\pdfoutput=1
\documentclass[structabstract]{aa}
\usepackage{graphicx}
\usepackage{txfonts}
\usepackage{amssymb}
\usepackage{natbib}
\usepackage{rotating}
\usepackage[rightcaption]{sidecap}
\bibpunct{(}{)}{;}{a}{}{,}

\begin{document}
\include{tables}
\title{High-\textit{J} CO survey of low-mass protostars \\ observed with
  \textit{Herschel}-HIFI\thanks{{\it Herschel} is an ESA space
    observatory with science instruments provided by European-led
    Principal Investigator consortia and with important participation
    from NASA.}}
\titlerunning{HIFI observations of high-$J$ CO in low-mass protostars}

\author{
     Umut~A.~Y{\i}ld{\i}z\inst{\ref{inst1}}
\and Lars~E.~Kristensen\inst{\ref{inst1},\ref{inst2}}
\and Ewine~F.~van~Dishoeck\inst{\ref{inst1},\ref{inst3}}
\and Irene~San~Jos{\'e}-Garc{\'i}a\inst{\ref{inst1}}
\and Agata~Karska\inst{\ref{inst3}}
\and Daniel~Harsono\inst{\ref{inst1}}
\and Mario~Tafalla\inst{\ref{inst4}}
\and Asuncion~Fuente\inst{\ref{inst5}}
\and Ruud~Visser\inst{\ref{inst6}}
\and Jes~K.~J{\o}rgensen\inst{\ref{inst7},\ref{inst8}}
\and Michiel~R.~Hogerheijde\inst{\ref{inst1}}
}

\institute{
Leiden Observatory, Leiden University, PO Box 9513, 2300 RA Leiden, The Netherlands\label{inst1} \email{yildiz@strw.leidenuniv.nl}
\and
Harvard-Smithsonian Center for Astrophysics, 60 Garden Street, Cambridge, MA 02138, USA\label{inst2}
\and
Max Planck Institut f\"{u}r Extraterrestrische Physik, Giessenbachstrasse 1, 85748 Garching, Germany\label{inst3}
\and
Observatorio Astron\'{o}mico Nacional (IGN), Calle Alfonso XII,3. 28014 Madrid, Spain\label{inst4}
\and
Observatorio Astron\'{o}mico Nacional, Apartado 112, 28803 Alcal\'{a} de Henares, Spain\label{inst5}
\and
Department of Astronomy, University of Michigan, 500 Church Street, Ann Arbor, MI 48109-1042, USA\label{inst6}
\and
Niels Bohr Institute, University of Copenhagen, Juliane Maries Vej 30, DK-2100 Copenhagen {\O}., Denmark\label{inst7}
\and
Centre for Star and Planet Formation, Natural History Museum of Denmark, University of Copenhagen, {\O}ster Voldgade 5-7, DK-1350 Copenhagen K., Denmark\label{inst8}
}

   \date{Draft: \today}

 
  \abstract
  {In the deeply embedded stage of star formation, protostars 
    start to heat and disperse their surrounding cloud cores. The
    evolution of these sources has traditionally been traced through
    dust continuum spectral energy distributions (SEDs), but the use of 
    CO excitation as an evolutionary probe has not yet been explored 
    due to lack of high-$J$ CO observations.
}
{The aim is to constrain the physical characteristics (excitation,
  kinematics, column density) of the warm gas in low-mass protostellar
  envelopes using spectrally-resolved \textit{Herschel} data of CO
  and compare those with the colder gas traced by
  lower excitation lines. 
  }
  {{\it Herschel}-HIFI observations of high-$J$ lines of $^{12}$CO,
    $^{13}$CO and C$^{18}$O (up to $J_{\rm u}=10$, $E_{\rm u}$ up to
    300 K) are presented toward 26 deeply embedded low-mass Class~0
    and Class~I young stellar objects, obtained as part of the \textit{Water
    In Star-forming regions with Herschel} (WISH) key program. This
    is the first large spectrally resolved high-$J$ CO survey
    conducted for these types of sources. Complementary lower-$J$~CO
    maps were observed using ground-based telescopes such as the
    JCMT and APEX and convolved to matching beam sizes.}
{ The $^{12}$CO~10--9 line is detected for all objects and can
  generally be decomposed into a narrow and broad component due to the
  quiescent envelope and entrained outflow material, respectively. 
  The $^{12}$CO excitation temperature increases with velocity from 
  $\sim$60~K up to $\sim$130 K. 
  The median excitation temperatures for $^{12}$CO, $^{13}$CO and
  C$^{18}$O derived from single-temperature fits to the $J_{\rm
    u}$=2--10 integrated intensities are $\sim$70~K,
  48~K and 37~K, respectively, with no significant difference between
  Class~0 and Class~I sources and no trend with $M_{\rm env}$ or $L_{\rm
    bol}$. Thus, in contrast with the continuum SEDs, the spectral
  line energy distributions (SLEDs) do not show an evolution during
  the embedded stage. In contrast, the integrated line intensities of
  all CO isotopologs show a clear decrease with evolutionary stage as
  the envelope is dispersed. Models of the collapse and evolution of
  protostellar envelopes reproduce well the C$^{18}$O results, but
  underproduce the $^{13}$CO and $^{12}$CO excitation temperatures,
  due to lack of UV heating and outflow components in those
  models. The H$_2$O $1_{10}-1_{01}$/CO 10--9 intensity ratio does not
  change significantly with velocity, in contrast with the H$_2$O/CO
  3--2 ratio, indicating that CO~10--9 is the lowest transition for
  which the line wings probe the same warm shocked gas as H$_2$O.
  Modeling of the full suite of C$^{18}$O lines indicates an abundance
  profile for Class 0 sources that is consistent with a freeze-out
  zone below 25~K and evaporation at higher temperatures, with but
  some fraction of the CO transformed into other species in the cold
  phase. In contrast, the observations for two Class I sources in Ophiuchus are
  consistent with a constant high CO abundance profile.
}
{
The velocity resolved line profiles trace the evolution from
  the Class 0 to the Class I phase through decreasing line
  intensities, less prominent outflow wings and increasing average CO
  abundances. However, the CO excitation temperature stays nearly constant.
The multiple components found here indicates that the 
analysis of spectrally unresolved data such as provided by SPIRE and
PACS must be done with caution.}

\maketitle



\section{Introduction}


Low-mass stars like our Sun form deep inside collapsing molecular
clouds by accreting material onto a central dense source.  As the
source evolves, gas and dust move from the envelope to the disk and
onto the star, resulting in a decrease of the envelope mass and a
shift in the peak of the continuum spectral energy distribution to
shorter wavelengths \citep[e.g.,][]{Lada99,Andre00,Young05}. At the
same time, jets and winds from the protostar entrain material and
disperse the envelope.
Spectral lines at submillimeter wavelengths trace this dense molecular
gas and reveal both the kinematic signature of collapse
\citep{Gregersen97,Myers00,Kristensen12} as well as the high velocity
gas in the outflows \citep{Arce07}. 

The most commonly used probe is CO because it is the second most
abundant molecule after H$_{2}$, has a simple energy level structure,
and all main isotopolog lines are readily detectable ($^{12}$CO,
$^{13}$CO, C$^{18}$O, C$^{17}$O).
Because of its small dipole moment, its rotational lines are easily
excited and therefore provide an excellent estimate of the gas column density
and the kinetic temperature. Although low excitation lines of CO have
been observed in protostars for decades
\citep[e.g.,][]{Hayashi94,Blake95,Bontemps96,Jorgensen02,Fuller02,Tachihara02,Hatchell05},
no systematic studies have been undertaken so far of the higher
excitation lines which probe the warm gas ($T$$>$100~K) during
protostellar evolution.  Ground-based observations of other molecules
exist as well and show in some (but not all) low-mass sources a
variety of complex organic species commonly ascribed to `hot cores'
where ices evaporate molecules back into the gas phase
\citep[e.g.,][]{vanDishoeck98,Ceccarelli07}. Quantifying these hot
core abundunces has been complicated by the lack of a good reference
of the H$_2$ column density in this warm $\geq$100~K gas.

With the launch of the {\it Herschel} Space Observatory
\citep{Pilbratt10} equipped with new efficient detectors, observations
of low-mass protostars in \mbox{higher-$J$} transitions of CO have
become possible. In this paper, high-$J$ refers to the lines $J_{\rm
  u}$$\geq$6 ($E_{\rm u}$$>$100~K) and low-$J$ refers to $J_{\rm
  u}$$\leq$5 ($E_{\rm u}$$<$100~K).  The Heterodyne Instrument for
Far-Infrared \citep[HIFI;][]{degraauw10} on {\it Herschel} offers a
unique opportunity to observe spectrally resolved high-$J$ CO lines of
various isotopologs with unprecendented sensitivity \citep[see][for
early results]{Yildiz10,Plume12}. Even higher transitions of CO up
to $J_{\rm u}$=50 are now routinely observed with the Photoconducting
Array and Spectrometer (PACS) \citep{Poglitsch10} and the Spectral and
Photometric Imaging Receiver (SPIRE) \citep{Griffin10} instruments on
{\it Herschel}, but those data are spectrally unresolved, detect
mostly $^{12}$CO, and probe primarily a hot shocked gas component
associated with the source
\citep[e.g.,][]{vanKempen10hh46,Herczeg12,Goicoechea12,Karska13,Manoj13,Green13}. To study the bulk of the protostellar system and disentangle the
various physical components, velocity resolved lines of isotopologs including
optically thin C$^{18}$O are needed.

In this paper, we use \textit{Herschel}-HIFI single pointing
observations of high-$J$ CO and its isotopologs up to the 10--9 ($E_{\rm
  u}/k$=300~K) transition from low-mass protostars obtained in the
``Water in Star-forming regions with {\it Herschel}'' (WISH) key
program \citep{vanDishoeck11}. The CO lines have been obtained as
complement to the large set of lines from H$_2$O, OH and other related
molecules in a sample of $\sim$80 low to high-mass protostars at
different evolutionary stages. This study focuses on low-mass
protostellar sources ($L_{\mathrm{bol}}$$<$100$~L_{\odot}$) ranging
from the most deeply embedded Class 0 phase to the more evolved Class
I stage. The {\it Herschel} CO data are complemented by ground-based
lower-$J$ transitions.

The warm gas probed by these high-$J$ CO lines is much more diagnostic
of the energetic processes that shape deeply embedded sources than the
low-$J$ lines observed so far.  Continuum data from submillimeter to
infrared wavelengths show that the temperature characterizing the peak
wavelength of the SED (the so-called bolometric temperature $T_{\rm
  bol}$; \citealt{MyersLadd93}) increases from about 25 K for the
earliest Class 0 sources to about 200--300 K for the more evolved
Class I sources, illustrating the increased dust temperatures as the
source evolves. At the same time, the envelope gradually decreases
with evolution from $\sim$1~M$_\odot$ to $<$0.05 M$_\odot$
\citep{Shirley00,Young05}.
Our CO data probe gas over the entire range of temperature 
and masses found in these protostellar envelopes. Hence we 
pose the following questions regarding the evolution
of the envelope and interaction with both the outflow and immediate
environment: ($i$) Does the CO line intensity decrease with
evolutionary stage from Class 0 to Class I in parallel with the dust?
($ii$) Does the CO excitation change with evolutionary stage, as does
the dust temperature?  ($iii$) How do the CO molecular line profiles
(i.e., kinematics) evolve through 2--1 up to 10--9. For example, what
fraction of emission is contained in the envelope and outflow components?
($iv$) What is the relative importance of the different energetic
processes in the YSO environment, e.g., passive heating of the
envelope, outflows, photon heating, and how is this quantitatively
reflected in the lines of the three CO isotopologs?  ($v$) Can our
data directly probe the elusive `hot core' and provide a column
density of quiescent warm ($T$$>$100~K) gas as reference for chemical
studies? How do those column densities evolve from Class 0 to Class I?

To address these questions, the full suite of lines and isotopologs is
needed. The $^{12}$CO line wings probe primarily the entrained outflow
gas. The $^{13}$CO lines trace the quiescent envelope but show excess
emission that has been interpreted as being due to UV-heated gas along
outflow cavity walls \citep{Spaans95,vanKempen09champ}. The C$^{18}$O
lines probe the bulk of the collapsing envelope heated by the
protostellar luminosity and can be used to constrain the CO abundance
structure. These different diagnostic properties of the CO and
isotopolog lines have been demonstrated through early
\textit{Herschel}-HIFI results of high-$J$ CO and isotopologs up to
10--9 by \citet{Yildiz10, Yildiz12} for three low-mass
protostars and by \citet{Fuente12} for one intermediate protostar.
Here we investigate whether the conclusions on column densities,
temperatures of the warm gas and CO abundance structure derived for
just a few sources hold more commonly in a large number of sources
covering different physical characteristics and evolutionary stages.

This paper presents \textit{Herschel}-HIFI CO and isotopolog spectra
for a sample of 26 low-mass protostars.  The \textit{Herschel} data are
complemented by ground-based spectra to cover as many lines as
possible from $J$=2--1 up to $J$=10--9, providing the most complete
survey of velocity resolved CO line profiles of these sources to date.
We demonstrate that the combination of low- and high-$J$ lines for the
various CO isotopologs is needed to get the complete picture.
The {\it Herschel} data presented here are also included in
the complementary paper by \citet{SanJoseGarcia13} comparing
high-$J$ CO from low- to high-mass YSOs. That paper investigates
trends across the entire mass spectrum, whereas this paper focusses on a
detailed analysis of the possible excitation mechanisms required to
explain the CO emission.

The outline of the paper is as follows. In
Section~\ref{4:sec:observations}, the observations and the telescopes
used to obtain the data are described. In Section~\ref{4:sec:results},
the \textit{Herschel} and complementary lines are presented and a
decomposition of the line profiles is made. In
Section~\ref{4:sec:rotdiags}, the data for each of the CO isotopologs
are analyzed, probing the different physical components. Rotational
excitation diagrams are constructed, column densities and abundances
are determined and kinetic temperatures in the entrained outflow gas
are constrained. The evolution of these properties from the Class~0 to
the Class~I sources is studied and compared with evolutionary models.
Section~\ref{4:sec:conclusions} summarizes the conclusions from this
work.

\begin{figure*}
\begin{center}
\begin{minipage}{4.0cm}
\includegraphics[width=4.cm]{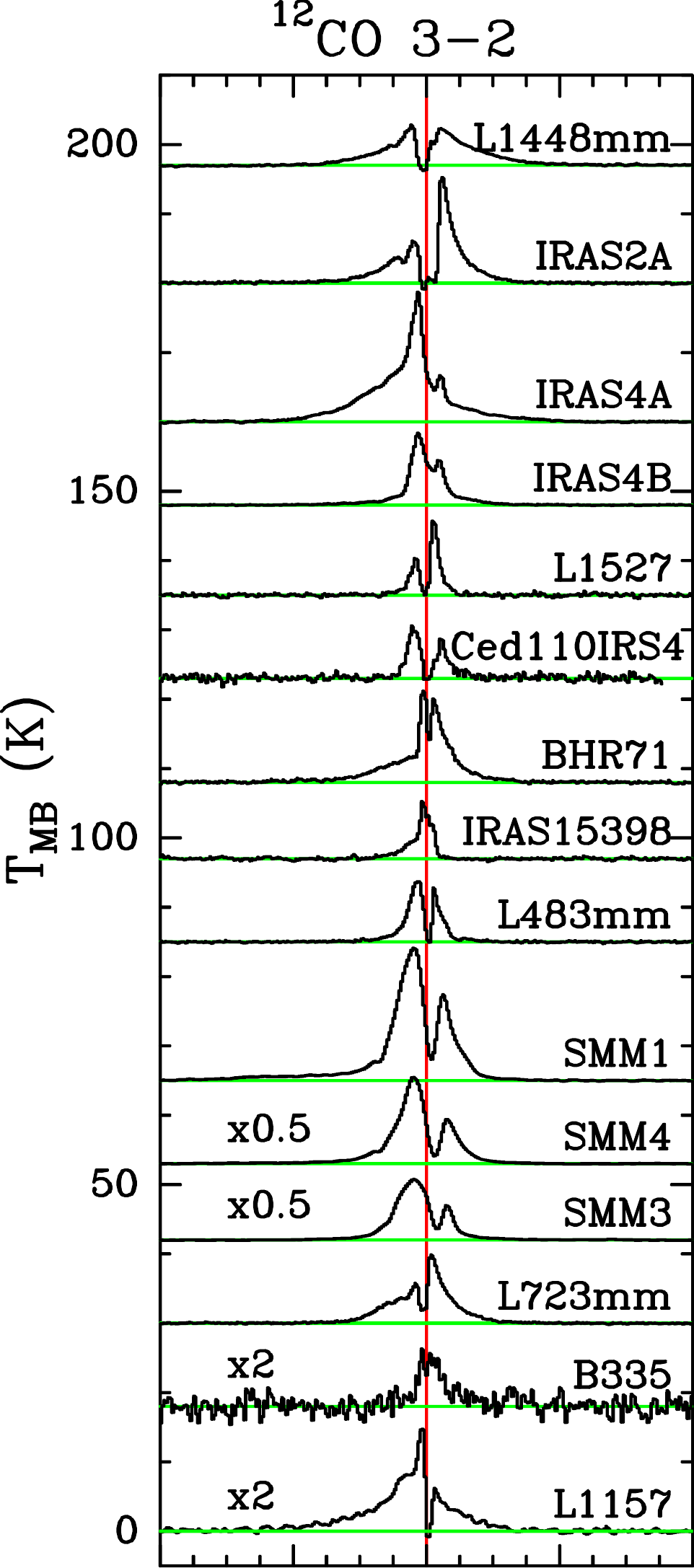}
\includegraphics[width=4.12cm]{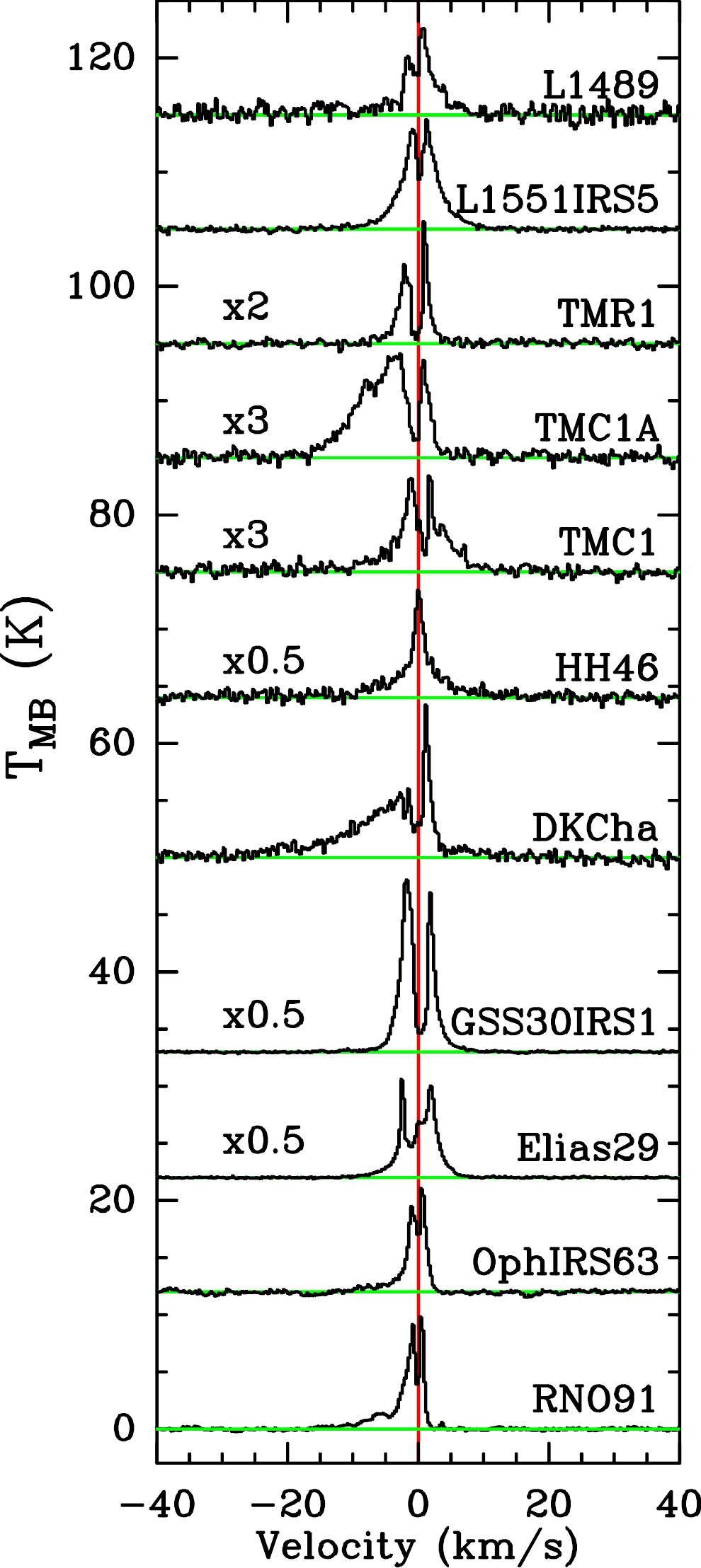}
\end{minipage}
\begin{minipage}{3.5cm}
\includegraphics[width=3.48cm]{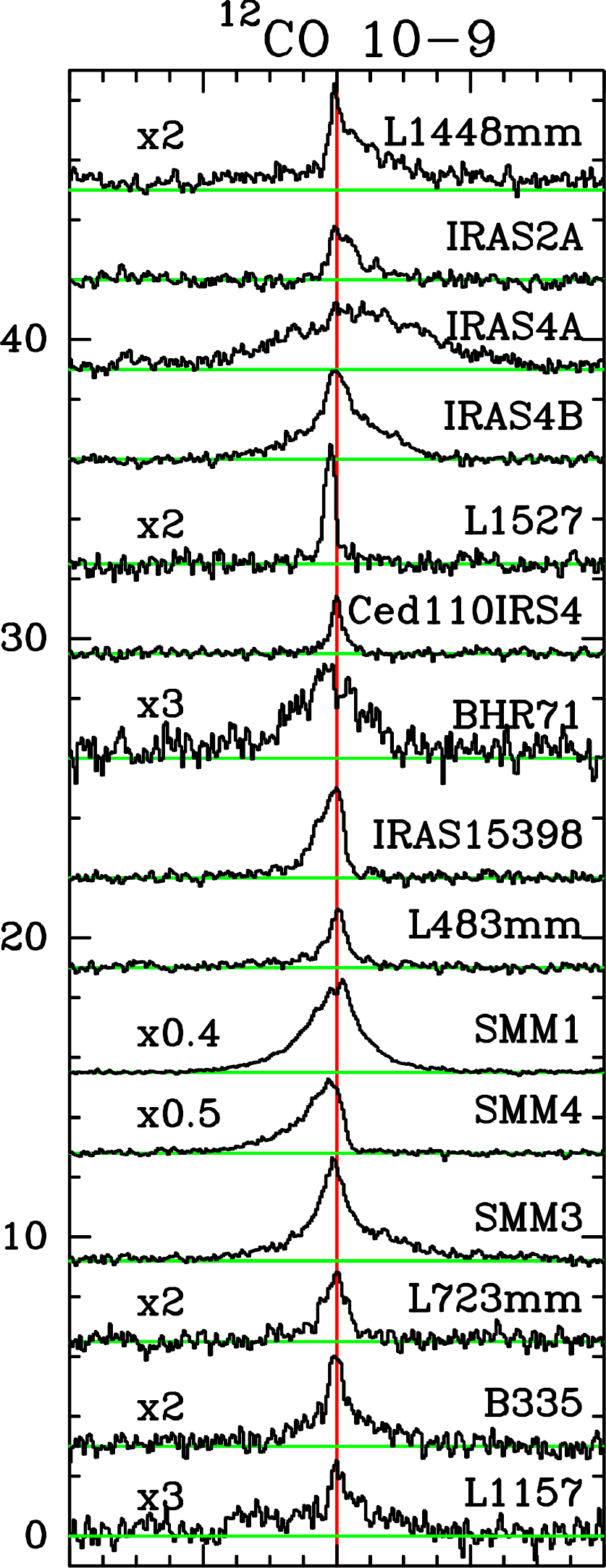}
\includegraphics[width=3.6cm]{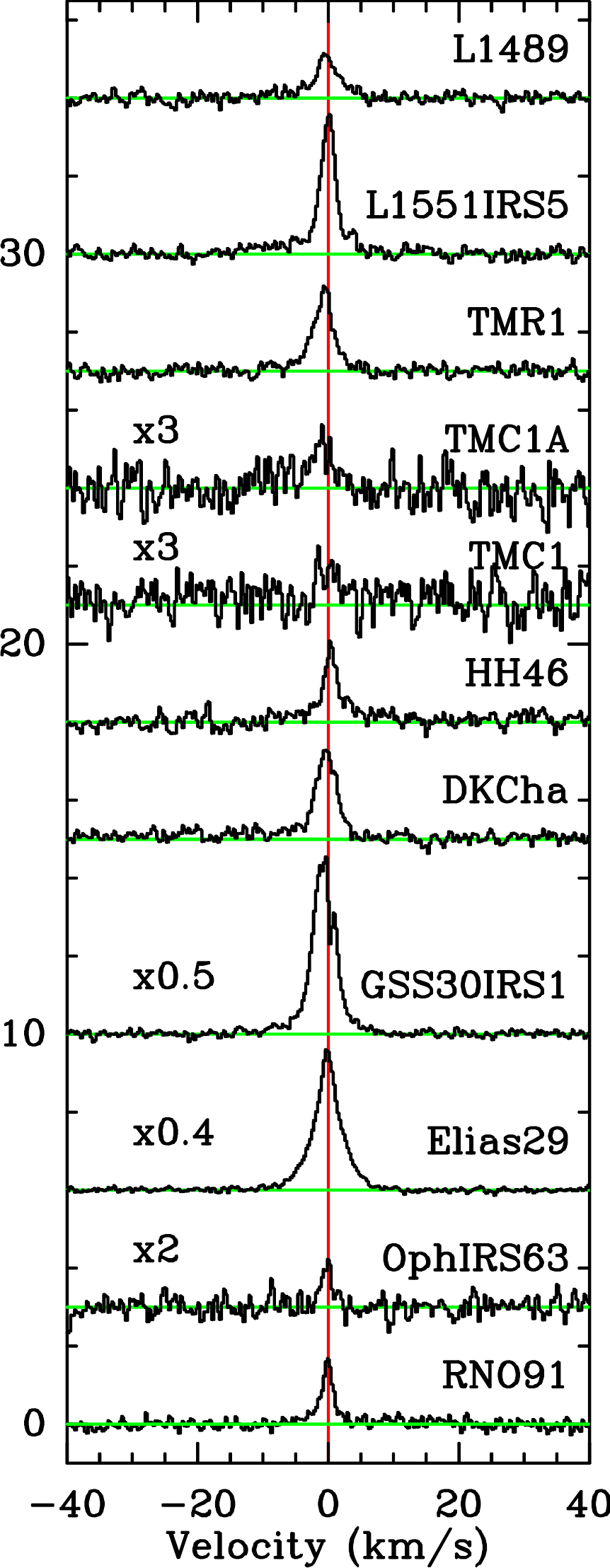}
\end{minipage}
\begin{minipage}{3.4cm}
\includegraphics[width=3.35cm]{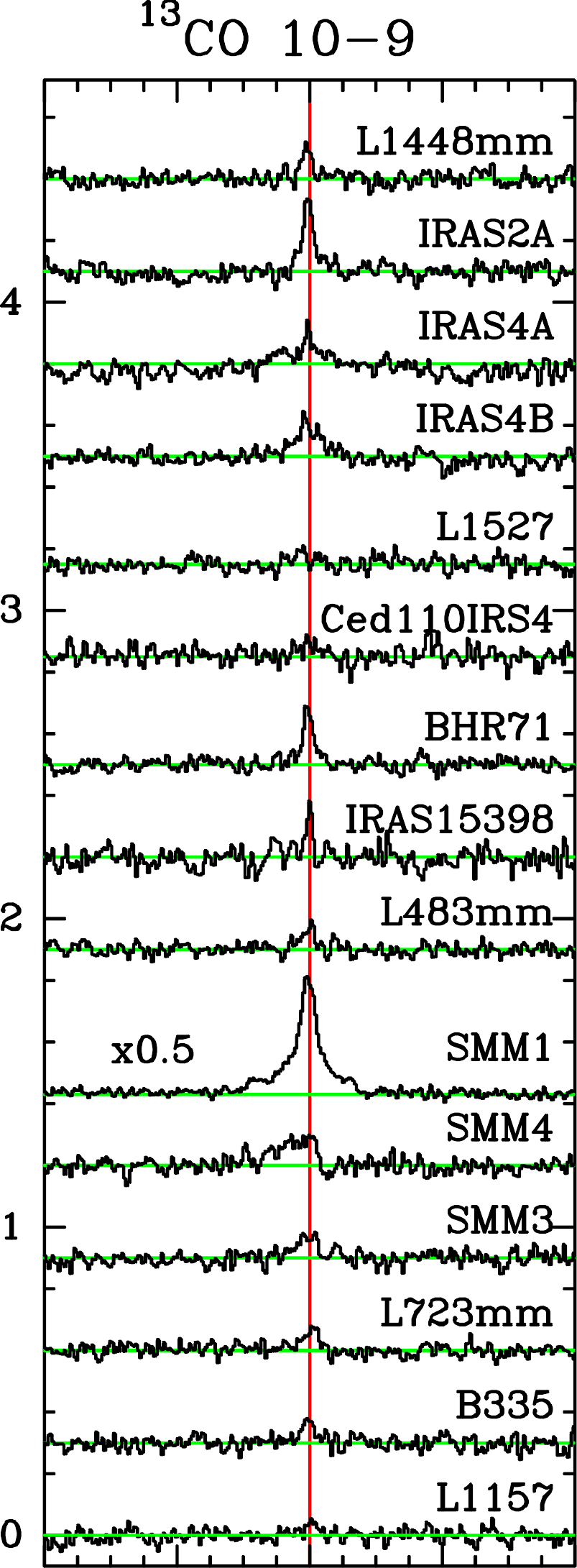}
\includegraphics[width=3.46cm]{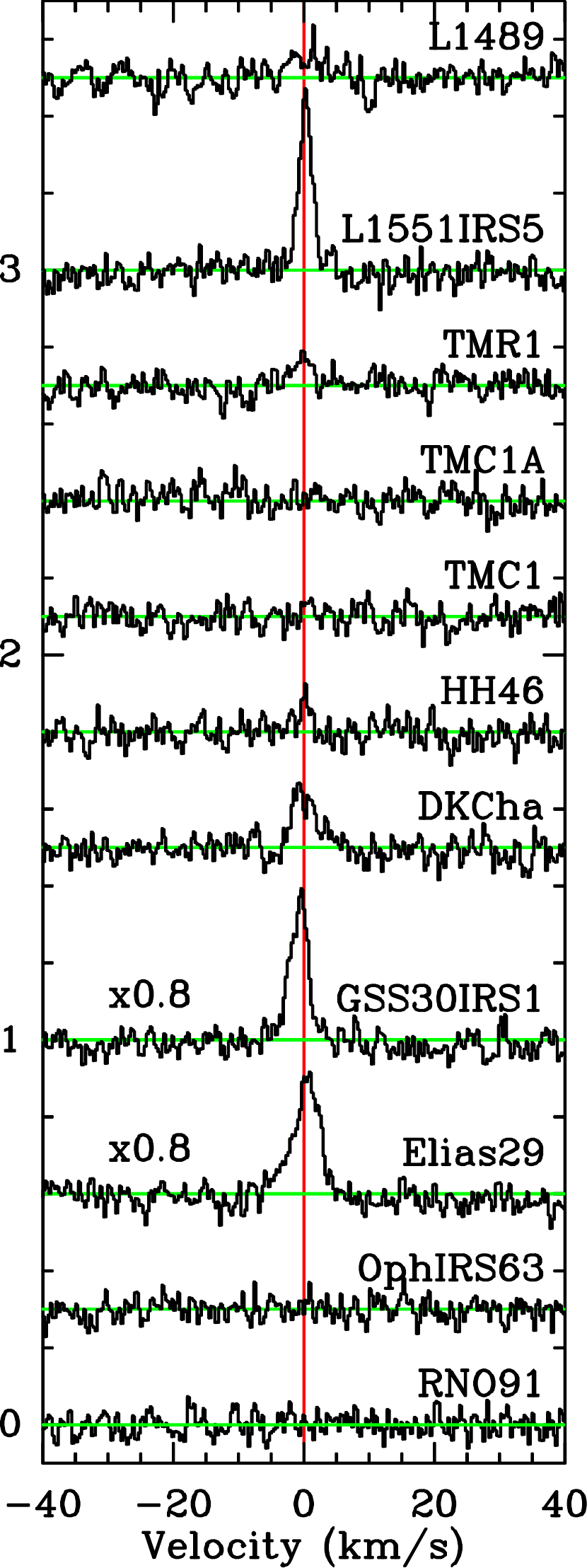}
\end{minipage}
\begin{minipage}{3.53cm}
    \hspace*{0.1cm}
\includegraphics[width=3.35cm]{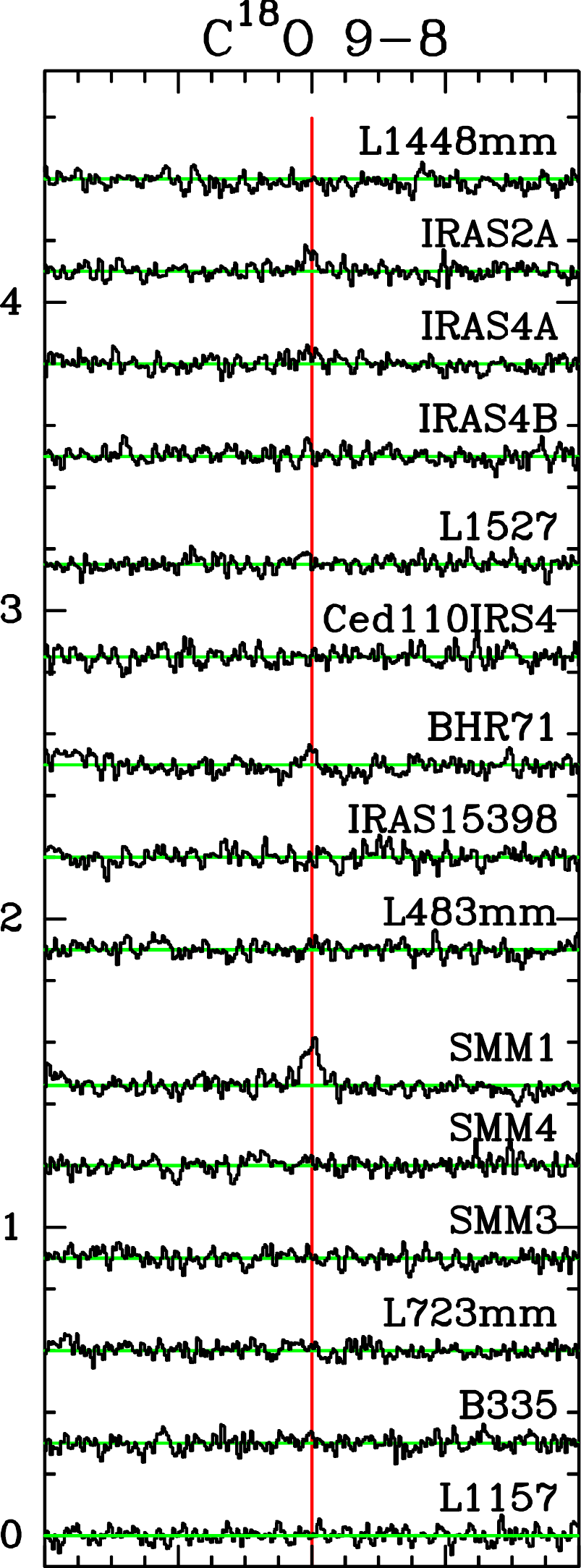}
\includegraphics[width=3.66cm]{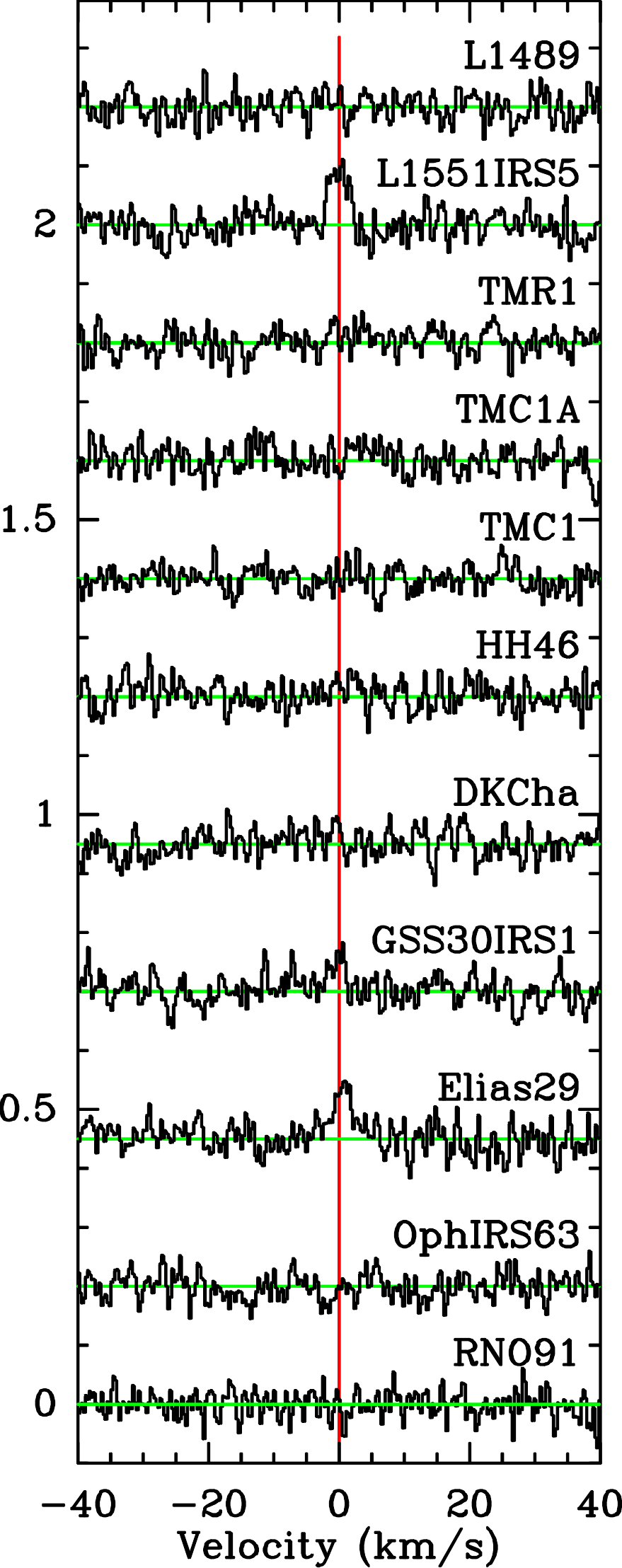}
\end{minipage}
\begin{minipage}{3.41cm}
    \hspace*{0.1cm}
\includegraphics[width=3.2cm]{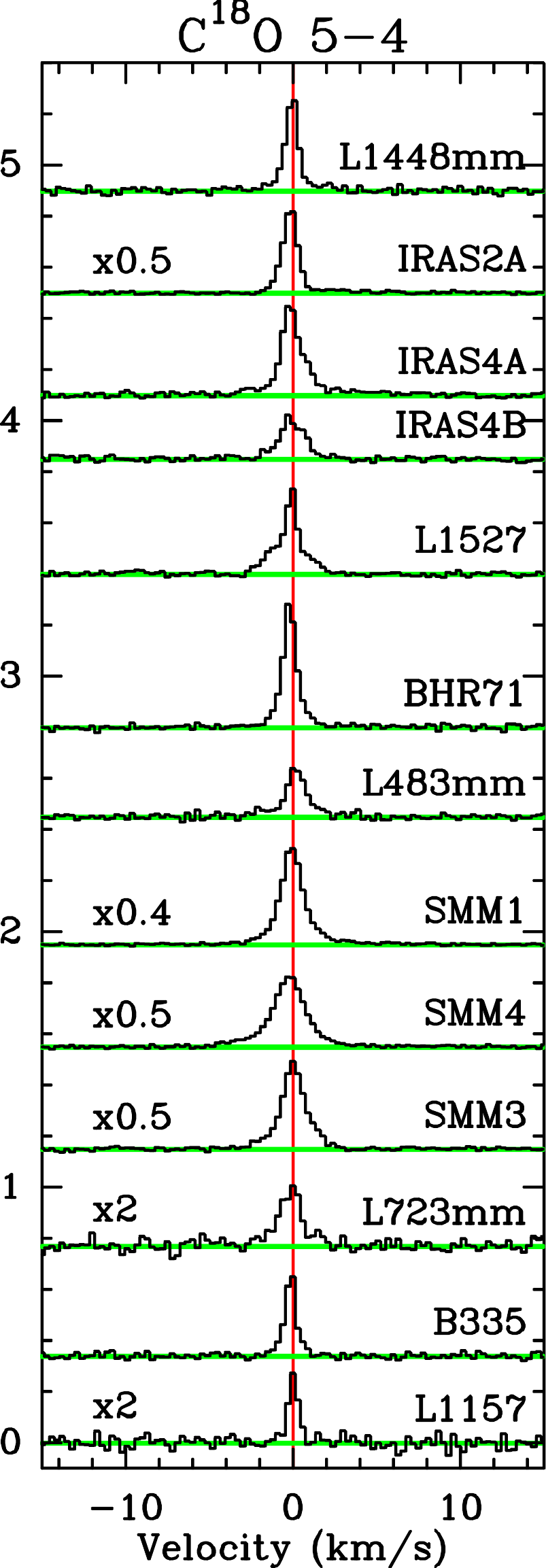}
\includegraphics[width=3.45cm]{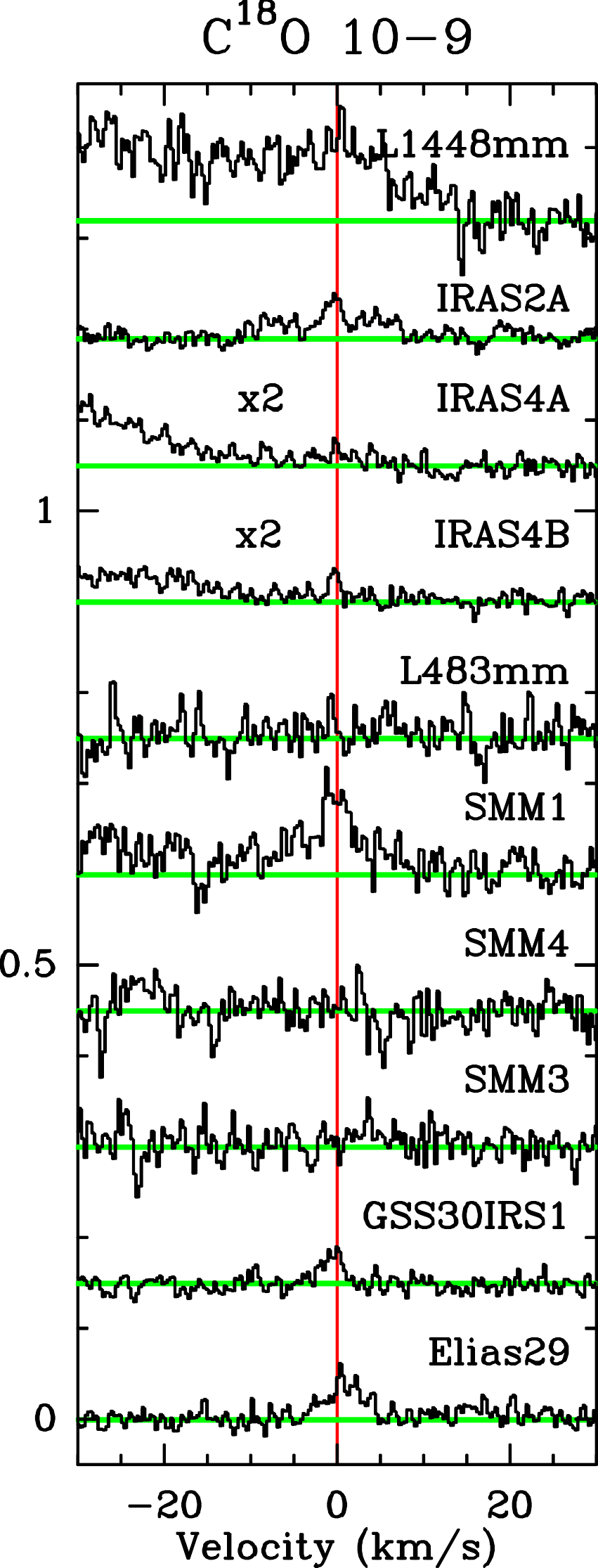}
\end{minipage}
\end{center}
    \caption{\small On source CO spectra convolved to a $\sim$20$''$
      beam. From left to right: $^{12}$CO~3--2, 10--9,
      $^{13}$CO~10--9, C$^{18}$O~9--8, 5--4 and 10--9,
      respectively. Only the CO~3--2 lines are observed with the JCMT,
      the rest of the data are from {\it Herschel}-HIFI. The spectra
      are plotted by shifting the source velocity ($V_{\rm lsr}$) to
      0~km~s$^{-1}$ (actual source velocities are given in Table
      \ref{tbl:overviewobs}). 
      The lines are shifted
      vertically. Intensity scale of some sources is multiplied
      by a constant value for easy viewing and marked if different from 1. 
      The top half of the figure shows
      the Class~0 sources whereas the bottom part displays the Class~I
      sources. The right-most column displays the C$^{18}$O 5--4 and
      C$^{18}$O~10--9 lines for the Class 0 sources only. The latter
      lines are very close to the H$_2$O 3$_{12}$-3$_{03}$ line
      resulting in an intensity rise on the blue side of the spectrum
      in some sources.}
    \label{fig:linesCO3-2andHIFI}
\end{figure*}

\section{Observations and complementary data}
\label{4:sec:observations}

\begin{table}[!t]
\caption{Overview of the observed transitions.}
\tiny
\begin{center}
\begin{tabular}{l r r r r l r r l r r r}
\hline \hline
Molecule & Trans.& $E_{\rm u}$ & $A_{\rm ul}$& Freq. \\ 
 & $J_{\mathrm{u}}$-$J_{\mathrm{l}}$ & [K] & [s$^{-1}$] & [GHz] \\ 
\hline
CO         & 2--1 & 16.6   & 6.910$\times$10$^{-7}$ & 230.538000  \\ 
           & 3--2 & 33.19  & 2.497$\times$10$^{-6}$ & 345.795989  \\
           & 4--3 & 55.3   & 6.126$\times$10$^{-6}$ & 461.040768  \\
           & 6--5 & 116.2  & 2.137$\times$10$^{-5}$ & 691.473076  \\
           & 7--6 & 154.9  & 3.422$\times$10$^{-5}$ & 806.651806  \\
           &10--9 & 304.2  & 1.006$\times$10$^{-4}$ &1151.985452  \\
 $^{13}$CO & 2--1 & 15.9   & 6.038$\times$10$^{-7}$ & 220.398684  \\
           & 3--2 & 31.7   & 2.181$\times$10$^{-6}$ & 330.587965  \\ 
           & 4--3 & 52.9   & 5.353$\times$10$^{-6}$ & 440.765174  \\
           & 6--5 & 111.1  & 1.868$\times$10$^{-5}$ & 661.067277  \\
           & 8--7 & 190.4  & 4.487$\times$10$^{-5}$ & 881.272808  \\
           &10--9 & 290.8  & 8.797$\times$10$^{-5}$ &1101.349597  \\
C$^{18}$O  & 2--1 & 15.8   & 6.011$\times$10$^{-7}$ & 219.560354  \\
           & 3--2 & 31.6   & 2.172$\times$10$^{-6}$ & 329.330553  \\
           & 5--4 & 79.0   & 1.062$\times$10$^{-5}$ & 548.831010  \\
           & 6--5 & 110.6  & 1.860$\times$10$^{-5}$ & 658.553278  \\
           & 9--8 & 237.0  & 6.380$\times$10$^{-5}$ & 987.560382  \\
           &10--9 & 289.7  & 8.762$\times$10$^{-5}$ &1097.162875  \\
\hline 
\end{tabular}
\end{center}
\tablefoot{The level energies, Einstein A coefficients, and line frequencies are from the LAMDA, JPL and CDMS databases \citep{Schoier2005LAMDA, Pickett1998JPL, Muller2005CDMS}. }
\label{tbl:overviewobslines}
\end{table}

An overview of observed spectral lines with their upper level energies, Einstein~$A$
coefficients, and rest frequencies are presented in Table~\ref{tbl:overviewobslines}.
The selection of the sources and their characteristics are described
in \citet{vanDishoeck11} and \citet{Kristensen12} together
with the targeted coordinates.  In total, 26 low-mass young stellar
objects were observed in CO of which 15 are Class~0 and 11 Class I
sources, with the boundary taken to be at $T_{\rm bol} =$ 70~K
\citep{MyersLadd93, Chen95}. In terms of envelope mass, the boundary
between the two classes is roughly at 0.5 M$_{\odot}$ \citep{Jorgensen02}.
All Class I sources have been vetted to be truly embedded `Stage I'
sources cf. \citet{Robitaille06} and not Class II edge-on disks or
reddened background stars \citep{vanKempen09_southc+,vanKempen09Oph}.
Throughout the paper, Class~0 sources are marked as red, and Class~I
sources are marked as blue in the figures.  In addition to
\textit{Herschel}-HIFI spectra, data come from the \mbox{12-m}
\mbox{sub-mm} Atacama Pathfinder Experiment Telescope,
APEX\footnote{This publication is based on data acquired with the
Atacama Pathfinder Experiment (APEX). APEX is a collaboration
between the Max-Planck-Institut f\"ur Radioastronomie, the European
Southern Observatory, and the Onsala Space Observatory.}  at Llano
de Chajnantor in Chile, and the 15-m James Clerk Maxwell Telescope,
JCMT\footnote{The JCMT is operated by The Joint Astronomy Centre on
behalf of the Science and Technology Facilities Council of the
United Kingdom, the Netherlands Organisation for Scientific
Research, and the National Research Council of Canada.} at Mauna
Kea, Hawaii. The overview of all the observations can be found in
Table \ref{tbl:overviewobssource}.

All data were acquired on the $T_{\rm A}^{*}$ antenna temperature 
scale, and were converted to main-beam brightness temperatures 
$T_{\rm MB} = T_{\rm A}^{*}/ \eta_{\rm MB}$ \citep{KutnerUlich81} 
by using the beam efficiencies ($\eta_{\rm MB}$) stated in each 
of the source Tables~C.(1--26) in the Appendix.

\textit{Herschel:} 
Spectral line observations of \mbox{$^{12}$CO~10--9}, \mbox{$^{13}$CO~10--9},
\mbox{C$^{18}$O~5--4, 9--8 and 10--9} were obtained with HIFI as part of the
WISH guaranteed time key program on {\it Herschel}.  Single pointing
observations at the source positions were
carried out between March~2010 and October~2011. 
An overview of the HIFI observations for each source is provided 
in Table \ref{tbl:HerschelObsids} (in the Appendix) with their corresponding 
{\it Herschel} Science Archive (HSA) obsids. The  lines were observed in 
dual-beam switch (DBS) mode using a switch of 3$\arcmin$.
The CO transitions were observed in combination with the water lines:
\mbox{$^{12}$CO~10--9} with \mbox{H$_2$O~3$_{12}$--2$_{21}$} (10~min);
\mbox{$^{13}$CO~10--9} with \mbox{H$_2$O~1$_{11}$--0$_{00}$} (40~min);
\mbox{C$^{18}$O~5--4} with \mbox{H$_{2}^{18}$O~1$_{10}$--1$_{01}$} (60~min); 
\mbox{C$^{18}$O~9--8} with \mbox{H$_2$O 2$_{02}$--1$_{11}$} (20 min); and 
\mbox{C$^{18}$O~10--9} with \mbox{H$_2$O~3$_{12}$--3$_{03}$} (30~min or 5 hours). 
Only a subset of the H$_2$O lines were observed toward all Class~I sources and
therefore C$^{18}$O~5--4 and part of the isotopolog \mbox{CO~10--9}
data are missing for these sources. For C$^{18}$O 9--8, IRAS~2A,
IRAS~4A, IRAS~4B, Elias~29, and GSS30~IRS1 were observed in very deep
integrations for 5 hours in the open-time program OT2$\_$rvisser$\_$2
(PI:~R.~Visser). Also, the C$^{18}$O 5--4 lines have very high $S/N$
because of the long integration on the H$_2^{18}$O line. Thus, the
noise level varies per source and per line.

The \textit{Herschel} data were taken using the wide-band spectrometer
(WBS) and high-resolution spectrometer (HRS) backends. Owing to the
higher noise ranging from a factor of 1.7 up to 4.7 of the HRS
compared with the WBS, mainly WBS data are presented here except for
the narrow \mbox{C$^{18}$O~5--4} lines where only the HRS
data are used.  The HIFI beam sizes are $\sim$20$\arcsec$ ($\sim$4000
AU for a source at $\sim$200 pc) at \mbox{1152 GHz} and 42$\arcsec$
($\sim$8400 AU) at \mbox{549 GHz}. The typical spectral resolution
ranges from 0.68~km~s$^{-1}$ (band 1) to 0.3~km~s$^{-1}$ (band 5) in
WBS, and 0.11~km~s$^{-1}$ (band 1) in HRS. Typical rms values range
from 0.1~K for $^{12}$CO~10-9 line to 9~mK for C$^{18}$O~10-9 in the longest
integration times.

Data processing started from the standard HIFI pipeline in the
\textit{Herschel} Interactive Processing Environment
(HIPE\footnote{HIPE is a joint development by the Herschel Science
Ground Segment Consortium, consisting of ESA, the NASA Herschel
Science Center, and the HIFI, PACS and SPIRE consortia.}) ver.
8.2.1 \citep{OttS10}, where the $V_{\mathrm{lsr}}$ precision is of the
order of a few m~s$^{-1}$.  Further reduction and analysis were performed
using the GILDAS-\verb1CLASS1\footnote{{http://www.iram.fr/IRAMFR/GILDAS/}}
software.  The spectra from the H- and V-polarizations were averaged to
obtain better $S/N$. In some cases a discrepancy of 30\% or more was
found between the two polarizations, in which case only the H band
spectra were used for analysis.
These sources are indicated in Tables~C.(1--26) in the online Appendix.
Significant emission contamination from one of the reference position 
was found at the $^{12}$CO~10--9 observation of IRAS~2A and IRAS~4A. In 
that case, only one reference position, which was clean, was used in 
order to reduce the data. On the other hand, even though the pointing 
accuracy is $\sim$2$\arcsec$, the H- and V-polarizations have slightly 
shifted pointing directions (Band~1:
$-$6\farcs2, $+$2\farcs2; Band~4: $-$1\farcs3,$-$3\farcs3; Band~5:
0\farcs0,$+$2\farcs8), which may give rise to different line profiles
in strong extended sources \citep{Roelfsema11}. No corrections were
made for these offsets. The HIFI beam efficiencies are 0.76, 0.74, and
0.64 for bands 1, 4, and 5, respectively \citep{Roelfsema11}.

\textit{APEX:} Maps of the $^{12}$CO~6--5, 7--6 and $^{13}$CO~6--5 lines
over a few arcmin region were observed with the CHAMP$^+$ instrument
\citep{Kasemann06, Guesten08} at the APEX telescope for all sources
visible from Chajnantor, whereas $^{13}$CO 8--7 and C$^{18}$O~6--5
lines were obtained for selected objects in staring mode.  The
CHAMP$^+$ instrument consists of two heterodyne receiver arrays, each
with seven pixel detector elements for simultaneous operations in the
\mbox{620--720 GHz} and \mbox{780--950} GHz frequency ranges. The APEX
beam sizes correspond to 8$\arcsec$ ($\sim$1600~AU for a source at
200~pc) at \mbox{809 GHz} and 9$\arcsec$ ($\sim$1800 AU) at \mbox{691
  GHz}. A detailed description of the instrument and observations of
several sources in the current sample have been presented in
\citet{vanKempen09champ2,vanKempen09champ,vanKempen09_southc+,Yildiz12}
and the remaining maps will be given in Y{\i}ld{\i}z et al. (in
prep.).  Here only the data for the central source positions are
considered. In addition, lower-$J$ transitions were observed for
southern sources using various receivers at APEX
\citep{vanKempen06DKCha,vanKempen09champ2,vanKempen09champ,vanKempen09_southc+}.
 
\textit{JCMT:} All sources visible from the JCMT were mapped by the
HARP \citep{Buckle09} instrument over an area of
2$\arcmin$$\times$2$\arcmin$ in $^{12}$CO, $^{13}$CO and C$^{18}$O~3--2
transitions. HARP consists of 16 SIS detectors with 4$\times$4 pixel
elements of 15$\arcsec$ each at 30$\arcsec$ separation. Other 2--1
lines were observed with the single pixel RxA instrument at a beam
size of $\sim$23$\arcsec$ by \citet{Jorgensen02}. Part of those
observations were fetched from the JCMT public archive\footnote{This
  research used the facilities of the Canadian Astronomy Data Centre
  operated by the National Research Council of Canada with the support
  of the Canadian Space Agency.}.

Since the observations involve a number of different telescopes and
frequencies, the beam sizes differ for each case. 
The maps obtained with the JCMT and APEX were resampled to a 
common resolution of 20$\arcsec$ so as to be directly comparable to the beam 
size of the HIFI CO 10--9 and 9--8 observations (20$\arcsec$ and 23$\arcsec$, respectively) 
as well as the JCMT CO 2--1 observations (22$\arcsec$).
The exception are a few CO~4--3 lines that are only available for a
single pointing in an 11$\arcsec$ beam size which are indicated in the
tables at the Appendix~C.  
The 20$\arcsec$ beam corresponds to a diameter of 2500~AU for a source
at 125~pc (closest distance), and 9000~AU for a source in 450~pc
(furthest distance), so the observing beam encloses both the bulk of
the dense envelope as well as outflow material.  The data reduction
and analysis for each source were finalized using the
GILDAS-\verb1CLASS1 software.  Calibration errors are estimated as
$\sim$20\% for the ground-based telescopes \citep[][for
JCMT]{Buckle09}, and $\sim$10\% for the HIFI lines
\citep{Roelfsema11}.

The full APEX and JCMT maps will be presented in Y{\i}ld{\i}z et
al. (in prep) where the outflows are studied in more detail.
The full set of lines for NGC~1333~IRAS~2A, IRAS~4A and IRAS~4B have
also been presented in \citet{Yildiz10, Yildiz12} (except
for the deeper C$^{18}$O~10--9 data), but for completeness
and comparison with the rest of the WISH sample, the data are included
in this paper.

\section{Results}
\label{4:sec:results}

\subsection{CO line gallery}

The $^{12}$CO, $^{13}$CO and C$^{18}$O spectra from $J$=2--1 up to
\mbox{$J$=10--9} for each source are provided in the online Appendix~C. 
This appendix contains figures of all the observed spectra and tables with the extracted information.
Summary spectra are presented in
Fig.~\ref{fig:linesCO3-2andHIFI} for the CO~3--2, 10--9,
$^{13}$CO~10--9, C$^{18}$O~5--4, 9--8, and 10--9 lines,
respectively. Emission is detected in almost all transitions with our
observing setup except some higher-$J$ isotopolog lines 
discussed below.  The high-$J$ CO lines observed with {\it Herschel}
are the first observations for these types of sources.
Decomposition of line profiles is discussed in detail in 
\citet{SanJoseGarcia13} and is only briefly summarized below.

\begin{figure}[!t]
    \centering
    \includegraphics[scale=0.3]{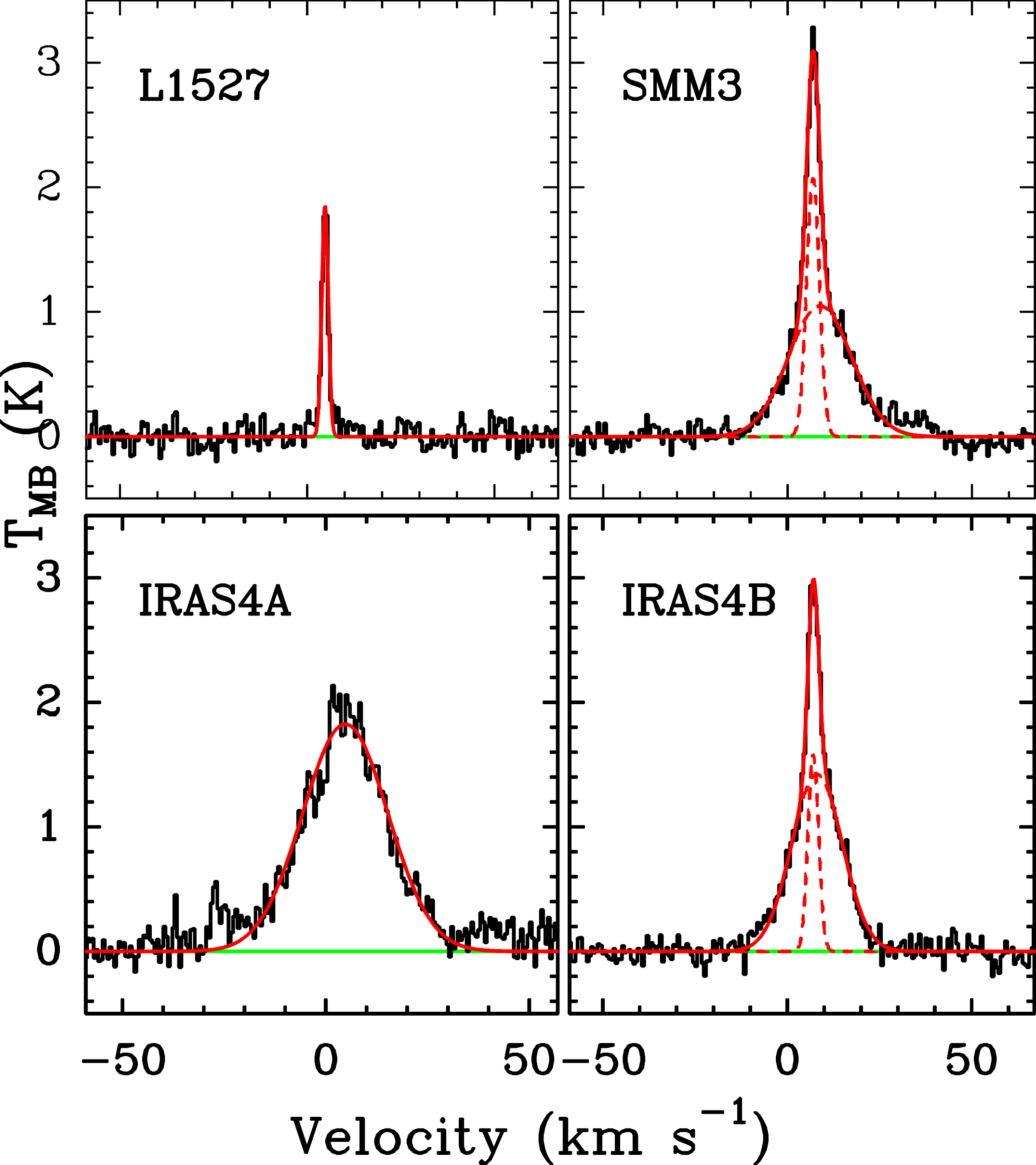}
    \caption{\small Gaussian decomposition of the
        $^{12}$CO~10--9 profile toward four sources. The
        profile toward two sources, L1527 and IRAS4A, can be
        decomposed into a single Gaussian, whereas SMM3, IRAS\,4B and
        all the remaining sources in the sample require two components
        (decomposition shown with the dashed red fit).}
    \label{fig:12COdecomposespectra}
\end{figure}

\begin{figure}[!t]
    \centering
\includegraphics[scale=0.6,angle=90]{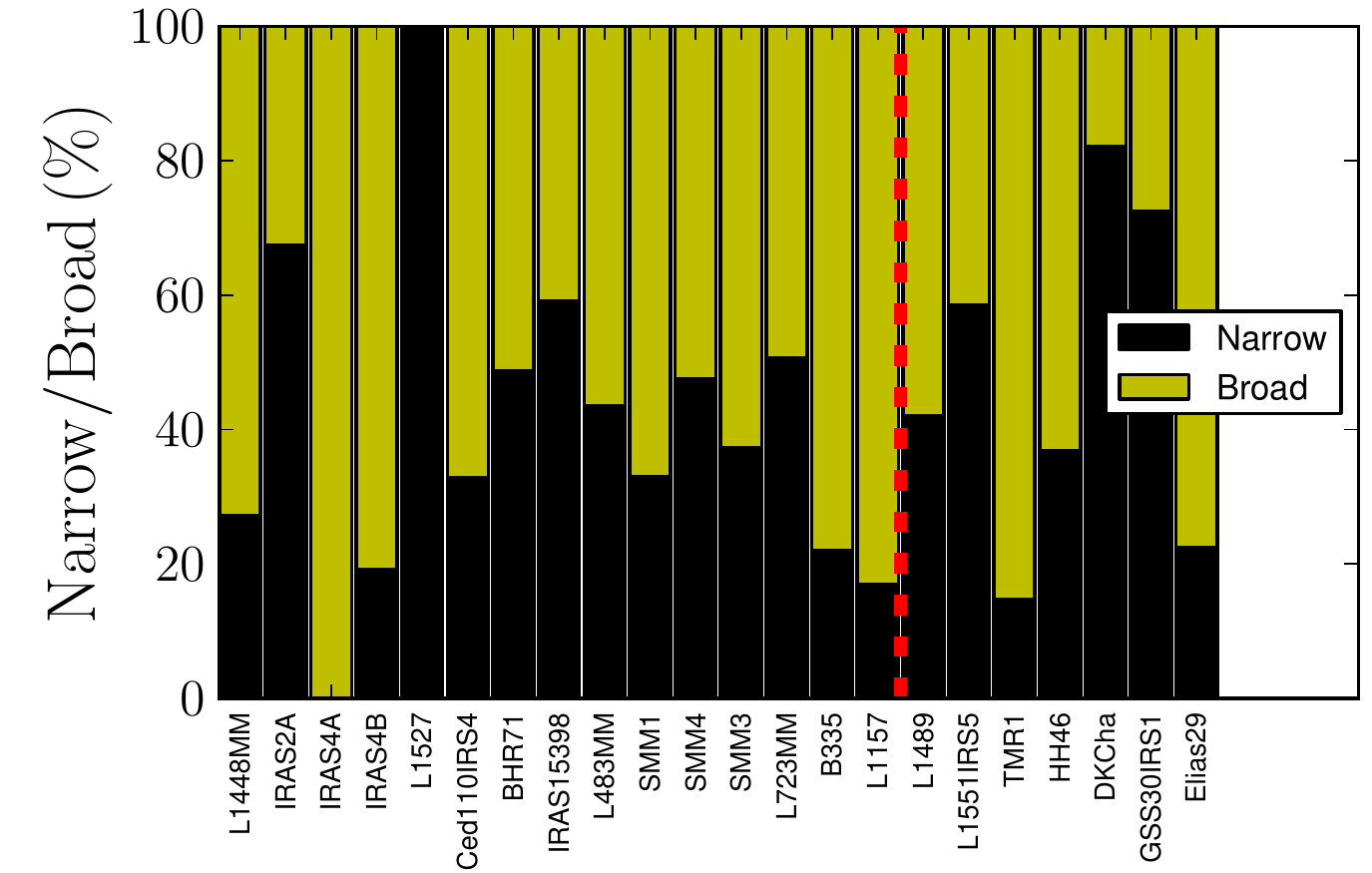}
\caption{\small Relative fraction of the integrated intensity of the
  narrow and broad components for each source. The $^{12}$CO~10--9
  profile decomposition is given in
  \citet{SanJoseGarcia13}. Yellow regions indicate the broad
  component fraction and black the narrow component fractions. The red
  dashed line divides the Class~0 (left) and Class~I (right) sources.}
    \label{fig:12COnarrowbroadpercent}
\end{figure}

\subsection{$^{12}$CO lines}
\label{4:sec:12COlines}

$^{12}$CO 10--9 emission is detected in all sources. 
Integrated and peak intensities are typically higher in the Class 0
sources compared with the Class I sources. Typical integrated
intensities at the source positions range from 1~K~km~s$^{-1}$ (in
Oph~IRS63) up to 82~K~km~s$^{-1}$ (in Ser-SMM1), whereas peak
intensities range from 0.6~K (in Oph~IRS63) up to 9.3~K (in
GSS30~IRS1). One striking result is that none of the $^{12}$CO~10--9
observations show self-absorption except for 
Ser-SMM1 and GSS30~IRS1, whereas all of the CO~3--2
  observations have strong self-absorption, which suggests optically
  thick line centers\footnote{The only exception is IRAS~15398 where
  the targeted position is off source, see
  \citet{Kristensen12}.}. The absorption components are
located at the source velocities as indicated by the peak of the
low-$J$~C$^{18}$O emission and are thus due to self-absorption from
the outer envelope.  Examining other available $^{12}$CO transitions
(lower than \mbox{$J$=10--9}) shows that the self-absorption
diminishes with increasing $J$ and disappears for 
all sources (again except Ser-SMM1 and GSS30~IRS1) at
around \mbox{$J$=10--9} (see Figures~C.(1--26) in Appendix).

\begin{table}[!t]
\caption{Relative fractions of integrated intensities calculated for broad and narrow components in $^{12}$CO~10--9 lines.}
\tiny
\begin{center}
\begin{tabular}{l c c c c c}
\hline \hline
Source & \multicolumn{2}{c} {$\int T_{\mathrm{MB}} \mathrm{d}V$ (K~km~s$^{-1}$) } & \multicolumn{2}{c} {$\%$\tablefootmark{a}} & \\ 
               & NC & BC & NC & BC &\\ 
\hline
L1448MM             & 12.7 & 33.9 & 27 & 73 \\
NGC1333-IRAS2A      & 6.4  & 3.1  & 67 & 33 \\
NGC1333-IRAS4A      & \dots  & 49.4 & \dots & 100 \\
NGC1333-IRAS4B      & 5.6 & 23.6 & 19 & 81 \\
L1527               & 4.2 & \dots & 100 & \dots \\
Ced110IRS4          & 1.8 & 3.7    & 33 & 67 \\
BHR71               & 7.8 & 8.2 & 49 & 51 \\
IRAS15398           & 9.9  & 6.8  & 59 & 41 \\
L483MM              & 4.7  & 6.3  & 44 & 56 \\
Ser SMM1            & 26.2 & 53.1 & 33 & 67  \\
Ser SMM4            & 31.7 & 35.0 & 47 & 53 \\
Ser SMM3            & 9.3  & 15.7 & 37 & 63 \\
L723MM              & 3.5  & 3.4  & 51 & 49 \\
B335                & 2.6  & 9.2  & 22 & 78 \\
L1157               & 1.4  & 6.8  & 17 & 83 \\
L1489               & 2.5  & 3.4  & 42 & 58 \\
L1551IRS5           & 8.5  & 6.0  & 59 & 41 \\
TMR1                & 1.3  & 7.6  & 15 & 85 \\
HH46                & 3.1  & 5.4  & 37 & 63 \\
DK~Cha              & 8.2 & 1.8 & 82 & 18 \\
GSS30IRS1           & 31.2 & 11.8 & 73 & 27 \\
Elias29             & 10.6 & 36.5 & 22 & 78 \\
TMC1A               & (1.4\tablefootmark{b}) & \dots & \dots & \dots \\
TMC1                & (2.9\tablefootmark{b}) & \dots & \dots & \dots \\
Oph~IRS63           & (1.1\tablefootmark{b}) & \dots & \dots & \dots \\
RNO91               & (5.2\tablefootmark{b}) & \dots & \dots & \dots \\
\hline 
\end{tabular}
\end{center}
\tablefoot{NC: Narrow Component, BC: Broad Component, 
\tablefoottext{a}{Relative percentages of the integrated intensities given in the 2$^{\rm nd}$ and 3$^{\rm rd}$ columns.}
\tablefoottext{b}{Due to low $S/N$ in their spectra, profiles could not be decomposed, however, total integrated intensities are given.}
}
\label{tbl:NarrowBroadRatios}
\end{table}

For the $^{12}$CO lines, more than two thirds of the sample can
readily be decomposed into two Gaussian components with 
line widths of $\leq$7.5~km~s$^{-1}$ (narrow)
and 11--25~km~s$^{-1}$ \citep[broad; see Fig.~\ref{fig:12COdecomposespectra};][for
details]{SanJoseGarcia13}.  
The narrow component is due to the quiescent envelope whereas the
broad component represents the swept-up outflow gas\footnote{The broad
CO component is not necessarily the same physical component as seen
in the broad H$_2$O profiles; this point will be further discussed
in Sect.~\ref{4:sec:cowater}.}. Fig.~\ref{fig:12COnarrowbroadpercent}
summarizes the relative fraction of each of the components in terms of
integrated intensities (also tabulated in
Table~\ref{tbl:NarrowBroadRatios}). For four sources in the sample,
i.e., TMC1A, TMC1, Oph~IRS63, and RNO91, the profiles could not be
decomposed due to low $S/N$ in their spectra.  The fraction of
emission contained in the narrow component ranges from close to 0\%
(IRAS~4A) to nearly 100\% (L1527),
with a median fraction of 42$\%$.
Particularly, for IRAS~4A, the narrow component is most likely hidden
under the strong broad component, whereas for L1527, outflows are in
the plane of the sky therefore the broad component is not
evident.  This decomposition demonstrates that the contributions from
these two components are generally comparable so care must be taken in
interpreting spectrally unresolved data from {\it Herschel}-SPIRE and
PACS, and, to some extent, near-IR transitions of the same molecules.

\begin{figure}[!t]
    \centering
    \includegraphics[scale=0.35]{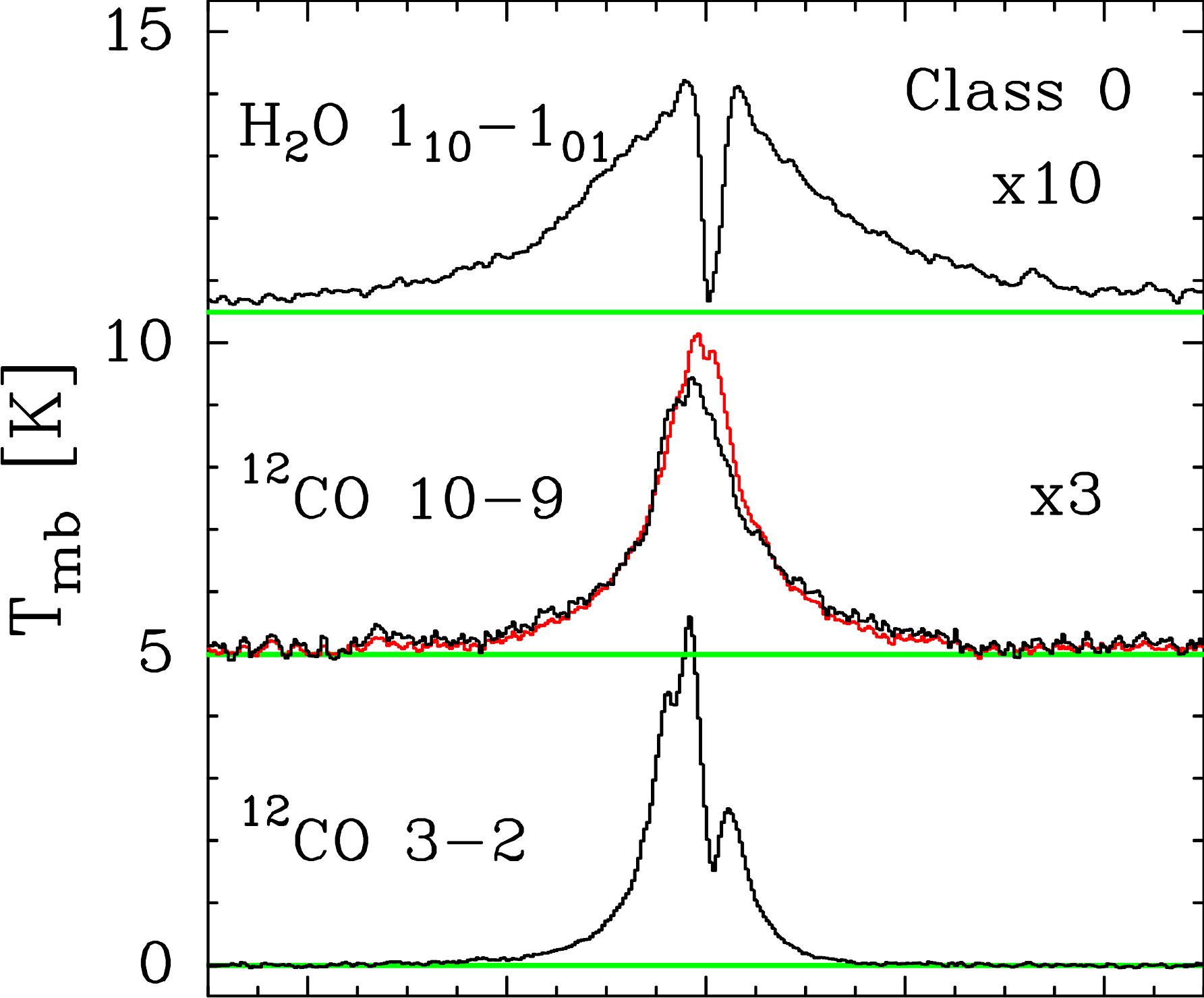}
    \includegraphics[scale=0.35]{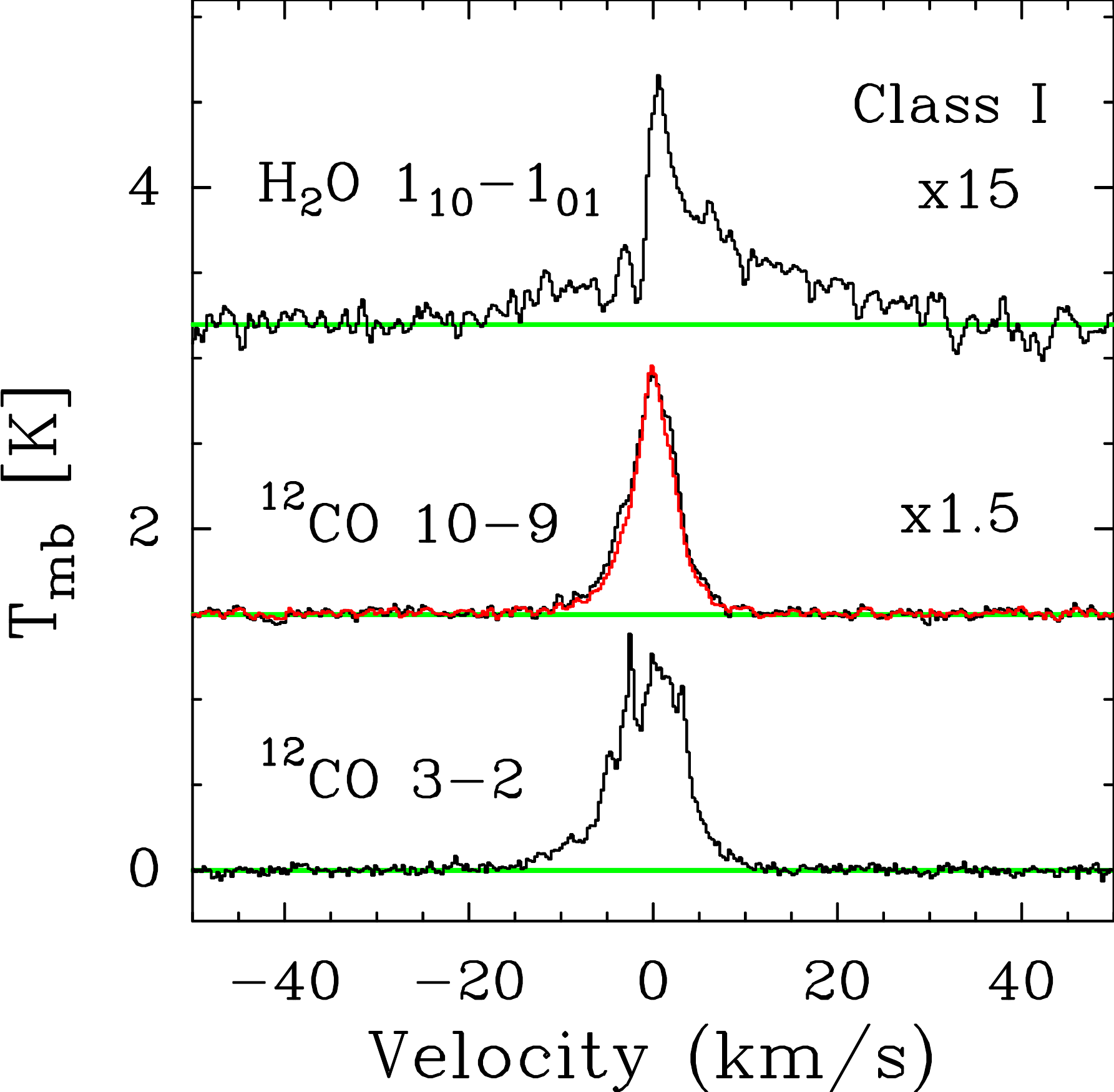}
    \caption{\small Composite H$_2$O~1$_{10}$-1$_{01}$, CO~10--9 and
      CO~3--2 spectra of Class~0 and Class~I sources averaged in order
      to compare the line profiles of two types of sources. All
      spectra are rescaled to a common distance of 200~pc, shifted to
      the central 0~km~s$^{-1}$ velocity, rebinned to a
      0.3~km~s$^{-1}$ velocity resolution. The CO spectra refer to a
      20$''$ beam, the H$_2$O spectra to a 40$''$ beam. The red
      spectra overlaid on top of the $^{12}$CO~10--9 are obtained by
      normalizing all the spectra to a common peak temperature first
      and then averaging them.  }
    \label{fig:composite}
\end{figure}

Figure \ref{fig:composite} presents the averaged
$^{12}$CO~3--2, 10--9, and H$_2$O~1$_{10}$-1$_{01}$ lines for the
Class~0 and Class~I sources in order to obtain a generic spectral structure
for one type of source. To compare with the H$_2$O spectra, a similar
 averaging procedure was followed as in \citet{Kristensen12},
where ground state ortho-water composite spectra observed with {\it
  Herschel}-HIFI at 557~GHz were presented in a beam of
40$\arcsec$. In this comparison, the IRAS~15398 (Class~0), TMC1 and
GSS30~IRS1 (Class~I) spectra have been excluded from the averaging
procedure.  The CO~10--9 line of IRAS~15398 is taken at a position
15$\arcsec$ offset from the source position, the TMC1 spectrum was too
noisy and the excitation of GSS30~IRS1 may not be representative of
Class~I sources \citep{Kristensen12}.  Therefore, 14~Class~0
and 9~Class~I spectra are scaled to a common distance of 200~pc and
averaged.  

It is seen that the broad CO outflow component is much more prominent
in the Class 0 than in the Class I sources (Figure
\ref{fig:composite}). For the Class~0 sources, the 10--9 line is
broader than the 3--2 lines. However, neither is as broad as the line
wings seen in H$_2$O 557 GHz lines, for which the average H$_2$O
spectra are taken from \citet{Kristensen12}.  The comparison
between CO and H$_2$O will be discussed further in
Sect.~\ref{4:sec:cowater}.

\begin{figure}[!t]
    \centering
\includegraphics[scale=0.32]{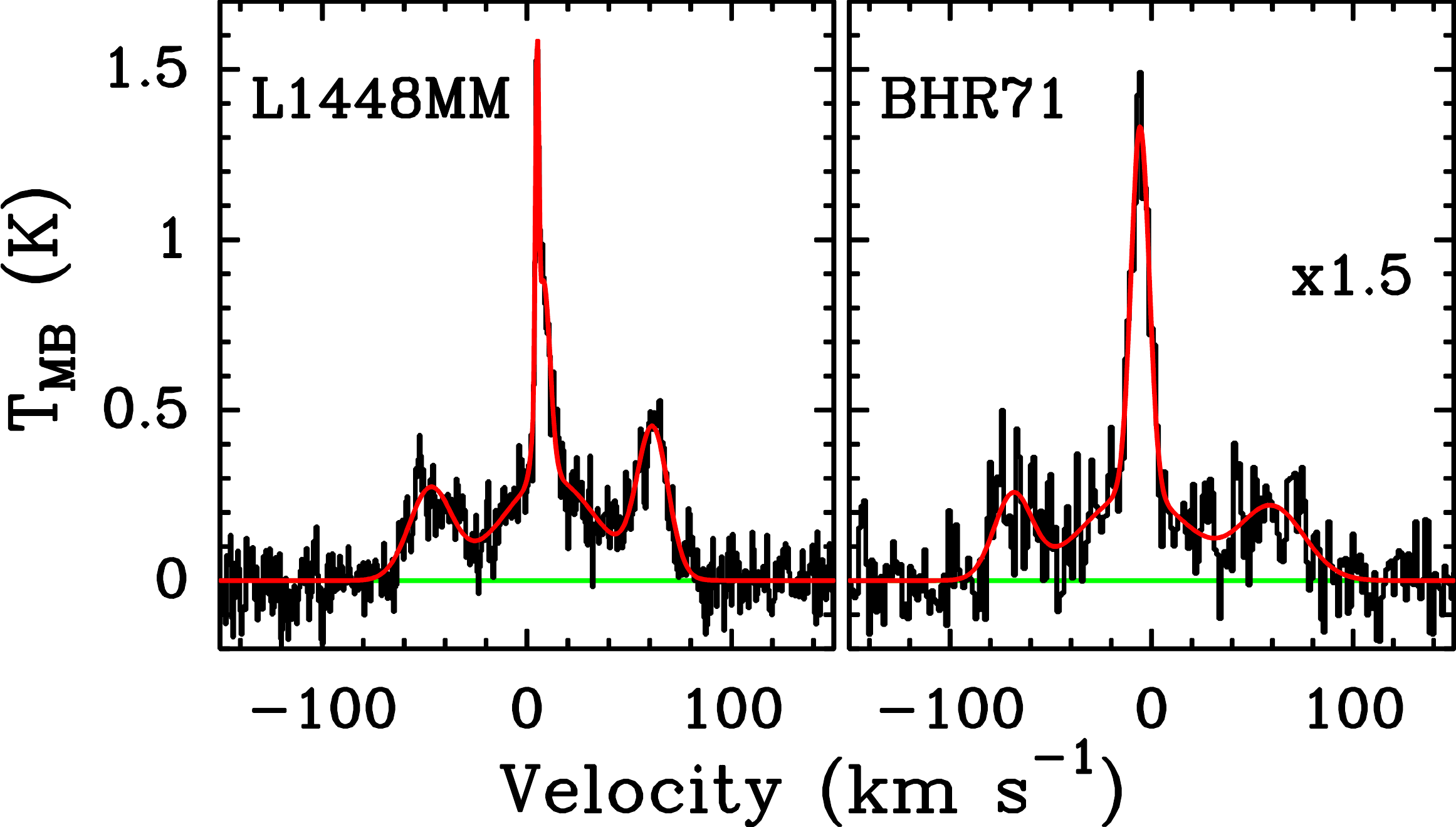}
    \caption{\small $^{12}$CO~10--9 spectra of L1448mm
      and BHR71, where bullet structures are
      shown. The green lines indicates the baseline and the red lines
      represent the Gaussian fits of the line profiles. Fit parameters are given in Table \ref{tbl:bullets}.}
    \label{fig:12CO109bullets}
\end{figure}

For a few sources, high velocity molecular emission features
associated with shock material moving at velocities up to hundred
km\,s$^{-1}$ have been observed \citep{Bachiller90,Tafalla10}.  For
species like SiO, their abundance is increased due to
shock-induced chemistry \citep{Bachiller97,Bourke97}.  These Extremely
High Velocity (EHV) components (or `bullet' emission) are also visible
in the higher-$J$~CO transitions, as well as in lower-$J$ transitions,
but the contrast in emission between bullet and broad outflow emission
is greatly enhanced at higher frequencies.  Bullets are visible
specifically in CO~6--5 and 10--9 data toward L1448mm and BHR71 at
$\sim\pm$60~km\,s$^{-1}$ (see Fig.~\ref{fig:12CO109bullets} for the
bullets and Table \ref{tbl:bullets} for the fit parameters).
These bullets are also seen in H$_2$O observations of the same
sources, as well as other objects \citep{Kristensen2011L1448, Kristensen12}.

\begin{table*}[!ht]
\caption{Fit parameters obtained from the bullet sources. }
\tiny
\begin{center}
\begin{tabular}{l c c c | c c c | c c c | c c c c c}
\hline \hline
                     & \multicolumn{3}{c} {EHV-B} &   \multicolumn{3}{c} {Broad} &  \multicolumn{3}{c} {Narrow} &  \multicolumn{3}{c}{EHV-R} & \\ 
Source        & $\int T_{\rm MB} \mathrm{d}V$ & $T_{\mathrm{peak}}$ & FWHM &  $\int T_{\rm MB} \mathrm{d}V$ & $T_{\mathrm{peak}}$ & FWHM &  $\int T_{\rm MB} \mathrm{d}V$ & $T_{\mathrm{peak}}$ & FWHM &   $\int T_{\rm MB} \mathrm{d}V$ & $T_{\mathrm{peak}}$ & FWHM\\
 & [K~km~s$^{-1}$] & [K] &   [km~s$^{-1}$]  & [K~km~s$^{-1}$] & [K] &  [km~s$^{-1}$]  & [K~km~s$^{-1}$] & [K] &   [km~s$^{-1}$]  & [K~km~s$^{-1}$] & [K] &   [km~s$^{-1}$] \\
\hline
L1448MM   & 6.9 & 0.26 & 24.6 & 17.5 & 0.30 & 54.3 & 6.5 & 1.10 & 7.5   & 8.1 & 0.43 & 17.8 \\
BHR71        & 4.2 & 0.16 & 25.2 & 10.8 & 0.18 & 55.6 & 4.2 & 0.71 & 10.0 & 3.9 & 0.14 & 27.1 \\
\hline 
\end{tabular}
\end{center}
\label{tbl:bullets}
\end{table*}

\subsection{$^{13}$CO lines}
$^{13}$CO emission is detected in all sources except for the 10--9
transition toward Oph~IRS63, RNO91, TMC1A, and TMR1. The 10--9
integrated and peak intensities are higher in Class~0 sources compared
with the Class~I sources. Typical integrated intensities range from
0.1~K (L1527) up to 3.4~K km~s$^{-1}$ (Ser-SMM1).  Peak intensities
range from 0.1~K (L1448 MM, L1527) up to 0.4~K (Ser-SMM1). 
All \mbox{$^{13}$CO~10--9} lines can be fitted by a single Gaussian (narrow component) except 
for Ser-SMM1 and IRAS~4A where two Gaussians (narrow and broader component) are needed. 

\begin{table*}
\caption{Source parameters.}
\small
\begin{center}
\begin{tabular}{l c c c c c c c c c}
\hline\hline
Source &d &  $\varv_{\rm LSR}$\tablefootmark{a} & $L_{\rm bol}$\tablefootmark{b} & $T_{\rm bol}$\tablefootmark{b} & $M_{\rm env~(10K)}$\tablefootmark{c} & $n_{\rm 1000AU}$\tablefootmark{c,d} & $n_{20\arcsec}$\tablefootmark{c,e}  \\
             & [pc] & [km s$^{-1}$] & [$L_\odot$] & [K] & [$M_\odot$] & [$10^5$~cm$^{-3}$]  & [$10^5$~cm$^{-3}$] \\ \hline
L1448-MM          & 235 & \phantom{1}$+$5.2 & \phantom{1}9.0 & \phantom{1}46 &	\phantom{1}3.9 & 39 & 11 \\ 
NGC1333-IRAS2A    & 235 & \phantom{1}$+$7.7 &                     35.7 & \phantom{1}50 &	\phantom{1}5.1 & 17 & 4.0 \\
NGC1333-IRAS4A    & 235 & \phantom{1}$+$7.0 & \phantom{1}9.1 & \phantom{1}33 &	\phantom{1}5.6 & 67  & 15 \\ 
NGC1333-IRAS4B    & 235 & \phantom{1}$+$7.1 & \phantom{1}4.4 & \phantom{1}28 &	\phantom{1}3.0 & 57  & 17 \\ 
L1527             & 140 & \phantom{1}$+$5.9 & \phantom{1}1.9 & \phantom{1}44 &	\phantom{1}0.9 & 8.1 & 6.0 \\
Ced110-IRS4       & 125 & \phantom{1}$+$4.2 & \phantom{1}0.8 & \phantom{1}56 &	\phantom{1}0.2 & 3.9 & 2.8 \\ 
BHR71             & 200 & \phantom{1}$-$4.4 &           14.8 & \phantom{1}44 &	\phantom{1}2.7 & 18  & 5.4 \\
IRAS15398         & 130 & \phantom{1}$+$5.1 & \phantom{1}1.6 & \phantom{1}52 &	\phantom{1}0.5 & 16  & 11\\
L483mm            & 200 & \phantom{1}$+$5.2 &           10.2 & \phantom{1}49 &	\phantom{1}4.4 & 5.1 & 2.8 \\
Ser-SMM1          & 230 & \phantom{1}$+$8.5 &           30.4 & \phantom{1}39 &          16.1 & 41  & 14 \\
Ser-SMM4          & 230 & \phantom{1}$+$8.0 & \phantom{1}1.9 & \phantom{1}26 &\phantom{1}2.1 & 54  & 23 \\
Ser-SMM3          & 230 & \phantom{1}$+$7.6 & \phantom{1}5.1 & \phantom{1}38 &	\phantom{1}3.2 & 11  & 5.5 \\
L723              & 300 &           $+$11.2 & \phantom{1}3.6 & \phantom{1}39 &	\phantom{1}1.3 & 8.0 & 2.2 \\
B335              & 250 & \phantom{1}$+$8.4 & \phantom{1}3.3 & \phantom{1}36 &	\phantom{1}1.2 & 15  & 4.3 \\
L1157             & 325 & \phantom{1}$+$2.6 & \phantom{1}4.7 & \phantom{1}46 &	\phantom{1}1.5 & 20  & 2.9\\ 
\hline
L1489             & 140 & \phantom{1}$+$7.2 & \phantom{1}3.8 &           200 &	\phantom{1}0.2 & 1.9 & 1.2 \\
L1551-IRS5        & 140 & \phantom{1}$+$6.2 &           22.1 & \phantom{1}94 &	\phantom{1}2.3 & 12  & 6.4 \\ 
TMR1              & 140 & \phantom{1}$+$6.3 & \phantom{1}3.8 &           133 &	\phantom{1}0.2 & 2.1 & 1.2 \\
TMC1A             & 140 & \phantom{1}$+$6.6 & \phantom{1}2.7 &           118 &	\phantom{1}0.2 & 2.2 & 1.3 \\
TMC1              & 140 & \phantom{1}$+$5.2 & \phantom{1}0.9 &           101 &	\phantom{1}0.2 & 1.8 & 1.2 \\
HH46-IRS          & 450 & \phantom{1}$+$5.2 &           27.9 &           104 &	\phantom{1}4.4 & 12  & 1.1 \\
DK~Cha            & 178 & \phantom{1}$+$3.1 &           35.4 &           569 &	\phantom{1}0.8 & 9.2 & 3.7 \\
GSS30-IRS1        & 125 & \phantom{1}$+$3.5 &           13.9 &           142 &	\phantom{1}0.6 & 1.7 & 1.1 \\
Elias 29          & 125 & \phantom{1}$+$4.3 &           14.1 &           299 &	\phantom{1}0.3 & 0.8 & 0.6 \\
Oph-IRS63         & 125 & \phantom{1}$+$2.8 & \phantom{1}1.0 &           327 &	\phantom{1}0.3 & 6.9 & 5.0 \\
RNO91             & 125 & \phantom{1}$+$0.5 & \phantom{1}2.6 &           340 &	\phantom{1}0.5 & 3.3 & 2.4 \\ 
\hline
\end{tabular}
\tablefoot{Sources above the horizontal line are Class 0, sources below are Class I.
	\tablefoottext{a}{Obtained from ground-based C$^{18}$O or C$^{17}$O observations.}
	\tablefoottext{b}{Measured using \textit{Herschel}-PACS data from the WISH and DIGIT key programmes \citep{Karska13,Green13}.}
	\tablefoottext{c}{Determined from DUSTY modeling of the sources; see \citet{Kristensen12}.}
	\tablefoottext{d}{Density at 1000~AU.}
	\tablefoottext{e}{Density at a region of 20$\arcsec$ diameter.}
	}
\end{center}
\label{tbl:overviewobs}
\end{table*}

\subsection{C$^{18}$O lines}

C$^{18}$O emission is detected in all sources up to $J$=5--4.  The
9--8 line is seen in several sources, mostly Class 0 objects (BHR71,
IRAS~2A, IRAS~4A, IRAS~4B, Ser-SMM1, L1551~IRS5).
The C$^{18}$O~10--9 line is detected after 5-hour integrations in IRAS~2A, IRAS~4A, IRAS~4B, Elias~29, GSS30~IRS1, and Ser-SMM1, with integrated intensities ranging from 0.05 (in IRAS~4A) up to
0.6~K~km~s$^{-1}$ (Ser-SMM1). 
Peak intensities range from 0.02~K up
to 0.07~K for the same sources.  The rest of the high-$J$ C$^{18}$O
lines do not show a detection but have stringent upper limits.
The high $S/N$ and high spectral resolution C$^{18}$O 5--4 data reveal
a weak, broad underlying component even in this minor isotopolog for
several sources \citep[see Fig.~1 in][]{Yildiz10}.  

The availability of transitions from 2--1 to 10--9 for many sources in
optically thin C$^{18}$O lines also gives an opportunity to revisit
source velocities, $V_{\rm lsr}$ that were previously obtained from
the literature. Table~\ref{tbl:overviewobs} presents the results with
nine sources showing a change in $V_{\rm lsr}$, compared with values
listed in \citet{vanDishoeck11} ranging from 0.2~km~s$^{-1}$ (IRAS~4A)
up to 1.0~km~s$^{-1}$ (L1551-IRS5).

\begin{center}
\begin{figure*}[!htb]
    \centering
    \includegraphics[scale=0.42]{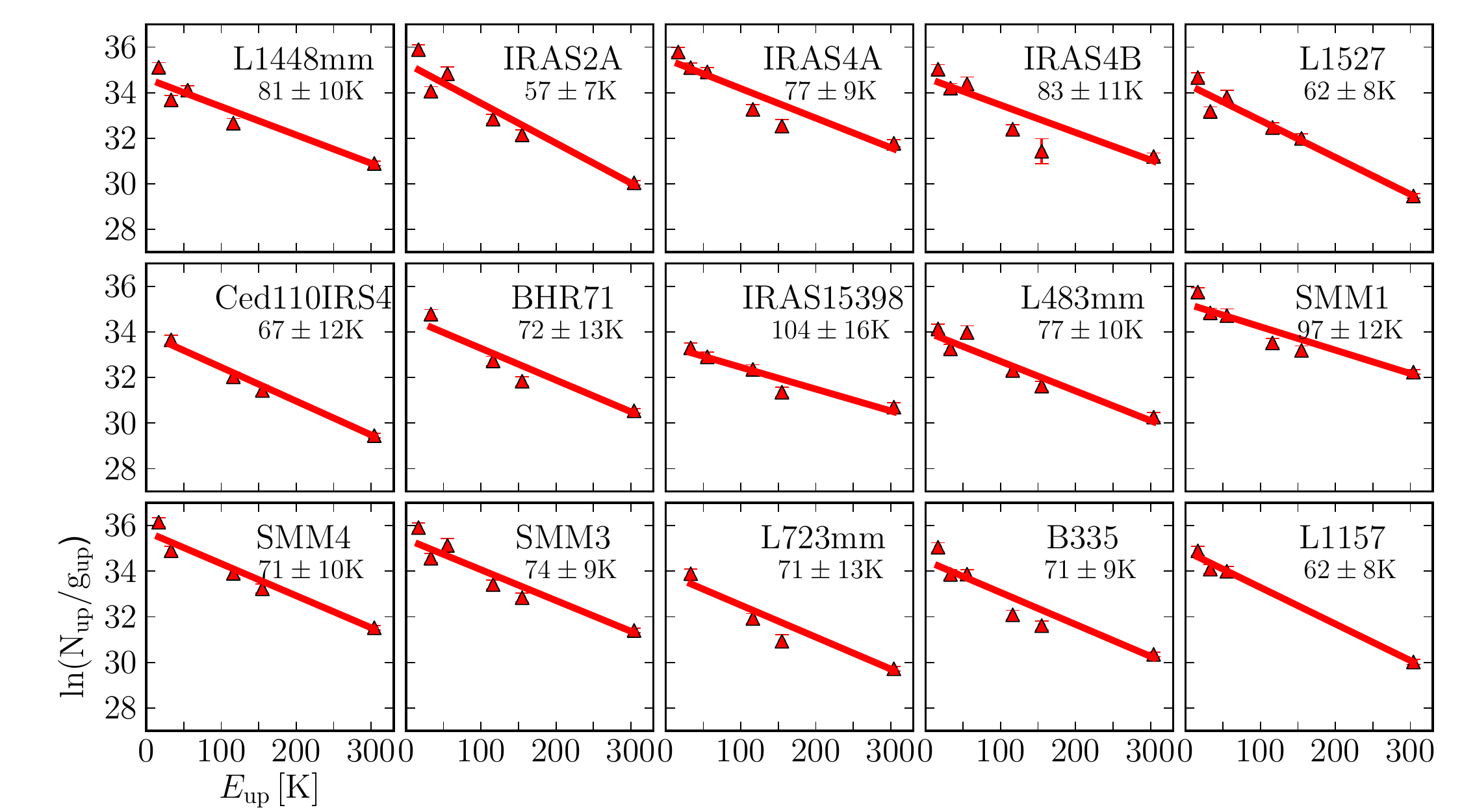}
    \includegraphics[scale=0.42]{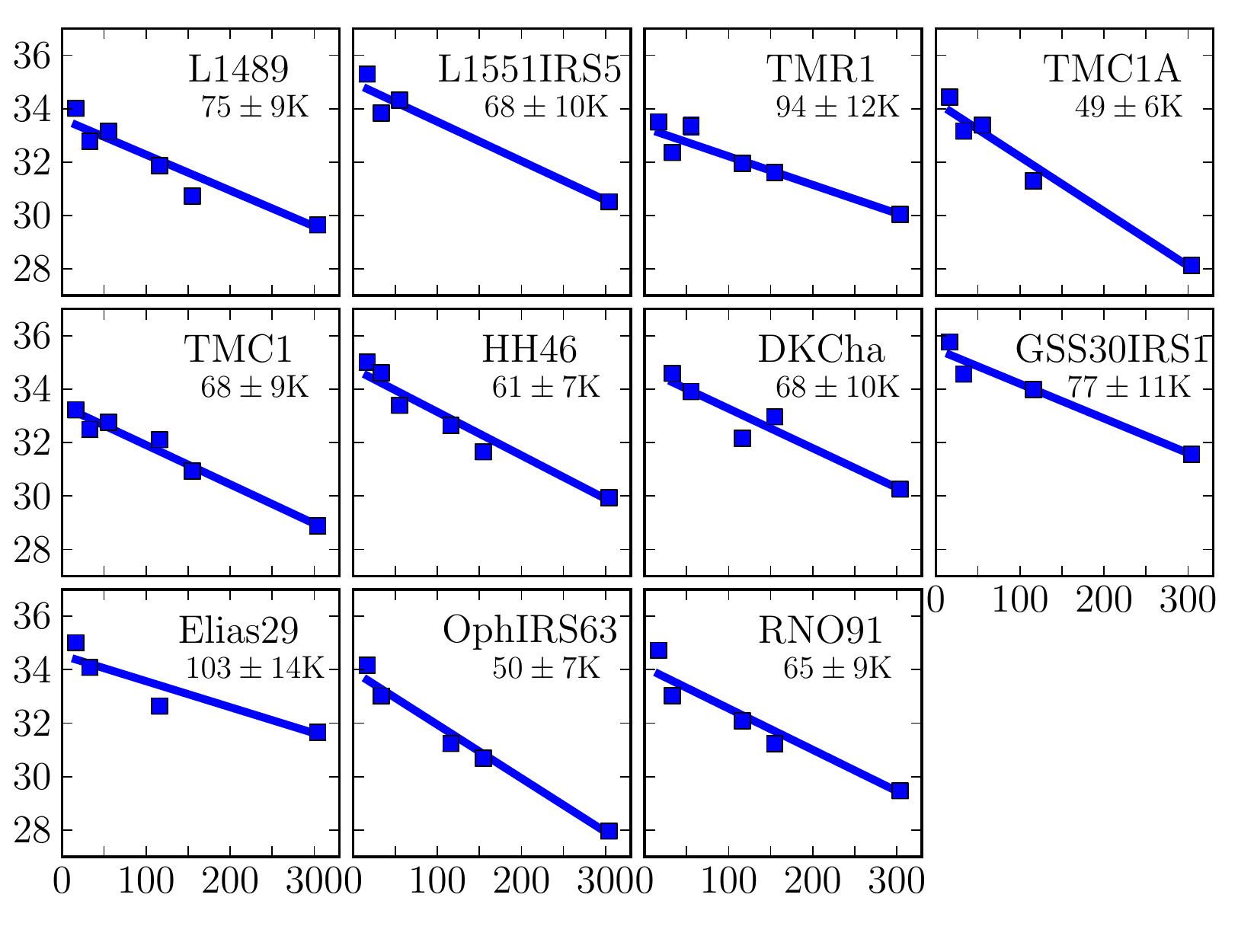}
    \caption{\small Rotational diagrams for $^{12}$CO lines using the
      integrated intensities. All data are convolved to a 20$\arcsec$
      beam and each plot shows the best single temperature fit to the
      observed transitions (see also Table
      \ref{tbl:rottempAndcolden}). {\it Left} panel (red): Class~0
      sources; {\it right} panel (blue): Class~I sources. }
    \label{fig:rotdiag12co}

\vspace{0.15in}
    \includegraphics[scale=0.42]{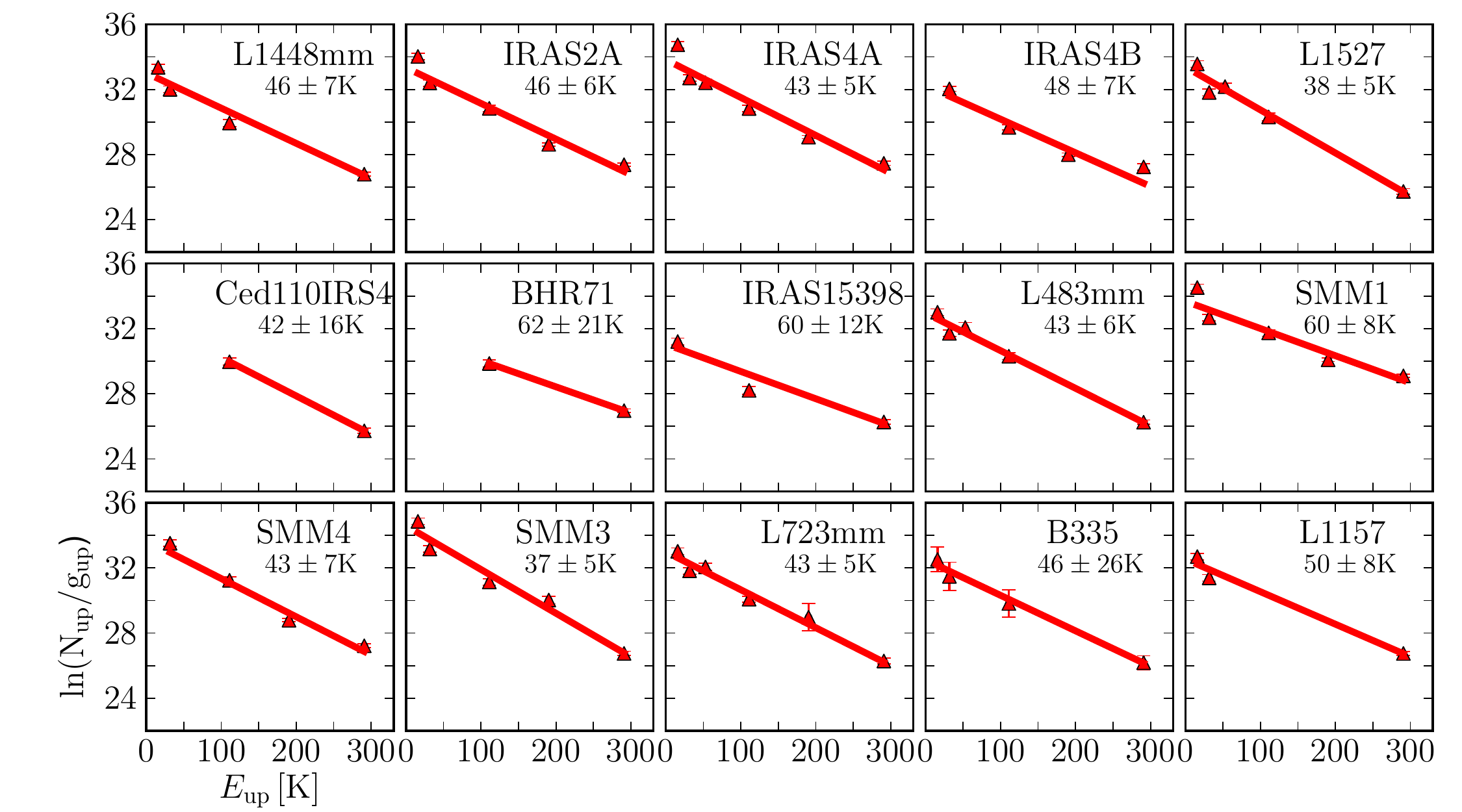}
    \includegraphics[scale=0.42]{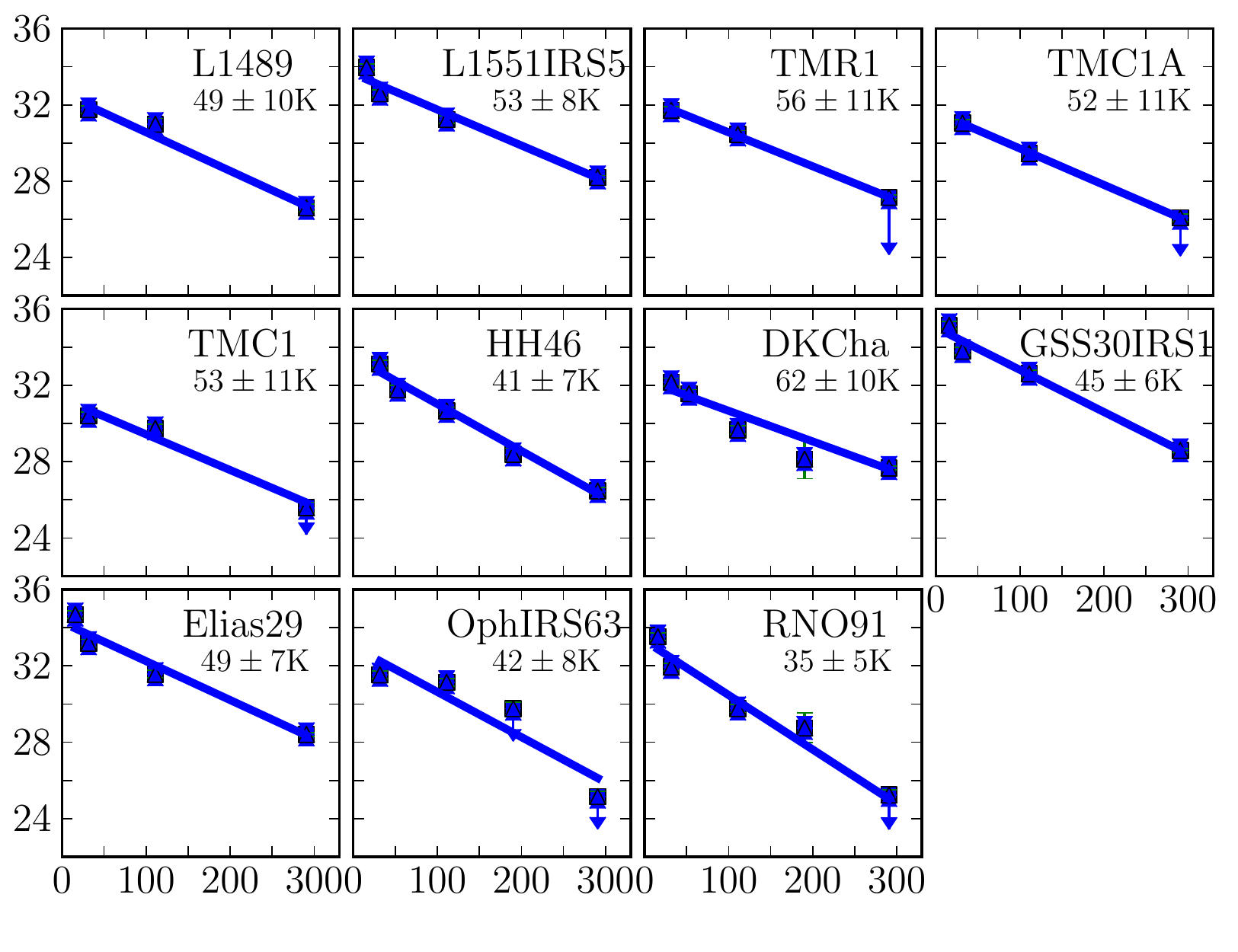}
    \caption{\small Same as Fig. \ref{fig:rotdiag12co} but for $^{13}$CO lines.}
    \label{fig:rotdiag13co}

\vspace{0.15in}
    \includegraphics[scale=0.42]{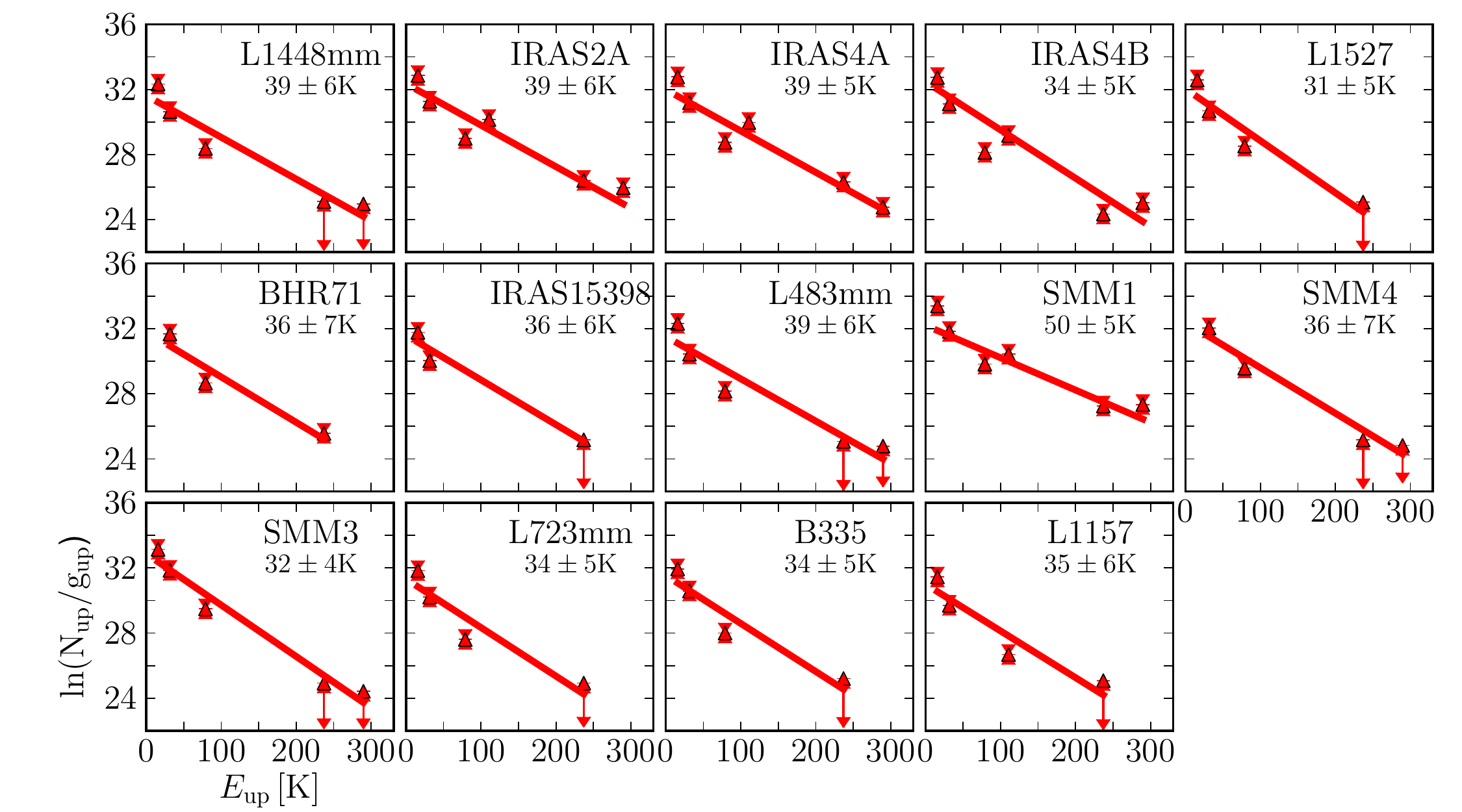}
    \includegraphics[scale=0.42]{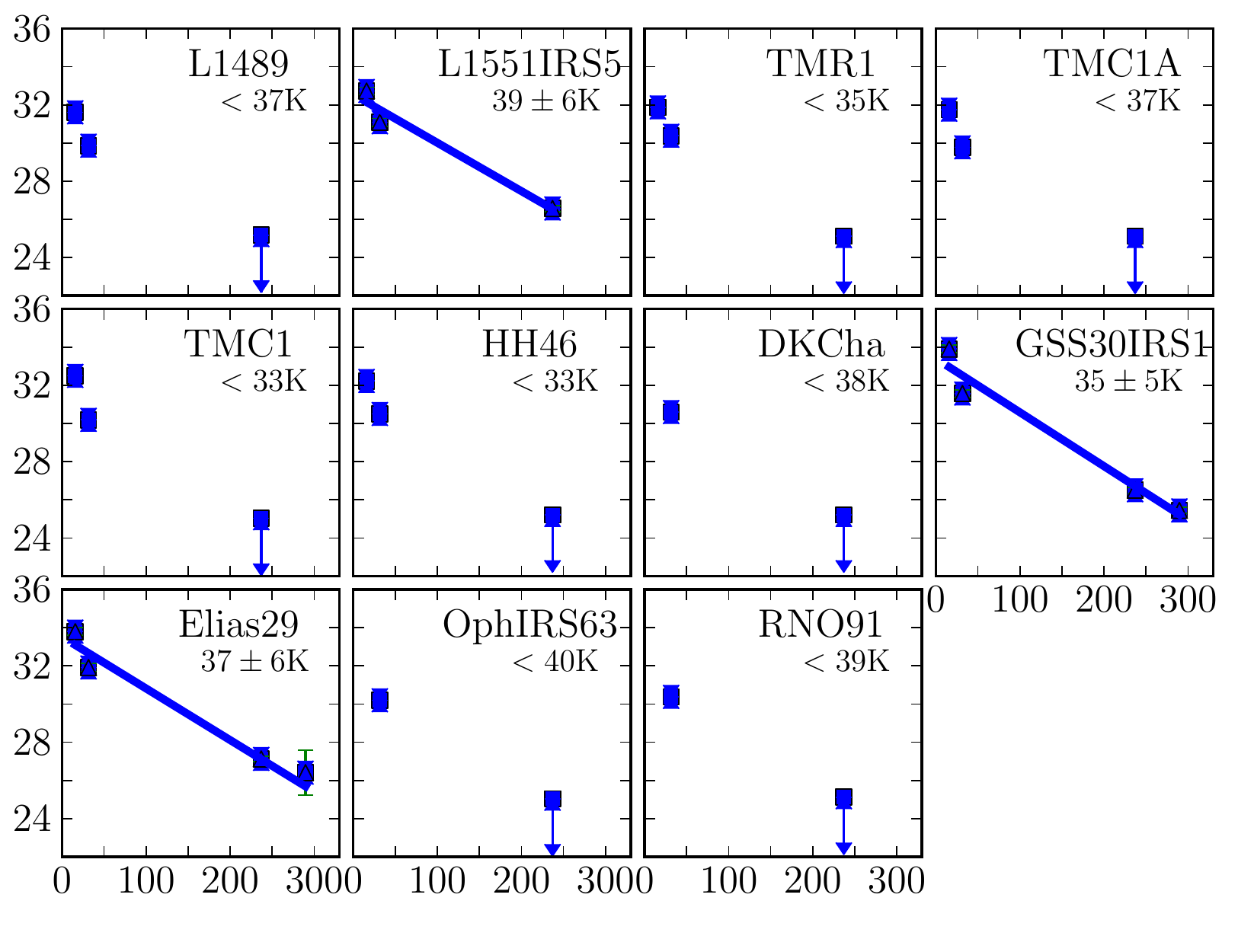}
    \caption{\small Same as Fig. \ref{fig:rotdiag12co} but for C$^{18}$O lines.}
    \label{fig:rotdiagc18o}
\end{figure*}
\end{center}

\section{Rotational diagrams}
\label{4:sec:rotdiags}

To understand the origin of the CO emission, rotational diagrams
provide a useful starting point to constrain the temperature of the
gas.  In our sample, we have CO and isotopolog emission lines of all
sources from $J$=2--1 up to 10--9 with upper level energies from
$E_{\rm up}$=5~K to $\sim$300~K. Rotational diagrams are constructed
assuming that the lines can be characterized by a single excitation
temperature $T_{\rm ex}$, also called rotational temperature $T_{\rm rot}$.
Typically, the isotopolog $^{13}$CO and C$^{18}$O lines are optically
thin, as well as the $^{12}$CO line wings 
\citep[see][]{vanKempen09champ,Yildiz12}, so no curvature should be
induced in their excitation diagrams due to optical depth effects.
However, the low-$J$ $^{12}$CO line profiles have strong
self-absorption and their cores are optically thick, 
leading to the column density of these levels being underestimated.
The \mbox{C$^{18}$O~5--4} line has a beam size of 42$\arcsec$ and may thus
contain unrelated cloud material, so its uncertainty is artificially
enhanced from 10\% to 20\% in order to reduce its weight in the fit
calculations.

Using the level energies, Einstein $A$ coefficients and line frequencies
from Table~1 and the cited databases,
rotational diagrams are constructed
where the column density for each level is plotted against its level
energy \citep{GoldsmithLanger99}. This temperature $T_{\rm rot}$ is basically defined from the Boltzmann
equation
\begin{equation}
\frac{N_{\rm u}}{N_{\rm l}} = \frac{g_{\rm u}}{g_{\rm l}} e^{\left(-{\Delta E/kT_{\rm rot}}\right)},
\end{equation}
where $N_{\rm u}$ and $N_{\rm l}$ are the column densities in the
upper and lower states, and $g_{\rm u}$ and $g_{\rm l}$ their statistical
weights equal to $2J_{\rm u}+1$ and $2J_{\rm l}+1$, respectively.
The CO column densities in individual levels are obtained from
\begin{equation}
\frac{N_{\rm u}}{g_{\rm u}} = \beta \frac{(\nu [\rm{GHz])^{2}}\, W [\mathrm{K\, km\, s^{-1}}]}{A_{\rm ul} [\rm s^{-1}]\, g_{\rm u}},
\end{equation}
where $\beta$ = 1937~cm$^{-2}$ and $W = \int T_{\rm mb} {\rm d}V$ is
the integrated intensity of the emission line.

The slope of the linear fit to the observations, $-(1/T_{\rm rot})$,
gives the rotational temperature, whereas the y-intercept gives the
total column density $ln(N_{\rm total}/Q(T_{\rm rot})$) where
$Q(T_{\rm rot})$ is the partition function referenced from CDMS for
the temperature given by the fit. 

The total integrated intensity $W$ for each line is measured
over the entire velocity range out to where line wings become equal to
the 1$\sigma$ noise. In L1448mm and BHR71, the bullet emission is not
included in the intensity calculation. In IRAS~2A, the emission in the
10--9 line is corrected for emission at one of the reference
positions, which results in a higher $T_{\rm rot}$ compared with
\citet{Yildiz12}.

\subsection{Rotational diagram results}

In Figs. \ref{fig:rotdiag12co}, \ref{fig:rotdiag13co}, and
\ref{fig:rotdiagc18o}, rotational diagrams are depicted for the
$^{12}$CO, $^{13}$CO and C$^{18}$O lines, respectively.  Extracted
excitation temperatures and column densities are presented in Table
\ref{tbl:rottempAndcolden}; $^{12}$CO column densities are not
provided because they are affected by optical depth effects. In all
sources the data can be fitted to a single temperature component from \mbox{$J$=2--1} up to
\mbox{10--9} with a range of uncertainty from 12\% to 21\% except $^{13}$CO 
temperatures in Ced110-IRS4 and BHR71 where only two observations are present. Curvature is present
for a number of sources which will be discussed in
Sect.~\ref{4:sec:twocomponents}.  The derived $^{12}$CO rotational
temperatures range from $\sim$50~K to $\sim$100~K. The median
temperatures for both Class~0 and Class~I sources are similar, $T_{\rm
  rot}$ = 71~K and 68~K, respectively.

\begin{figure}[!t]
    \centering
    \includegraphics[scale=0.4]{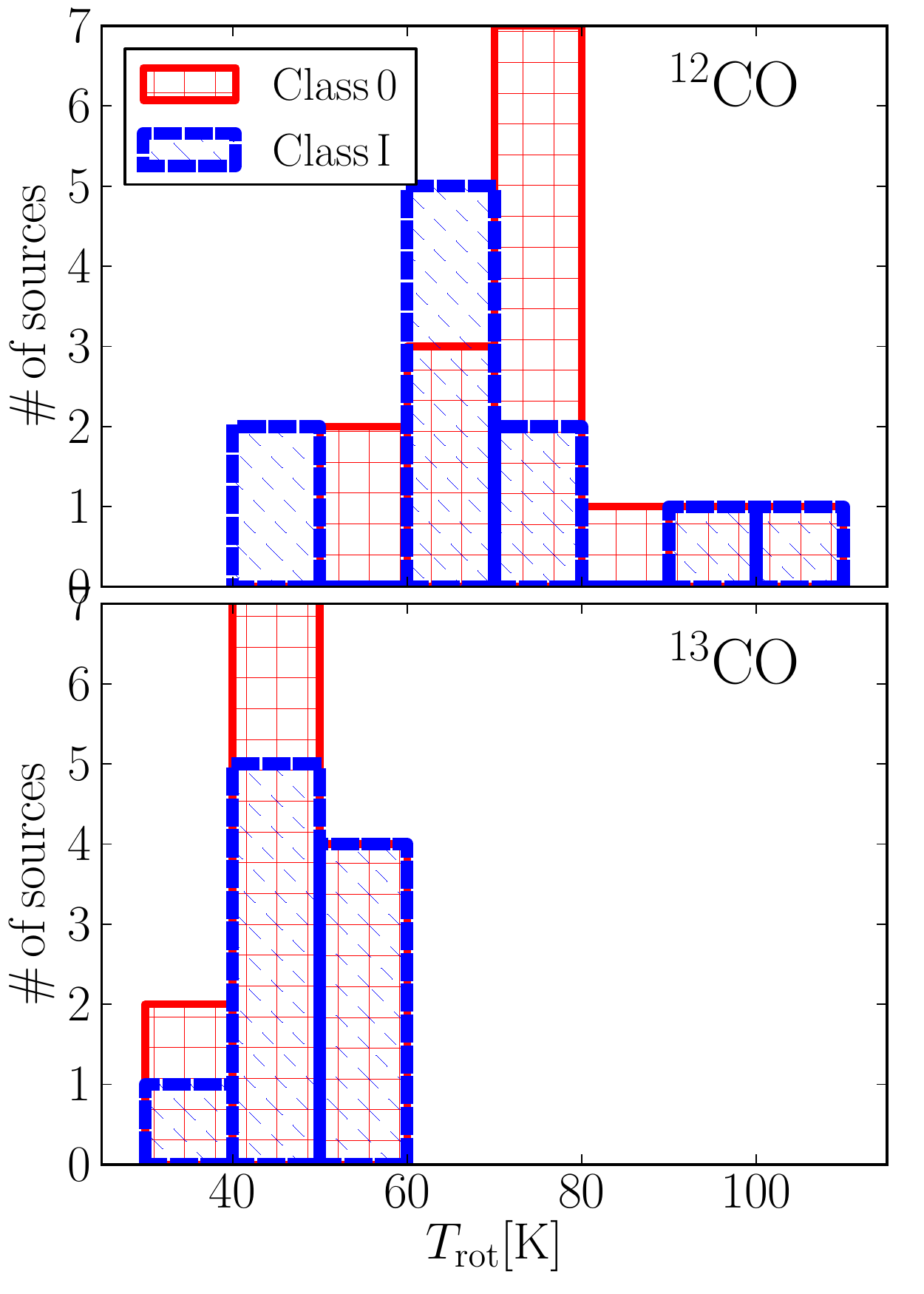}
    \caption{\small Distribution of rotation temperatures ($T_{\rm rot}$) 
calculated from $^{12}$CO and $^{13}$CO line observations. 
The median temperatures are listed in Table \ref{tbl:rotdiaglist}.}    
\label{fig:Texhistogram}
\end{figure}

For $^{13}$CO, the temperatures range from $T_{\rm rot}$ $\sim$35~K to
$\sim$60~K, with a median $T_{\rm rot}$=46~K and 49~K for Class~0 and
Class~I sources, respectively.  For C$^{18}$O, the median temperature
for Class~0 sources is $T_{\rm rot}$ = 36~K.  For Class~I sources,
either lack of observational data or non-detections make it harder to
obtain an accurate temperature. Nevertheless, upper limits are still
given. The median $T_{\rm rot}$ for the sources with $\geq$3 data
points is 37~K. A summary of the median rotational temperatures is
given in Table~\ref{tbl:rotdiaglist}.  Fig.~\ref{fig:Texhistogram}
presents the $^{12}$CO and $^{13}$CO temperatures in histogram mode
for the Class~0 and Class~I sources, with no statistically significant
differences between them. Note that this analysis assumes that all
lines have a similar filling factor in the $\sim$20$''$ beam; if the
higher-$J$ lines would have a smaller filling factor than the
lower-$J$ lines the inferred rotational temperatures would be lower
limits. 
 
\begin{figure*}
\sidecaption
   \includegraphics[scale=0.5]{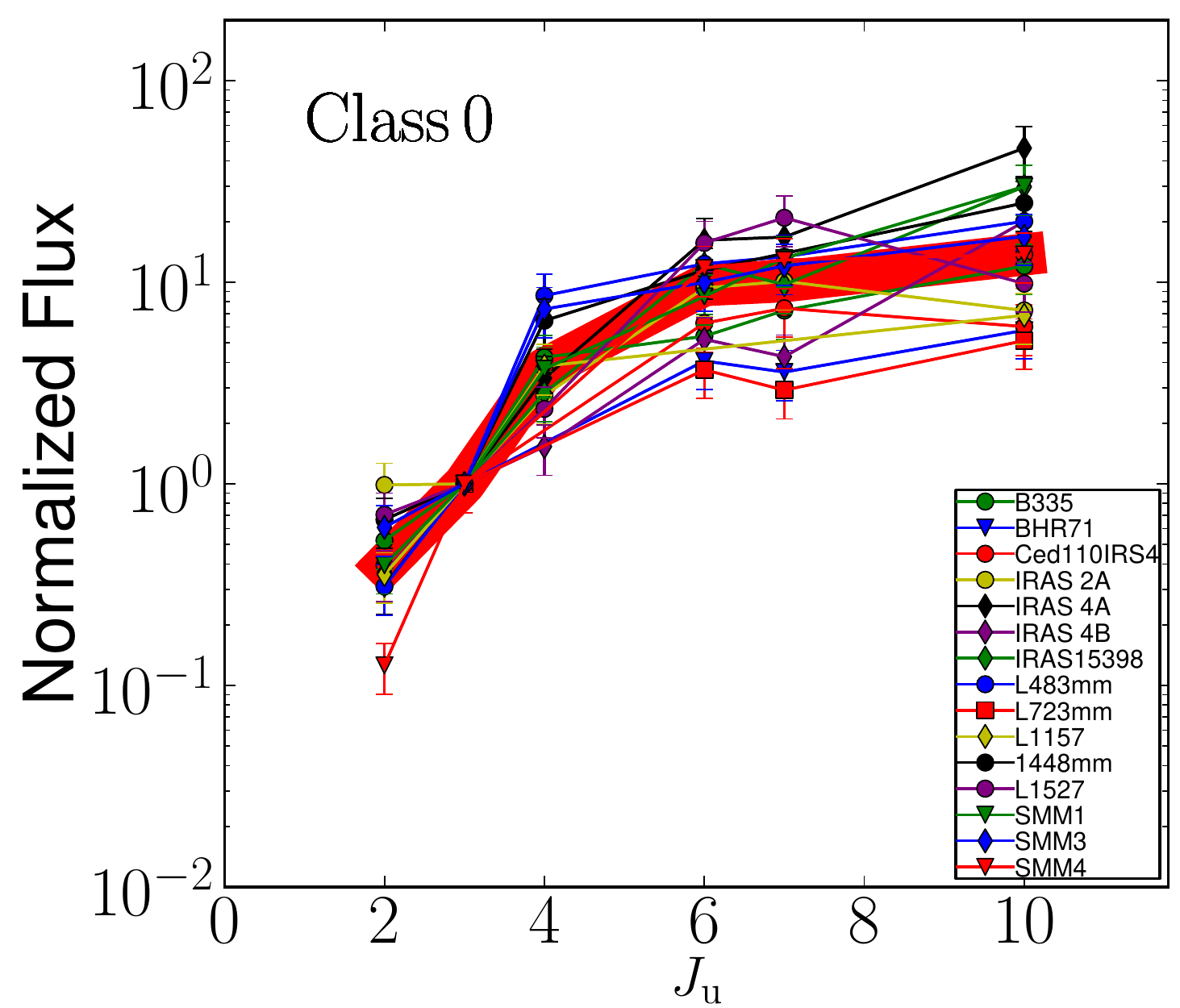}
   \includegraphics[scale=0.5]{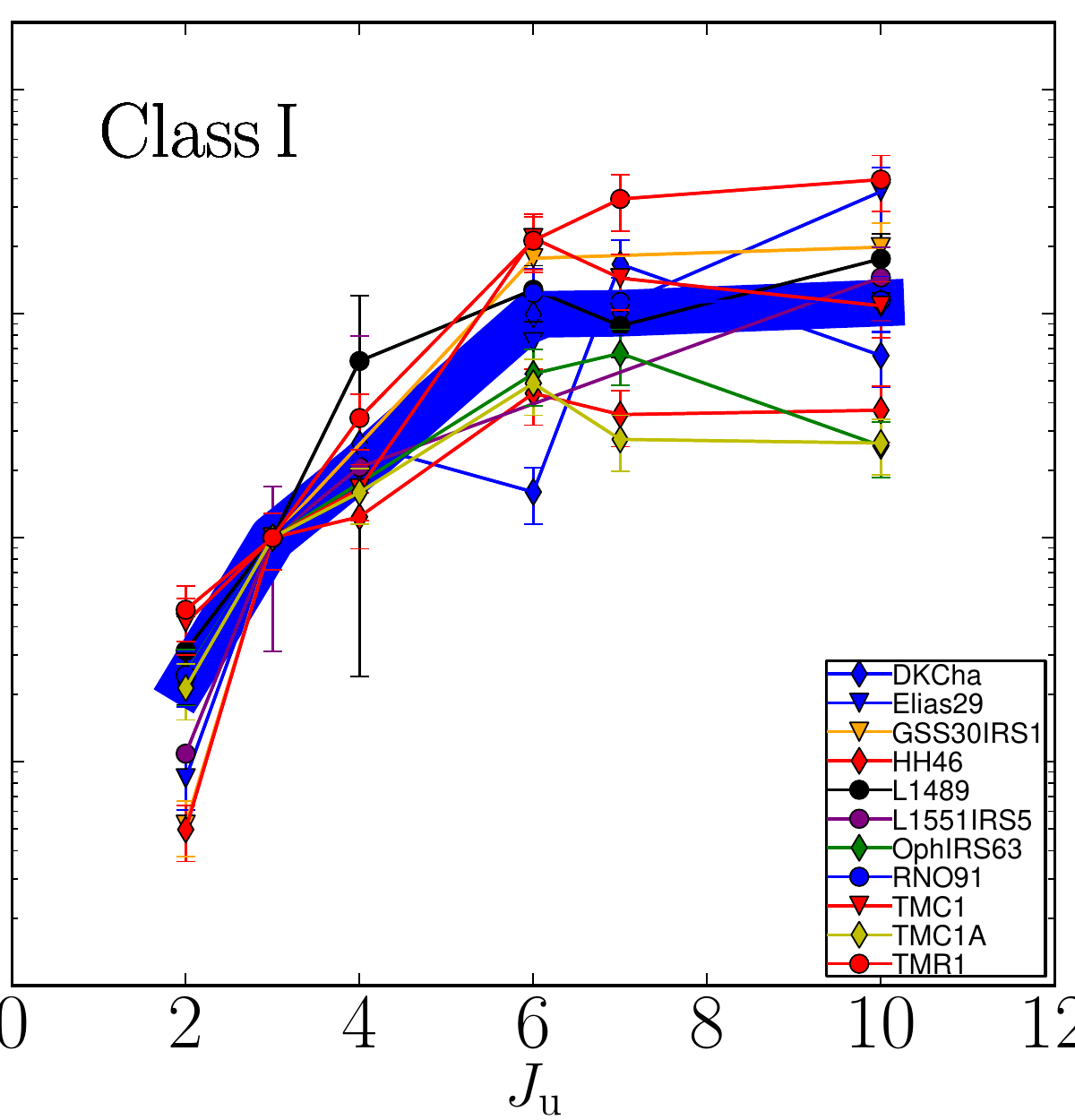}
   \caption{$^{12}$CO spectral line energy distribution (SLED) for the
     observed transitions. All of the fluxes are normalized to their
     own CO~3-2 flux for Class~0 ({\it left}) and Class~I ({\it
       right}) sources separately. The beams are $\sim$20'' beam. The
     thick lines are the median values of each transitions. }
\label{fig:JupvsIntensity}
\end{figure*}

\begin{table*}
\caption{Extracted rotational temperatures and column densities}
\small
\begin{center}
\begin{tabular}{l c c c c c c c c c c}
\hline\hline
 & \multicolumn{3}{c} {Rotational Temperature [K]} & \multicolumn{2}{c} {H$_2$ Column Density\tablefootmark{a} [10$^{21}$ cm$^{-2}$]} & \multicolumn{2}{c} {Kinetic Temperature\tablefootmark{b} [K]} & $X_{\rm constant}$\tablefootmark{c}\\
\hline
Source              & $^{12}$CO & $^{13}$CO & C$^{18}$O & from $^{13}$CO & from C$^{18}$O & Blue Wing & Red Wing & $\times$10$^{-8}$\\ 
\hline
L1448-MM         & 81  $\pm$ 10 & 46 $\pm$  7  & 39 $\pm$  6  & 5.0  $\pm$ 0.7 & 4.4 $\pm$ 0.6 & 60--75   & 100--200 & 2.1 \\
NGC1333-IRAS2A   & 57  $\pm$ 7  & 46 $\pm$  6  & 39 $\pm$  6  & 6.8  $\pm$ 0.9 & 9.2 $\pm$ 1.3 & \dots    & \dots    & 6.0 \\
NGC1333-IRAS4A   & 77  $\pm$ 9  & 43 $\pm$  5  & 39 $\pm$  5  & 10.6 $\pm$ 1.3 & 6.5 $\pm$ 0.8 & 70--80   & 80--110  & 1.4 \\
NGC1333-IRAS4B   & 83  $\pm$11  & 48 $\pm$  7  & 34 $\pm$  5  & 2.4  $\pm$ 0.4 & 9.0 $\pm$ 1.4 & 110--170 & 110--280 & 1.5 \\
L1527            & 62  $\pm$ 8  & 38 $\pm$  5  & 31 $\pm$  5  & 6.0  $\pm$ 0.8 & 5.5 $\pm$ 0.8 & 80--130  & 80--110  & 14.0 \\
Ced110-IRS4      & 66  $\pm$ 9  & 42 $\pm$ 16  & \dots        & 3.0  $\pm$ 1.1 & \dots   & \dots          & \dots    & 17.0 \\
BHR71            & 72  $\pm$13  & 62 $\pm$ 21  & 36 $\pm$  7  & 1.7  $\pm$ 0.6 & 5.0 $\pm$ 1.0 & 75--105  & 70--140  & 8.5 \\
IRAS15398        &104  $\pm$16  & 60 $\pm$ 12  & 35 $\pm$  5  & 0.9  $\pm$ 0.2 & 3.8 $\pm$ 0.7 & \dots    & \dots    & 3.5 \\
L483             & 77  $\pm$10  & 43 $\pm$  6  & 39 $\pm$  6  & 4.5  $\pm$ 0.6 & 4.0 $\pm$ 0.6 & 105--300 & 100--240 & 6.0 \\
Ser SMM1         & 97  $\pm$12  & 60 $\pm$  8  & 50 $\pm$  5  & 12.7 $\pm$ 1.7 & 10.1$\pm$ 1.1 & 90--120  & 100--230 & 2.2 \\
Ser SMM4         & 71  $\pm$10  & 43 $\pm$  7  & 36 $\pm$  7  & 9.2  $\pm$ 1.6 & 8.8 $\pm$ 1.6 & 75--120  & 50--120  & 6.0 \\
Ser SMM3         & 74  $\pm$9   & 37 $\pm$  5  & 32 $\pm$  4  & 19.8 $\pm$ 2.8 & 12.7$\pm$ 1.8 & 75--175  & 80--260  & 10.0 \\
L723             & 71  $\pm$13  & 43 $\pm$  5  & 34 $\pm$  5  & 4.6  $\pm$ 0.6 & 2.8 $\pm$ 0.4 & \dots    & \dots    & 4.4 \\
B335             & 71  $\pm$9   & 46 $\pm$  6  & 34 $\pm$  5  & 3.0  $\pm$ 1.7 & 3.6 $\pm$ 0.5 & \dots    & \dots    & 9.0 \\
L1157            & 62  $\pm$8   & 50 $\pm$  8  & 35 $\pm$  6  & 3.4  $\pm$ 0.5 & 2.1 $\pm$ 0.3 & 75--130  & 85--120  & 2.4 \\
\hline
L1489            & 75  $\pm$ 9  & 49 $\pm$ 10  & $<$37        & 3.5  $\pm$ 0.7 & $<$3.7   & \dots         & \dots    & 17.0 \\
L1551-IRS5       & 68  $\pm$10  & 53 $\pm$  8  & 39 $\pm$ 6   & 10.7 $\pm$ 1.5 & 11.3 $\pm$ 1.8 & 70--130 & 70--130  & 8.0 \\
TMR1             & 94  $\pm$12  & 56 $\pm$  11 & $<$35        & 3.1  $\pm$ 0.6 &$<$ 4.9   & \dots         & \dots    & 38.0 \\
TMC1A            & 49  $\pm$ 6  & 52 $\pm$  11 & $<$37        & 1.4  $\pm$ 0.3 & $<$3.4    & \dots        & \dots    & 17.0 \\
TMC1             & 68  $\pm$ 9  & 53 $\pm$ 11  & $<$33        & 1.1  $\pm$ 0.2 & $<$6.0   & \dots         & \dots    & 50.0 \\
HH46-IRS         & 61  $\pm$7   & 41 $\pm$ 7   & $<$33        & 7.1  $\pm$ 1.1 & $<$8.0   & 100--130      & 105--300 & 9.0 \\
DK~Cha           & 68  $\pm$10  & 62 $\pm$ 10  & $<$38        & 3.2  $\pm$ 0.5 & $<$3.6   & \dots         & \dots    & 7.0 \\
GSS30-IRS1       & 77  $\pm$ 11 & 45 $\pm$  6  & 35 $\pm$ 5   & 37.0 $\pm$ 5.1 & 24.0 $\pm$ 3.7 & 160--300& 130--240 & 45.0 \\
Elias 29         & 103 $\pm$14  & 49 $\pm$  7  & 39 $\pm$ 6   & 19.1 $\pm$ 2.7 & 31.1 $\pm$ 4.4 & 340--400& 340--400 & 50.0 \\
Oph-IRS63        & 50  $\pm$ 7  & 42 $\pm$  8  & $<$40        & 4.6  $\pm$ 0.9 & $<$2.4  & \dots          & \dots    & 40.0 \\
RNO91            & 65  $\pm$ 9  & 35 $\pm$  5  & $<$39        & 5.2  $\pm$ 0.8 & $<$2.9  & \dots          & \dots    & 12.0 \\
\hline
\end{tabular}
\tablefoot{Rotational temperatures and column densities calculated from $^{12}$CO, $^{13}$CO and C$^{18}$O observations. Kinetic temperatures calculated from blue and red line wings.
\tablefoottext{a}{In H$_2$ column densities. Derived using a constant abundance ratio of CO/H$_2$=10$^{-4}$.}
\tablefoottext{b}{Obtained from CO~10--9/3--2 line intensity ratios as explained in Sect. \ref{sect:kintemp}}.
\tablefoottext{c}{Constant abundance fits as explained in Sect.~\ref{4:sec:coabundance}}.
}
\end{center}
\label{tbl:rottempAndcolden}
\end{table*}

For the case of Serpens-SMM1, our inferred rotational temperatures of
97$\pm$12 and 60$\pm$8 K compare well with those of 103$\pm$15 and
76$\pm$6 K found by \citet{Goicoechea12} from {\it Herschel-}SPIRE
data for $^{12}$CO and $^{13}$CO, respectively. The SPIRE values were
obtained from a fit to the $J_{\rm u}$=4--14 levels, with the beam changing
by a factor of $\sim$3 from $\sim$47$\arcsec$ to $\sim$13$\arcsec$ across the
ladder.

The $^{12}$CO and \mbox{$^{13}$CO~2--1} lines included in
Figs.~\ref{fig:rotdiag12co} and \ref{fig:rotdiag13co} 
are observed in a similar
$\sim$20$\arcsec$ beam as the higher transitions, but they most likely
include cold cloud emission as well. Removing those
lines from the rotational diagrams increases the
temperatures around 10-15~$\%$.
Similarly, the \mbox{C$^{18}$O~5--4} line is observed in a 42$\arcsec$ beam
and removing this line from the fit increases the temperatures around
5--10$\%$, which is still within the error bars. In practice, we did
not discard the \mbox{C$^{18}$O~5--4} observations but increased their
uncertainty to give them less weight in the calculations. The effect 
of multiple velocity components in discussed in Sect.~\ref{4:sec:velresdiag}.

The $^{13}$CO and C$^{18}$O column densities
are converted to $^{12}$CO column densities by using $^{12}$C/$^{13}$C = 65
\citep[based on][]{LangerPenzias90,Vladilo93} and $^{16}$O/$^{18}$O =
550 \citep{WilsonRood94} and then using CO/H$_2$~=~10$^{-4}$ to obtain
the H$_2$ column density (tabulated in Table ~\ref{tbl:rottempAndcolden}). 
The median H$_2$ column densities for Class~0 
sources are 4.6$\times$10$^{21}$~cm$^{-2}$  and  5.2$\times$10$^{21}$~cm$^{-2}$ 
for $^{13}$CO and C$^{18}$O data, respectively. For Class~I sources the 
value is 4.6$\times$10$^{21}$~cm$^{-2}$ for $^{13}$CO, with the caveat 
that only few Class~I sources have been measured in C$^{18}$O.
The agreement between the two isotopologs indicates that the lines are
not strongly affected by optical depths. Since this conversion uses a
CO/H$_2$ abundance ratio close to the maximum, the inferred H$_2$
column densities should be regarded as minimum values. In particular,
for Class 0 sources freeze-out and other chemical processes will lower
the CO/H$_2$ abundance (see Sect.~\ref{4:sec:envelope})
\citep{Jorgensen02}. Thus, the actual difference in column densities
between the Class~0 and Class~I stages is larger than is shown in
Table~\ref{tbl:rottempAndcolden}.

\begin{table}[!t]
\caption{Median rotational temperatures and colum densities of Class~0 
and Class~I sources calculated from $^{12}$CO, $^{13}$CO and C$^{18}$O.}
\tiny
\begin{center}
\begin{tabular}{l c c c c c c c}
\hline
\hline
 & \multicolumn{3}{c} {Rotational Temperature [K]} & \multicolumn{2}{c} {Column Density\tablefootmark{a} [10$^{21}$~cm$^{-2}$]} \\
\hline
 &  {$^{12}$CO} & {$^{13}$CO}  & {C$^{18}$O} & {$^{13}$CO}  & {C$^{18}$O} \\
\hline  
Class~0& 71$\pm$13 & 46$\pm$6   & 36$\pm$7 & 4.6$\pm$0.6 & 5.2$\pm$0.9 \\
Class~I & 68$\pm$9    & 49$\pm$10 & \dots          & 4.6$\pm$0.9 & \dots \\
\hline
\end{tabular}
\end{center}
\tablefoot{Median's error is taken as global error value.
\tablefoottext{a}{Derived using a constant abundance ratio of CO/H$_2$=10$^{-4}$.}
}
\label{tbl:rotdiaglist}
\end{table}

\subsection{CO ladders}
\label{4:sec:sled}

Spectral Line Energy Distribution (SLED) plots are another way of
representing the CO ladder where the integrated flux is plotted
against upper level rotational quantum number, $J_{\rm up}$. In
Fig.~\ref{fig:JupvsIntensity}, $^{12}$CO line fluxes for the observed
transitions are plotted.  Since the CO~3--2 lines are available for
all sources, the fluxes are all normalized to their own CO~3--2
flux.  The thick blue and red lines are the median values of all
sources for each transition for each class. It is readily seen that
the Class~0 and Class~I sources in our sample have similar excitation
conditions, but that the Class~I sources show a larger spread at
high-$J$ and have higher error bars due to the weaker absolute
intensities. 
Similarly, as can be inferred from Fig.~\ref{fig:Texhistogram}, 
the $^{13}$CO SLEDs do not show any significant difference between 
the two classes. Thus, although the continuum SEDs show a significant
evolution from Class~0 to Class I with $T_{\rm bol}$ increasing from
$<$30 K to more than 500~K, this change is not reflected in the line
SLEDs of the $^{12}$CO or the optically thinner $^{13}$CO excitation. 
This limits the usage of CO SLEDs as an evolutionary probe. One
of the explanations for this lack of evolution is  that $T_{\rm bol}$ 
depends on thermal emission from both the dust in the envelope
and the extincted stellar flux, whereas the SLED only traces the
temperature of the gas in the envelope and/or outflow, but has no 
stellar component. Moreover, the infrared dust emission comes from 
warm (few hundred K) optically thick dust very close to the protostar, 
whereas, the CO originates further out the envelope.

\subsection{Two temperature components?}
\label{4:sec:twocomponents}

\begin{figure}[!t]
    \centering
	\includegraphics[scale=0.45]{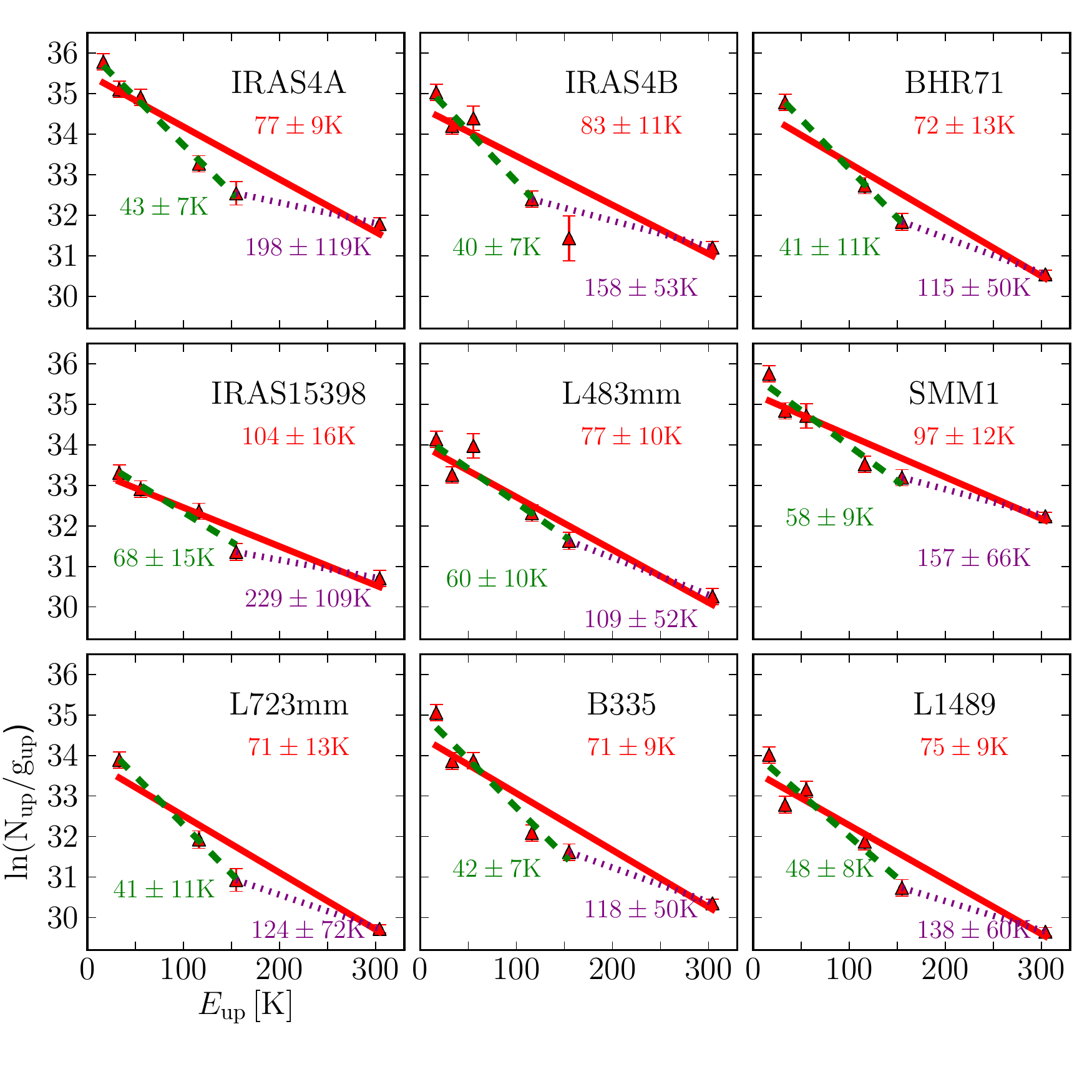}
    \caption{\small Two temperature components fitted for selected sources 
    to the $^{12}$CO lines, from 2--1 to 7--6 (cold), from 7--6 to 10--9 
    (warm), and 2--1 to 10--9 (global).} 
    \label{fig:TexMultiComp}
\end{figure}

Unresolved line observations of higher-$J$~CO transitions ($J$=13 up
to 50) by {\it Herschel}-PACS \citep{Herczeg12,Karska13,Manoj13}
typically show two temperature components with $\sim$300~K and
$\sim$900~K.  \citet{Goicoechea12} found three temperature components
from combined SPIRE and PACS data, with the lower temperature of
$\sim$100 K fitting lines up to $J_{\rm u}$=14, similar to that found
in our data. The question addressed here is if the higher 300~K
component only appears for lines with $J_{\rm u}$$>$10 or whether it
becomes visible in our data. One third of our sample shows a positive
curvature in the $^{12}$CO rotation diagrams (Fig.~\ref{fig:rotdiag12co}), 
specifically IRAS~4A, IRAS~4B, BHR71, IRAS~15398, L483mm, Ser-SMM1, 
L723mm, B335, and L1489 (see Fig.~\ref{fig:TexMultiComp}).

The curvature in $^{12}$CO rotation diagrams is treated by dividing
the ladder into two components, where the first fit is from 2--1 to
7--6 for the colder component and the second fit from 7--6 to 10--9
for the warmer component. The fit from 2--1 to 10--9 is named as
global. The median $T_{\rm rot}$ is 43~K for the colder component and
138~K for the warmer component for these nine sources. 

Close inspection of the SPIRE data by \citet{Goicoechea12} shows a
slight curvature for low-$J$ in their cold component as well. The
``two-component'' decomposition is perhaps a generic feature for
Class~0 low-mass protostars, which implies that the CO~10--9
transition is at the border of the transitions for the cold
  ($T_{\rm rot}$$<$100~K) and warm ($T_{\rm rot}$$\sim$300~K)
  component associated with the currently shocked gas
\citep{Karska13}.  $T_{\rm rot}$ clearly increases when higher
  rotational levels are added (Fig.~\ref{fig:TexMultiComp}), and for
  the brightest sources in NGC1333 and Serpens, $T_{\rm
    rot}$$>$150~K. See Sect.~\ref{4:sec:velresdiag} for further discussion.

\subsection{Velocity resolved diagrams}
\label{4:sec:velresdiag}

 \begin{figure}[!t]
    \centering
    \includegraphics[scale=0.6]{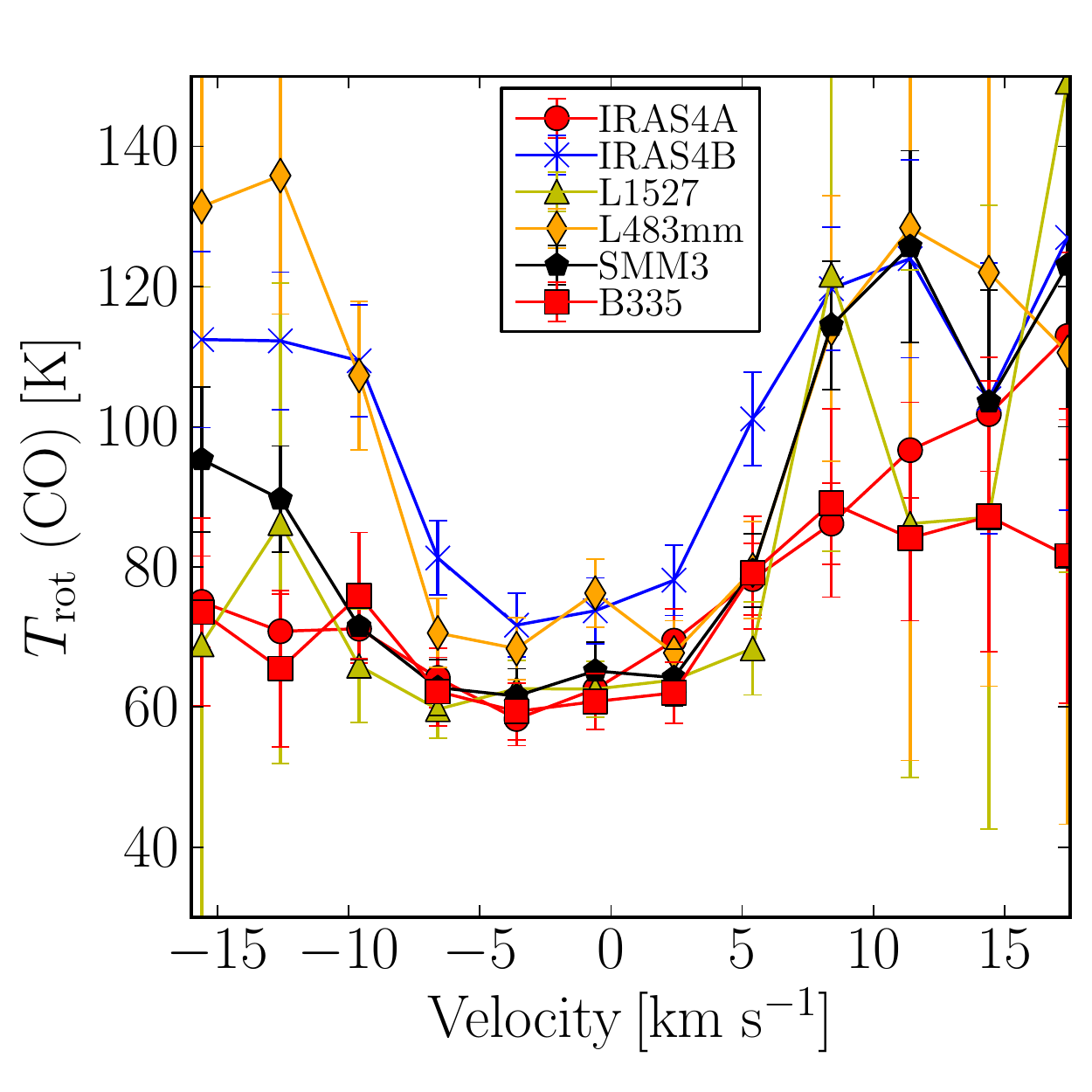}
    \caption{\small Rotational temperatures calculated channel by
      channel for $^{12}$CO. Each spectrum is shifted to $V_{\rm
        lsr}$=0~km~s$^{-1}$ and rebinned to 3~km~s$^{-1}$ velocity
      resolution.}
    \label{fig:VvsTex}
\end{figure}

To investigate whether the $^{12}$CO narrow and broad components have
different temperatures, Fig.~\ref{fig:VvsTex} presents excitation
temperatures calculated channel by channel for a few sources with high
$S/N$. Each spectrum is shifted to $V_{\rm lsr}$=0~km~s$^{-1}$ and
rebinned to 3~km~s$^{-1}$ velocity resolution. 
It is obvious from Fig.~\ref{fig:linesCO3-2andHIFI} that the line
wings are more prominent in the CO~10--9 transitions, specifically for
Class~0 sources, as is reflected also in the increasing line widths
with the increasing rotation level \citep[see Figs.~C.(1--26) in the
  Appendix;][]{SanJoseGarcia13}. Figure~\ref{fig:VvsTex} shows that
in the optically thin line wings, the excitation temperatures are a
factor of 2 higher than in the line centers implying that the wings of
the \mbox{higher-$J$~CO} lines are associated with the warmer material
described in Sect.~\ref{4:sec:twocomponents}. Since the presence of 
self-absorption at line centers
of the lower-$J$ lines reduces their emission, the excitation
temperatures at low velocities are further decreased if this
absorption is properly corrected for.

Another way to illustrate the change with velocity is to look at the
\mbox{CO~10--9} and \mbox{CO~3--2} spectra for each source as shown in
Fig.~\ref{fig:12CO_10-9--12CO_3-2_ontop} (in the Appendix).
Fig.~\ref{fig:12CO_10-9--12CO_3-2_ratios} shows the blue and red
line-wing ratios for the 14 sources with the highest $S/N$.  For all
sources, the line ratios increase with increasing velocity, consistent
with Fig.~\ref{fig:VvsTex}.

In summary, while a single rotational temperature provides a decent
fit to the bulk of the CO and isotopolog data, both the $^{12}$CO
integrated intensity rotational diagrams and the velocity resolved
diagrams indicate the presence of a second, highly excited component for Class
0 sources. This warmer and/or denser component is most likely associated with the
broad line wings, as illustrated by the 10--9/3--2 line ratios,
whereas the colder component traces the narrow quiescent
envelope gas. On average, the integrated intensities have roughly equal
contributions from the narrow and broad components
(Fig.~\ref{fig:12COnarrowbroadpercent}) so the single rotational
temperatures are a weighted mean of the cold and warmer values.

\subsection{Kinetic temperature}
\label{sect:kintemp}

\begin{figure*}[!th]
    \centering
\includegraphics[scale=0.65]{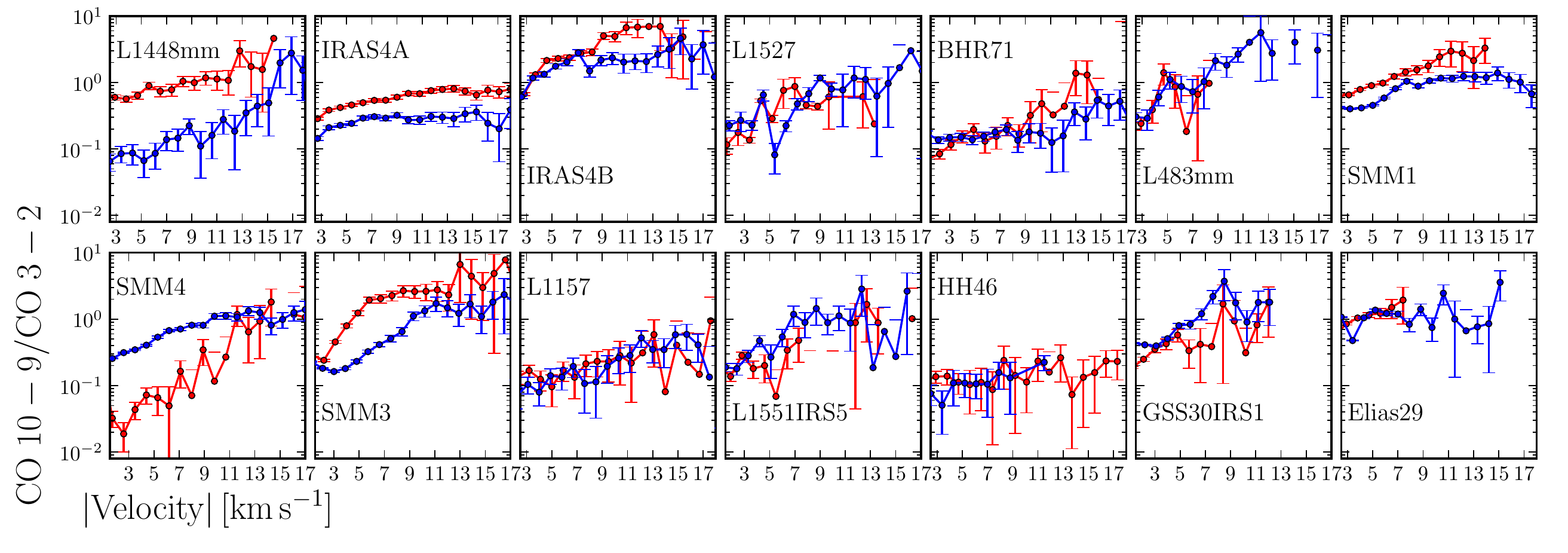}
\caption{\small Ratios of the \mbox{CO~10--9/3--2} line wings for 14
  protostars as function of velocity offset from the central
  emission. Spectra are shifted to $V_{\rm lsr}$=0~km~s$^{-1}$. The
  wings start at $\pm$2.5~km-s$^{-1}$ from $V_{\rm lsr}$ in order to
  prevent adding the central self-absorption feature in the CO~3--2 
  lines.}
    \label{fig:12CO_10-9--12CO_3-2_ratios}
\end{figure*}

Having lines from low-to-high-$J$~CO provides information about the
physical conditions in the different parts of the envelope. The
critical densities for the different transitions are $n_{\rm
  cr}$=4.2$\times$10$^{5}$ cm$^{-3}$, 1.2$\times$10$^{5}$~cm$^{-3}$,
and 2$\times$10$^{4}$~cm$^{-3}$ for CO~10--9, 6--5 and 3--2
transitions at $\sim$50--100~K, using the CO collisional rate
coefficients by \citet{Yang10}. For densities higher than $n_{\rm
  cr}$, the emission is thermalized and therefore a clean temperature
diagnostic, however, for lower densities the precise value of the
density plays a role in the analysis. In the high density case
($n$$>$$n_{\rm cr}$), the kinetic temperature is equal to the rotation
temperature.  By using two different $^{12}$CO lines, kinetic
temperatures can be calculated if the density is known independently
using the \verb1RADEX1 non-LTE excitation and radiative transfer
program \citep{vanderTak07}.  The analysis below for CO~10--9/3--2
assumes that the emission originates from the same gas.

\begin{figure}[!t]
    \centering
	\includegraphics[scale=0.65]{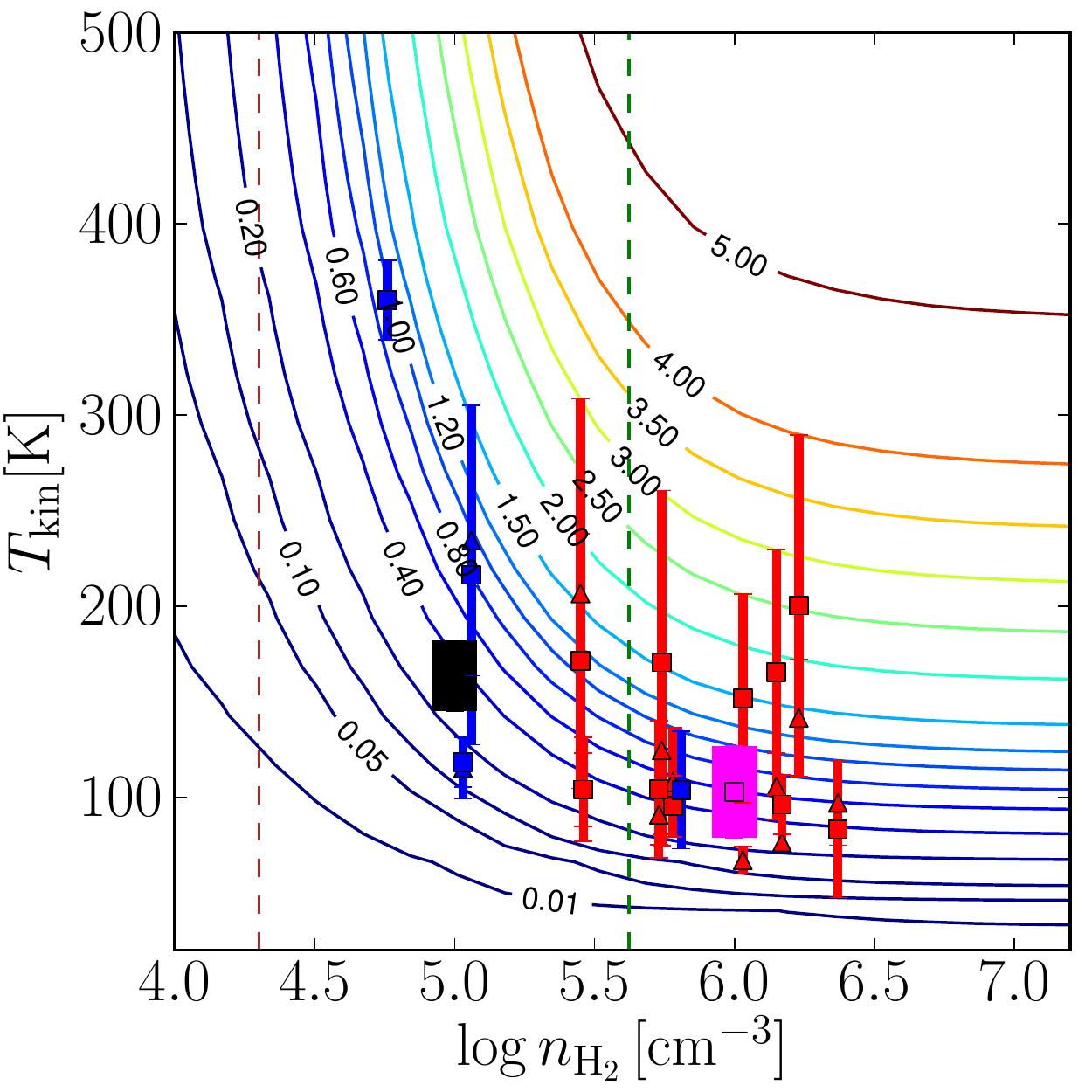}
        \caption{\small Model CO~10--9/3--2 line intensity ratios as
          function of temperature and density, obtained for a CO
          column density of 10$^{16}$ cm$^{-2}$ with line-widths of 10
          km~s$^{-1}$, representative of the observed CO intensity and
          line width. Note that the bars are the ranges of lower and
          upper limits as seen from the observations in
          Fig. \ref{fig:12CO_10-9--12CO_3-2_ratios}.  Red markers are
          for Class~0 and blue markers are for Class~I sources. Blue
          wing ratios are indicated with triangles and red wing ratios
          have square symbols. Thick magenta and black lines indicate
          the ratios from the composite spectra for Class~0 and
          Class~I sources, respectively.
          Vertical dashed lines indicate the limits for $n_{\rm cr}$
          for CO~3--2 (brown) and CO~10--9 (green). In the relevant
        density range, higher ratios are indicative of higher kinetic
        temperatures.}
    \label{fig:radex}
\end{figure}

The resulting model line ratios are presented in Fig.~\ref{fig:radex}
for a grid of temperatures and densities.  Densities for each source
are calculated from the envelope parameters determined by modeling of
the submillimeter continuum emission and spectral energy distribution
\citep{Kristensen12}. A spherically symmetric envelope model
with a power-law density structure is assumed \citep{Jorgensen02}.
The 20$\arcsec$ diameter beam covers a range of radii from
$\sim$1250~AU (e.g., Ced110IRS4, Oph sources) up to 4500~AU (HH46) and
the densities range from 5.8$\times$10$^{4}$~cm$^{-3}$ (Elias~29) up
to 2.3$\times$10$^{6}$~cm$^{-3}$ (Ser-SMM4). The densities for all
sources at the 10$\arcsec$ radius are given at the last column in
Table~\ref{tbl:overviewobs}. Note that these are lower limits since
the densities increase inward of $10''$. The envelope densities are
used here as a proxy for the densities at the outflow walls where the
entrainment occurs.

The majority of the Class~0 sources have densities that are similar or
higher than the critical densities of the high-$J$~CO lines, with the
possible exceptions of Ced110-IRS4, L483mm, L723mm, and
L1157. However, in Class~I sources, the majority of the densities are
lower than the critical densities with the exception of
L1551~IRS5. 
The inferred kinetic temperatures from the \mbox{CO~10--9/3--2} blue
and red line-wings, which generally increase with velocity,
are presented in Table \ref{tbl:rottempAndcolden} and range mostly from
70~K to 250~K.  The ratios for individual sources are included in
Fig. \ref{fig:radex} at the $10''$ radius density of the sources.

The thick magenta and black bars indicate the
average values for the composite Class~0 and
Class~I sources (Sect.~\ref{4:sec:cowater}).  
For average densities at a 10$\arcsec$ radius of
$\sim$10$^{6}$~cm$^{-3}$ and $\sim$10$^{5}$~cm$^{-3}$ for Class~0 and
Class~I sources, respectively, the \mbox{CO~10--9/CO~3--2} line ratios
would imply kinetic temperatures of around 80--130~K for
Class~0 and 140--180~K for Class~I sources (Fig.~\ref{fig:radex}),
assuming the two lines probe the same physical component. If
  part of the CO 10--9 emission comes from a different physical component,
these values should be regarded as upper limits.

\section{Correlations with physical properties}

\subsection{Integrated intensities}

\begin{figure*}[!ht]
\sidecaption
    \centering
    \includegraphics[scale=0.6]{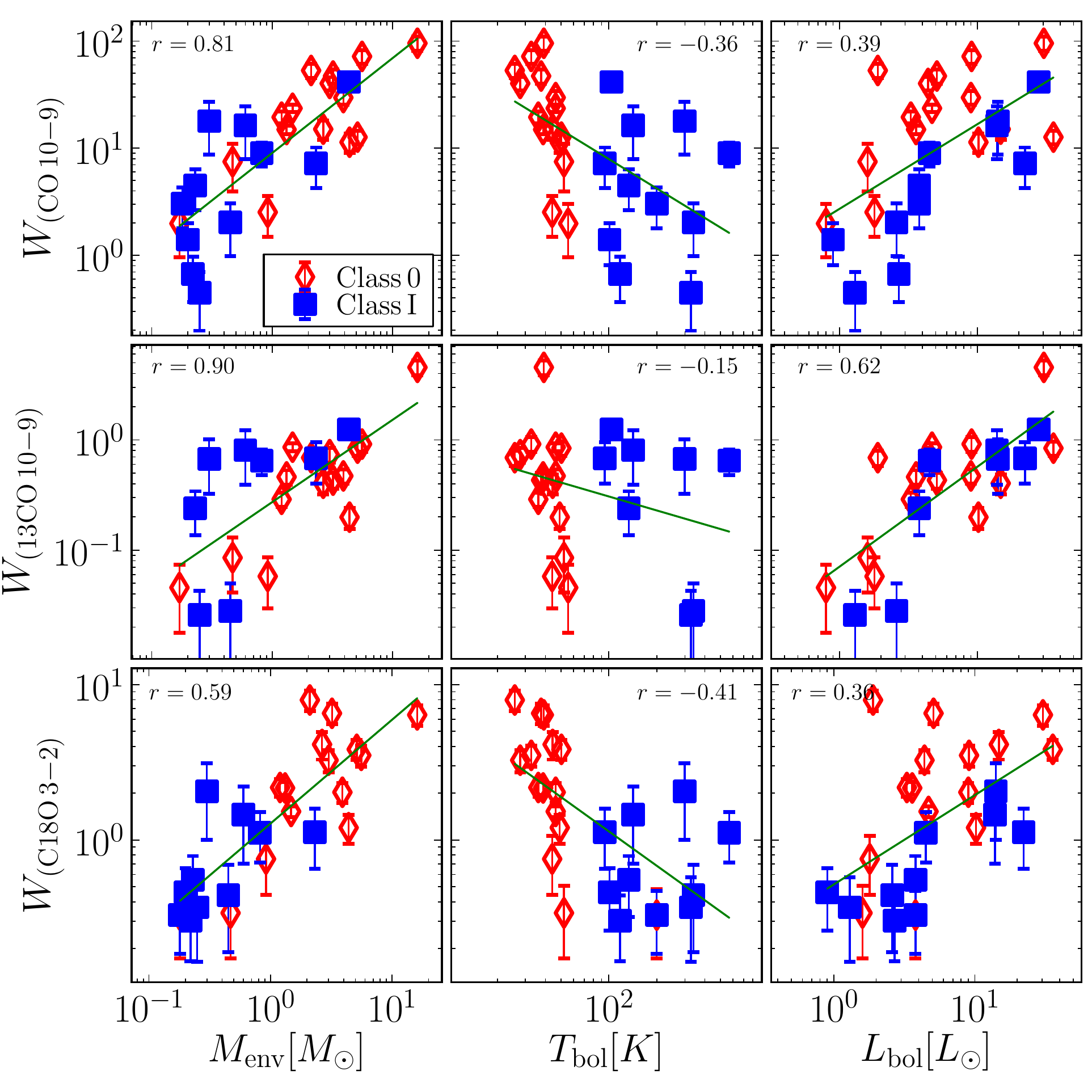}
     \caption{\small From top to bottom, $^{12}$CO~10--9, $^{13}$CO~10--9, 
     and C$^{18}$O~3--2 integrated intensity $W$ normalized at 200~pc are 
     plotted against various physical properties:  envelope mass at 10~K 
     radius, $M_{\rm env}$; bolometric temperature, $T_{\rm bol}$; and 
     bolometric luminosity, $L_{\rm bol}$. Green lines are the best fit to
     the data points and $r$ values in each of the plots are Pearson 
     correlation coefficient.}
     \label{fig:WLbolTbolMenv}
\end{figure*}

Figure \ref{fig:WLbolTbolMenv} shows the integrated intensities $W$ of
the CO~10--9 lines plotted against $L_{\rm bol}$, $M_{\rm env}$, and
$T_{\rm bol}$. The intensities are scaled to a
common distance of 200~pc.  The bolometric luminosity, $L_{\rm bol}$
and bolometric temperature $T_{\rm bol}$ of the sources have been
measured using data from infrared to millimeter wavelengths including
new {\it Herschel} far-infrared fluxes, and are presented in
\citet{Kristensen12}.
These are commonly used evolutionary tracers in order to distinguish
young stellar objects. The envelope mass, $M_{\rm env}$ is calculated 
from the \verb1DUSTY1 modeling by \citet{Kristensen12}.

In Fig. \ref{fig:WLbolTbolMenv}, the green lines are the best
power-law fits to the entire data set. Clearly, the CO 10--9 lines
are stronger for the Class 0 sources which have higher $M_{\rm
  env}$ and $L_{\rm bol}$ and lower $T_{\rm bol}$, for all isotopologs.
The same correlation is seen for other (lower-$J$) CO and isotopolog
lines, such as CO~2--1, 3--2, 4--3, 6--5 and 7--6; examples for
$^{13}$CO~10--9 and C$^{18}$O~3--2 are included in
Fig.~\ref{fig:WLbolTbolMenv}.
The Pearson correlation coefficients for $^{12}$CO 10--9 are $r$=0.39 (1.87$\sigma$),
0.81 (4.05$\sigma$), and --0.36 (--1.68$\sigma$) for $L_{\rm bol}$, $M_{\rm env}$, and $T_{\rm bol}$, 
respectively. The coefficients, $r$, for  $M_{\rm env}$ in $^{13}$CO 10--9 and C$^{18}$O~3--2
are 0.90 (4.49$\sigma$) and 0.59 (2.97$\sigma$).  
Those correlations indicate that there is a strong
correlation between the intensities and envelope mass, $M_{\rm env}$, 
consistent with the lines becoming weaker as the envelope is dissipated. 
Together with the high-$J$~CO lines, the C$^{18}$O low-$J$ lines are
also good evolutionary tracers in terms of $M_{\rm env}$ and $T_{\rm bol}$.  
Adding intermediate and high-mass WISH sources to extend the
correlation to larger values of $L_{\rm bol}$ and $M_{\rm env}$ shows
that these sources follow the same trend with similar slopes with a
strong correlation \citep{SanJoseGarcia13}. The scatter in the 
correlation partly reflects the fact that the CO abundance is not
constant throughout the envelope and changes with evolutionary stage
(see Sect.~\ref{4:sec:envelope}).

\subsection{Excitation temperature and comparison with evolutionary models}
\begin{figure*}[!ht]
\sidecaption
  \hspace*{0.2cm}
  \includegraphics[scale=0.592]{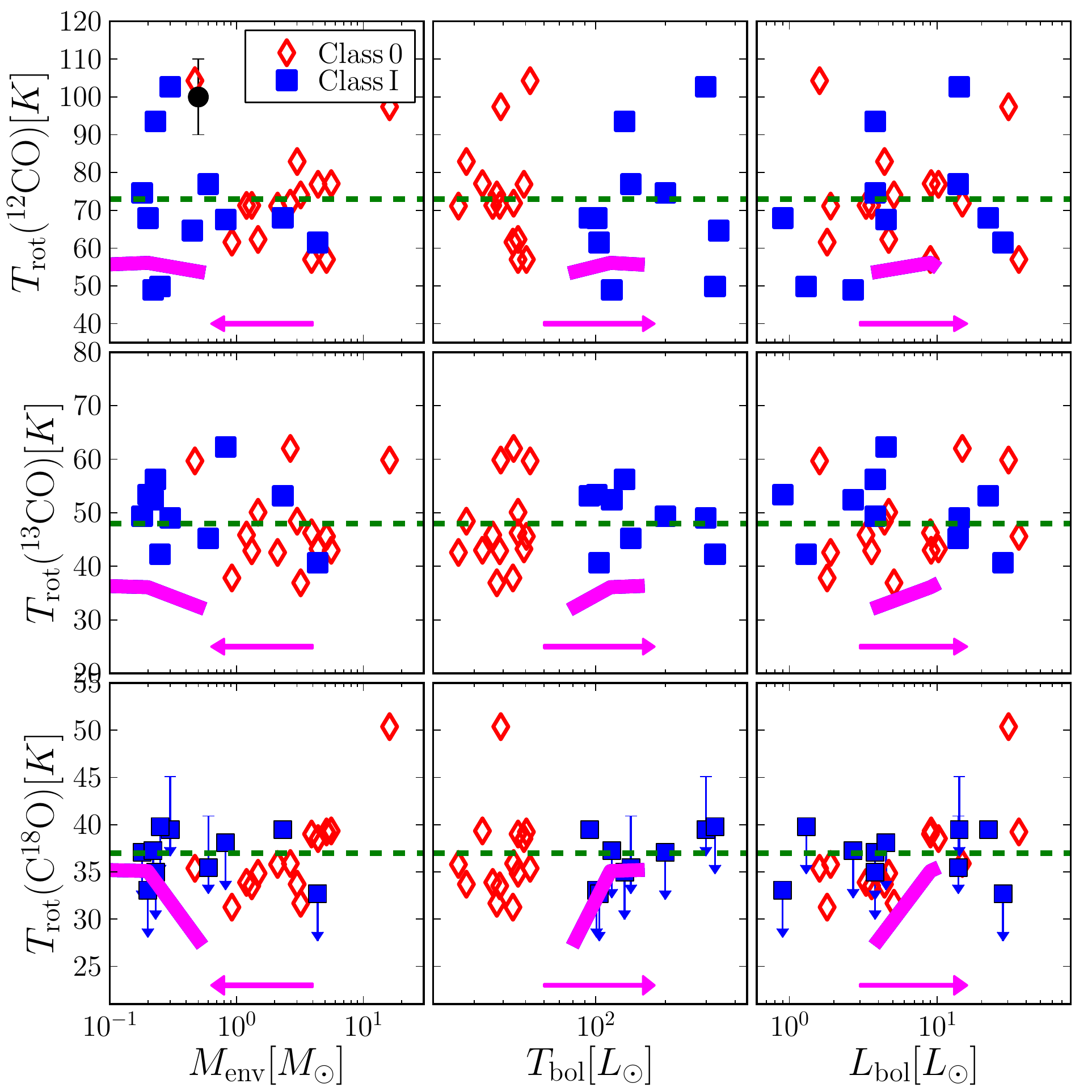}
  \caption{\small Calculated rotational temperatures, $T_{\rm rot}$,
    plotted $M_{\rm env}$, $T_{\rm bol}$, and $L_{\rm bol}$,
    for $^{12}$CO ({\it top}), $^{13}$CO ({\it middle}), C$^{18}$O lines ({\it
    bottom}). The median excitation temperatures of $\sim$70~K,
    48~K and 37~K for $^{12}$CO, $^{13}$CO and C$^{18}$O,
    respectively, are indicated with the green dashed lines. Typical
    error bar for each of the $T_{\rm rot}$ values is shown in the
    upper left plot, represented in black circle data point. In the C$^{18}$O plots, blue arrows 
    indicate the upper limits for a number of sources. These figures are
    compared with disk evolution models of \citet[][magenta solid lines]{Harsono13}, 
    which represents all three models discussed in the text. Magenta arrows 
    also show the direction of time. The figures show
    that the excitation temperature does not change with the
    increasing luminosity, envelope mass or density, confirming that
    Class~0 and I sources have similar excitation conditions. }
  \label{fig:LbolvsTex}
\end{figure*}

Figure \ref{fig:LbolvsTex} presents the derived rotational
temperatures for $^{12}$CO, $^{13}$CO and C$^{18}$O versus 
$M_{\rm env}$, $T_{\rm bol}$, and $L_{\rm bol}$.  In contrast with the
integrated intensities, no systematic trend is seen for any
parameter. As noted in Sect. \ref{4:sec:sled}, this lack of change in
excitation temperature with evolution is in stark contrast with the
evolution of the continuum SED as reflected in the range of $T_{\rm bol}$. 

To investigate whether the lack of evolution in excitation temperature
is consistent with our understanding of models of embedded protostars,
a series of collapsing envelope and disk formation models with time
has been developed by \citet{Harsono13}, based on the
formulation of \citet{Visser09} and \citet{Visser10}.  Three different
initial conditions are studied.  The total mass of the envelope is
taken as 1~M$_{\odot}$ initially in all cases, but different
assumptions about the sound speed $c_S$ and initial core angular
momentum $\Omega$ result in different density structures as a function
of time. The three models have $\Omega$=10$^{-14}$, 10$^{-14}$, 
10$^{-13}$ s$^{-1}$ and $c_S$=0.26, 0.19, 0.26 km~s$^{-1}$, 
respectively, covering the range of parameters expected for
low-mass YSOs. The luminosity of the source changes with time from
$<$1 to $\sim$5--10 L$_\odot$. The dust temperature is computed at
each time step using a full 2D radiative transfer model and the gas
temperature is taken equal to the dust temperature.

Given the model physical structures, the CO excitation is then
computed as a function of time (evolution) through full 2D non-LTE
excitation plus radiative transfer calculations.  The line fluxes are
computed for $i$=45$\degr$ inclination but do not depend strongly on
the value of $i$.  The resulting line intensities are convolved to a
20$''$ beam and CO rotational temperatures are computed using the 2--1
up to 10--9 lines. Details are provided in \citet{Harsono13}. The 
resulting model excitation temperatures are plotted against
model $L_{\rm bol}$, $T_{\rm bol}$, and $M_{\rm env}$ values as a
function of time, with envelope mass decreasing with time. Note 
that because these models were run for a single $M_{\rm env}$=1~M$_{\odot}$, 
they do not cover the full range of observed $M_{\rm env}$.

The first conclusion from this comparison is that the model rotational
temperatures hardly show any evolution with time consistent with the
observations, in spite of the envelope mass changing by two orders of
magnitude. Although the density decreases to values below the critical
densities, the temperature increases throughout the envelope so the
rotational temperatures stay constant.

The second conclusion is that the model rotational temperatures are
generally below the observed temperatures, especially for
  $^{12}$CO ($\sim$55~K) and $^{13}$CO ($\sim$36~K). For $^{12}$CO
this could be due to the fact that outflow emission is not included in
the models, which accounts typically for more than half of the line
intensity and has a higher rotational temperature (see
Sect.~\ref{4:sec:velresdiag}). Indeed, the rotational
temperature of the narrow component is around 60--70 K, closer to
the model values.
For $^{13}$CO, the model $T_{\rm rot}$ values could be lower
than the observed values because UV photon heating contributes along
the outflow walls \citep[see][for quantitative
discussion]{Visser12,Yildiz12}, although this effect may be small
within $\sim$1000~AU radius of the source position itself \citep{Yildiz12}. 
The model C$^{18}$O rotational temperatures are close to the observed values
(typically 28--35~K), illustrating that the envelope models
are an accurate representation of the observations.

\subsection{High-$J$ CO vs. water}
\label{4:sec:cowater}

\begin{figure}[!t]
    \centering
\includegraphics[scale=0.45]{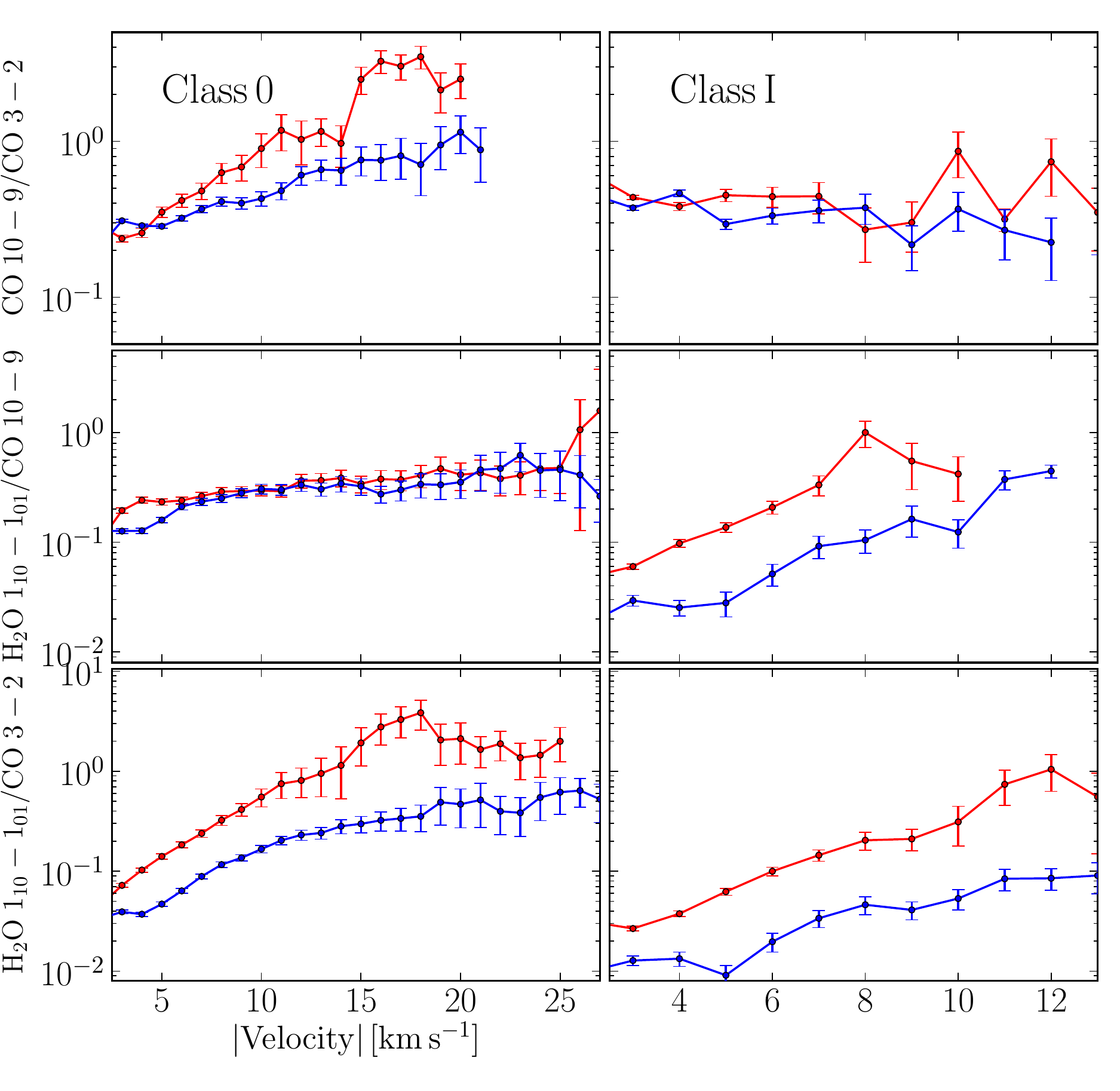}
\caption{\small Blue and red line wing ratios of CO~10--9/3--2 ({\it
    top panel}), H$_2$O~1$_{10}$-1$_{01}$/CO~10--9 ({\it middle
    panel}), and H$_2$O~1$_{10}$-1$_{01}$/CO~3--2 ({\it bottom panel})
  for the composite Class~0 and Class~I source spectra. The H$_2$O/CO
  10--9 and H$_2$O/CO 3--2 ratios are a factor of two lower than shown in
  Fig.~\ref{fig:composite} due to the beam size difference of
  20$\arcsec$ (CO~10--9 lines) to 40$\arcsec$
  (H$_2$O~1$_{10}$-1$_{01}$ lines).}
    \label{fig:compositeratios}
\end{figure}

Do the average spectra of the Class~0 and Class~I sources show any
evolution and how does this compare with water?  In
Fig.~\ref{fig:composite}, stacked and averaged \mbox{$^{12}$CO~3--2},
\mbox{10--9}, and \mbox{H$_2$O~1$_{10}$-1$_{01}$} spectra for the Class 0 and
I sources have been  presented.
Consistent with the discussion in Sect.~\ref{4:sec:12COlines} and
\citet{SanJoseGarcia13}, Class 0 sources have broader line
widths than Class~I sources, showing the importance of protostellar
outflows in Class~0 sources.  In general, Class~I sources show weaker
overall emission except for the bright sources GSS30~IRS1 and
Elias~29, consistent with the trend in Fig.~\ref{fig:WLbolTbolMenv}.
Figure~\ref{fig:compositeratios} shows the
H$_2$O~1$_{10}$-1$_{01}$/CO~10--9, H$_2$O~1$_{10}$-1$_{01}$/CO~3--2,
and CO~10--9/CO~3--2 line ratios of the averaged spectra.  
The \mbox{CO~10--9/CO~3--2} line wings show increasing ratios from
$\sim$0.2 to $\sim$1.0--3.0 for the averaged Class~0 spectrum, but this trend is
weaker  for the Class~I sources which have a near constant ratio of 0.3.  
The implied kinetic temperature has been discussed in Sect. 4.5.

\citet{Franklin08} examined the H$_2$O abundance as a function of
velocity by using CO~1--0 as a reference frame. Here it is
investigated how H$_2$O/CO line ratios change with increasing $J$ by
using CO~3--2 and 10--9 as reference frames
(Fig. \ref{fig:compositeratios}).  In this figure, the CO
3--2 line has been convolved to a 40$\arcsec$ beam using the JCMT data
\citep[also done by][]{Kristensen12}.
For the 10--9 line, no map is available so the emission is taken to
scale linearly with the beam size, as appropriate for outflow line
wings assuming a 1D structure \citep{Tafalla10}.  
Thus, the 10--9 intensities are a factor of two lower than those shown
in, for example, Fig.~\ref{fig:composite} where data
in a 20$\arcsec$ beam were used.
The line wing ratios are computed up to
the velocity where the CO emission reaches down to $\sim$2$\sigma$
noise limit, even though the H$_2$O line wings extend further.

Consistent with \citet{Kristensen12}, an increasing trend of
H$_2$O/CO~3--2 line ratios with velocity is found for both Class~0 and
Class~I sources. 
However, the H$_2$O/CO~10--9 ratios show little variation with velocity 
for Class~0 sources, and the ratio is constant within the error bars.
For the Class I sources an increasing trend in H$_2$O/CO~10--9 line 
ratios is still seen.

Because of the similarity of the CO~10--9 line wings with those of
water, it is likely that they are tracing the same warm gas. This is
in contrast with the 3--2 line, which probes the colder entrained gas.
The conclusion that H$_2$O and high-$J$~CO emission go together (but
not low-$J$ CO) is consistent with recent analyses
\citep{Santangelo12,Vasta12,Tafalla13} of WISH data at outflow
positions offset from the source.  The CO 10--9 line seems to be the
lowest $J$ transition whose line wings probe the warm shocked gas
rather than the colder entrained outflow gas (see also
Sect.~\ref{4:sec:twocomponents}); HIFI observations of higher-$J$ lines up
to \mbox{$J$=16--15} should show an even closer correspondence between
H$_2$O and CO (Kristensen et al., in prep.). That paper will also
present H$_2$O/CO abundance ratios since deriving those from the data
requires further modeling because the H$_2$O lines are subthermally
excited and optically thick.

\begin{table}
\caption{Summary of C$^{18}$O abundance profiles.}
\scriptsize
\begin{center}
\begin{tabular}{l l l l l l l }
\hline \hline
Source & Profile      & $X_{\rm in}$ & $T_{\rm ev}$ & $X_{\rm D}$ & $n_{\rm de}$ & $X_{0}$ \\
       &  &            & [K]          &                     & [cm$^{-3}$]   &              \\
\hline
IRAS~2A\tablefootmark{a} & Drop &  1.5$\times$10$^{-7}$ & 25 & $\sim$ 4$\times$10$^{-8}$ & 7$\times$10$^4$ & 5$\times$10$^{-7}$ \\
IRAS~4A\tablefootmark{b} & Drop & $\sim$1$\times$10$^{-7}$ & 25 & 5.5$\times$10$^{-9}$ & 7.5$\times$10$^4$ & 5$\times$10$^{-7}$ \\
IRAS~4B\tablefootmark{b} & Jump & 3$\times$10$^{-7}$ & 25 & 1$\times$ 10$^{-8}$ & \dots & \dots \\
Ser-SMM1                 & Drop &  1.5$\times$10$^{-7}$  & 25 & 1.0$\times$10$^{-8}$ & 1.4$\times$10$^5$ & 5$\times$10$^{-7}$ \\
Elias29                  & Constant & 5$\times$10$^{-7}$ & \dots &  5$\times$10$^{-7}$ & \dots & 5$\times$10$^{-7}$    \\
GSS30IRS1                & Jump &  4$\times$10$^{-7}$ & 25 & 2$\times$ 10$^{-8}$ & \dots &  \dots \\
\hline
\end{tabular}
\end{center}
\tablefoot{$X_{\rm in}$ is the abundance of inner envelope, $T_{\rm ev}$ 
is the evaporation temperature, $X_{\rm D}$ drop zone abundance, $n_{\rm de}$ 
is the desorption density, and $X_{0}$ is the abundance of the outermost 
part of the envelope. See Fig. B1 in \citet{Yildiz10} for definition.
Results from 
\tablefoottext{a} \citet{Yildiz10}
\tablefoottext{b} \citet{Yildiz12}
}
\label{tbl:abundancestable}
\end{table}

\section{CO abundance and warm inner envelope}
\label{4:sec:envelope}
\subsection{CO abundance profiles}
\label{4:sec:coabundance}

The wealth of high quality C$^{18}$O lines probing a wide range of
temperatures allows the CO abundance structure throughout the
quiescent envelope to be constrained. The procedure has been described
in detail in \citet{Yildiz10, Yildiz12}. Using the density and temperature 
structure of each envelope as derived by \citet[][their Table C.1; see 
Sect.~\ref{sect:kintemp}]{Kristensen12}, the CO abundance profile can 
be inferred by comparison with the C$^{18}$O data. The \verb1Ratran1 
\citep{Hogerheijde00} radiative-transfer modeling code is used to compute 
line intensities for a given trial abundance structure.

Six sources with clear detections of C$^{18}$O~9--8 and 10--9  have been modeled.
The outer radius of the models is important for the lower-$J$ lines
and for Class~0 sources. It is taken to be the radius where either the
density $n$ drops to $\sim$1.0$\times$10$^{4}$~cm$^{-3}$, or the
temperature drops below 8--10~K, whichever is reached first. In some
Class 0 sources (e.g., IRAS~4A, IRAS~4B), however, the density is still high
even at temperatures of $\sim$8~K; here the temperature was taken to
be constant at 8~K and the density was allowed to drop until
$\sim$10$^{4}$~cm$^{-3}$.  The turbulent velocity (Doppler-$b$
parameter) is set to 0.8~km~s$^{-1}$, which is representative of the
observed C$^{18}$O line widths for most sources \citep{Jorgensen02}
except for Elias~29 where 1.5~km~s$^{-1}$ is adopted. The model
emission is convolved with the beam in which the line has been
observed.

\begin{figure}[!t]
    \centering
    \includegraphics[scale=0.45]{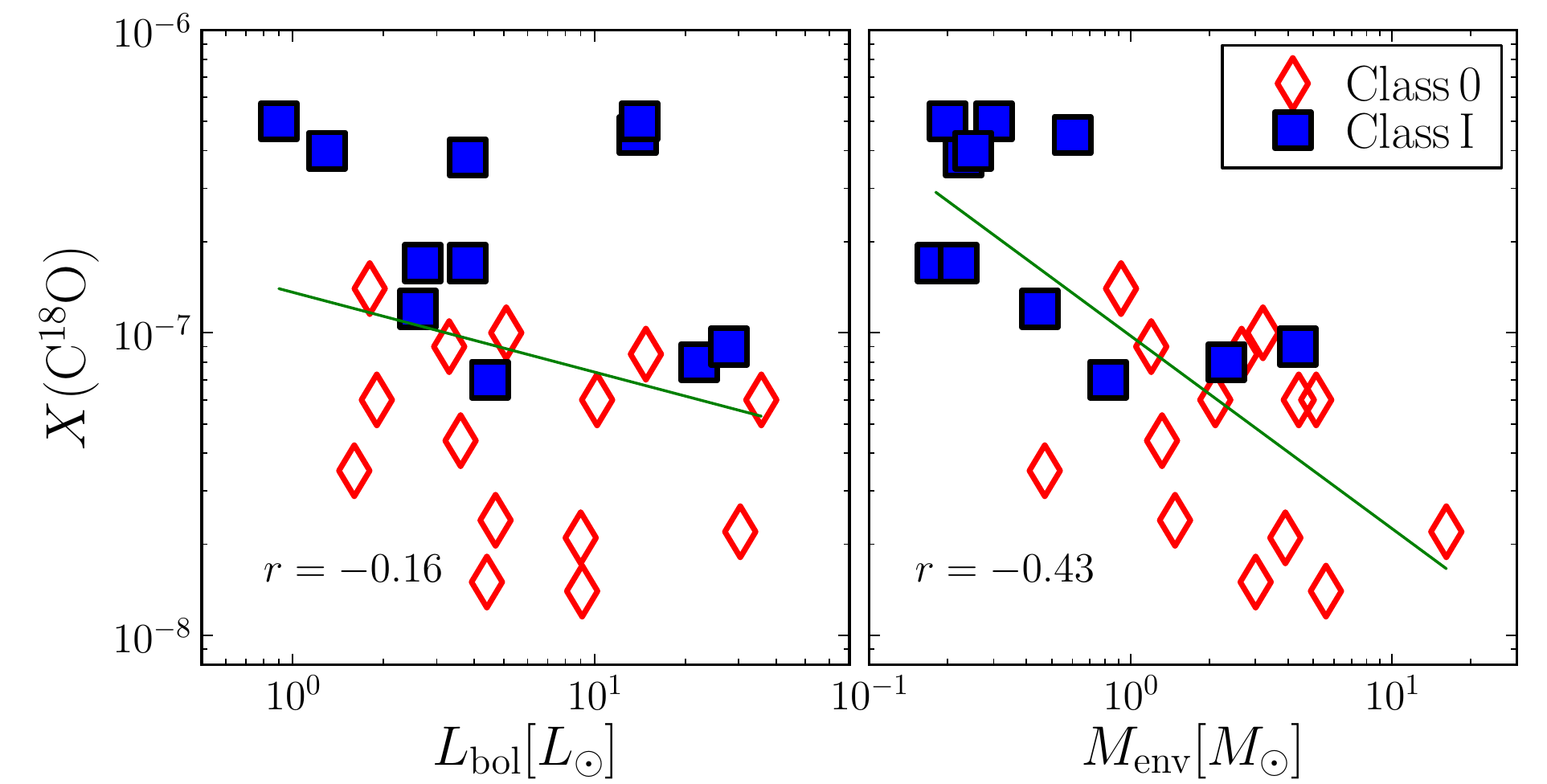}
    \caption{\small Constant abundance profiles are fitted to the lower-$J$ 
    C$^{18}$O~3--2 are shown as function of $L_{\rm bol}$ and $M_{\rm env}$. 
    Green lines are the best fit to the data points and $r$ values in each 
    of the plots are Pearson correlation coefficient.}
     \label{fig:constC18Omodels}
\end{figure}

First, constant abundance profiles are fitted to the lower-$J$
\mbox{C$^{18}$O~3--2}, together with the 2--1 transitions, if
available and they are tabulated in Table \ref{tbl:rottempAndcolden}.
In Fig.~\ref{fig:constC18Omodels}, these abundances are
plotted as function of bolometric luminosities and envelope
masses. Consistent with \citet{Jorgensen02}, Class~0 sources with
higher envelope mass have lower average abundances in their envelopes
than Class~I sources, by more than an order of magnitude. This result
is firm for \mbox{lower-$J$} transitions; however, in order to fit
higher-$J$ lines simultaneously, it is necessary to introduce a more
complex `drop' abundance profile with a freeze-out zone
\citep{Jorgensen04}.  The inner radius is determined by where the dust
temperature falls below the CO evaporation temperature of 25~K. The
outer radius is determined by where the density becomes too low for
freeze-out to occur within the lifetime of the core.

Following \citet{Yildiz10,Yildiz12} for the
NGC~1333~IRAS~2A, IRAS~4A and IRAS~4B protostars, 
such a drop-abundance profile provides a better fit to the C$^{18}$O
data than a constant or `anti-jump' abundance. In these
profiles, $X_{\rm in}$ is defined as the abundance of the inner
envelope down to the evaporation temperature of CO, $T_{\rm ev}$.
The outer abundance $X_0$ is set to 5$\times$10$^{-7}$ below at a
certain desorption density, $n_{\rm de}$, corresponding to the maximum
expected CO abundance of 2.7$\times$10$^{-4}$. The drop
abundance zone is defined as the freeze-out region in the envelope
between the limit of $T_{\rm ev}$ and $n_{\rm de}$ (see Fig.~B.1 in
\citealt{Yildiz10}.) Best fit abundances for different sources are
summarized in Table \ref{tbl:abundancestable}.  As in our previous
work and in \citet{Fuente12} and \citet{Albi10}, the CO abundance in
the inner envelope is below the canonical value of 2.7$\times$10$^{-4}$
\citep{Lacy94} by a factor of a few for the Class 0 sources, probably
due to processing of CO to other species on the grains during the cold
phase.

Only two of the Class~I sources (GSS30~IRS1 and Elias~29) have been
observed in deep integrations of C$^{18}$O~10--9 and therefore they
are the only Class~I sources modeled in detail. These sources are
located in the Ophiuchus molecular cloud, where two low-density
foreground sheets contribute to the lowest $J$~1--0 and 2--1 lines
\citep[e.g.,][]{Loren89,vanKempen09Oph}. To take this into account, 
a single slab foreground cloud is added in front of the protostars with 
15~K temperature, 1.5$\times$10$^{4}$~cm$^{-3}$ H$_2$ density, and 
10$^{16}$~cm$^{-2}$ CO column density.  Best-fit models for the 
C$^{18}$O~3--2, 5--4, 9--8and 10--9 lines toward GSS30~IRS1 and Elias~29 
can then be well fit with a constant CO abundance close to the canonical 
value and require at most only a small freeze-out zone (for the case of 
GSS30~IRS1). \citet{Jorgensen05freeze} argue that the size of the freeze-out 
zone evolves during protostellar evolution, i.e., for Class~0 sources the
drop-zone should be much larger than for Class~I sources. This is
indeed consistent with the results found here. Interestingly,
however, the constant or inner abundances $X_{\rm in}$ are high,
consistent with the maximum CO gas phase abundance. If these Class I
sources went through a previous Class 0 phase with a more massive
and colder envelope, apparently less CO ice has been converted to
other species during this phase than found for the NGC~1333
sources. One possible explanation is that the dust in this part of
Ophiuchus is on average warmer due to the UV radiation from the
nearby massive stars. Higher dust temperatures decrease the
efficiency of CO hydrogenation because the hydrogen atom has only a
short residence time on the grain \citep{Fuchs09}. Alternatively, 
the low CO abundance component may have been incorporated to the disk 
and/or dispersed. Finally, there may be a PDR contribution to the 
observed intensities \citep{Liseau2012,Bjerkeli12}.

\begin{figure}[!t]
    \centering
    \includegraphics[scale=0.5]{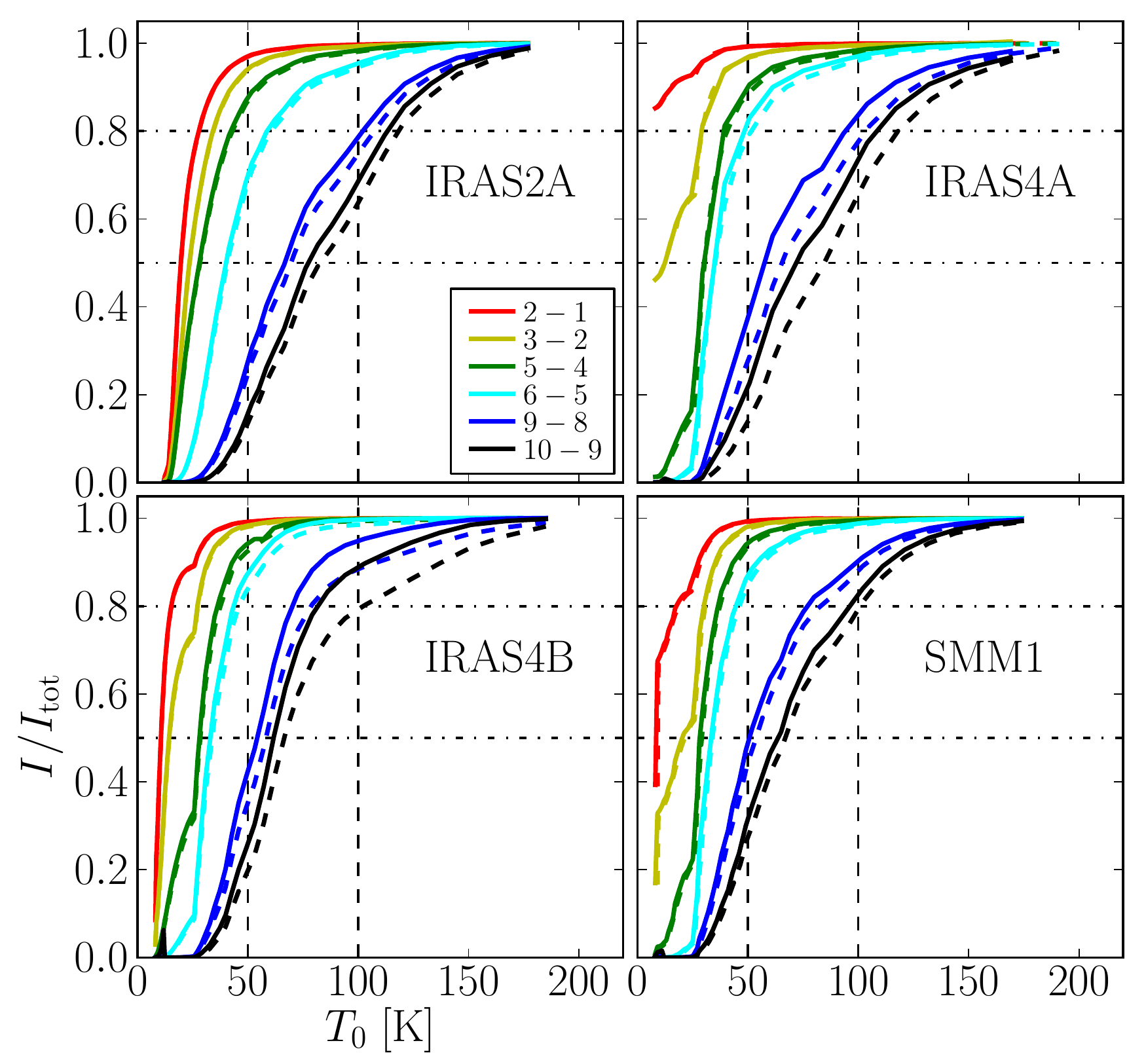}
    \caption{\small Cumulative intensity $I/I_{\rm tot}$ for various
      C$^{18}$O lines as function of envelope radius (as indicated by
      the temperature $T_o$) with the dust opacity included (solid
      curves), and dust opacity off (dashed curves). The dash-dotted
      lines indicates the fractions of 50 and 80$\%$, respectively. }
     \label{fig:dust-nodustcumulative}
\end{figure}

\subsection{Warm inner envelope}

The higher-$J$~C$^{18}$O transitions such as 9--8 and 10--9 
($E_{\rm u}$ up to $\sim$300~K) in principle trace directly the warmer gas in
the inner envelope. Can we now use these data to put limits on the
amount of warm $>$100 K gas, which could then be used as a reference
for determining abundances of complex molecules?

In principle, one could argue that simply summing the observed column density
of molecules in each level should provide the total warm column
density, as done by \citet{Plume12} for the case of Orion and in
Table~\ref{tbl:rottempAndcolden} based on the rotational
diagrams. However, in the low-mass sources considered here freeze-out
also plays a role and CO can be converted to other species on the
grains. Thus, to convert to $N$(H$_2$) in the warm gas, one needs to
use the values of $X_{\rm in}$ that have been derived in Table 7.
However, these values are derived in the context of a physical model
of the source, so in principle one simply recovers the input model.

Besides the complication of the changing CO abundance with radius,
there are two other effects that make a direct observational
determination of the warm H$_2$ column densities far from simple. The
first issue is the fact that not all emission in the 9--8 or 10--9
lines arises from gas at $>$100~K even though $E_{\rm u}$=237--290 K.
Figure \ref{fig:dust-nodustcumulative} shows the cumulative C$^{18}$O
line intensities (solid curves) as functions of radius (or,
equivalently, temperature) for four source models. $I_{\rm tot}$ are 
the intensities measured from the best-fit abundance models as
tabulated in Table \ref{tbl:abundancestable}. Thus, the curves
represent the fraction of line intensities which have their origins in
gas at temperatures below $T_0$ for the different transitions.
As expected, about 90--95$\%$ of the C$^{18}$O emission in the
lower-$J$ transitions up to $J_{\rm u}\leq$3 comes from gas at
$<$40~K.  However, even for the 9--8 transition, 30--50$\%$ comes from
gas at less than 50~K whereas only $\sim$10--20$\%$ originates at
temperatures above 100~K. For the 10--9 transition, $\sim$20--40$\%$
of the emission comes from $>$100~K. Thus, these are additional
correction factors that would have to be applied to obtain the columns
of gas $>$100 K.

\begin{figure}[!t]
    \centering
  \includegraphics[scale=0.5]{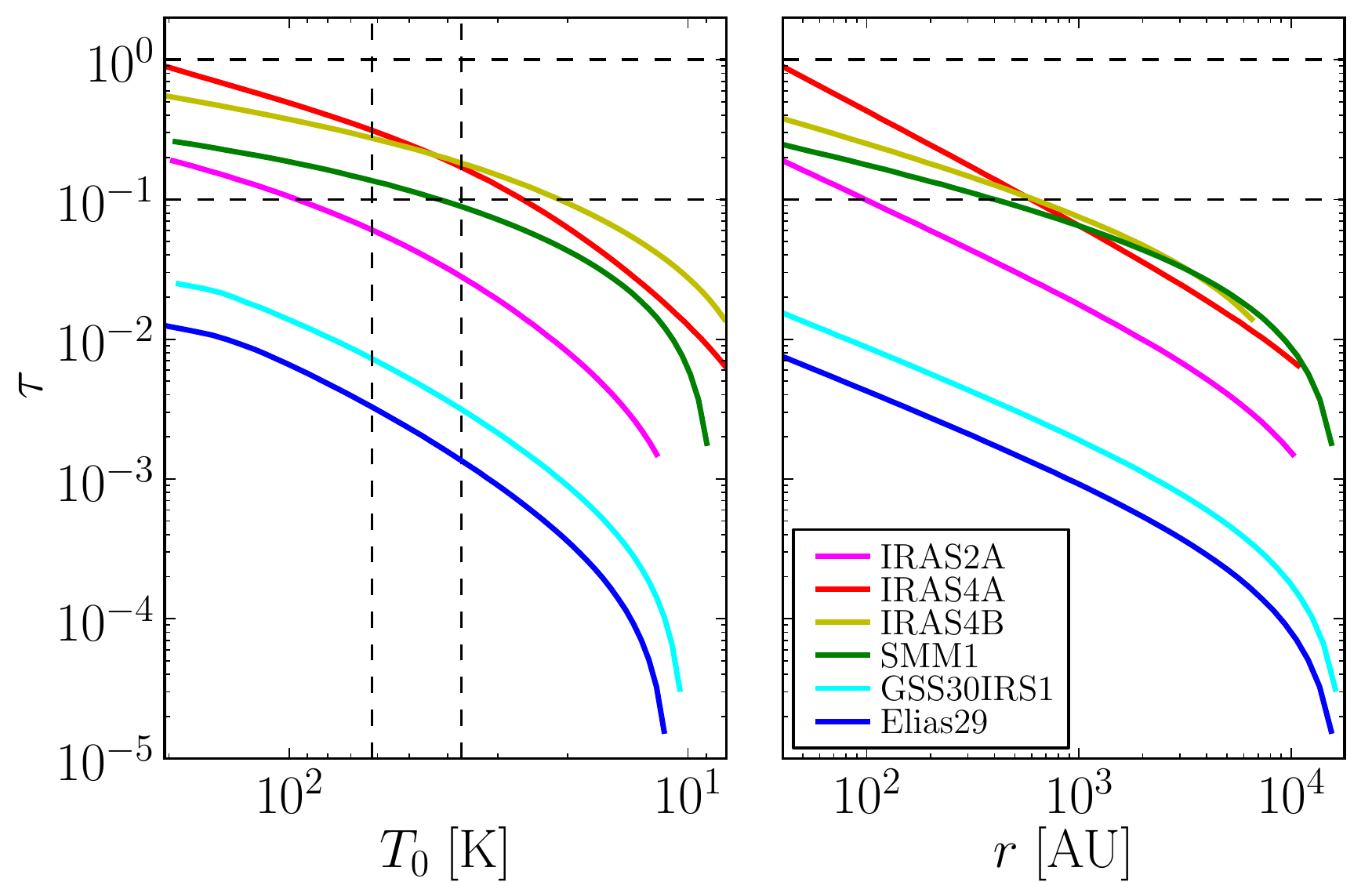}
\caption{\small Dust optical depth at 1 THz as function of envelope radius
  and temperature. Horizontal dashed lines indicate the $\tau$=1 and 
  $\tau$=0.1 values whereas vertical dash-dotted lines are radii, where the 
  temperatures are between 40~K and 60~K.}
    \label{fig:opacities}
\end{figure}

A second potential issue is that the dust continuum at 1~THz may
become optically thick so that warm C$^{18}$O emission cannot escape.
Figure \ref{fig:opacities} shows the dust opacity as function of
radius throughout the envelope, obtained by multiplying the column
densities with the $\kappa_{\rm dust}$(1~THz) \citep[][their Table 1,
column 5]{OsseHen94} and integrating from the outer edge to each
radius to find the optical depth ($\tau$).  It is seen that
the dust is optically thin ($\tau<$1 at 1 THz) throughout the envelopes of
our sources except for IRAS~4A, where it reaches unity in the
innermost part of the envelope. To what extent is the fraction of
high-$J$ emission coming from $>$100 K affected by the dust?  In
Fig.~\ref{fig:dust-nodustcumulative} (dashed curves), the dust
emission has been turned off in the \verb1Ratran1 models. For the
lower-$J$, lower-frequency transitions, almost no difference is found;
however, for the higher-$J$ transitions the cumulative intensities are
somewhat higher (up to a factor of 2) when no dust is present.

In summary, there are various arguments why warm H$_2$ column
densities cannot be inferred directly from the high-$J$ C$^{18}$O data.
Moreover, all of these arguments use a simple spherically symmetric
physical model of the source to quantify the effects. It is known that
such models fail on the smaller scales (less than a few hundred AU)
due to the presence of a (pseudo)disk and outflow cavities
\citep[e.g.,][]{Jorgensen05}. Spatially resolved data are needed to
pin down the structure of the inner envelopes and properly interpret
the origin of the high-$J$ C$^{18}$O emission.

\section{Conclusions}
\label{4:sec:conclusions}
We have presented the first large-scale survey of spectrally resolved
low- to high-$J$~CO and isotopolog lines ($2 \leq J_{\rm up} \leq 10$)
in 26 low-mass young stellar objects by using data from {\it
  Herschel}-HIFI, APEX, and JCMT telescopes. Velocity resolved data are
key to obtain the complete picture of the protostellar envelope and
the interaction of the protostar with the environment and follow the
evolution from the Class 0 to Class I phase.

\begin{itemize}

\item The $^{12}$CO line profiles can be decomposed into narrow and
  broad components, with the relative fractions varying from zero to
  nearly 100\%, with a median of 42\% of the $J$=10--9 emission in the
  narrow component. The average Class 0 profile shows a broader, more
  prominent wing than the average Class I profile.

\item The 10--9 together with 3--2 intensities correlate strongly with
  total luminosity $M_{\rm env}$ and are inversely proportional with
  bolometric temperature $T_{\rm bol}$, illustrating the dissipation
  of the envelope with evolution.

\item Rotation diagrams are constructed for each source in
  order to derive rotational temperatures and column
  densities. Median temperatures of $T_{\rm rot}$ are 70~K, 48~K, and
  37~K, for the $^{12}$CO, $^{13}$CO and C$^{18}$O transitions,
  respectively, for integrated intensities over the entire line ptofiles. 
  The excitation temperatures and SLEDs are very similar for Class~0
  and Class~I sources in all three isotopologs and do not show any 
  trend with $M_{\rm env}$, $L_{\rm bol}$ or $T_{\rm bol}$, in
  contrast with the continuum SEDs.  

\item The $^{12}$CO~10--9/CO~3--2 intensity ratio as well as
  the overall rotational temperature generally increase with
  velocity. The central narrow component due to the quiescent envelope
  has a lower temperature than the broad outflow component.

\item Models of the CO emission from collapsing envelopes reproduce
  the lack of evolution found in the observed excitation
  temperatures. They agree quantitatively for C$^{18}$O but
  underproduce the $^{12}$CO and $^{13}$CO excitation temperatures,
  pointing to the combined effects of outflows and photon-heating in
  boosting these temperatures.
 
\item Comparison of the $^{12}$CO profiles with those of H$_2$O shows
  that the H$_2$O~1$_{10}$-1$_{01}$/CO~10--9 intensity ratio is nearly
  constant with velocity for Class~0 sources, contrary to the case for
  low-$J$ CO lines. Combined with other findings, this suggests that
  the CO~10--9 line has contributions from the warmer ($\sim$300~K)
  shocked gas seen in PACS data containing also water, rather
  than the colder ($\sim$100~K) entrained outflow gas traced by
  low-$J$~CO lines.

\item The C$^{18}$O line intensities give average CO
  abundances that increase by more than an order of magnitude with
  decreasing envelope mass from the Class 0 to the Class I phase,
  consistent with earlier findings. Modeling of the full set of
  higher-$J$~C$^{18}$O lines within a given physical model shows
  further evidence for a freeze-out zone (`drop' abundance profile) in
  the envelopes of Class~0 sources with an inner CO abundance that is
  a factor of a few lower than the canonical CO abundance probably due
  to processing of CO on grains into other molecules. For two Class I
  sources in Ophiuchus, no freeze-out zone is needed and the data are
  consistent with a constant high abundance value at a level that is
  close to the maximum gas-phase CO abundance. The lack of processing
  for these sources may de due to the higher dust temperatures in
  Ophiuchus, evolutionary effects, or a PDR contribution.

\item The warm ($>$100 K) H$_2$ column densities cannot be derived
  directly from C$^{18}$O~9--8 or 10--9 lines because of contributions
  to the emission from colder gas in the envelope, dust extinction at
  high frequency, and more generally a lack of knowledge of the source
  structure on a few hundred AU scales.
\end{itemize}

Overall, our data show that the evolution from the Class 0 to the
Class I phase is traced in the decrease of the line intensities
reflecting envelope dissipation, a less prominent broad wing
indicating a decrease in the outflow power, and an increase in the
average CO abundance, reflecting a smaller freeze-out zone. The CO
excitation temperature from $J_{\rm u}$=2 to 10 shows little evolution
between these two classes, however.

The next step in the study of CO in low mass protostars is clearly to
obtain higher spectral and spatial resolution data with instruments
like ALMA that recover the full range of spatial scales from $<$100 AU
to $>$1000~AU in both low- and high$-J$ CO and isotopolog lines.


\begin{acknowledgements}

The authors are grateful to the rest of the WISH team for stimulating
discussions and to Joseph Mottram, Carolyn M$^{\rm c}$Coey and 
the anonymous referee for valuable comments on the manuscript. They 
also thank the APEX, JCMT and {\it Herschel} staff as well 
as NL and MPIfR observers for carrying out
the observations.  Astrochemistry in Leiden is supported by the
Netherlands Research School for Astronomy (NOVA), by a Spinoza grant
and grant 614.001.008 from the Netherlands Organisation for Scientific
Research (NWO), and by the European Community's Seventh Framework
Programme FP7/2007-2013 under grant agreement 238258 (LASSIE).  The
authors are grateful to many funding agencies and the HIFI-ICC staff
who has been contributing for the construction of \textit{Herschel}
and HIFI for many years. HIFI has been designed and built by a
consortium of institutes and university departments from across
Europe, Canada and the United States under the leadership of SRON
Netherlands Institute for Space Research, Groningen, The Netherlands
and with major contributions from Germany, France and the US.
Consortium members are: Canada: CSA, U.Waterloo; France: CESR, LAB,
LERMA, IRAM; Germany: KOSMA, MPIfR, MPS; Ireland, NUI Maynooth; Italy:
ASI, IFSI-INAF, Osservatorio Astrofisico di Arcetri- INAF;
Netherlands: SRON, TUD; Poland: CAMK, CBK; Spain: Observatorio
Astron{\'o}mico Nacional (IGN), Centro de Astrobiolog{\'i}a
(CSIC-INTA). Sweden: Chalmers University of Technology - MC2, RSS $\&$
GARD; Onsala Space Observatory; Swedish National Space Board,
Stockholm University - Stockholm Observatory; Switzerland: ETH Zurich,
FHNW; USA: Caltech, JPL, NHSC.Construction of CHAMP$^{+}$ is a
collaboration between the Max-Planck-Institut fur Radioastronomie
Bonn, Germany; SRON Netherlands Institute for Space Research,
Groningen, the Netherlands; the Netherlands Research School for
Astronomy (NOVA); and the Kavli Institute of Nanoscience at Delft
University of Technology, the Netherlands; with support from the
Netherlands Organization for Scientific Research (NWO) grant
600.063.310.10.

\end{acknowledgements}
\bibliographystyle{aa}
\bibliography{bibdata}

\appendix

\onecolumn
\section{CO~10--9/3--2 spectra}

\begin{figure*}[!th]
    \centering
\includegraphics[scale=0.67]{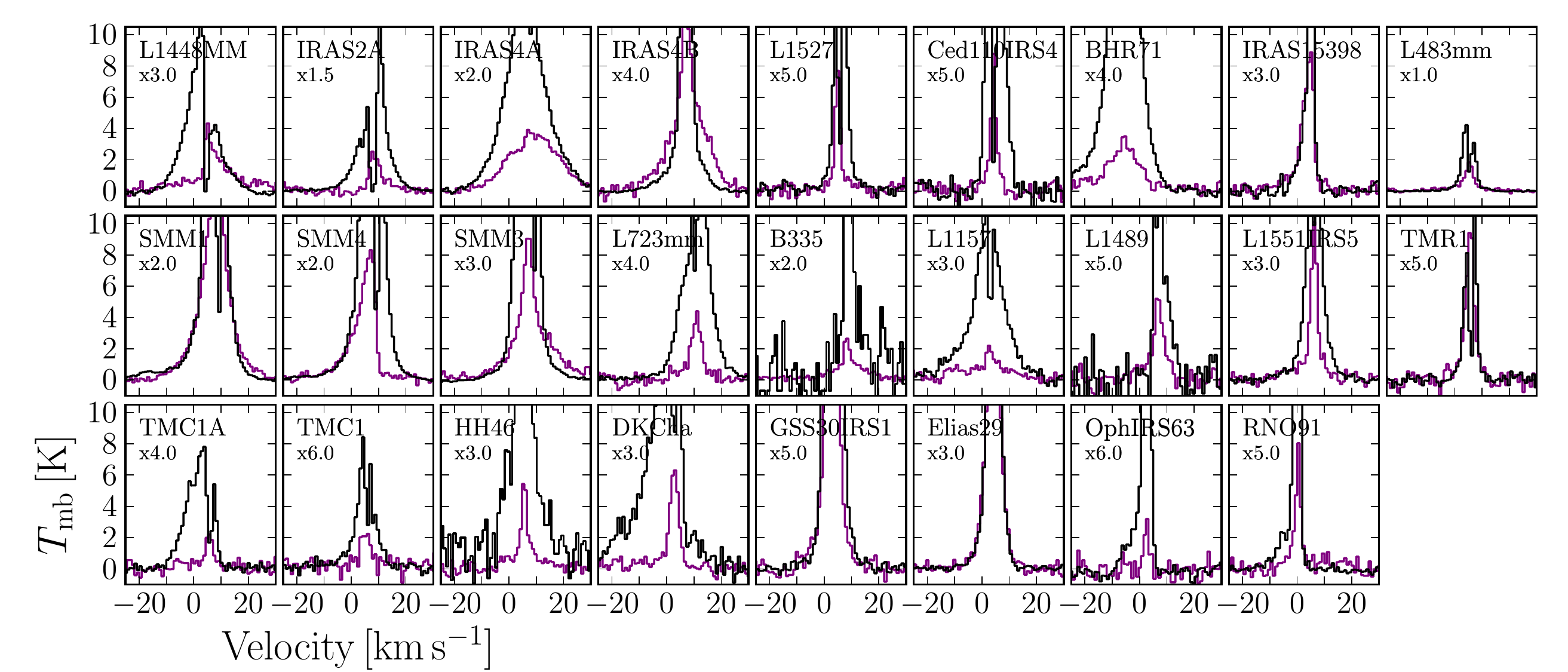}
\caption{\small Zoom-in spectra of the line wings of $^{12}$CO~3--2
  (black) and $^{12}$CO~10--9 (magenta) overplotted for each
  source. The spectra are multiplied by a constant factor indicated in
  the figure in order to show the line wings in more detail. They have
  been rebinned to 0.6--1~km~s$^{-1}$, to enhance
  signal-to-noise.}
    \label{fig:12CO_10-9--12CO_3-2_ontop}
\end{figure*}

\section{Observations Matrix}
\begin{sidewaystable*}[!th]
\caption{Observed sources and lines.}
\small
\begin{center}
\begin{tabular}{@{}l c c c c c c | c c c c c c | c c c c c c c c c }
\hline\hline
                     & \multicolumn{6}{c} {12CO} &   \multicolumn{6}{c} {13CO} &  \multicolumn{6}{c} {C18O} & \\ 
Source                         & 2--1 & 3--2 & 4--3 & 6--5 & 7--6 & 10--9 & 2--1 & 3--2 & 4--3 & 6--5 & 8--7 & 10--9 & 2--1 & 3--2 & 5--4 & 6--5 & 9--8 & 10--9 \\ \hline
L1448-MM               & JCMT & JCMT & JCMT & \dots\tablefootmark{a} & \dots\tablefootmark{a} & HIFI & JCMT & JCMT & \dots & APEX & \dots & HIFI & JCMT & JCMT & HIFI & \dots & HIFI & HIFI  \\
\object{NGC1333-IRAS2A}& JCMT & JCMT & JCMT & APEX & APEX & HIFI & \dots   & JCMT & JCMT & APEX & \dots   & HIFI & JCMT & JCMT & HIFI  & APEX & HIFI & HIFI  \\
\object{NGC1333-IRAS4A}& JCMT & JCMT & JCMT & APEX & APEX & HIFI & JCMT & JCMT & JCMT & APEX & HIFI    & HIFI & JCMT & JCMT & HIFI  & APEX & HIFI & HIFI  \\
\object{NGC1333-IRAS4B}& JCMT & JCMT & JCMT & APEX & APEX & HIFI & \dots   & JCMT & \dots   & APEX & HIFI    & HIFI & JCMT & JCMT & HIFI  & APEX & HIFI & HIFI  \\
\object{L1527}      & JCMT & JCMT & JCMT & APEX & APEX & HIFI & \dots   & JCMT & JCMT & APEX & HIFI     & HIFI & JCMT & JCMT & HIFI  & \dots   & HIFI & \dots    \\
\object{Ced110-IRS4}   & \dots   & APEX & APEX & APEX & APEX & HIFI & \dots   & \dots   & \dots   & APEX & \dots   & HIFI & \dots   & \dots   & \dots & \dots   & HIFI & \dots    \\
\object{BHR71}         & \dots   & APEX & \dots   & APEX & APEX & HIFI & \dots   & \dots   & \dots   & APEX & \dots   & HIFI & \dots   & APEX & HIFI  & \dots   & HIFI & \dots    \\
\object{IRAS15398}     & \dots   & JCMT & JCMT & APEX & APEX & HIFI & \dots   & JCMT & \dots   & APEX & \dots   & HIFI & JCMT & JCMT & HIFI & \dots   & HIFI & \dots    \\
\object{L483mm}        & JCMT & JCMT & JCMT & APEX & APEX & HIFI & JCMT & JCMT & JCMT & APEX & \dots   & HIFI & JCMT & JCMT & HIFI  & \dots   & HIFI & HIFI  \\
\object{Ser-SMM1}      & JCMT & JCMT & JCMT & APEX & APEX & HIFI & JCMT & JCMT & \dots   & APEX & HIFI    & HIFI & JCMT & JCMT & HIFI  & APEX & HIFI & HIFI  \\
\object{Ser-SMM4}      & JCMT & JCMT & JCMT & APEX & APEX & HIFI & \dots   & JCMT & \dots   & APEX & HIFI    & HIFI & \dots   & JCMT & HIFI  & \dots   & HIFI & HIFI  \\
\object{Ser-SMM3}      & JCMT & JCMT & JCMT & APEX & APEX & HIFI & JCMT & JCMT & \dots   & APEX & APEX & HIFI & JCMT & JCMT & HIFI  & \dots   & HIFI & HIFI  \\
\object{L723}          & \dots   & JCMT & \dots   & APEX & APEX & HIFI & JCMT & JCMT & JCMT & APEX & APEX & HIFI & JCMT & JCMT & HIFI  & \dots   & HIFI & \dots    \\
\object{B335}          & JCMT & JCMT & JCMT & APEX & APEX & HIFI & JCMT & JCMT & \dots   & APEX & \dots   & HIFI & JCMT & JCMT & HIFI  & \dots   & HIFI & \dots    \\
\object{L1157}         & JCMT & JCMT & JCMT & JCMT & \dots   & HIFI & JCMT & JCMT & \dots   & JCMT & \dots   & HIFI & JCMT & JCMT & HIFI  & \dots   & HIFI & \dots    \\
\hline
\object{L1489}        & JCMT & JCMT & JCMT & APEX & APEX & HIFI & \dots   & JCMT & \dots   & APEX & \dots   & HIFI & JCMT & JCMT & \dots   & \dots   & HIFI & \dots    \\
\object{L1551-IRS5}   & JCMT & JCMT & JCMT & APEX & \dots   & HIFI & JCMT & JCMT & \dots   & APEX & \dots   & HIFI & JCMT & JCMT & \dots   & \dots   & HIFI & \dots    \\
\object{TMR1}         & JCMT & JCMT & JCMT & APEX & APEX & HIFI & \dots   & JCMT & \dots   & APEX & \dots   & HIFI & JCMT & JCMT & \dots   & \dots   & HIFI & \dots    \\
\object{TMC1A}        & JCMT & JCMT & JCMT & APEX & APEX & HIFI & \dots   & JCMT & \dots   & APEX & \dots   & HIFI & JCMT & JCMT & \dots   & \dots   & HIFI & \dots    \\
\object{TMC1}         & JCMT & JCMT & JCMT & APEX & APEX & HIFI & \dots   & JCMT & \dots   & APEX & \dots   & HIFI & JCMT & JCMT & \dots   & \dots   & HIFI & \dots    \\
\object{HH46-IRS}     & APEX & APEX & APEX & APEX & APEX & HIFI & \dots   & APEX & APEX & APEX & APEX & HIFI & JCMT & APEX & \dots   & APEX & HIFI & \dots    \\
\object{DK~Cha}       & \dots   & APEX & APEX & APEX & APEX & HIFI & \dots   & \dots   & APEX & APEX & APEX & HIFI & APEX & APEX & \dots   & APEX & HIFI & \dots    \\
\object{GSS30-IRS1}   & JCMT & JCMT & \dots   & APEX & \dots   & HIFI & JCMT & JCMT & \dots   & APEX & \dots   & HIFI & JCMT & JCMT & \dots   & \dots   & HIFI & \dots    \\
\object{Elias 29}     & JCMT & JCMT & \dots   & APEX & \dots   & HIFI & JCMT & JCMT & \dots   & APEX & \dots   & HIFI & JCMT & JCMT & HIFI & \dots   & HIFI & \dots    \\
\object{Oph-IRS63}    & JCMT & JCMT & \dots   & APEX & APEX & HIFI & \dots   & JCMT & \dots   & APEX & APEX & HIFI & JCMT & JCMT & \dots   & \dots   & HIFI & \dots    \\
\object{RNO91}        & JCMT & JCMT & \dots   & APEX & APEX & HIFI & JCMT & JCMT & \dots   & APEX & APEX & HIFI & JCMT & JCMT & \dots   & \dots   & HIFI & \dots    \\
\hline
\end{tabular}
\end{center}
\tablefoot{Sources above the horizontal line are Class 0, sources below are Class I. HIFI indicates {\it Herschel}-HIFI observations.
\tablefoottext{a}{Gomez-Ruiz et al. in prep.}
}
\label{tbl:overviewobssource}
\end{sidewaystable*}

\Online

\onecolumn

\section{CO spectra of sample}

\onecolumn
\subsection{L1448MM}
\begin{figure*}[htb]
    \centering
    \includegraphics[scale=0.3]{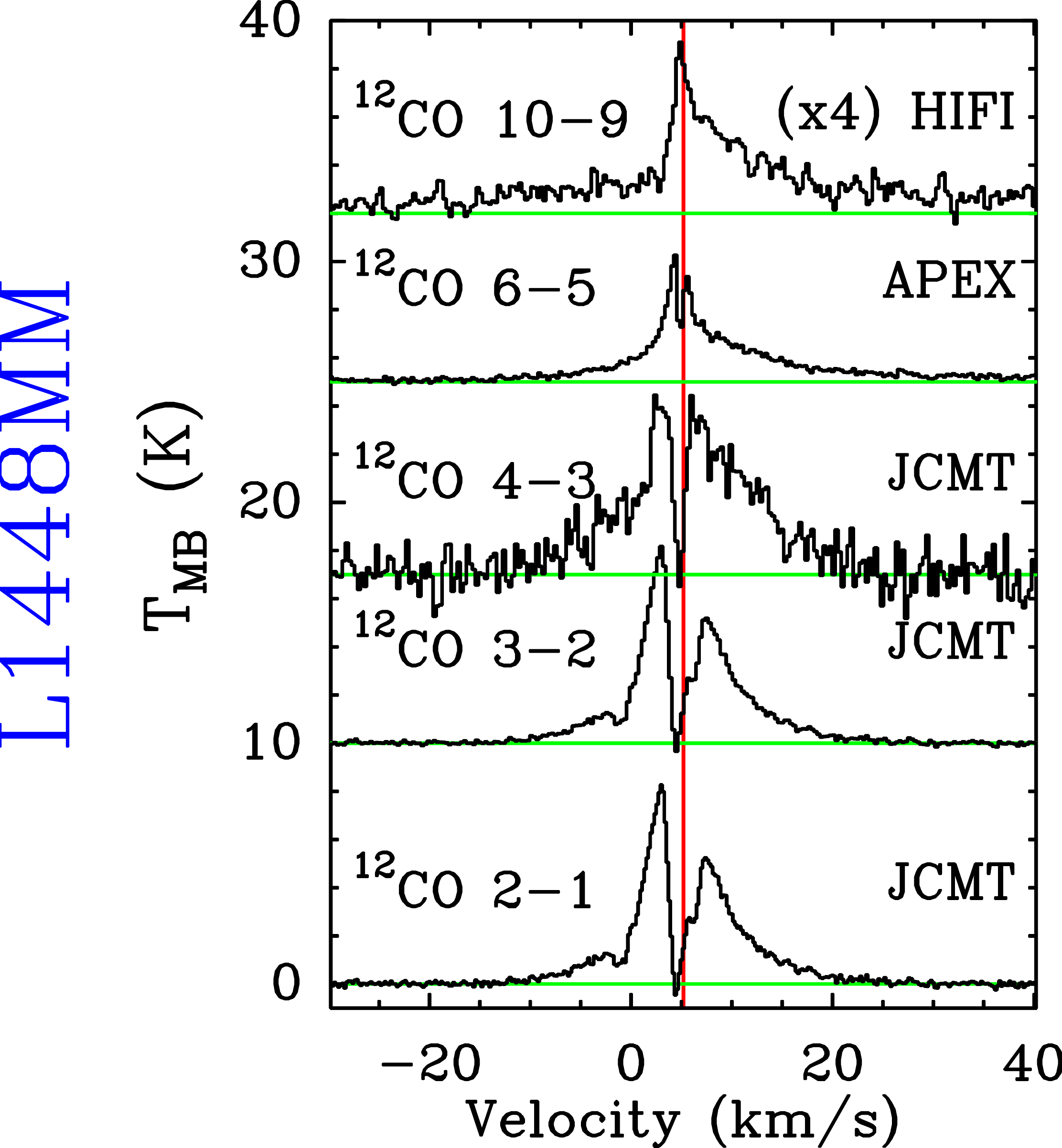}
    \includegraphics[scale=0.3]{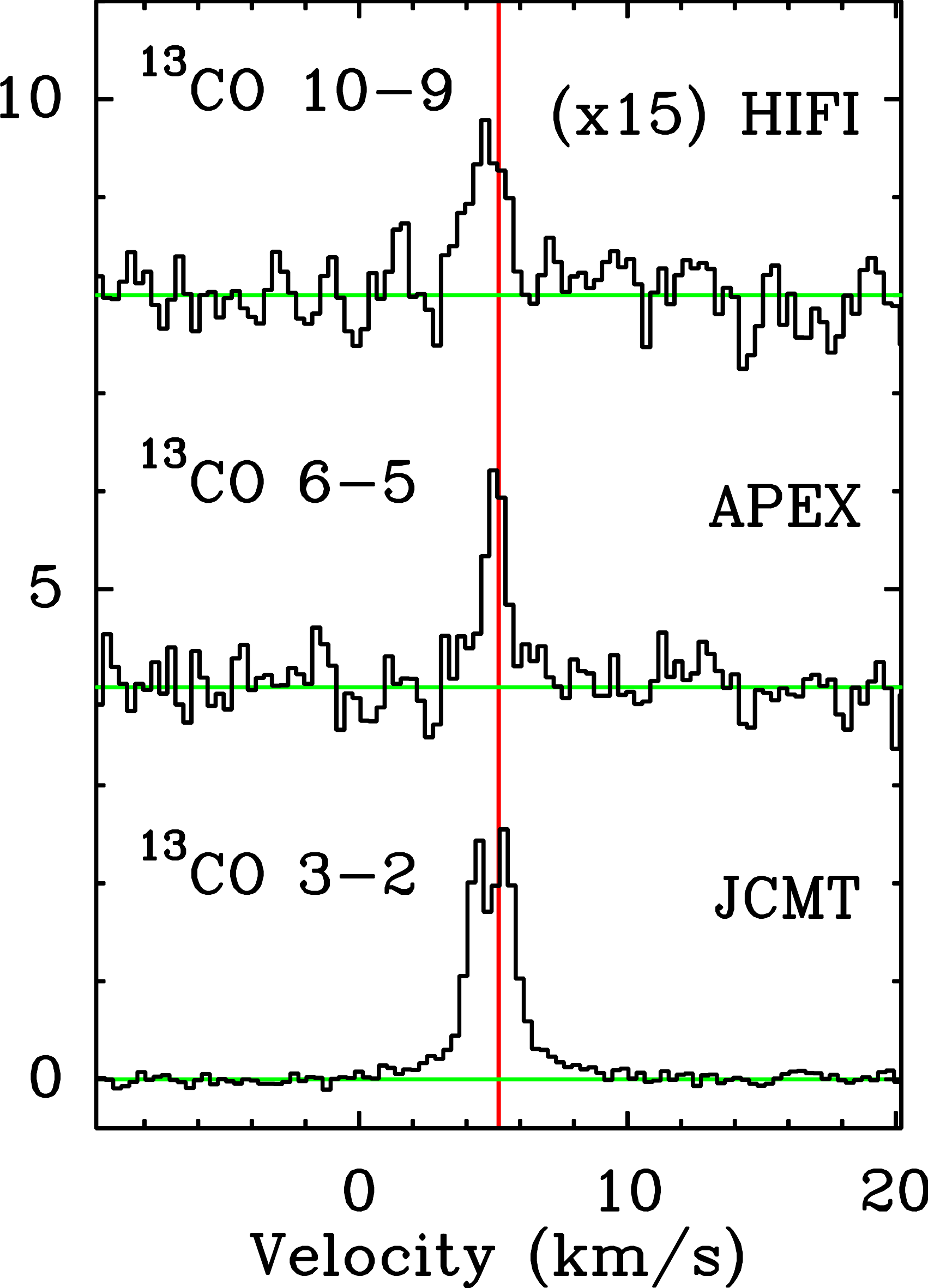}
    \includegraphics[scale=0.3]{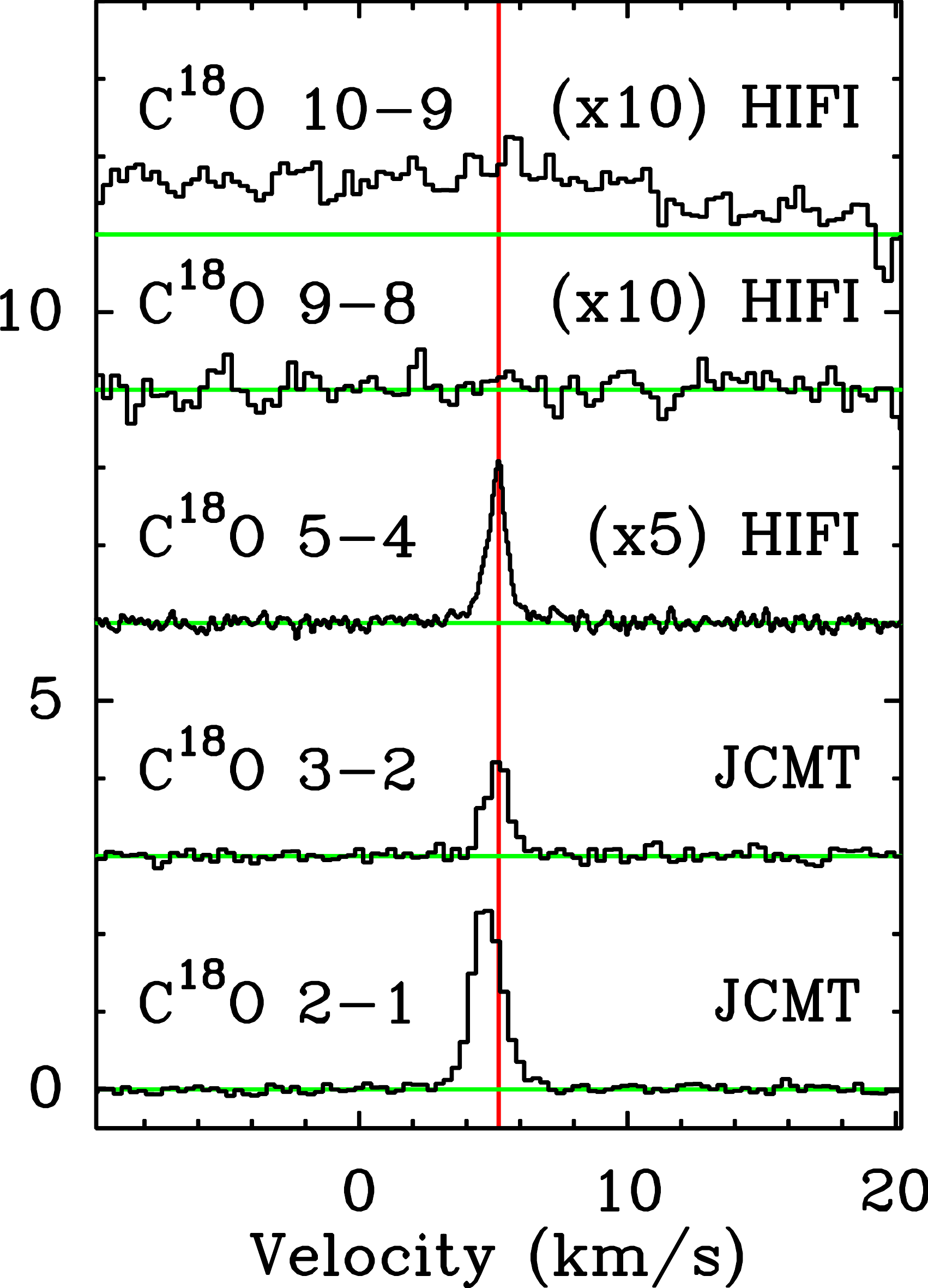}
    \caption{\small Observed $^{12}$CO, $^{13}$CO, and C$^{18}$O transitions for L1448MM}
    \label{fig:linesL1448MM}
\end{figure*}

\begin{table*}[!ht]
\caption{Observed line intensities for L1448MM in all observed transitions. }
\normalsize
\begin{center}
\begin{tabular}{l l l r r r r r r r r r}
\hline \hline
Mol.  & Transition & Telescope & Efficiency & $\int T_{\rm MB} \mathrm{d}V$ & $T_{\mathrm{peak}}$ & $rms$ \\
 &  & &   $\eta$ &[K km s$^{-1}$] & [K] &   [K]\\
\hline
CO        & 2--1 & JCMT-RxA       & 0.69 & 59.71\phantom{0} & 8.26\phantom{0}  & 0.10\phantom{0} \\
          & 3--2 & JCMT-HARPB     & 0.63 & 32.15\phantom{0} & 4.20\phantom{0}  & 0.085 \\
          & 4--3 & JCMT           & 0.38 & 87.49\phantom{0} & 7.45\phantom{0}  & 0.77\phantom{0} \\
          & 6--5\tablefoottext{a} & APEX-CHAMP$^+$ & 0.46  & 46.15\phantom{0}  & 5.67\phantom{0}  & 0.12\phantom{0} \\
          &10--9 & {\it Herschel}-HIFI\tablefootmark{b} & 0.64 & 21.53\phantom{0} & \dots \phantom{0}  & 0.15\phantom{0} \\
$^{13}$CO & 2--1 & JCMT-RxA       & 0.74 & 9.84\phantom{0}  & 6.50\phantom{0}   & 0.066 \\
          & 3--2 & JCMT-HARPB     & 0.63 & 5.74\phantom{0}  & 2.70\phantom{0}  & 0.065 \\
          & 6--5 & APEX-CHAMP$^+$ & 0.48 & 2.88\phantom{0}  & 2.69\phantom{0}  & 0.34\phantom{0} \\
          & 10--9& {\it Herschel}-HIFI\tablefootmark{b}    & 0.74 & 0.34\phantom{0} & 0.13\phantom{0} & 0.019 \\
C$^{18}$O & 2--1 & JCMT-RxA       & 0.69 & 3.53\phantom{0}  & 2.35\phantom{0}  & 0.061 \\
          & 3--2 & JCMT-HARPB     & 0.63 & 1.47\phantom{0}  & 1.48\phantom{0}  & 0.10\phantom{0} \\
          & 5--4 & {\it Herschel}-HIFI\tablefootmark{b}  & 0.76 & 0.42\phantom{0} & 0.45\phantom{0} & 0.018 \\
          & 9--8 & {\it Herschel}-HIFI\tablefootmark{c}  & 0.74 & $<$0.052 & \dots\phantom{0} & 0.023 \\
          &10--9 & {\it Herschel}-HIFI\tablefootmark{c}  & 0.74 & $<$0.055 & \dots\phantom{0} & 0.024 \\
\hline 
\end{tabular}
\end{center}
\tablefoot{
\tablefoottext{a}{Gomez-Ruiz et al. in prep.}
\tablefoottext{b}{Only H-polarization observation is used.}
\tablefoottext{c}{H- and V-polarization observations averaged.}
}
\label{tbl:linesL1448MM}
\end{table*}

\newpage
\clearpage

\onecolumn
\subsection{IRAS2A}
\begin{figure*}[htb]
    \centering
    \includegraphics[scale=0.3]{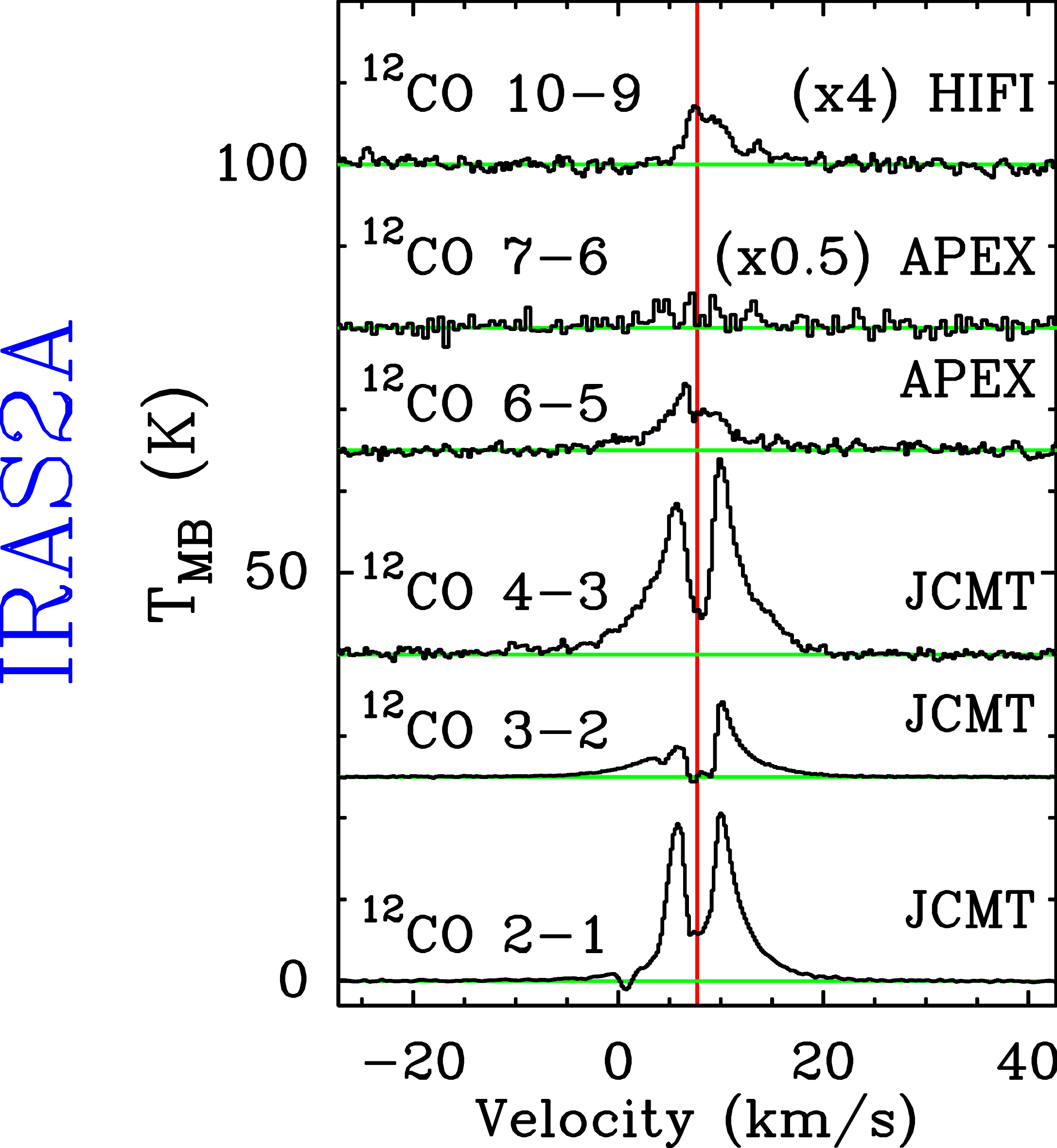}
    \includegraphics[scale=0.3]{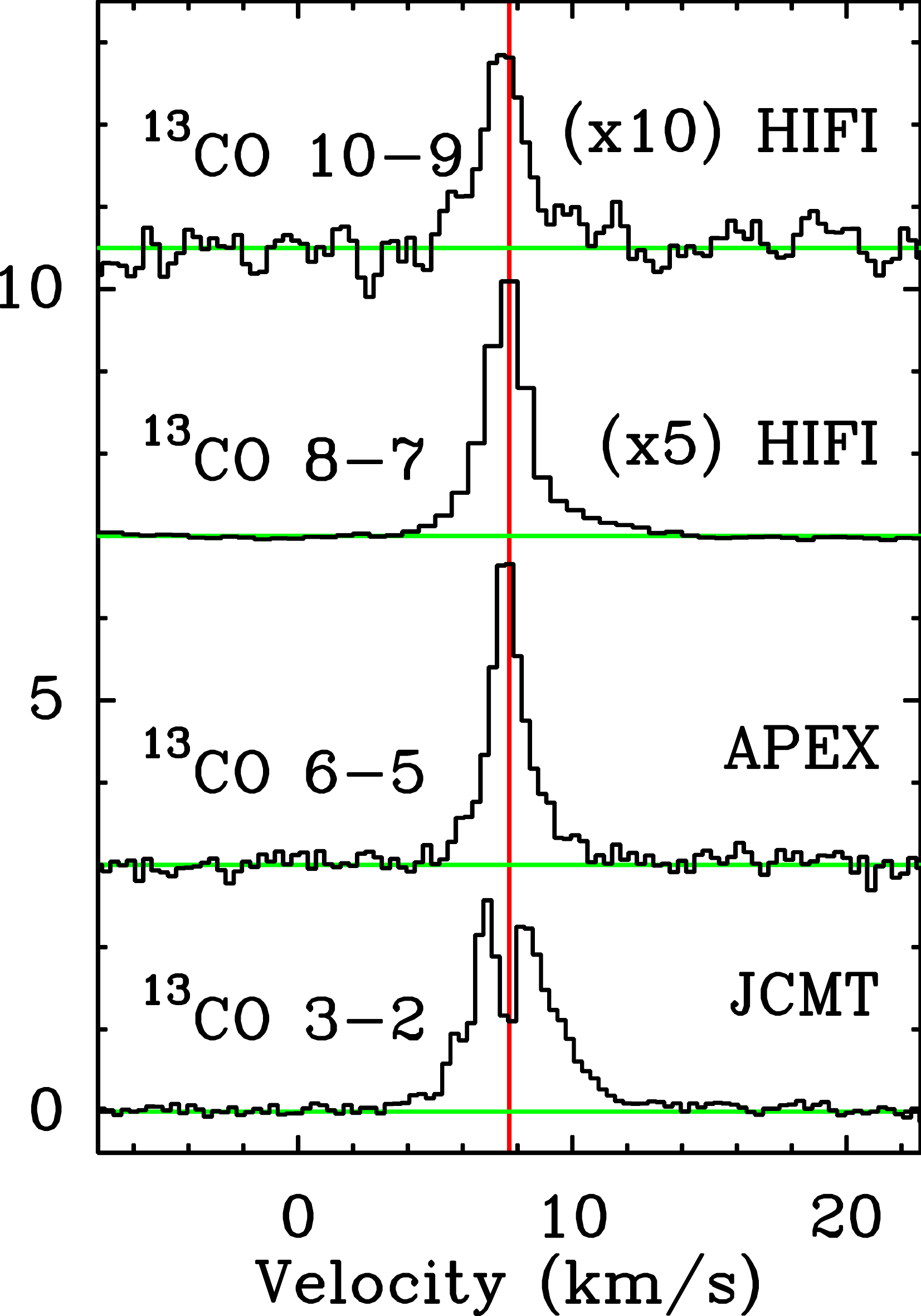}
    \includegraphics[scale=0.3]{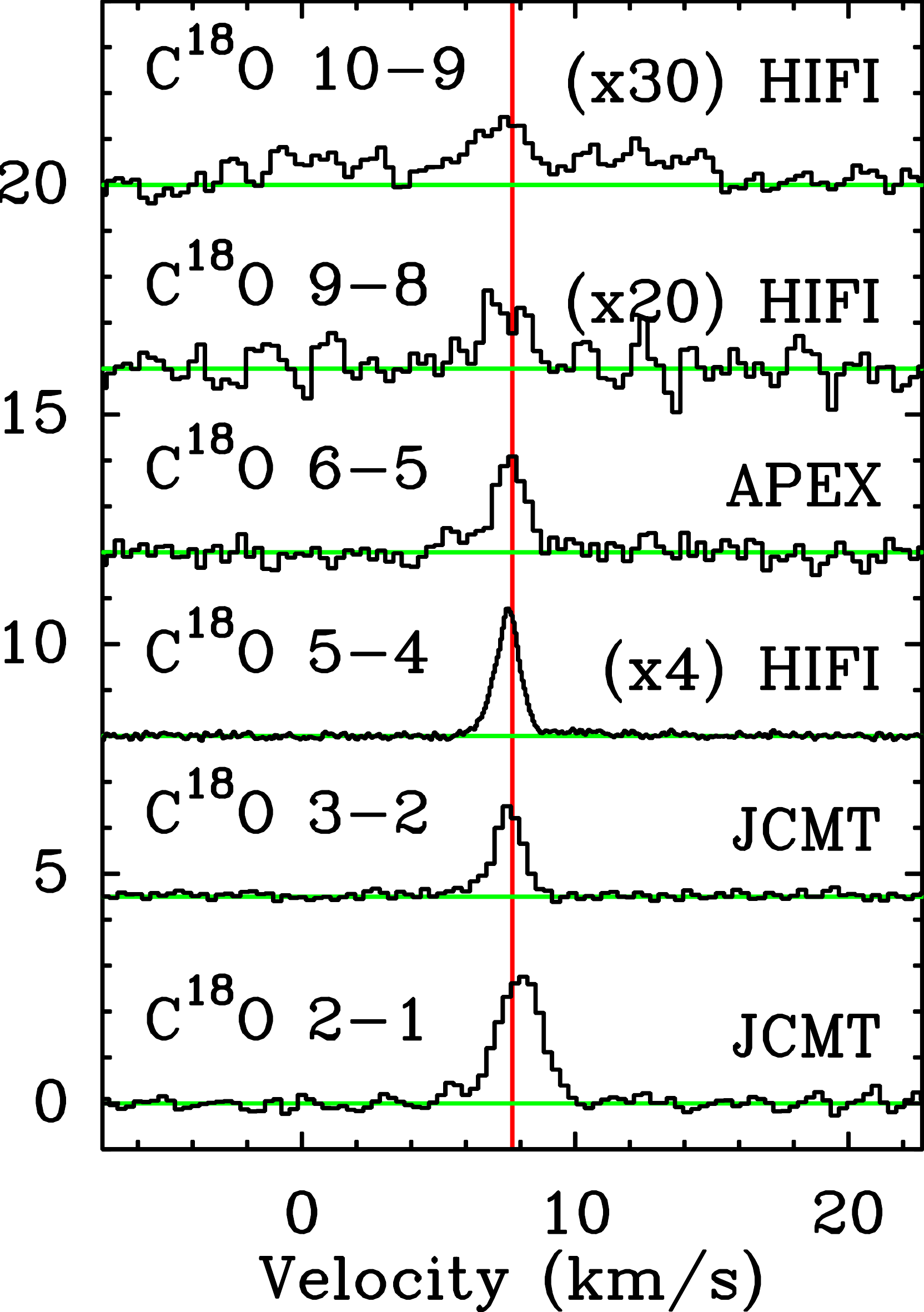}
    \caption{\small Observed $^{12}$CO, $^{13}$CO, and C$^{18}$O transitions for IRAS 2A.}
    \label{fig:linesIRAS2A}
\end{figure*}

\begin{table*}[!ht]
\caption{Observed line intensities for IRAS2A in all observed transitions.}
\normalsize
\begin{center}
\begin{tabular}{l l l r r r r r r r r r}
\hline \hline
Mol.  & Transition & Telescope & Efficiency & $\int T_{\rm MB} \mathrm{d}V$ & $T_{\mathrm{peak}}$ & $rms$ \\
 &  & &   $\eta$ &[K km s$^{-1}$] & [K] &   [K]\\
\hline
CO        & 2--1 & JCMT-RxA        & 0.69   & 129.70\phantom{0} & 20.54\phantom{0} & 0.083 \\
          & 3--2 & JCMT-HARPB      & 0.63   & 46.93\phantom{0}  & 9.38\phantom{0}  & 0.061 \\
          & 4--3 & JCMT\tablefootmark{a}    & 0.38      & 180.00\phantom{0} &  23.67\phantom{0}  & 0.55\phantom{0} \\
          & 6--5 & APEX-CHAMP$^+$  & 0.52   & 55.02\phantom{0}  & 8.89\phantom{0}  & 0.60\phantom{0} \\
          & 7--6 & APEX-CHAMP$^+$  & 0.49   & 37.39\phantom{0}  & 6.35\phantom{0}  & 1.29\phantom{0} \\
          &10--9 & {\it Herschel}-HIFI\tablefootmark{b} & 0.64 & 9.20\phantom{0}   & 1.80\phantom{0}  & 0.14\phantom{0} \\
$^{13}$CO & 2--1 & JCMT-RxA        & 0.69   & 19.25\phantom{0}  & 3.95\phantom{0}  & 0.10\phantom{0} \\
          & 3--2 & JCMT-HARPB      & 0.63   & 8.48\phantom{0}   & 2.66\phantom{0}  & 0.078 \\
          & 6--5 & APEX-CHAMP$^+$  & 0.52   & 7.08\phantom{0}   & 3.95\phantom{0}  & 0.15\phantom{0} \\
          & 8--7 & {\it Herschel}-HIFI\tablefootmark{c} & 0.75 & 1.37\phantom{0}   & 0.66\phantom{0} & 0.005 \\
          & 10--9& {\it Herschel}-HIFI\tablefootmark{b} & 0.74 & 0.61\phantom{0}   & 0.30\phantom{0} & 0.030 \\
C$^{18}$O & 2--1 & JCMT-RxA        & 0.69   & 6.00\phantom{0}   & 2.81\phantom{0}  & 0.15\phantom{0} \\
          & 3--2 & JCMT-HARPB      & 0.63   & 2.78\phantom{0}   & 2.10\phantom{0}  & 0.10\phantom{0} \\
          & 5--4 & {\it Herschel}-HIFI\tablefootmark{b} & 0.76 &  0.78\phantom{0}  &  0.67\phantom{0} & 0.013 \\
          & 6--5 & APEX-CHAMP$^+$  & 0.56   & 3.58\phantom{0}   & 2.09\phantom{0}  & 0.27\phantom{0} \\
          & 9--8 & {\it Herschel}-HIFI\tablefootmark{c} & 0.74 & 0.19\phantom{0}   & 0.090 & 0.020 \\
          &10--9 & {\it Herschel}-HIFI\tablefootmark{b} & 0.74 & 0.15\phantom{0}   & 0.046 & 0.009 \\
\hline 
\end{tabular}
\end{center}
\tablefoot{
\tablefoottext{a}{Taken in 11$\arcsec$ beam.}
\tablefoottext{b}{Only H-polarization observation is used.}
\tablefoottext{c}{H- and V-polarization observations averaged.}
}
\label{tbl:linesIRAS2A}
\end{table*}

\newpage
\clearpage

\onecolumn
\subsection{IRAS4A}
\begin{figure*}[htb]
    \centering
    \includegraphics[scale=0.3]{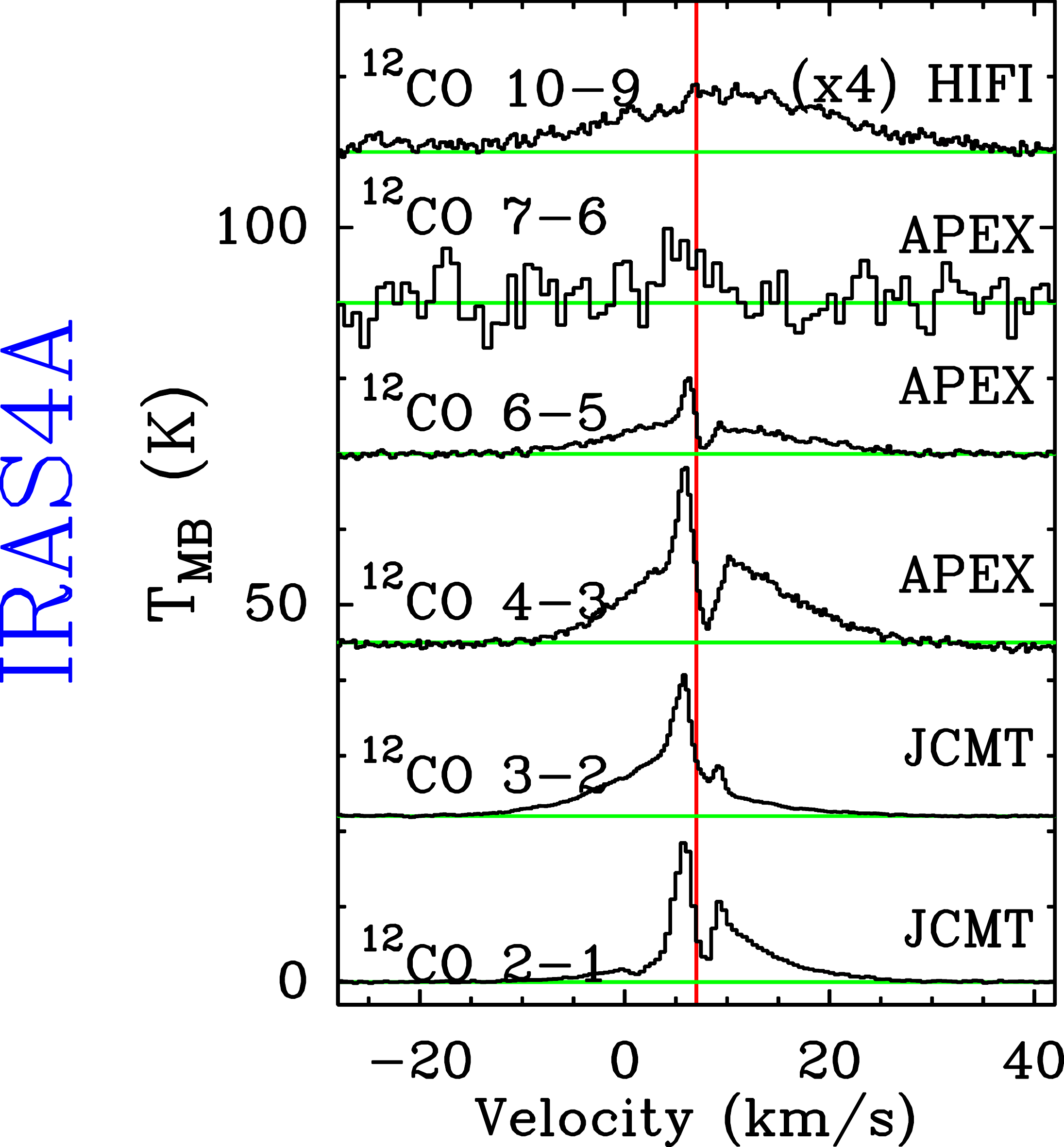}
    \includegraphics[scale=0.3]{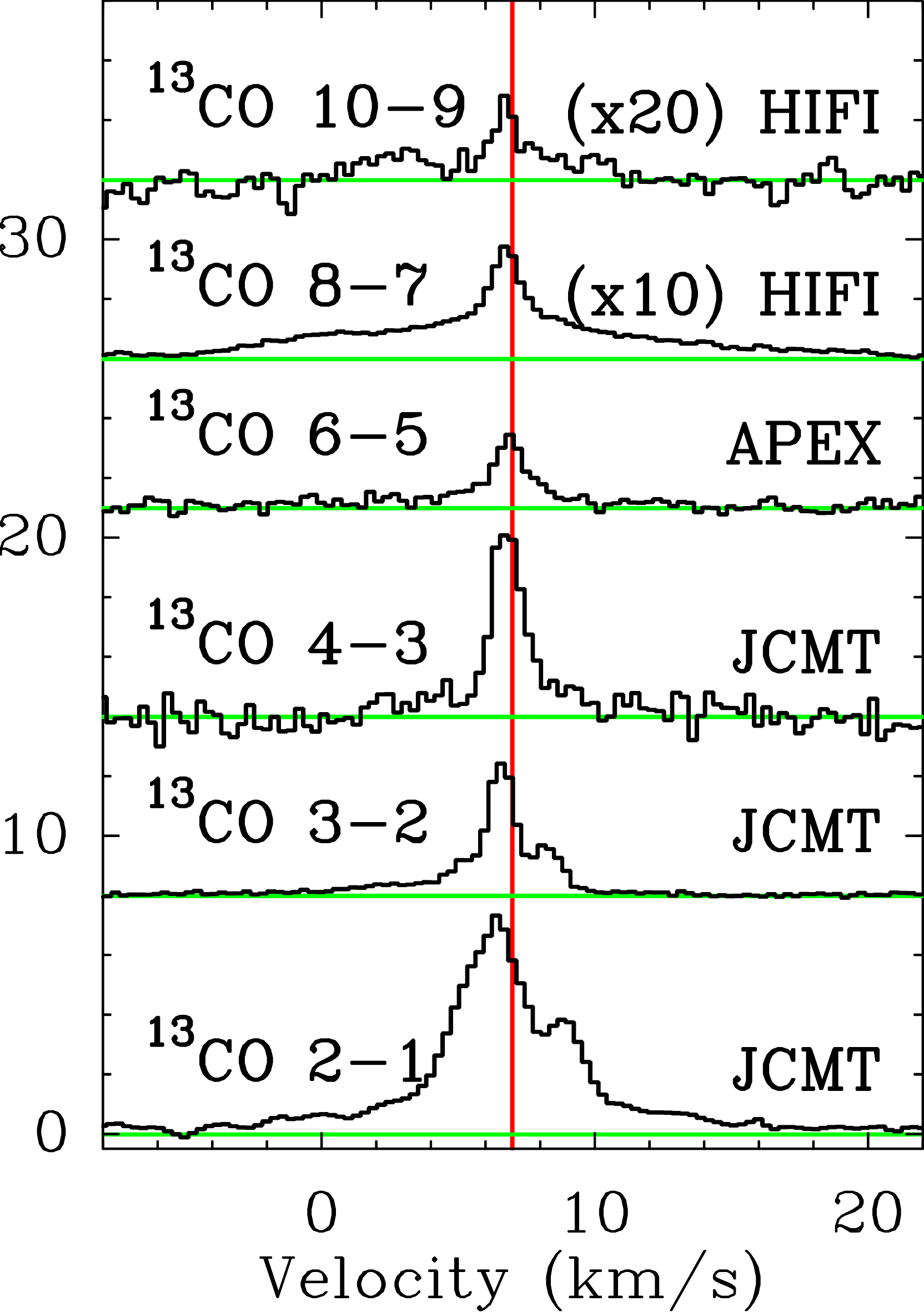}
    \includegraphics[scale=0.3]{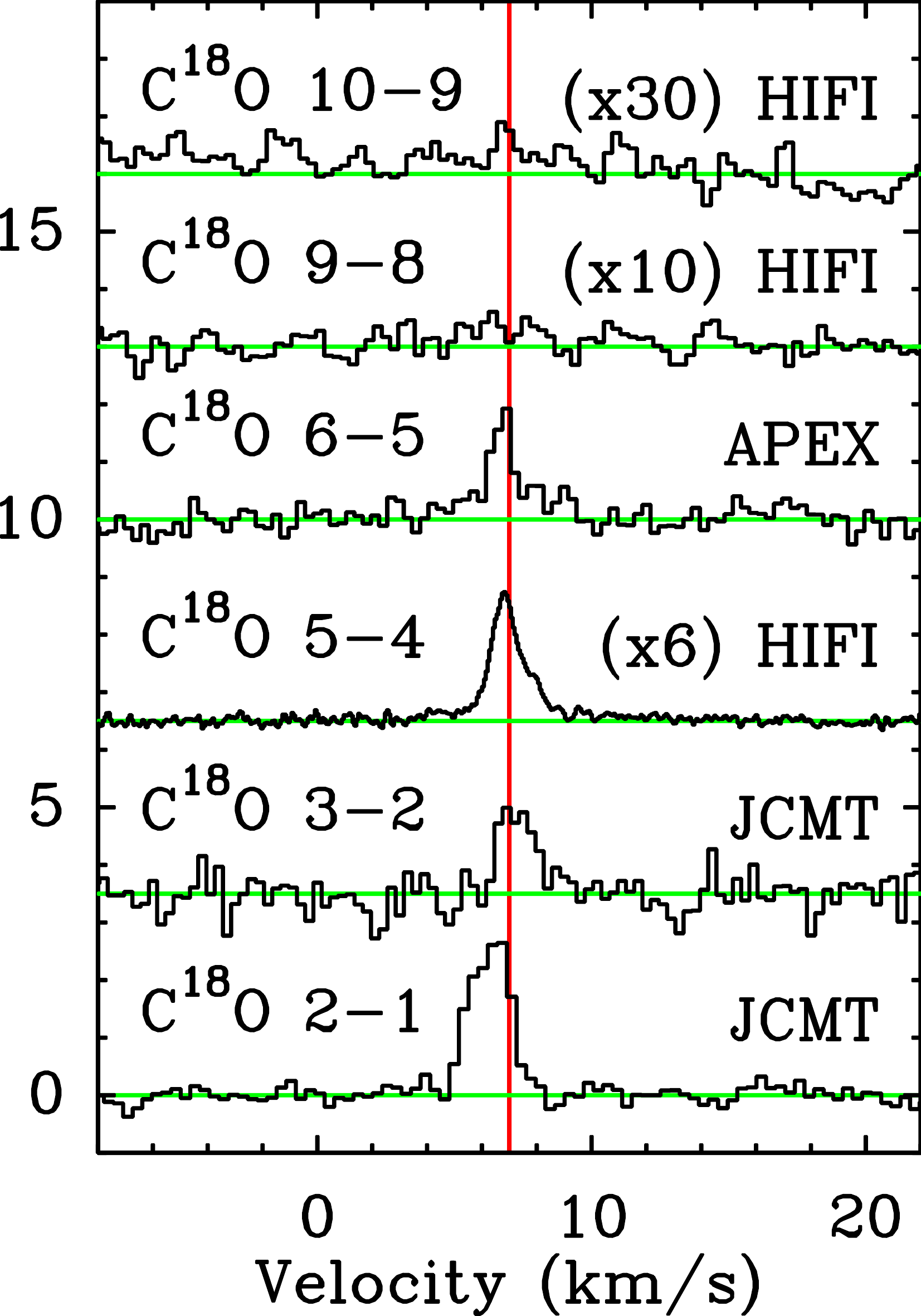}
    \caption{\small Observed $^{12}$CO, $^{13}$CO, and C$^{18}$O transitions for IRAS 4A.}
    \label{fig:linesIRAS4A}
\end{figure*}

\begin{table*}[!ht]
\caption{Observed line intensities for IRAS4A in all observed transitions.}
\normalsize
\begin{center}
\begin{tabular}{l l l r r r r r r r r r}
\hline \hline
Mol.  & Transition & Telescope & Efficiency & $\int T_{\rm MB} \mathrm{d}V$ & $T_{\mathrm{peak}}$ & $rms$ \\
 &  & &   $\eta$ &[K km s$^{-1}$] & [K] &   [K]\\
\hline
 CO       & 2--1 & JCMT-RxA        & 0.69   & 116.49\phantom{0} & 18.43\phantom{0} & 0.11\phantom{0} \\
          & 3--2 & JCMT-HARPB      & 0.63   & 132.56\phantom{0} & 18.85\phantom{0} & 0.070 \\
          & 4--3 & JCMT\tablefootmark{a}    & 0.38 & 194.25\phantom{0} &  23.78\phantom{0} & 0.92\phantom{0} \\
          & 6--5 & APEX-CHAMP$^+$  & 0.48   & 84.23\phantom{0}  & 10.03\phantom{0} & 0.27\phantom{0} \\
          & 7--6 & APEX-CHAMP$^+$  & 0.45   & 55.00\phantom{0}  & 17.04\phantom{0} & 4.39\phantom{0} \\
          &10--9 & {\it Herschel}-HIFI\tablefootmark{b} & 0.64 & 52.16\phantom{0} &  2.31\phantom{0} & 0.18\phantom{0} \\
$^{13}$CO & 2--1 & JCMT-RxA        & 0.74   & 39.30\phantom{0}  & 7.57\phantom{0}  & 0.15\phantom{0} \\
          & 3--2 & JCMT-HARPB      & 0.63   & 11.43\phantom{0}  & 4.49\phantom{0}  & 0.078 \\
          & 4--3 & JCMT\tablefootmark{a}    & 0.38 & 15.19\phantom{0} & 7.35\phantom{0} & 0.78\phantom{0} \\
          & 6--5 & APEX-CHAMP$^+$  & 0.52   & 7.03\phantom{0}   & 2.61\phantom{0}  & 0.20\phantom{0} \\
          & 8--7 & {\it Herschel}-HIFI\tablefootmark{c} & 0.75 & 2.14\phantom{0} & 0.38\phantom{0} & 0.009 \\
          & 10--9& {\it Herschel}-HIFI\tablefootmark{c} & 0.74 & 0.66\phantom{0}  & 0.16\phantom{0} & 0.022 \\
C$^{18}$O & 2--1 & JCMT-RxA        & 0.69   & 4.90\phantom{0}   & 2.70\phantom{0}  & 0.23\phantom{0} \\
          & 3--2 & JCMT-HARPB      & 0.63   & 2.54\phantom{0}   & 2.30\phantom{0}  & 0.22\phantom{0} \\
          & 5--4 & {\it Herschel}-HIFI\tablefootmark{d} & 0.76 & 0.61\phantom{0}  & 0.37\phantom{0}& 0.010 \\
          & 6--5 & APEX-CHAMP$^+$  & 0.48   & 3.00\phantom{0}   & 2.00\phantom{0}  & 0.26\phantom{0} \\
          & 9--8 & {\it Herschel}-HIFI\tablefootmark{c} & 0.74 & 0.17\phantom{0} & 0.050     & 0.023 \\
          &10--9 & {\it Herschel}-HIFI\tablefootmark{e} & 0.74 & 0.045 & 0.023   & 0.008 \\
\hline 
\end{tabular}
\end{center}
\tablefoot{The values given here are calculated for 20$\arcsec$ beam, therefore values are slightly different than \citet{Yildiz12}.
\tablefoottext{a}{Taken in 11$\arcsec$ beam.}
\tablefoottext{b}{Herschel observation corrected for the chopped emission and only H polarization observation is used.}
\tablefoottext{c}{H- and V-polarization observations averaged.}
\tablefoottext{d}{Only H-polarization observation is used.}
\tablefoottext{e}{Observed by open time program OT2$\_$rvisser$\_$2. H- and V-polarization observations averaged. }
}
\label{tbl:linesIRAS4A}
\end{table*}

\newpage

\onecolumn
\subsection{IRAS4B}
\begin{figure*}[htb]
    \centering
    \includegraphics[scale=0.3]{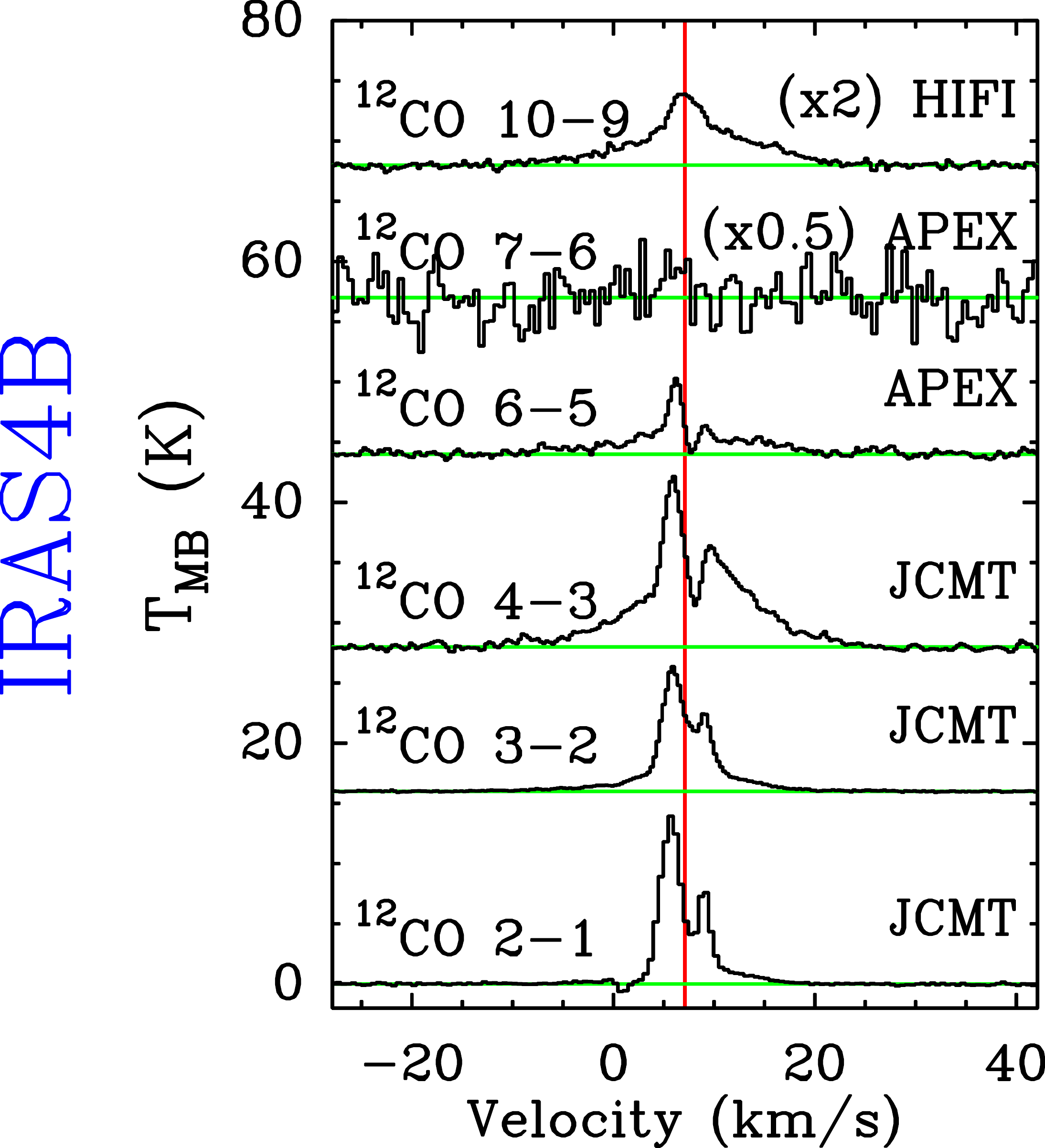}
    \includegraphics[scale=0.3]{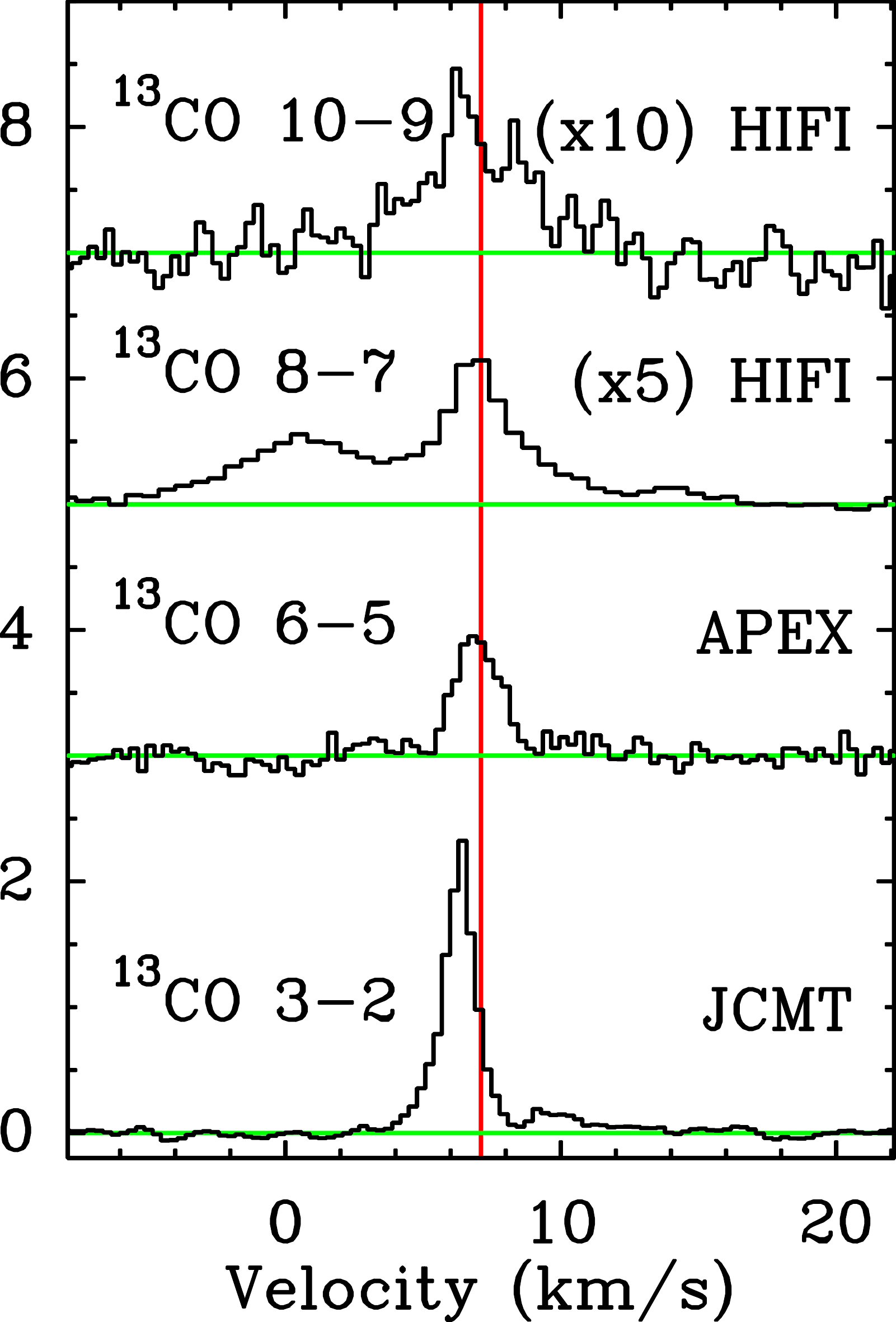}
    \includegraphics[scale=0.3]{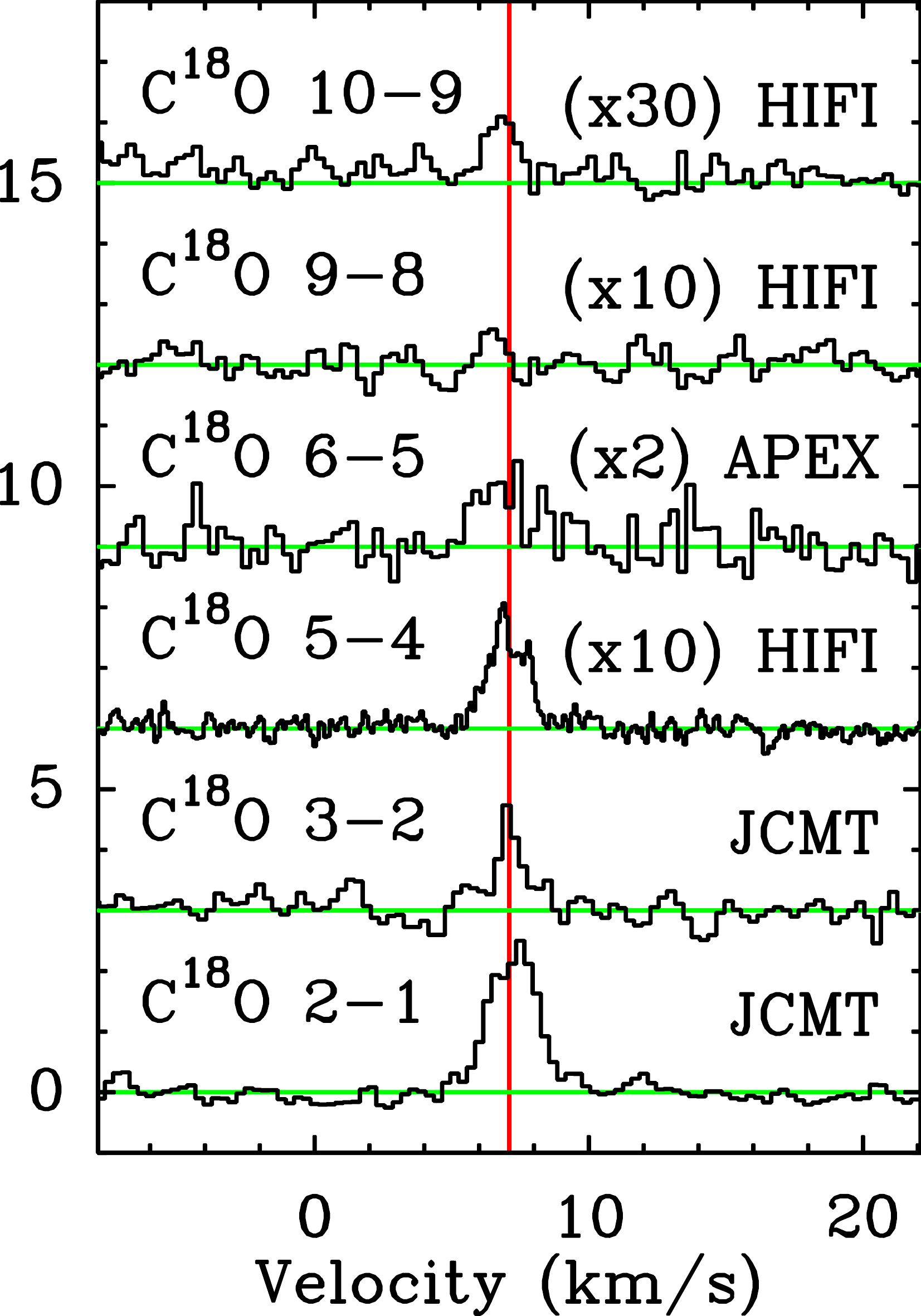}
    \caption{\small Observed $^{12}$CO, $^{13}$CO, and C$^{18}$O transitions for IRAS 4B.}
    \label{fig:linesIRAS4B}
\end{figure*}

\begin{table*}[!ht]
\caption{Observed line intensities for IRAS4B in all observed transitions.}
\normalsize
\begin{center}
\begin{tabular}{l l l r r r r r r r r r}
\hline \hline
Mol.  & Transition & Telescope & Efficiency & $\int T_{\rm MB} \mathrm{d}V$ & $T_{\mathrm{peak}}$ & $rms$ \\
 &  & &   $\eta$ &[K km s$^{-1}$] & [K] &   [K]\\
\hline
CO        & 2--1 & JCMT-RxA        & 0.69   & 54.70\phantom{0} & 13.94\phantom{0} & 0.070 \\
          & 3--2 & JCMT-HARPB      & 0.63   & 53.89\phantom{0} & 10.39\phantom{0} & 0.034 \\
          & 4--3 & JCMT\tablefootmark{a}    & 0.38 & 115.19\phantom{0} & 14.38\phantom{0} & 0.26\phantom{0} \\
          & 6--5 & APEX-CHAMP$^+$  & 0.48   & 35.17\phantom{0} & 6.29\phantom{0}  &  0.29\phantom{0} \\
          & 7--6 & APEX-CHAMP$^+$  & 0.45   & $<$35.00\phantom{0} & \dots\phantom{0}  & 4.51\phantom{0} \\
          &10--9 & {\it Herschel}-HIFI\tablefootmark{b} & 0.64 & 29.17\phantom{0} &  2.96\phantom{0} &  0.10\phantom{0} \\
$^{13}$CO & 3--2 & JCMT-HARPB      & 0.63   & 5.90\phantom{0}  & 2.32\phantom{0}  &  0.023 \\
          & 6--5 & APEX-CHAMP$^+$  & 0.52   & 2.20\phantom{0}  & 1.03\phantom{0}   & 0.10\phantom{0} \\
          & 8--7 & {\it Herschel}-HIFI\tablefootmark{c} & 0.75 & 0.73\phantom{0}  & 0.25\phantom{0} & 0.009 \\
          & 10--9& {\it Herschel}-HIFI\tablefootmark{c} & 0.74 & 0.53\phantom{0}  & 0.15\phantom{0} & 0.021 \\
C$^{18}$O & 2--1 & JCMT-RxA        & 0.69   & 5.30\phantom{0}  & 2.50\phantom{0}   & 0.16\phantom{0} \\
          & 3--2 & JCMT-HARPB      & 0.63   & 2.36\phantom{0}  & 1.70\phantom{0}   & 0.30\phantom{0} \\
          & 5--4 & {\it Herschel}-HIFI\tablefootmark{d} & 0.76 & 0.33\phantom{0} & 0.21\phantom{0} & 0.014 \\
          & 6--5 & APEX-CHAMP$^+$  & 0.48   &  1.35\phantom{0} & 0.82\phantom{0} & 0.22\phantom{0} \\
          & 9--8 & {\it Herschel}-HIFI\tablefootmark{c} & 0.74 & $<$0.024  & \dots\phantom{0} & 0.022 \\
          &10--9 & {\it Herschel}-HIFI\tablefootmark{e} & 0.74 & 0.059     & 0.04\phantom{0} & 0.009 \\
\hline 
\end{tabular}
\end{center}
\tablefoot{The values given here are calculated for 20$\arcsec$ beam, therefore values are slightly different than \citet{Yildiz12}.
\tablefoottext{a}{Taken in 11$\arcsec$ beam.}
\tablefoottext{b}{{\it Herschel} observation corrected for the chopped emission and only H polarization observation is used.}
\tablefoottext{c}{H- and V-polarization observations averaged.}
\tablefoottext{d}{Only H-polarization observation is used.}
\tablefoottext{e}{Observed by open time program OT2$\_$rvisser$\_$2. H- and V-polarization observations averaged. }
}
\label{tbl:linesIRAS4B}
\end{table*}

\newpage
\clearpage

\onecolumn
\subsection{L1527}
\begin{figure*}[htb]
    \centering
    \includegraphics[scale=0.3]{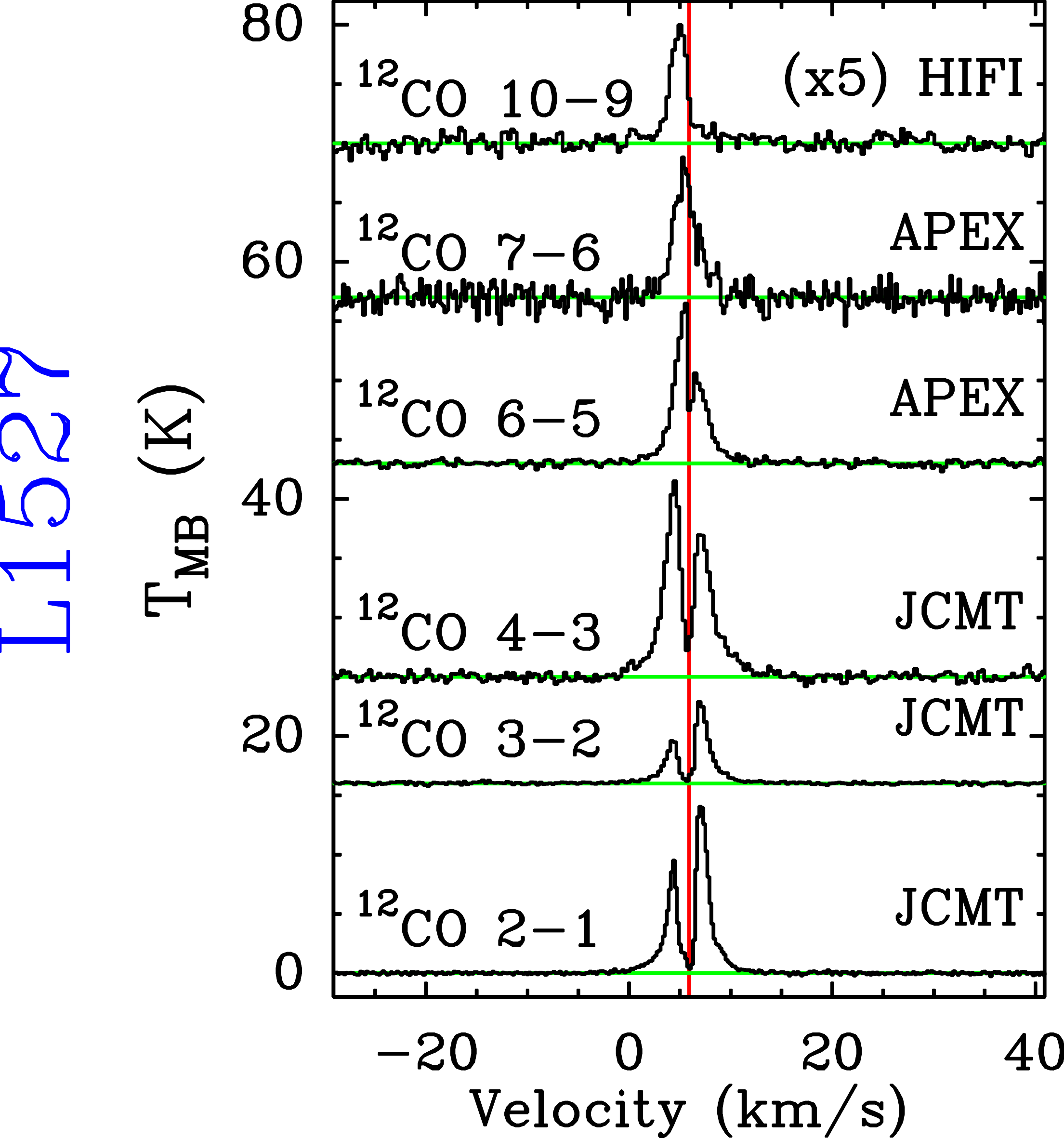}
    \includegraphics[scale=0.3]{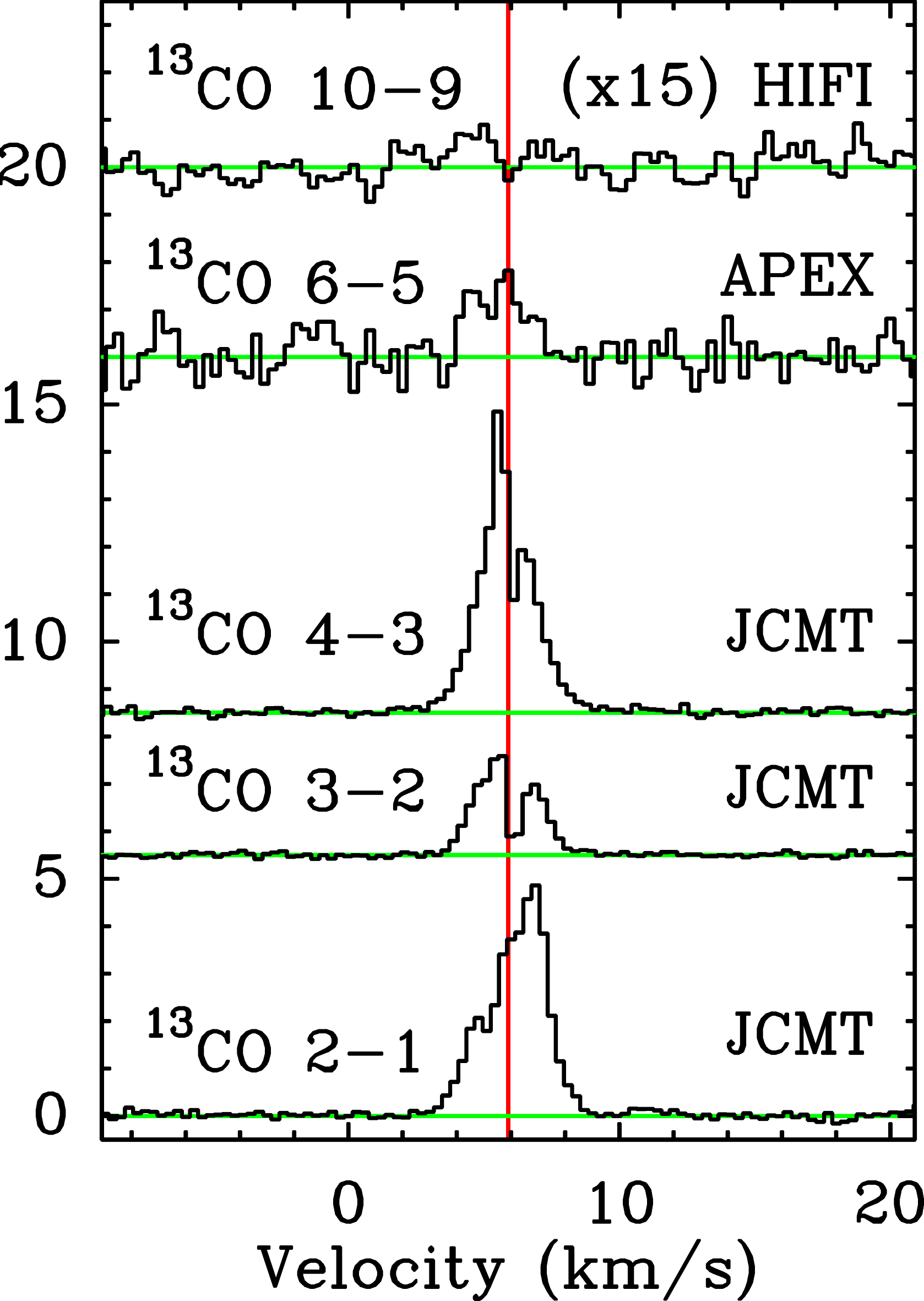}
    \includegraphics[scale=0.3]{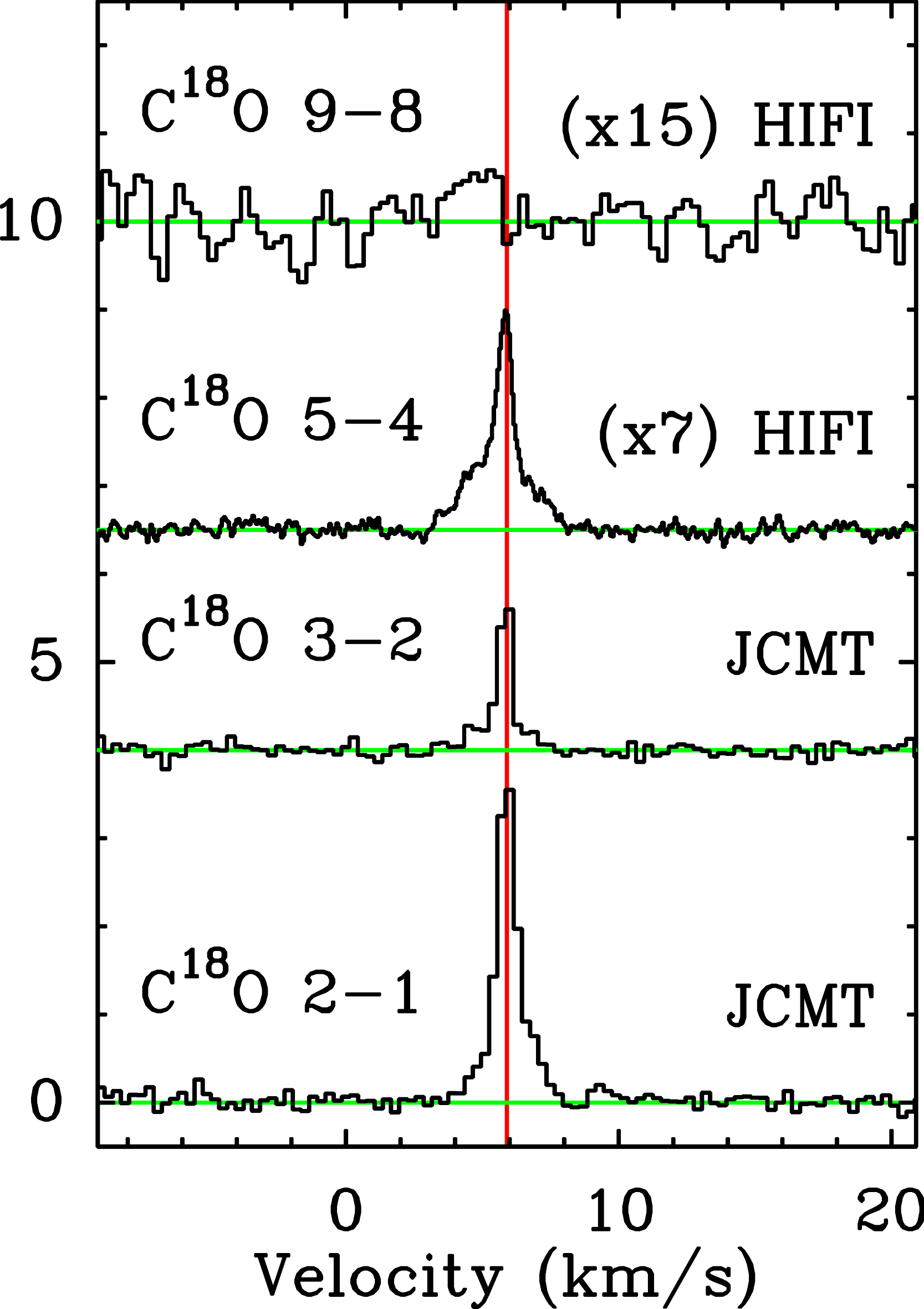}
    \caption{\small Observed $^{12}$CO, $^{13}$CO, and C$^{18}$O transitions for L1527.}
    \label{fig:linesL1527}
\end{figure*}

\begin{table*}[!ht]
\caption{Observed line intensities for L1527 in all observed transitions.}
\normalsize
\begin{center}
\begin{tabular}{l l l r r r r r r r r r}
\hline \hline
Mol.  & Transition & Telescope & Efficiency & $\int T_{\rm MB} \mathrm{d}V$ & $T_{\mathrm{peak}}$ & $rms$ \\
 &  & &   $\eta$ &[K km s$^{-1}$] & [K] &   [K]\\
\hline
CO        & 2--1 & JCMT-RxA       & 0.69   &  38.14\phantom{0} & 14.04\phantom{0} & 0.086 \\
          & 3--2 & JCMT-HARPB     & 0.63   &  19.38\phantom{0}   &  6.90\phantom{0}  & 0.12\phantom{0} \\
          & 4--3 & JCMT\tablefootmark{a}   & 0.38   &  63.64\phantom{0}   &  16.39\phantom{0}  & 0.46\phantom{0} \\
          & 6--5 & APEX-CHAMP$^+$ & 0.52   &  38.02\phantom{0}   &  14.88\phantom{0}  & 0.33\phantom{0} \\
          & 7--6 & APEX-CHAMP$^+$ & 0.49   &  31.90\phantom{0}   &  11.82\phantom{0}  & 0.85\phantom{0} \\
          &10--9 & {\it Herschel}-HIFI\tablefootmark{b} & 0.64   &  5.17\phantom{0} & 2.00\phantom{0}  & 0.12\phantom{0} \\
$^{13}$CO & 2--1 & JCMT-RxA       & 0.69   &  12.12\phantom{0}   &  4.92\phantom{0}  & 0.085 \\
          & 3--2 & JCMT-HARPB     & 0.63   &  4.72\phantom{0}    &  2.71\phantom{0}  & 0.073 \\
          & 4--3 & JCMT\tablefootmark{a}   & 0.38   &  12.12\phantom{0}   &  6.40\phantom{0}  & 0.089 \\
          & 6--5 & APEX-CHAMP$^+$ & 0.48   &  4.72\phantom{0}    &  2.35\phantom{0}  & 0.51\phantom{0} \\
          & 10--9&  {\it Herschel}-HIFI\tablefootmark{b}        & 0.76 & 0.12\phantom{0} &  0.06\phantom{0}  & 0.019 \\
C$^{18}$O & 2--1 & JCMT-RxA       & 0.69   &  4.55\phantom{0}    &  3.89\phantom{0}  & 0.11\phantom{0} \\
          & 3--2 & JCMT-HARPB     & 0.63   &  1.54\phantom{0}    &  1.90\phantom{0}  & 0.10\phantom{0} \\
          & 5--4 & {\it Herschel}-HIFI\tablefootmark{b}         & 0.76 & 0.49\phantom{0} & 0.35\phantom{0} & 0.009 \\
          & 9--8 & {\it Herschel}-HIFI\tablefootmark{c}         & 0.74 & $<$0.050 & \dots\phantom{0}  & 0.022 \\
\hline 
\end{tabular}
\end{center}
\tablefoot{
\tablefoottext{a}{Taken in 11$\arcsec$ beam.}
\tablefoottext{b}{Only H-polarization observation is used.}
\tablefoottext{c}{H- and V-polarization observations averaged.}
}
\label{tbl:linesL1527}
\end{table*}

\newpage

\onecolumn
\subsection{Ced110IRS4}
\begin{figure*}[htb]
    \centering
    \includegraphics[scale=0.3]{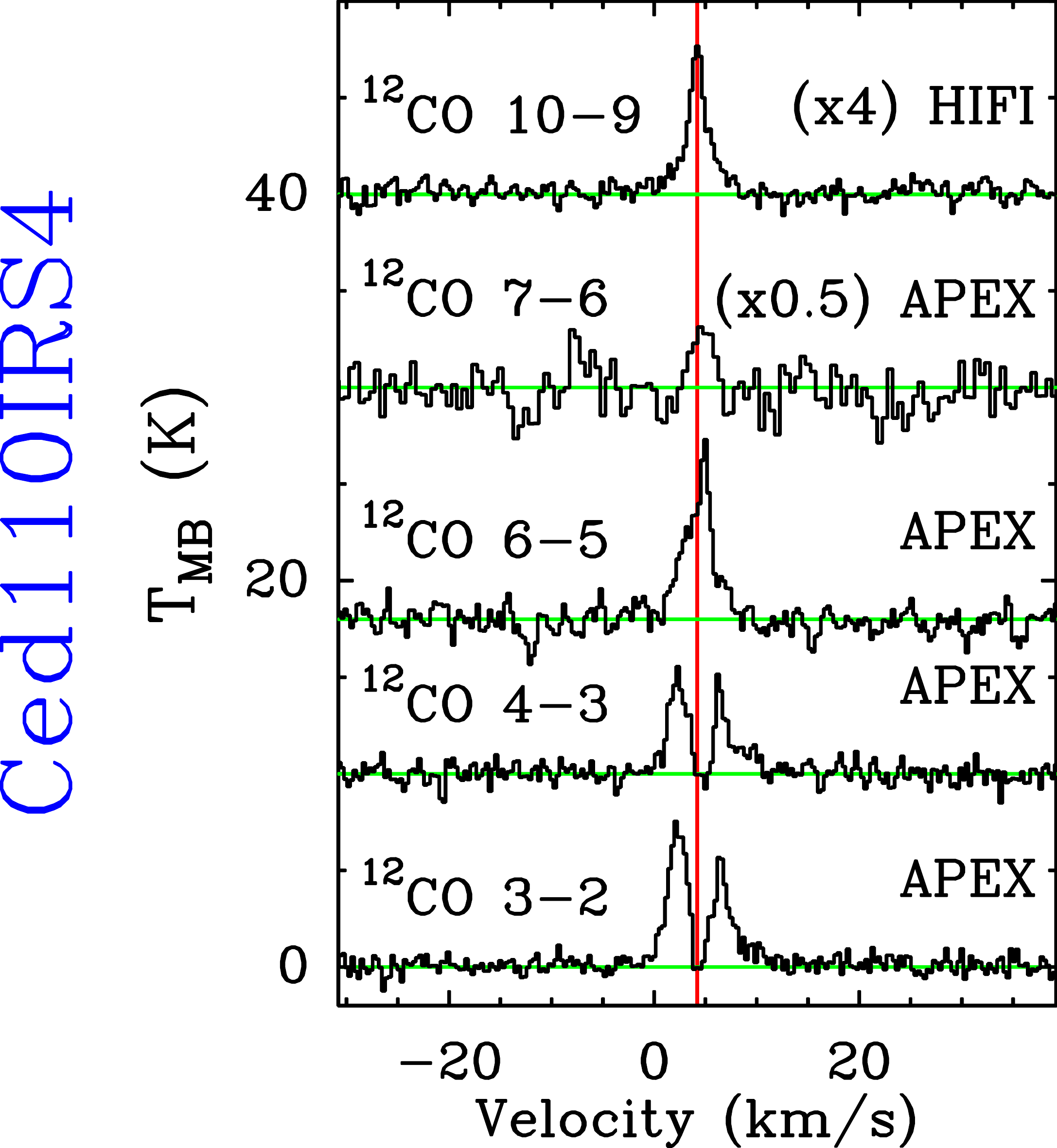}
    \includegraphics[scale=0.3]{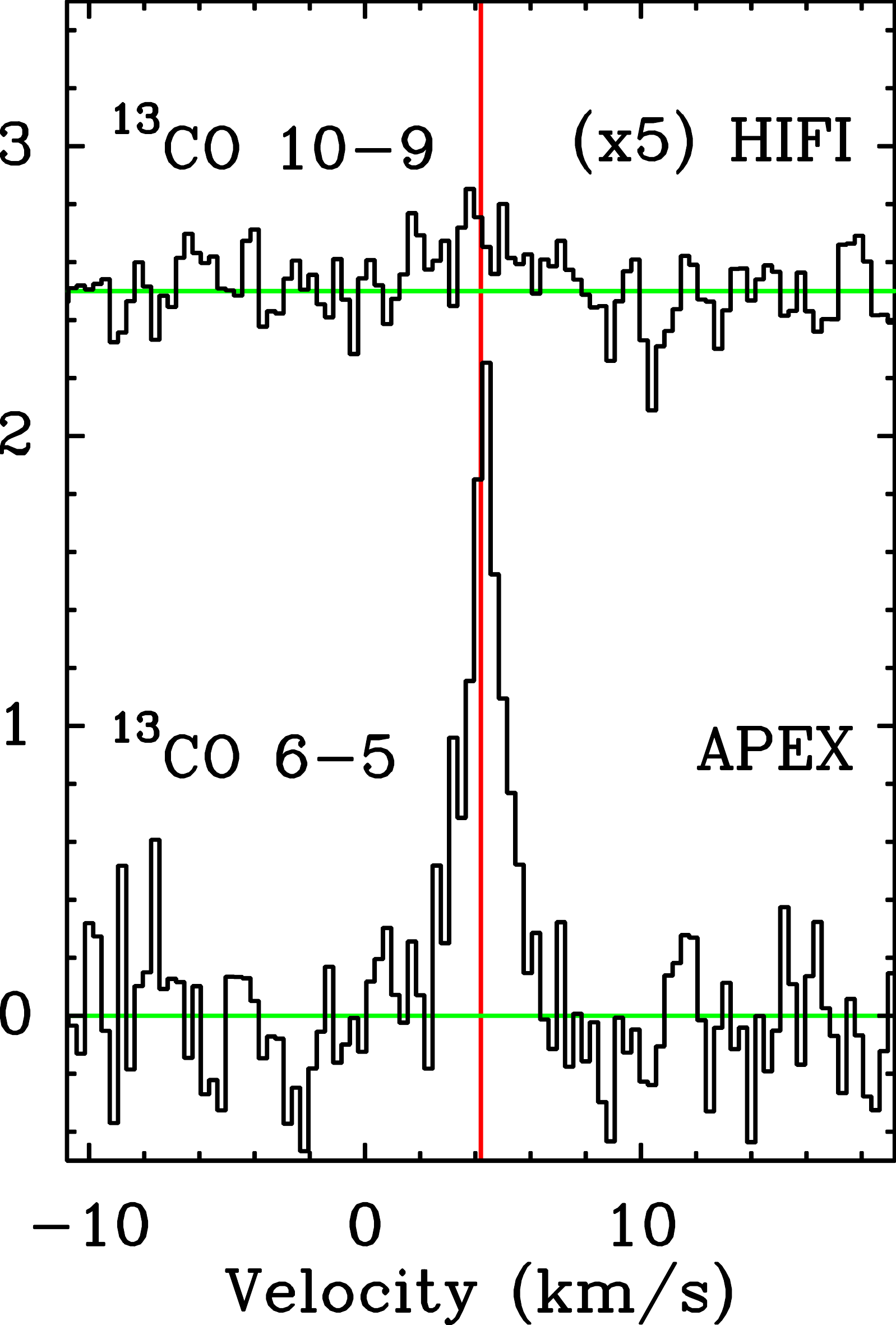}
    \includegraphics[scale=0.3]{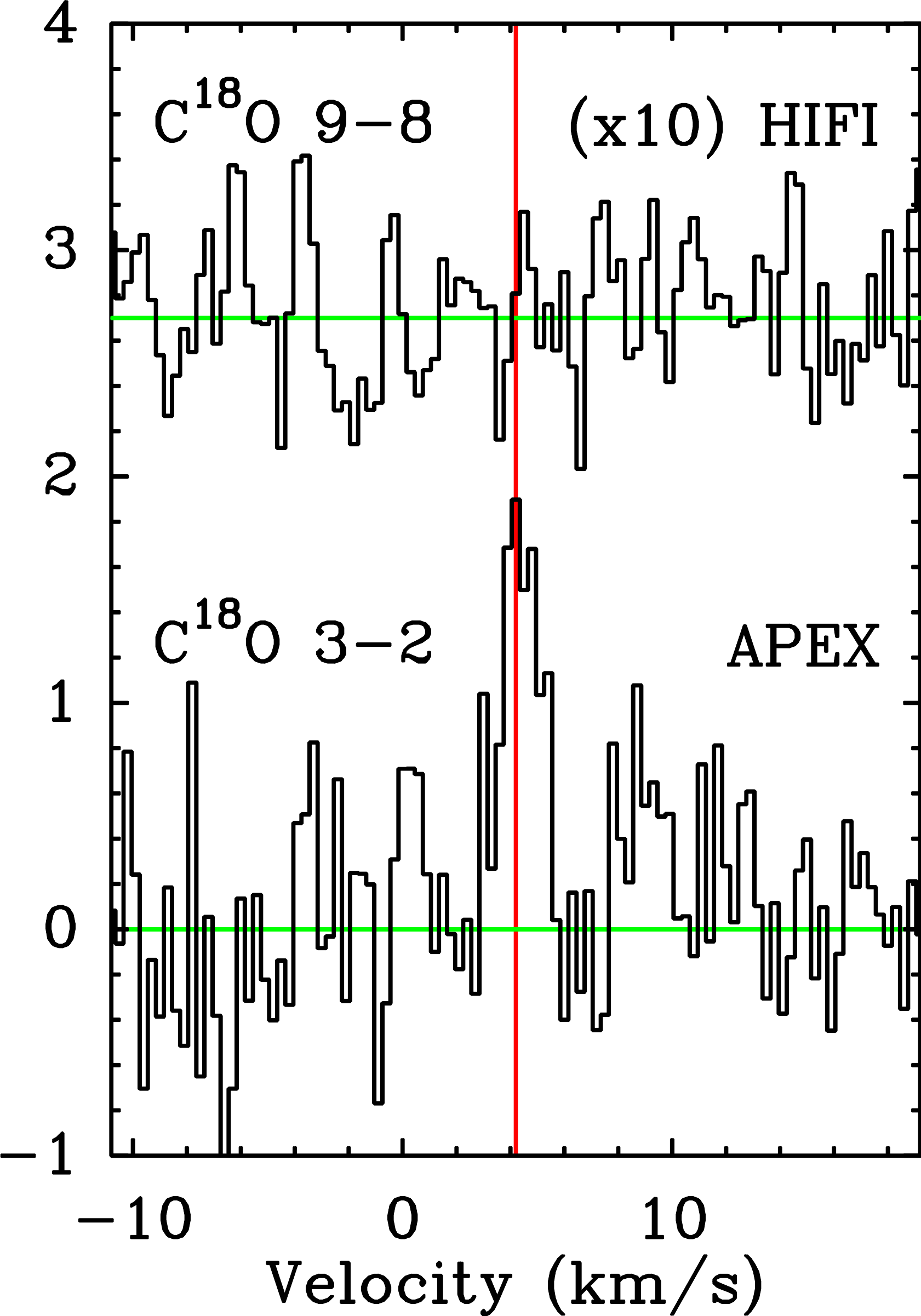}
    \caption{\small Observed $^{12}$CO, $^{13}$CO, and C$^{18}$O transitions for Ced110-IRS4.}
    \label{fig:linesCed110IRS4}
\end{figure*}

\begin{table*}[!ht]
\caption{Observed line intensities for Ced110-IRS4 in all observed transitions.}
\normalsize
\begin{center}
\begin{tabular}{l l l r r r r r r r r r}
\hline \hline
Mol.  & Transition & Telescope & Efficiency & $\int T_{\rm MB} \mathrm{d}V$ & $T_{\mathrm{peak}}$ & $rms$ \\
 &  & &   $\eta$ &[K km s$^{-1}$] & [K] &   [K]\\
\hline
CO        & 3--2 & APEX           & 0.73 & 31.14\phantom{0} & 7.70\phantom{0}  & 0.35\phantom{0} \\
          & 4--3 & APEX           & 0.60 & 20.00\phantom{0} & 5.70\phantom{0} & 0.62\phantom{0} \\
          & 6--5 & APEX-CHAMP$^+$ & 0.45 & 24.42\phantom{0} & 10.36\phantom{0} & 0.71\phantom{0} \\
          & 7--6 & APEX-CHAMP$^+$ & 0.42 & 18.23\phantom{0} & 6.82\phantom{0}  & 1.49\phantom{0} \\
          &10--9 & {\it Herschel}-HIFI\tablefootmark{a}& 0.66 & 5.08\phantom{0} & 1.92\phantom{0} & 0.12\phantom{0} \\
$^{13}$CO & 6--5 & APEX-CHAMP$^+$ & 0.45 & 2.98\phantom{0} & 2.70\phantom{0}  & 0.30\phantom{0} \\
          & 10--9& {\it Herschel}-HIFI\tablefootmark{a} & 0.74 & 0.12\phantom{0}  & 0.08\phantom{0} & 0.027 \\
C$^{18}$O & 3--2 & APEX--2a       & 0.73 & 3.10\phantom{0} & 1.90 \phantom{0} & 0.47\phantom{0} \\
          & 9--8 & {\it Herschel}-HIFI\tablefootmark{a} & 0.74 & $<$0.059 & \dots\phantom{0} & 0.026 \\
\hline 
\end{tabular}
\end{center}
\tablefoot{
\tablefoottext{a}{H- and V-polarization observations averaged.}
}
\label{tbl:linesCed110IRS4}
\end{table*}

\newpage
\clearpage

\onecolumn
\subsection{BHR71}
\begin{figure*}[htb]
    \centering
    \includegraphics[scale=0.3]{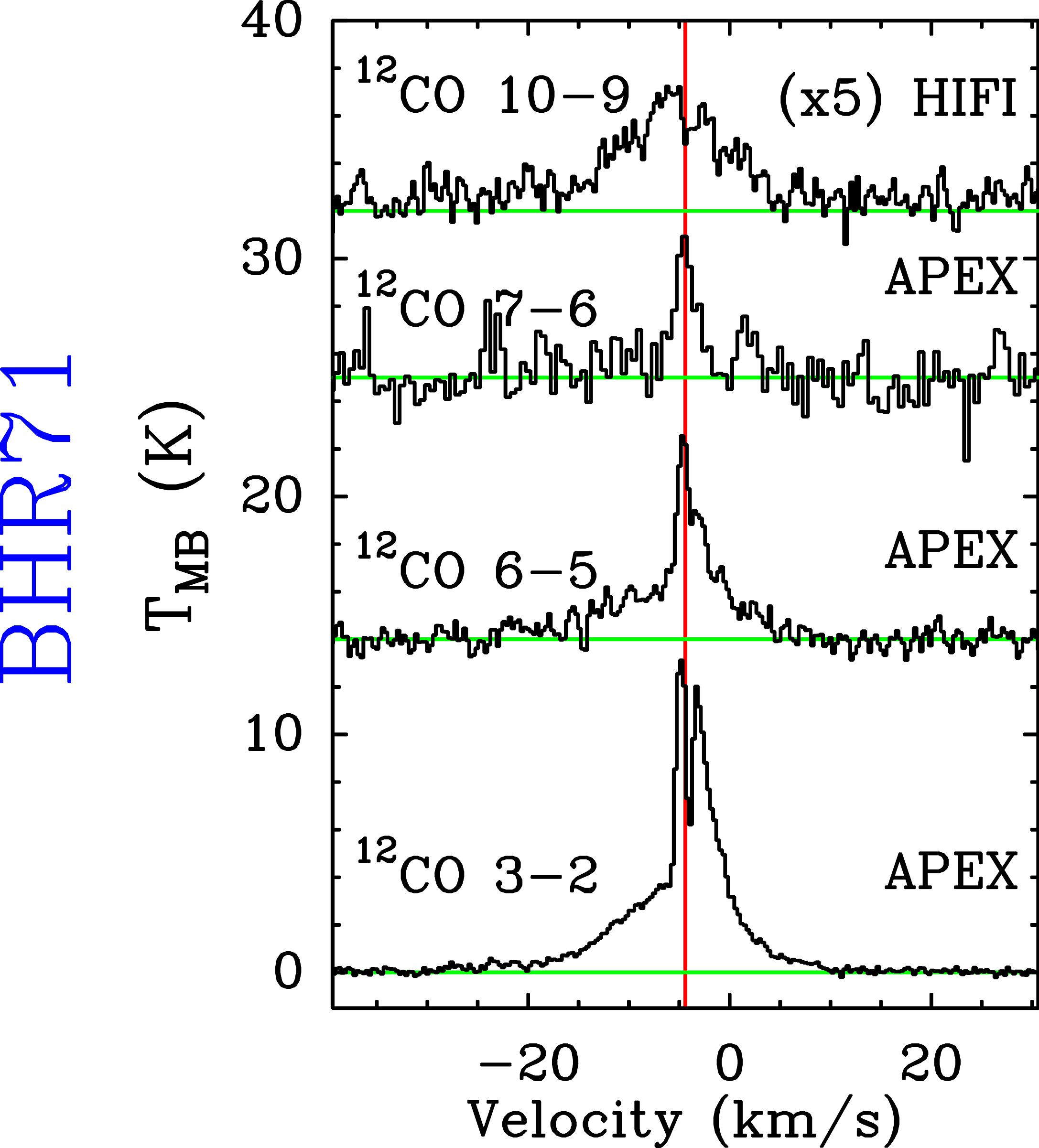}
    \includegraphics[scale=0.3]{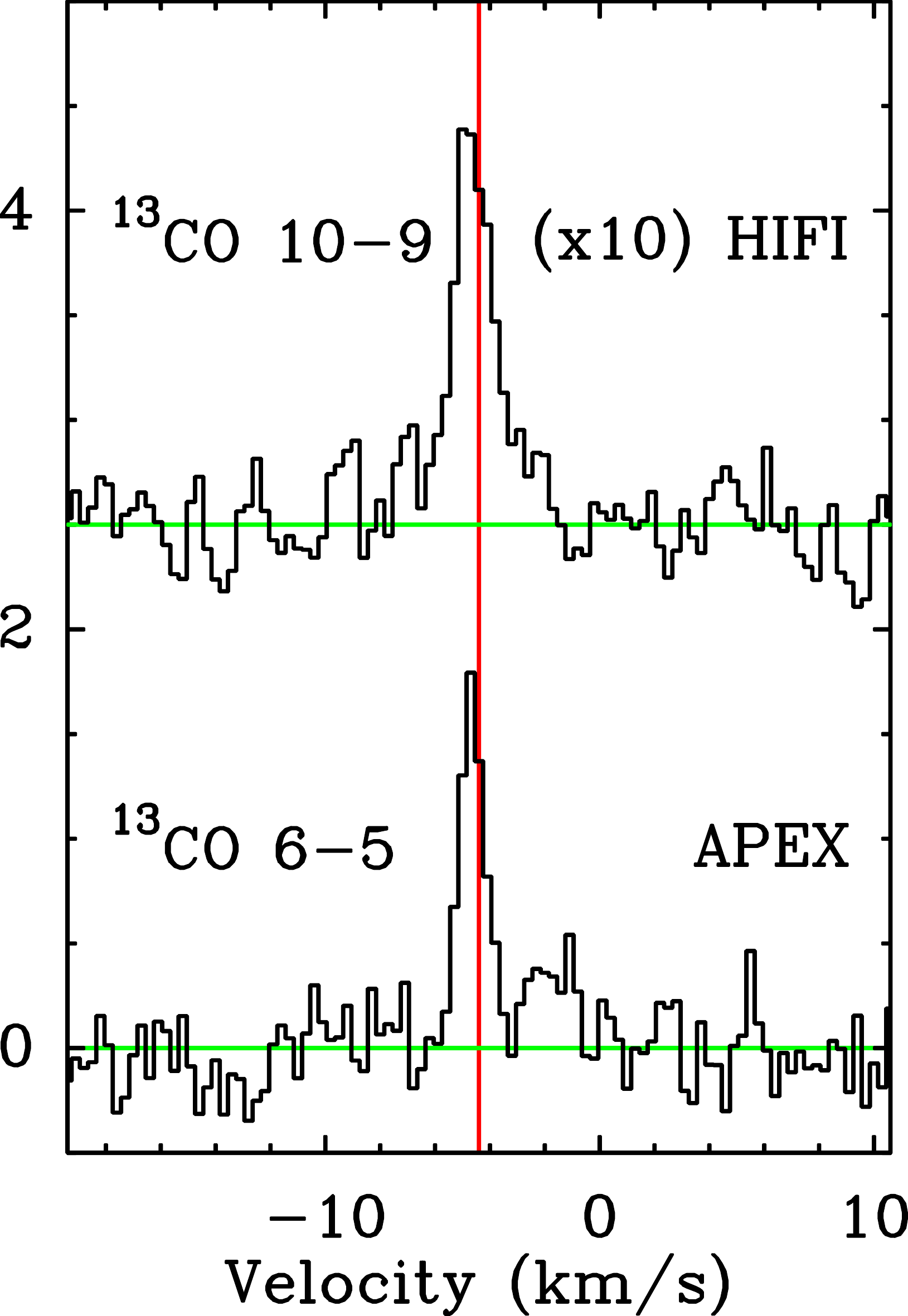}
    \includegraphics[scale=0.3]{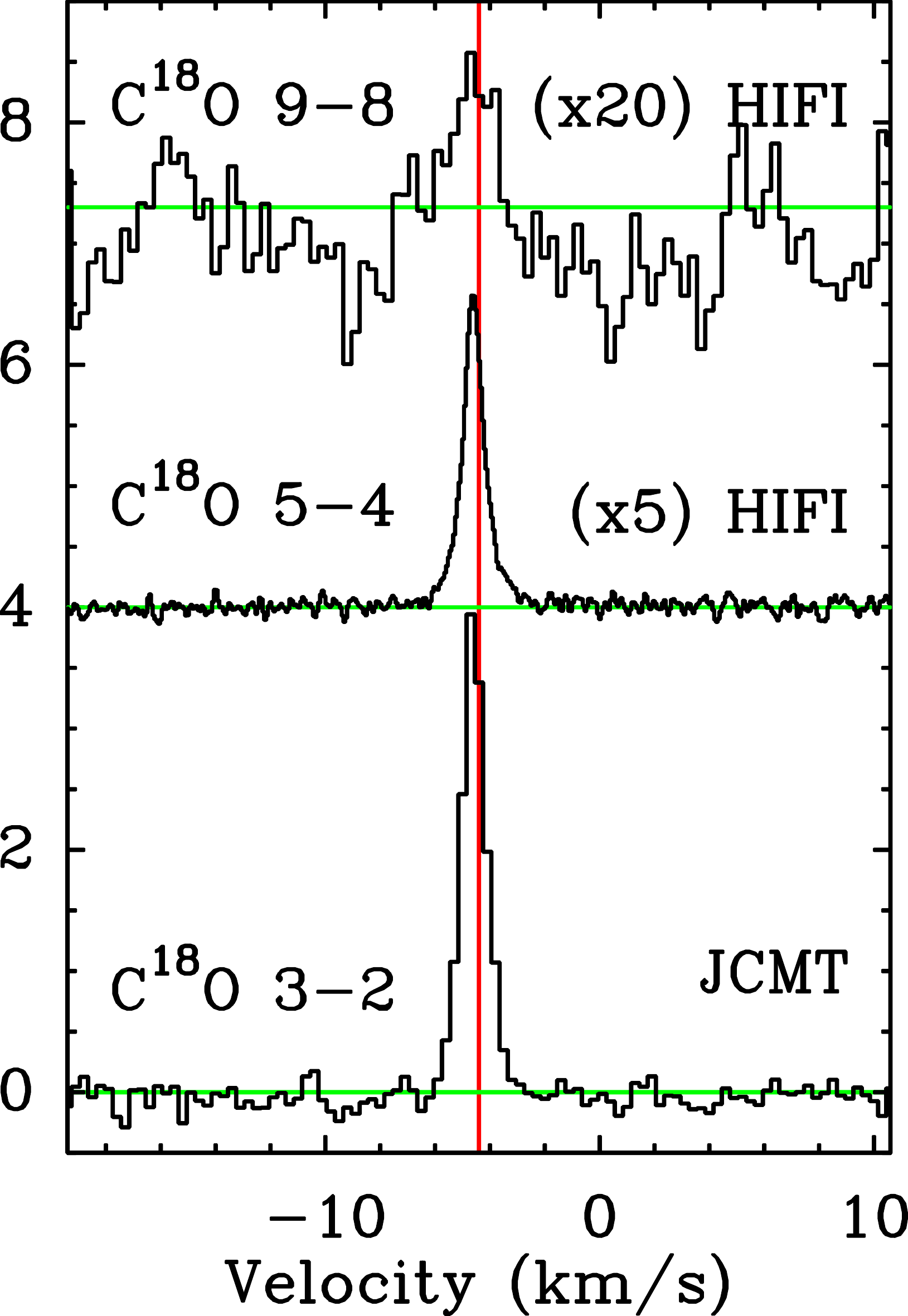}
    \caption{\small Observed $^{12}$CO, $^{13}$CO, and C$^{18}$O transitions for BHR71.}
    \label{fig:linesBHR71}
\end{figure*}

\begin{table*}[!ht]
\caption{Observed line intensities for BHR71 in all observed transitions.}
\normalsize
\begin{center}
\begin{tabular}{l l l r r r r r r r r r}
\hline \hline
Mol.  & Transition & Telescope & Efficiency & $\int T_{\rm MB} \mathrm{d}V$ & $T_{\mathrm{peak}}$ & $rms$ \\
 &  & &   $\eta$ &[K km s$^{-1}$] & [K] &   [K]\\
\hline
CO        & 3--2 & APEX           & 0.73   &  96.49\phantom{0}   & 14.46\phantom{0}  & 0.20\phantom{0} \\
          & 6--5 & APEX-CHAMP$^+$ & 0.46   &  49.14\phantom{0}   &  8.72\phantom{0}  & 0.52\phantom{0} \\
          & 7--6 & APEX-CHAMP$^+$ & 0.49   &  27.26\phantom{0}   &  6.83\phantom{0}  & 1.31\phantom{0} \\
          &10--9 & {\it Herschel}-HIFI\tablefootmark{a} & 0.64 & 15.08\phantom{0} & 1.00\phantom{0} & 0.15\phantom{0} \\
$^{13}$CO & 6--5 & APEX-CHAMP$^+$ & 0.45   &  2.67\phantom{0} &  1.92\phantom{0}  & 0.23\phantom{0} \\
          &10--9 & {\it Herschel}-HIFI\tablefootmark{b}  & 0.74   &  0.40\phantom{0} &  0.20\phantom{0} & 0.017 \\
C$^{18}$O & 3--2 & APEX-2a        & 0.73   &  4.12\phantom{0}  & 4.05\phantom{0}   & 0.11\phantom{0} \\
          & 5--4 & {\it Herschel}-HIFI\tablefootmark{a} & 0.76   &  0.56\phantom{0} &  0.51\phantom{0} & 0.010 \\
          & 9--8 & {\it Herschel}-HIFI\tablefootmark{b} & 0.74   &  0.08\phantom{0} &  0.07\phantom{0} & 0.022 \\
\hline 
\end{tabular}
\end{center}
\tablefoot{
\tablefoottext{a}{Only H-polarization observation is used.}
\tablefoottext{b}{H- and V-polarization observations averaged.}
}
\label{tbl:linesBHR71}
\end{table*}

\newpage
\clearpage

\onecolumn
\subsection{IRAS153981}
\begin{figure*}[htb]
    \centering
    \includegraphics[scale=0.3]{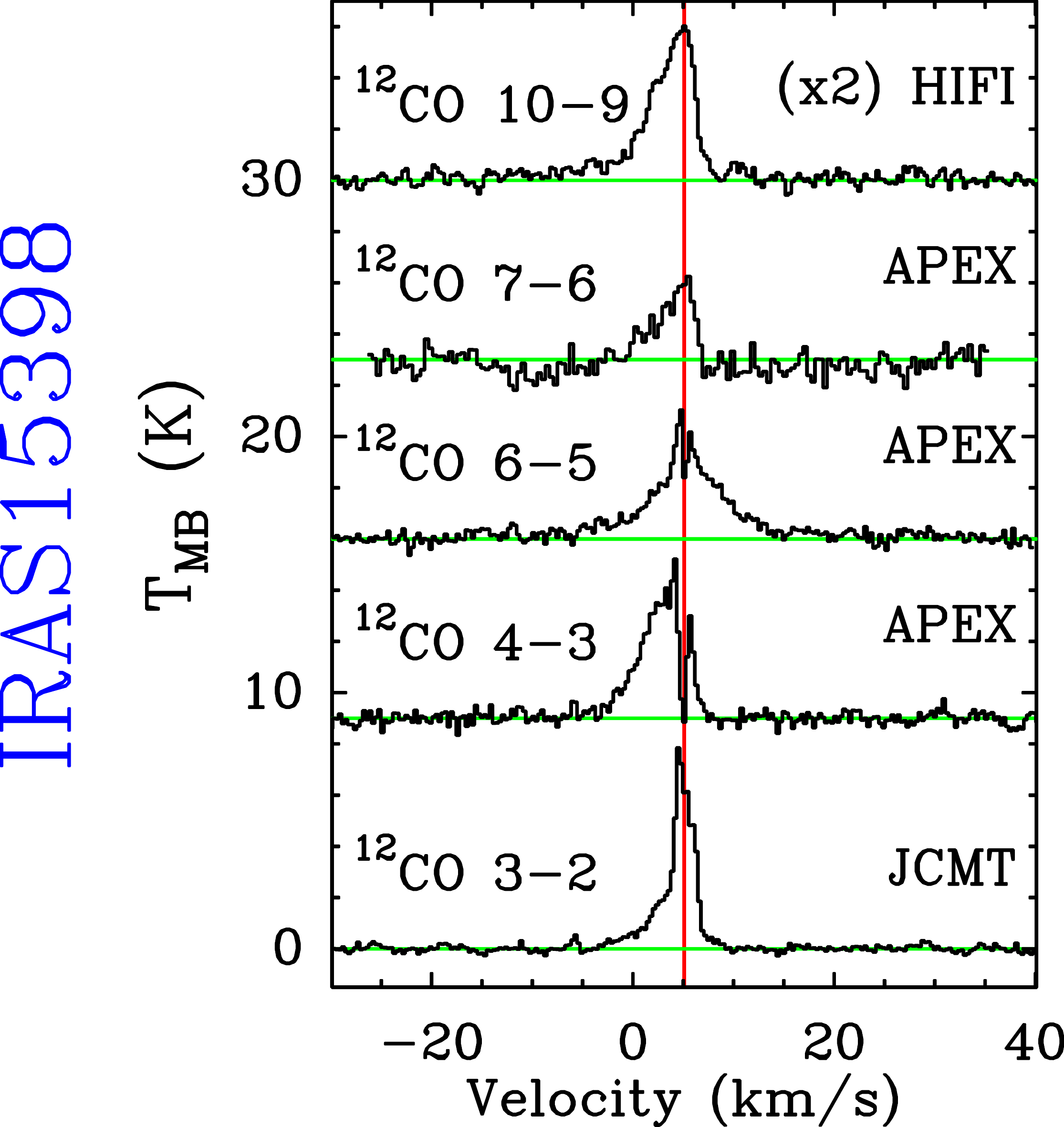}
    \includegraphics[scale=0.3]{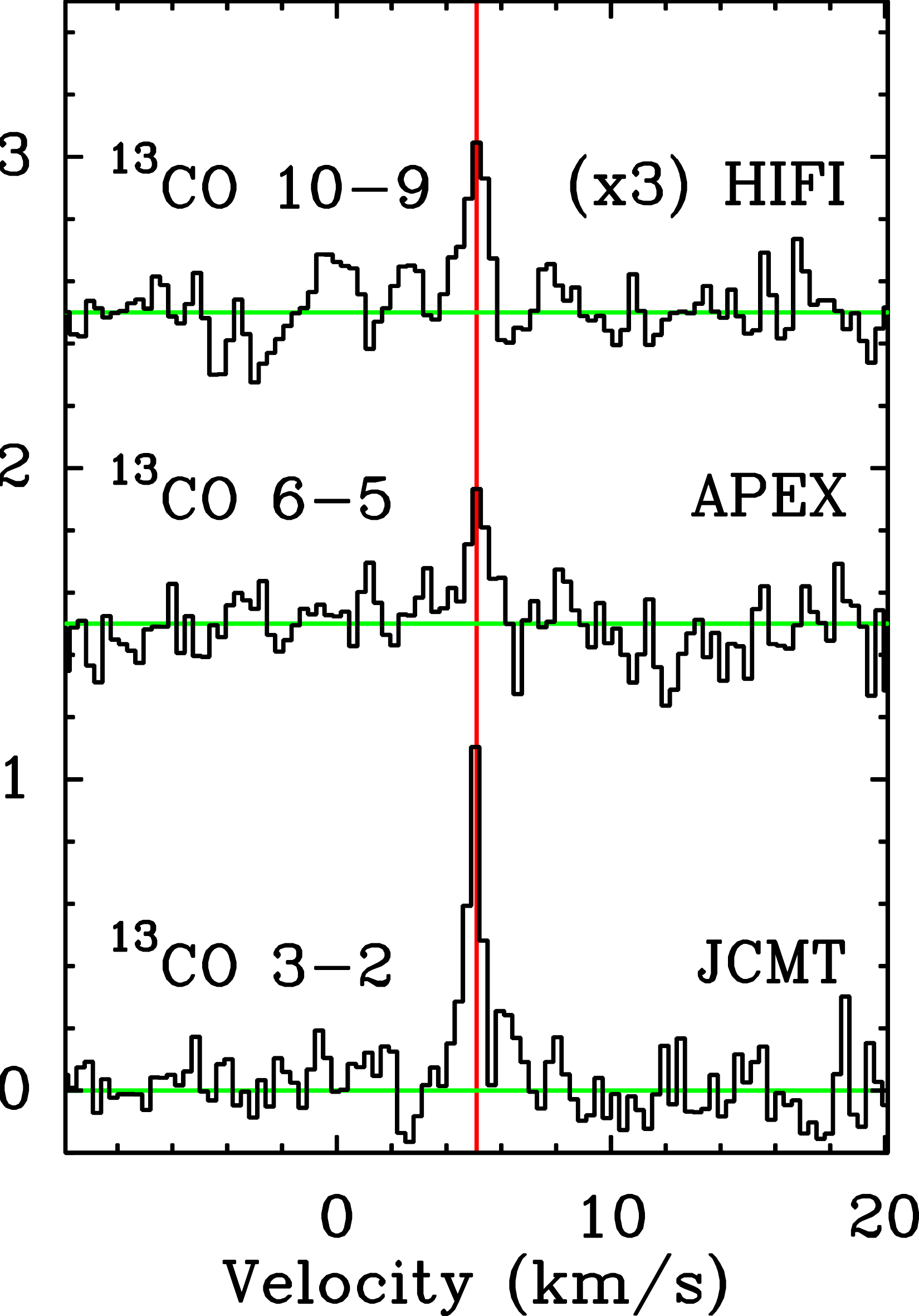}
    \includegraphics[scale=0.3]{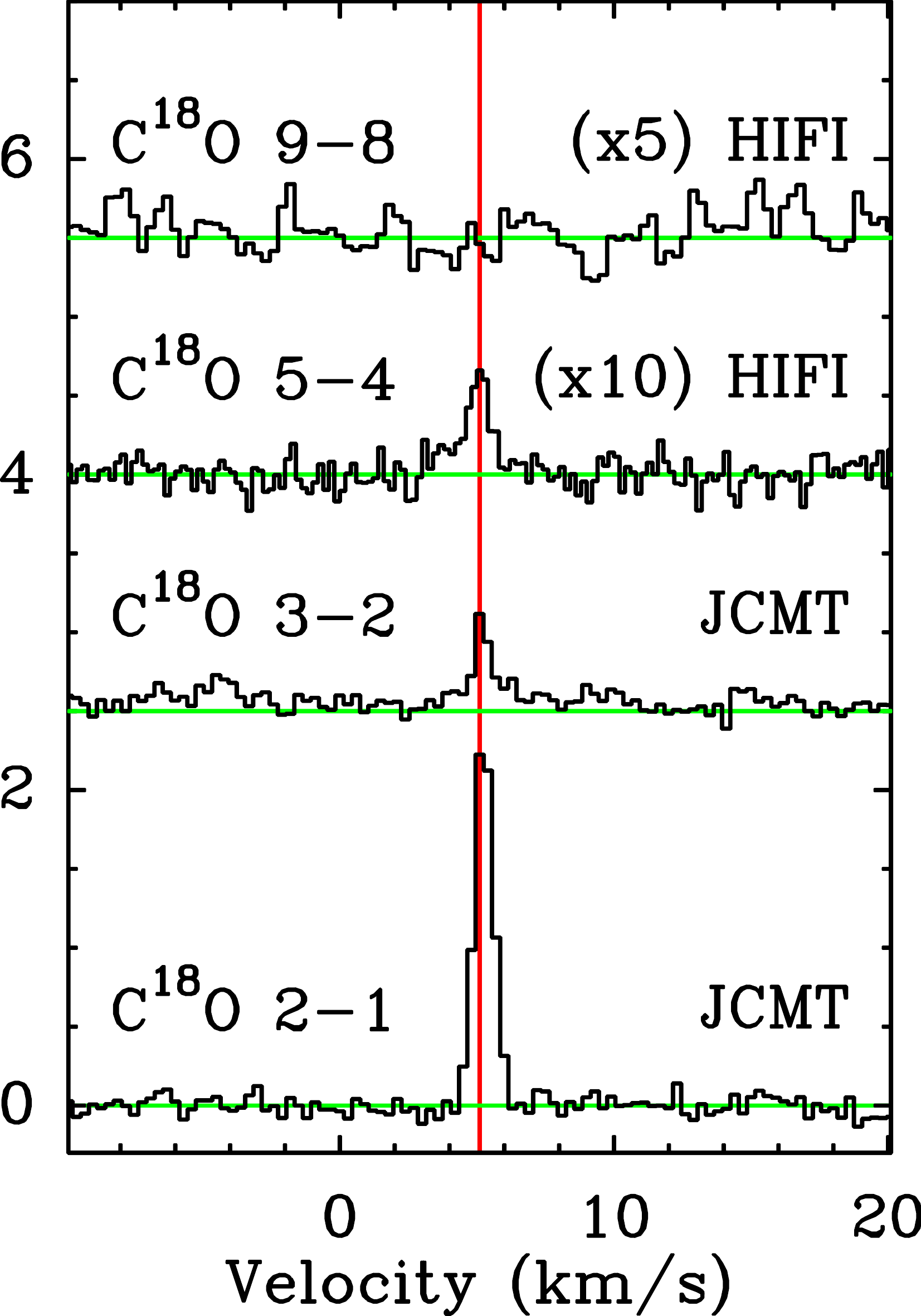}
    \caption{\small Observed $^{12}$CO, $^{13}$CO, and C$^{18}$O transitions for IRAS15398.}
    \label{fig:linesIRAS15398}
\end{figure*}

\begin{table*}[!ht]
\caption{Observed line intensities for IRAS15398 in all observed transitions.}
\normalsize
\begin{center}
\begin{tabular}{l l l r r r r r r r r r}
\hline \hline
Mol.  & Transition & Telescope & Efficiency & $\int T_{\rm MB} \mathrm{d}V$ & $T_{\mathrm{peak}}$ & $rms$ \\
 &  & &   $\eta$ &[K km s$^{-1}$] & [K] &   [K]\\
\hline
CO        & 3--2 & APEX           & 0.73   &  22.00\phantom{0}   &  8.54\phantom{0}  & 0.15\phantom{0} \\
          & 4--3 & APEX           & 0.60   &  26.26\phantom{0}   &  6.76\phantom{0}  & 0.32\phantom{0} \\
          & 6--5 & APEX-CHAMP$^+$ & 0.48   &  33.91\phantom{0}   &  5.33\phantom{0}  & 0.25\phantom{0} \\
          & 7--6 & APEX-CHAMP$^+$ & 0.49   &  16.88\phantom{0}   &  3.90\phantom{0}  & 0.64\phantom{0} \\
          &10--9 & {\it Herschel}-HIFI\tablefootmark{a} & 0.64 & 17.74\phantom{0} & 3.01\phantom{0} & 0.10\phantom{0} \\
$^{13}$CO & 3--2 & JCMT-HARPB     & 0.63   &  1.15\phantom{0} &  1.27\phantom{0}  & 0.12\phantom{0} \\
          & 6--5 & APEX-CHAMP$^+$ & 0.48   &  0.51\phantom{0} &  0.51\phantom{0}  & 0.12\phantom{0} \\
          & 10--9&  {\it Herschel}-HIFI\tablefootmark{a} & 0.74 & 0.20\phantom{0} & 0.18\phantom{0}  & 0.027 \\
C$^{18}$O & 2--1 & JCMT-RxA       & 0.69   &  2.03\phantom{0} &  2.52\phantom{0}  & 0.093 \\
          & 3--2 & JCMT-HARPB     & 0.63   &  0.80\phantom{0} &  0.72\phantom{0}  & 0.075 \\
          & 5--4 & {\it Herschel}-HIFI\tablefootmark{a} & 0.76 &  0.085   &  0.065 & 0.085 \\
          & 9--8 & {\it Herschel}-HIFI\tablefootmark{a} & 0.74 &  $<$0.055 & \dots\phantom{0}  & 0.024 \\
\hline 
\end{tabular}
\end{center}
\tablefoot{
\tablefoottext{a}{H- and V-polarization observations averaged.}
}
\label{tbl:linesIRAS15398}
\end{table*}

\newpage

\onecolumn
\subsection{L483mm}
\begin{figure*}[htb]
    \centering
    \includegraphics[scale=0.3]{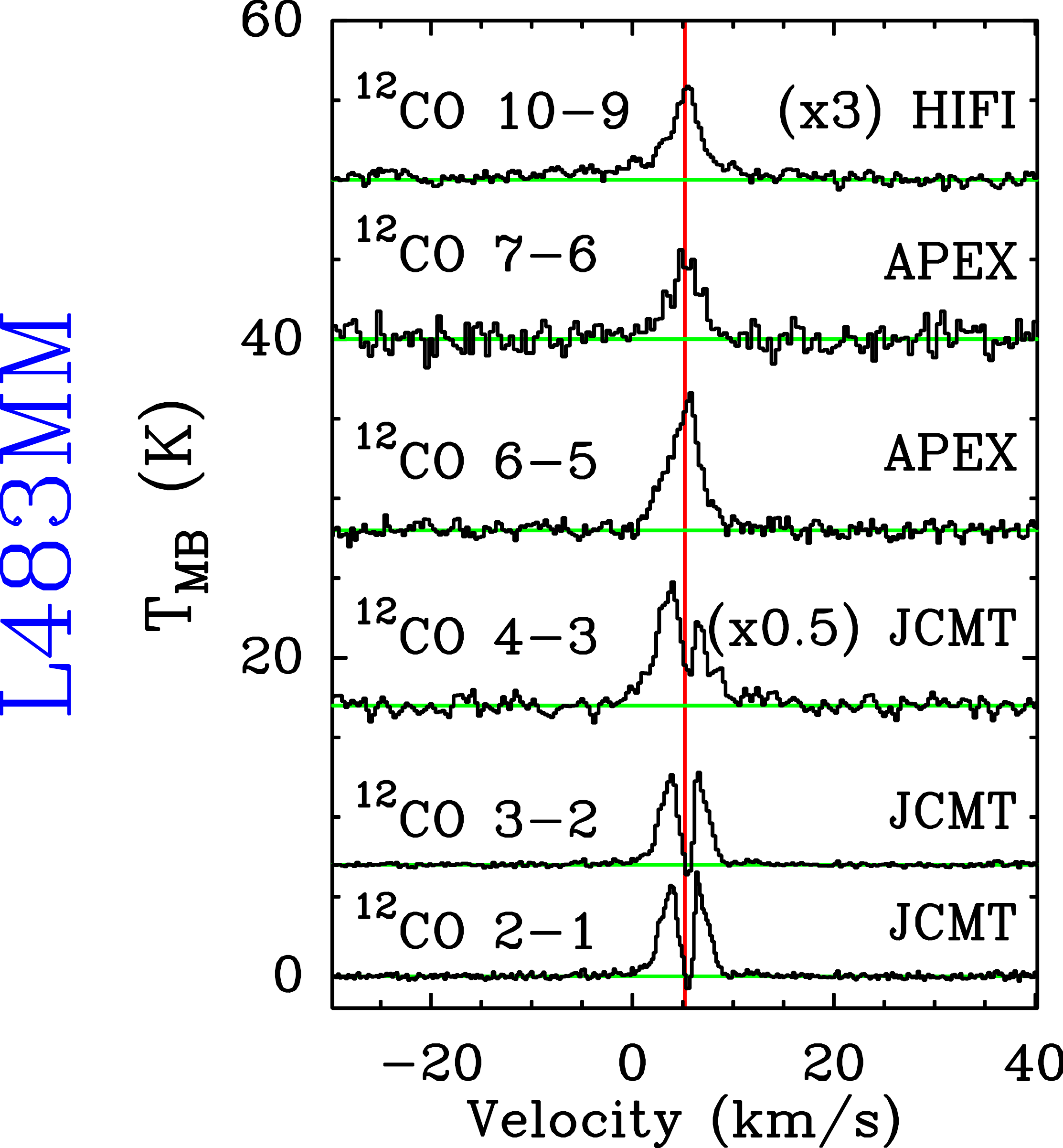}
    \includegraphics[scale=0.3]{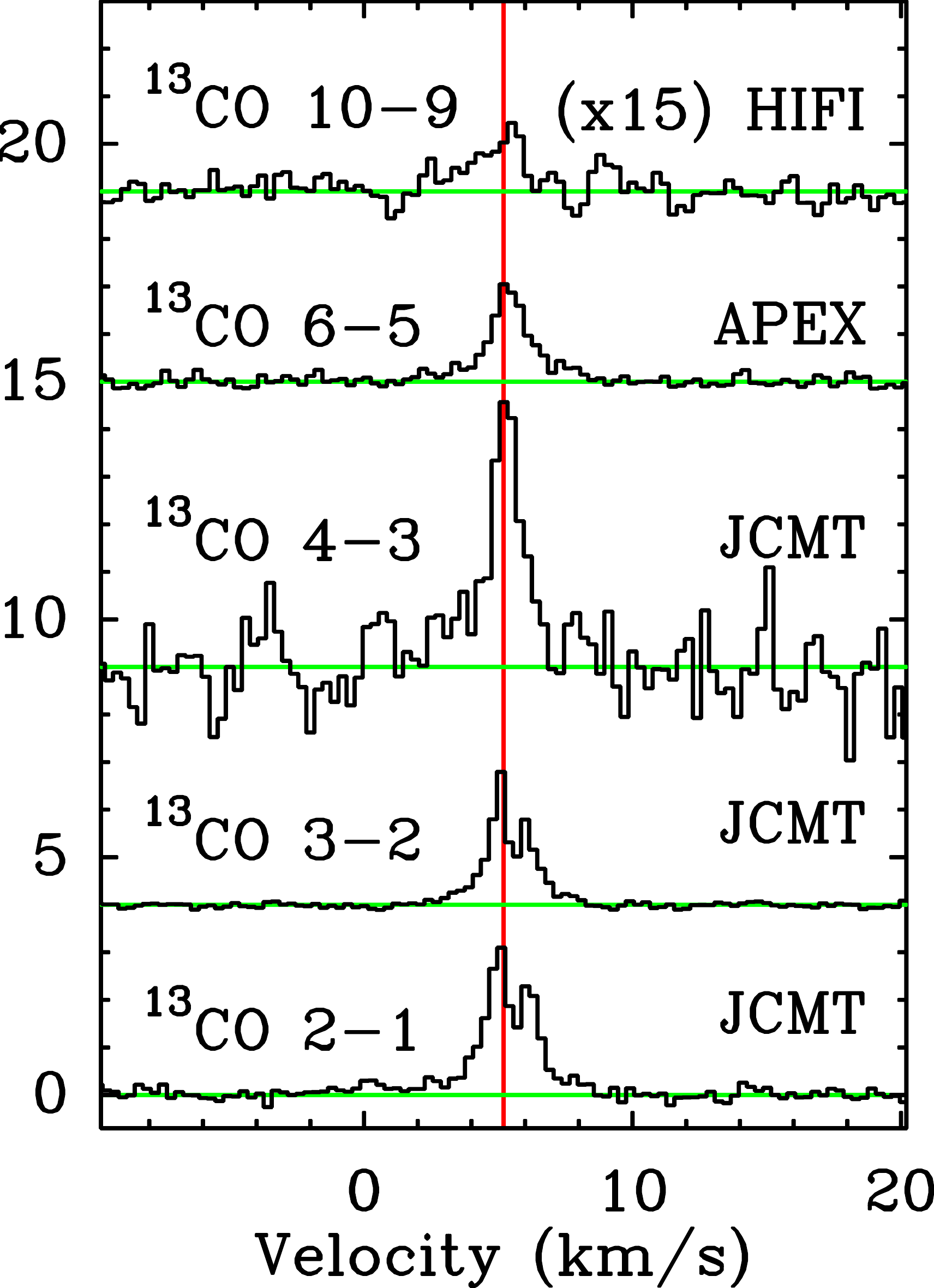}
    \includegraphics[scale=0.3]{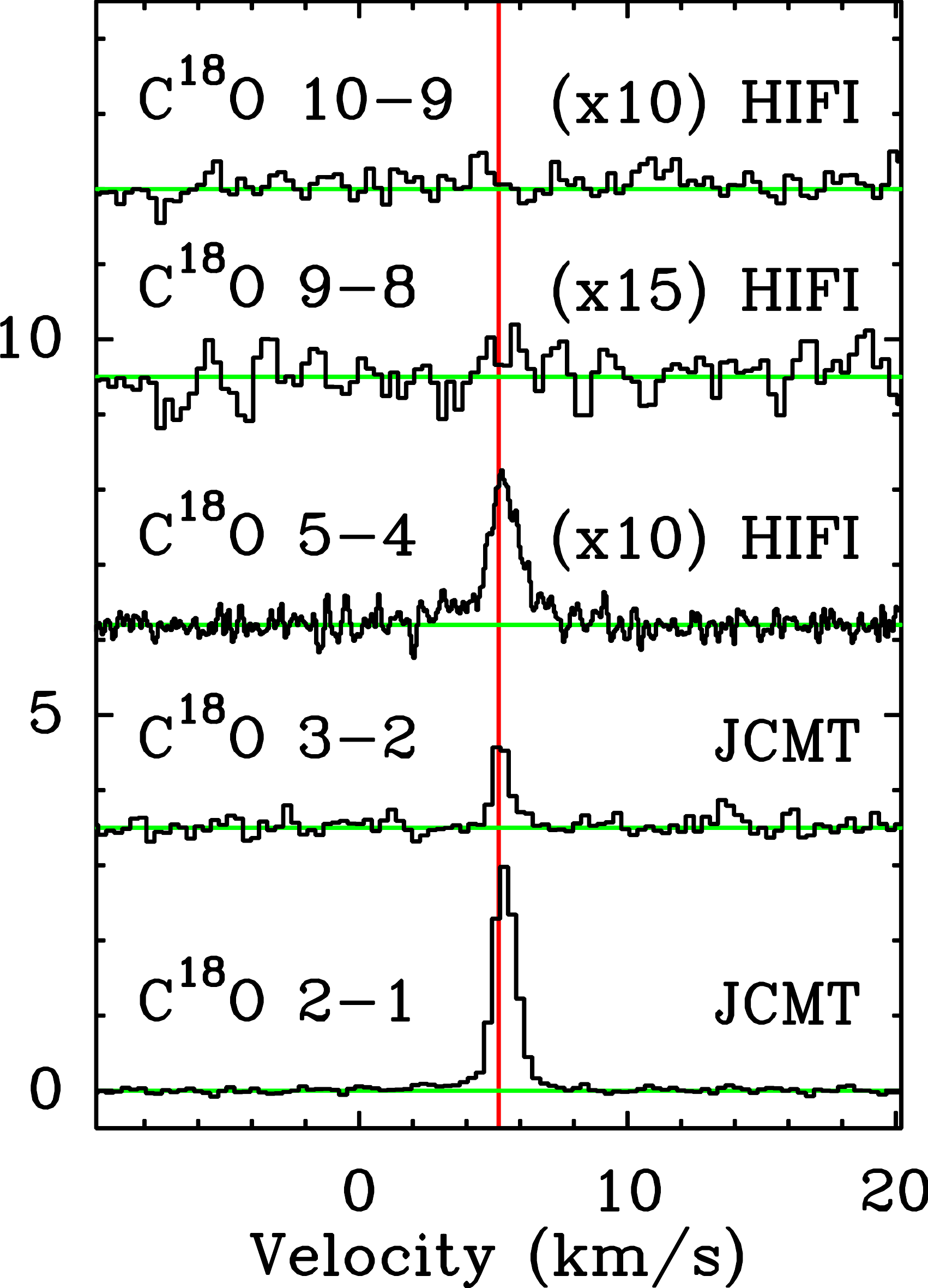}
    \caption{\small Observed $^{12}$CO, $^{13}$CO, and C$^{18}$O transitions for L483mm.}
    \label{fig:linesL483mm}
\end{figure*}

\begin{table*}[!ht]
\caption{Observed line intensities for L483mm in all observed transitions.}
\normalsize
\begin{center}
\begin{tabular}{l l l r r r r r r r r r}
\hline \hline
Mol.  & Transition & Telescope & Efficiency & $\int T_{\rm MB} \mathrm{d}V$ & $T_{\mathrm{peak}}$ & $rms$ \\
 &  & &   $\eta$ &[K km s$^{-1}$] & [K] &   [K]\\
\hline
CO        & 2--1 & JCMT-RxA       & 0.69   &  22.40\phantom{0}  & 6.53\phantom{0}  & 0.13\phantom{0} \\
          & 3--2 & JCMT-HARPB     & 0.63   &  20.98\phantom{0}  & 4.77\phantom{0}  & 0.060 \\
          & 4--3 & JCMT           & 0.38   &  76.09\phantom{0}  & 18.61\phantom{0} & 0.94\phantom{0} \\
          & 6--5 & APEX-CHAMP$^+$ & 0.45   &  32.40\phantom{0}  & 8.92\phantom{0}  & 0.49\phantom{0} \\
          & 7--6 & APEX-CHAMP$^+$ & 0.49   &  22.12\phantom{0}  & 6.29\phantom{0}  & 1.11\phantom{0} \\
          &10--9 & {\it Herschel}-HIFI\tablefootmark{a} & 0.64 & 11.39\phantom{0} & 2.00\phantom{0} & 0.10\phantom{0} \\
$^{13}$CO & 2--1 & JCMT-RxA       & 0.74   &  6.98\phantom{0}   & 3.23\phantom{0}  & 0.17\phantom{0} \\
          & 3--2 & JCMT-HARPB     & 0.63   &  4.28\phantom{0}   & 2.94\phantom{0}  & 0.071 \\
          & 4--3 & JCMT\tablefootmark{b}   &  0.38   &  10.87\phantom{0}    &  6.64\phantom{0} & 0.93\phantom{0} \\
          & 6--5 & APEX-CHAMP$^+$ & 0.46   &  4.17\phantom{0}   &  2.16\phantom{0}  & 0.13\phantom{0} \\
          & 10--9&  {\it Herschel}-HIFI\tablefootmark{a}       & 0.74   &  0.20\phantom{0} & 0.10\phantom{0}  & 0.017 \\
C$^{18}$O & 2--1 & JCMT-RxA       & 0.69   &  3.52\phantom{0}   & 3.10\phantom{0}  & 0.042 \\
          & 3--2 & JCMT-HARPB     & 0.63   &  1.20\phantom{0}   & 1.39\phantom{0}  & 0.15\phantom{0} \\
          & 5--4 & {\it Herschel}-HIFI\tablefootmark{c} & 0.76 & 0.34\phantom{0}  & 0.20\phantom{0}  & 0.012 \\
          & 9--8 & {\it Herschel}-HIFI\tablefootmark{a} & 0.74 & $<$0.050 &  \dots\phantom{0}  & 0.022 \\
          &10--9 & {\it Herschel}-HIFI\tablefootmark{a} & 0.74 & $<$0.046 &  \dots\phantom{0}  & 0.020 \\
\hline 
\end{tabular}
\end{center}
\tablefoot{
\tablefoottext{a}{H- and V-polarization observations averaged.}
\tablefoottext{b}{Taken in 11$\arcsec$ beam.}
\tablefoottext{c}{Only H-polarization observation is used.}
}
\label{tbl:linesL483mm}
\end{table*}

\newpage
\clearpage

\onecolumn
\subsection{SMM1}
\begin{figure*}[htb]
    \centering
    \includegraphics[scale=0.3]{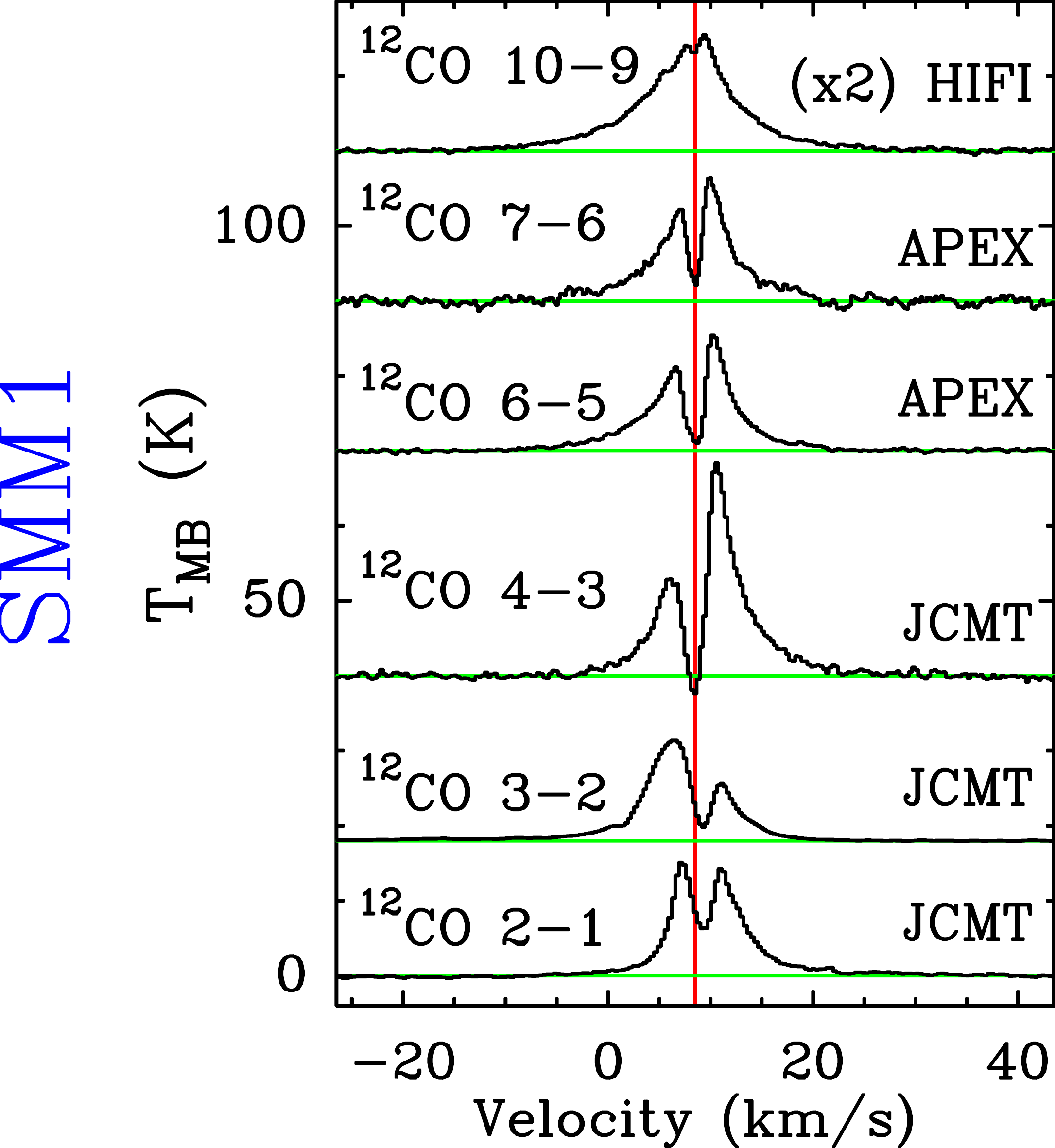}
    \includegraphics[scale=0.3]{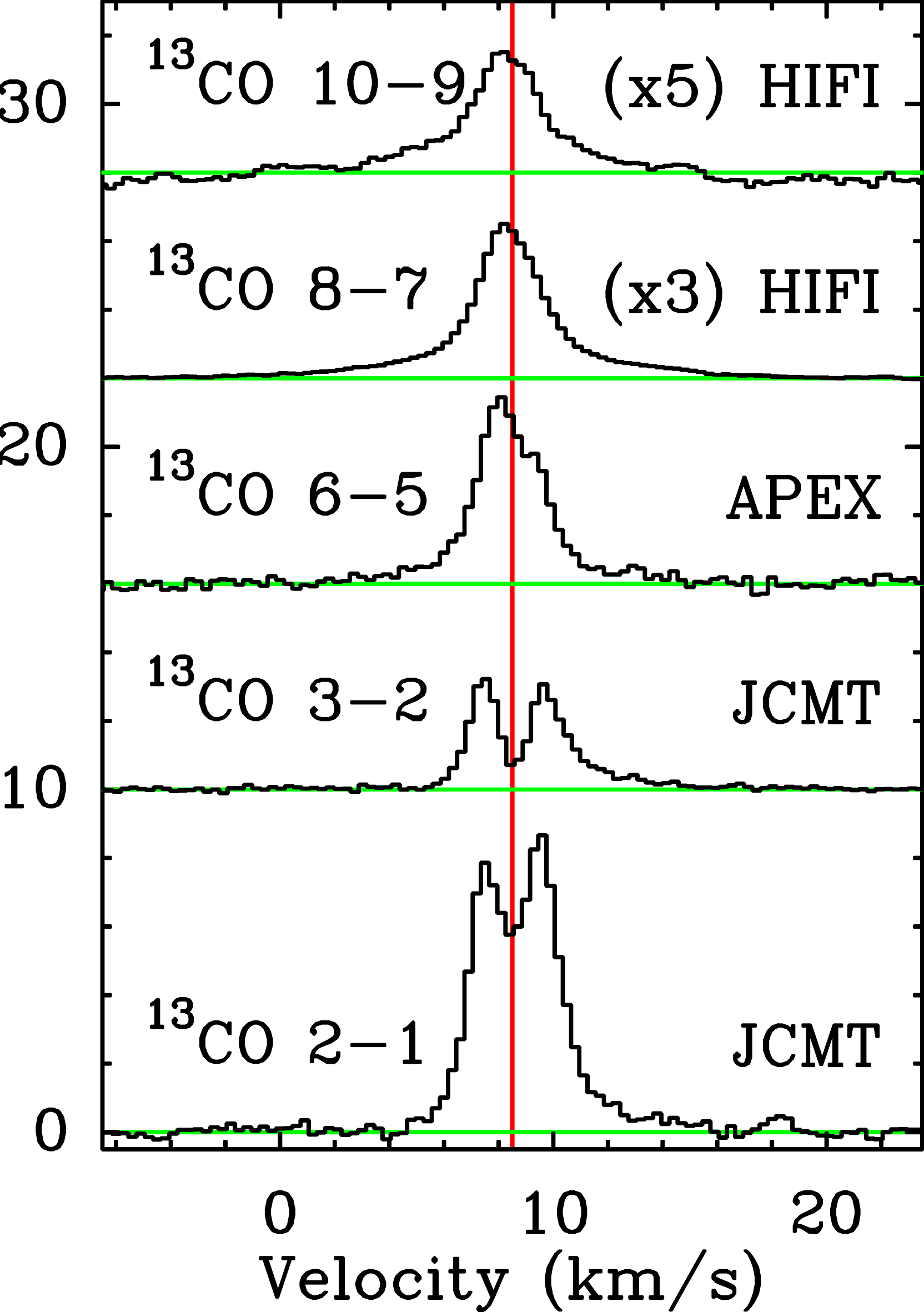}
    \includegraphics[scale=0.3]{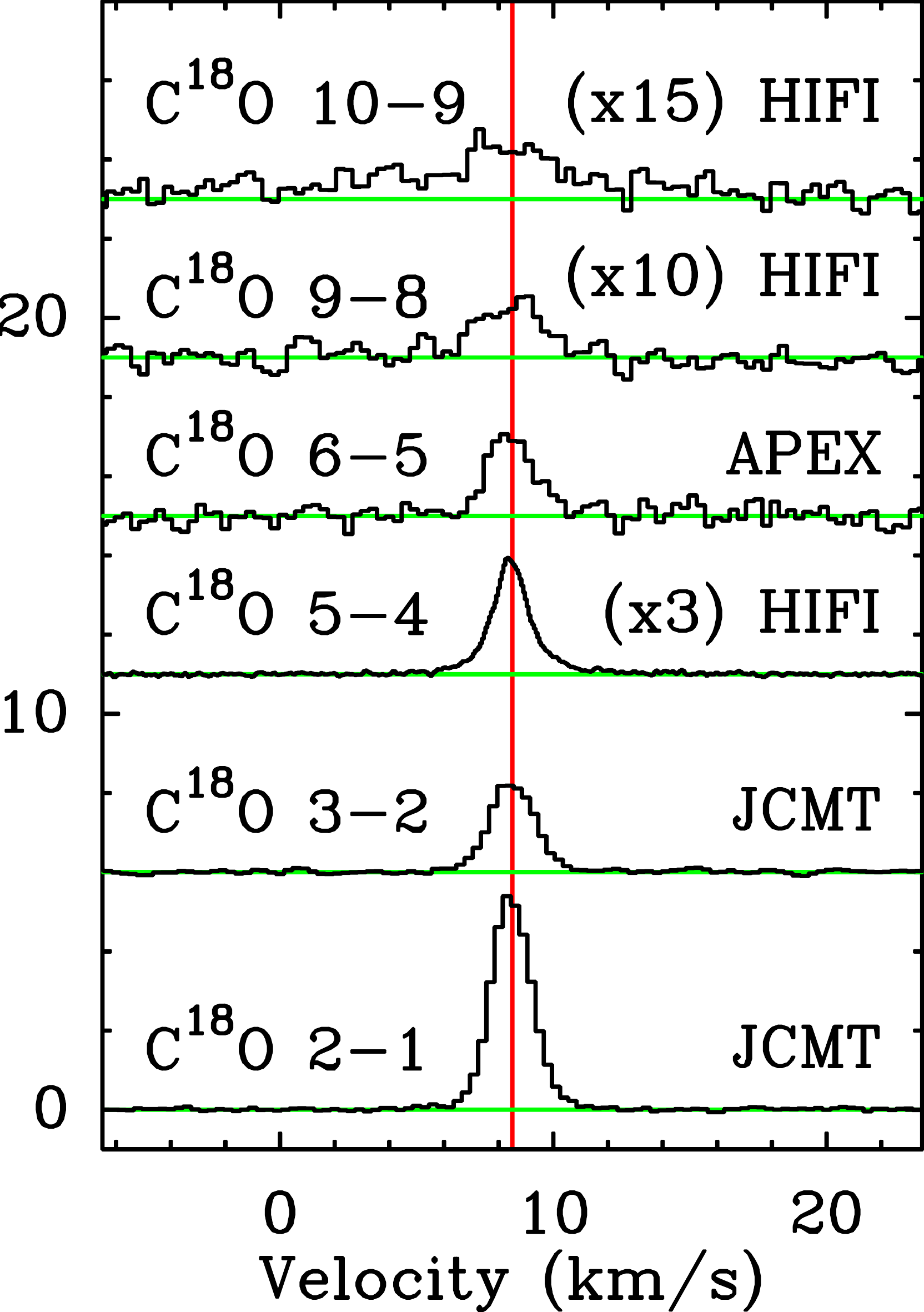}
    \caption{\small Observed $^{12}$CO, $^{13}$CO, and C$^{18}$O transitions for SerSMM1.}
    \label{fig:linesSMM1}
\end{figure*}

\begin{table*}[!ht]
\caption{Observed line intensities for SerSMM1 in all observed transitions.}
\normalsize
\begin{center}
\begin{tabular}{l l l r r r r r r r r r}
\hline \hline
Mol.  & Transition & Telescope & Efficiency & $\int T_{\rm MB} \mathrm{d}V$ & $T_{\mathrm{peak}}$ & $rms$ \\
 &       &      &   $\eta$   &[K km s$^{-1}$]           & [K]            & [K]\\
\hline
CO        & 2--1 & JCMT-RxA         & 0.69   &  112.49\phantom{0} & 15.09\phantom{0} & 0.19\phantom{0} \\
          & 3--2 & JCMT-HARPB       & 0.63   &  101.82\phantom{0} & 13.45\phantom{0} & 0.023 \\
          & 4--3 & JCMT             & 0.38   &  159.90\phantom{0} & 29.03\phantom{0} & 0.39\phantom{0} \\
          & 6--5 & APEX-CHAMP$^+$   & 0.52   &  108.36\phantom{0} & 15.49\phantom{0} & 0.17\phantom{0} \\
          & 7--6 & APEX-CHAMP$^+$   & 0.40   &  105.61\phantom{0} & 16.41\phantom{0} & 0.59\phantom{0} \\
          &10--9 & {\it Herschel}-HIFI\tablefootmark{a} & 0.64    & 82.28\phantom{0} & 7.74\phantom{0} & 0.09\phantom{0} \\
$^{13}$CO & 2--1 & JCMT-RxA         & 0.74   &  31.78\phantom{0}  & 8.87\phantom{0}  & 0.17\phantom{0} \\
          & 3--2 & JCMT-HARPB       & 0.63   &  11.18\phantom{0}  & 3.44\phantom{0}  & 0.074 \\
          & 6--5 & APEX-CHAMP$^+$   & 0.46   &  17.53\phantom{0}  & 5.58\phantom{0}  & 0.18\phantom{0} \\
          & 8--7 & {\it Herschel}-HIFI\tablefootmark{b}  & 0.75   & 5.93\phantom{0}  & 1.49\phantom{0}& 0.007 \\
          & 10--9& {\it Herschel}-HIFI\tablefootmark{b}  & 0.74   & 3.44\phantom{0}  & 0.66\phantom{0} & 0.027 \\
C$^{18}$O & 2--1 & JCMT-RxA         & 0.69   &  10.31\phantom{0}  & 5.49\phantom{0}  & 0.06\phantom{0} \\
          & 3--2 & JCMT-HARPB       & 0.63   &  4.83\phantom{0}   & 2.20\phantom{0}  & 0.049 \\
          & 5--4 & {\it Herschel}-HIFI\tablefootmark{b}    & 0.76 & 1.80\phantom{0}  & 0.97\phantom{0} & 0.010 \\
          & 6--5 & APEX-CHAMP$^+$   & 0.56   &  4.75\phantom{0}   & 2.10\phantom{0}  & 0.26\phantom{0} \\
          & 9--8 & {\it Herschel}-HIFI\tablefootmark{a}   & 0.74  & 0.44\phantom{0}  & 0.18\phantom{0} & 0.019 \\
          &10--9 & {\it Herschel}-HIFI\tablefootmark{a}   & 0.74  & 0.59\phantom{0}  & 0.12\phantom{0} & 0.017 \\
\hline 
\end{tabular}
\end{center}
\tablefoot{
\tablefoottext{a}{H- and V-polarization observations averaged.}
\tablefoottext{b}{Only H-polarization observation is used.}
}
\label{tbl:linesSMM1}
\end{table*}

\newpage

\onecolumn
\subsection{SMM4}
\begin{figure*}[htb]
    \centering
    \includegraphics[scale=0.3]{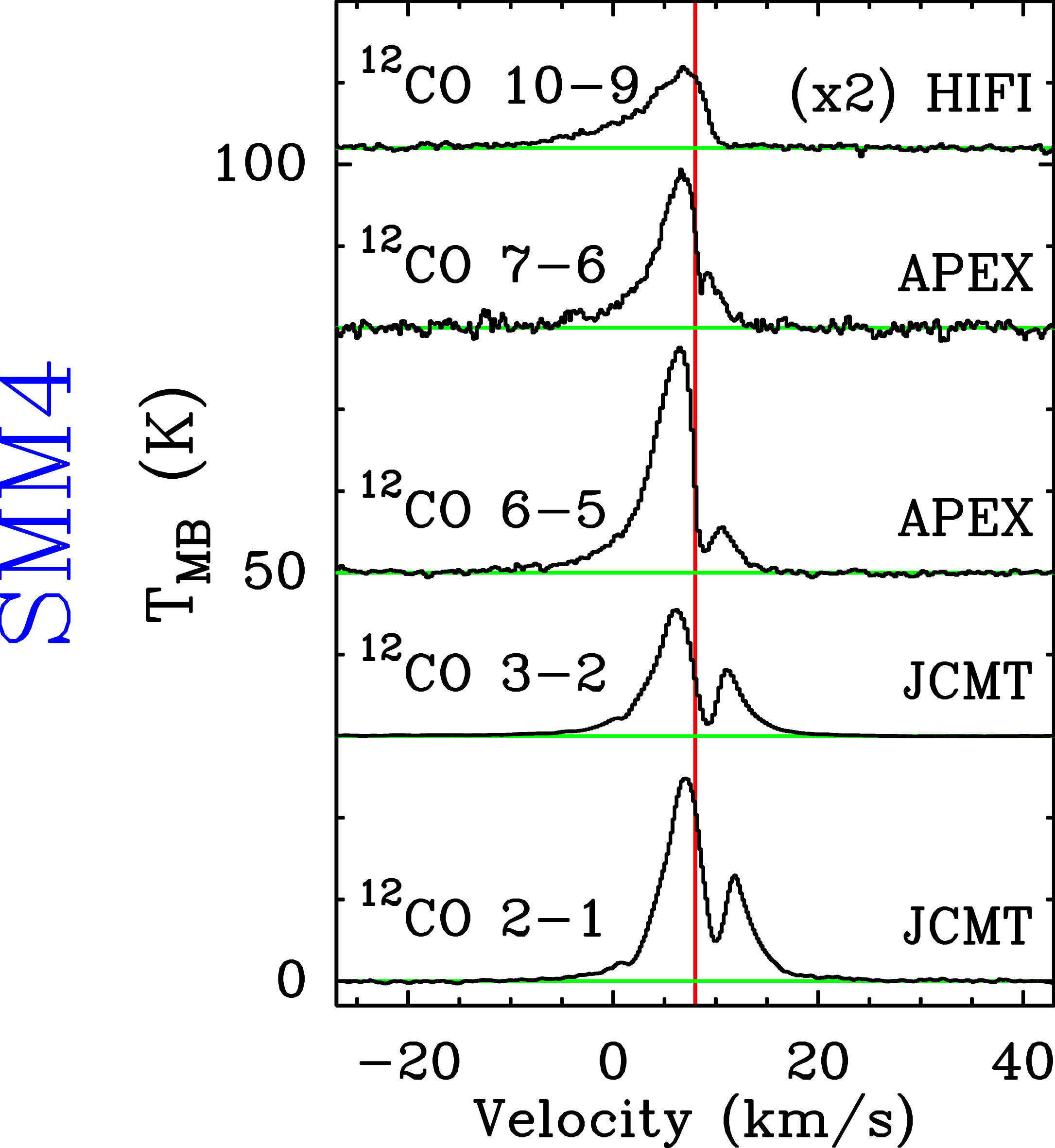}
    \includegraphics[scale=0.3]{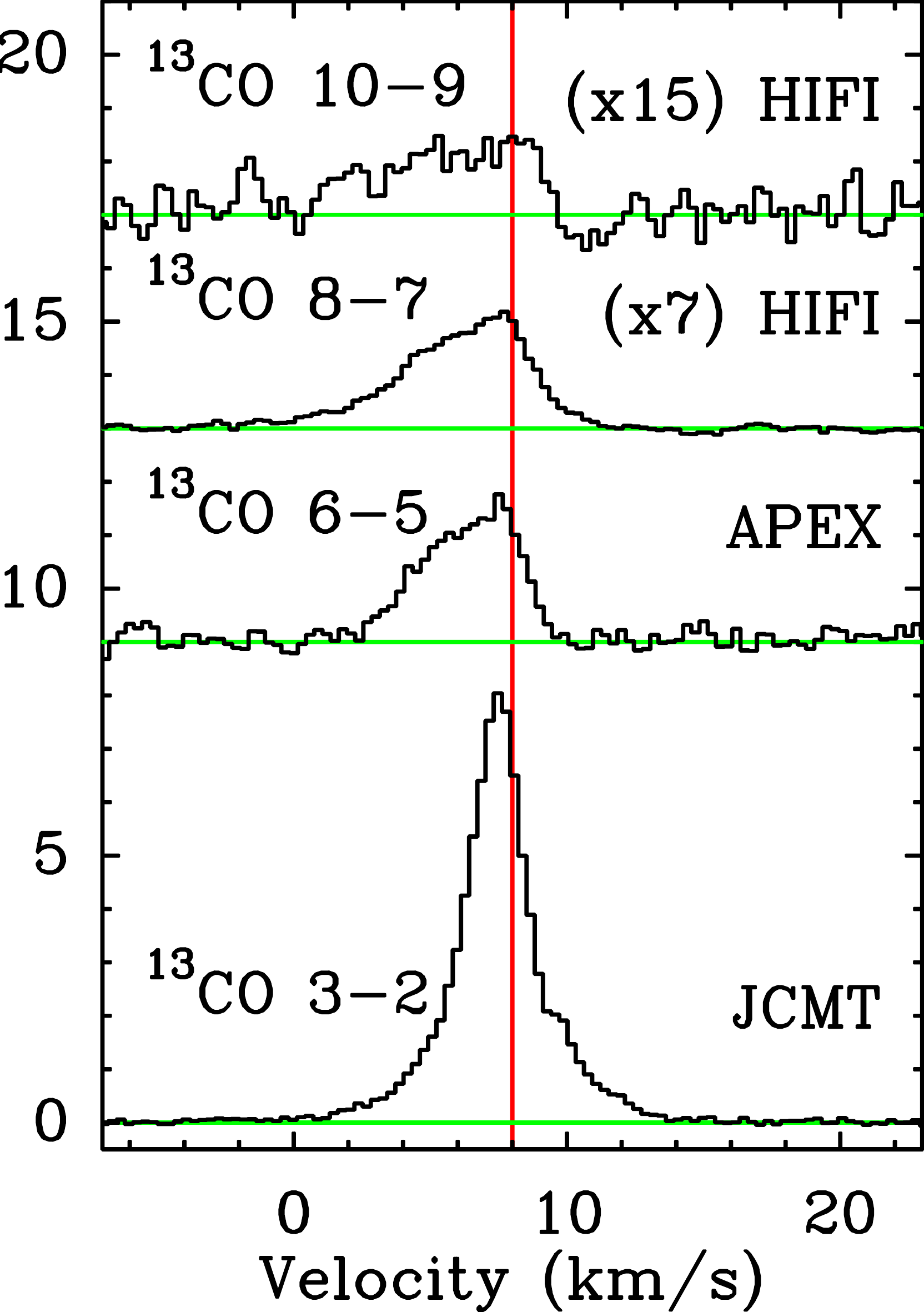}
    \includegraphics[scale=0.3]{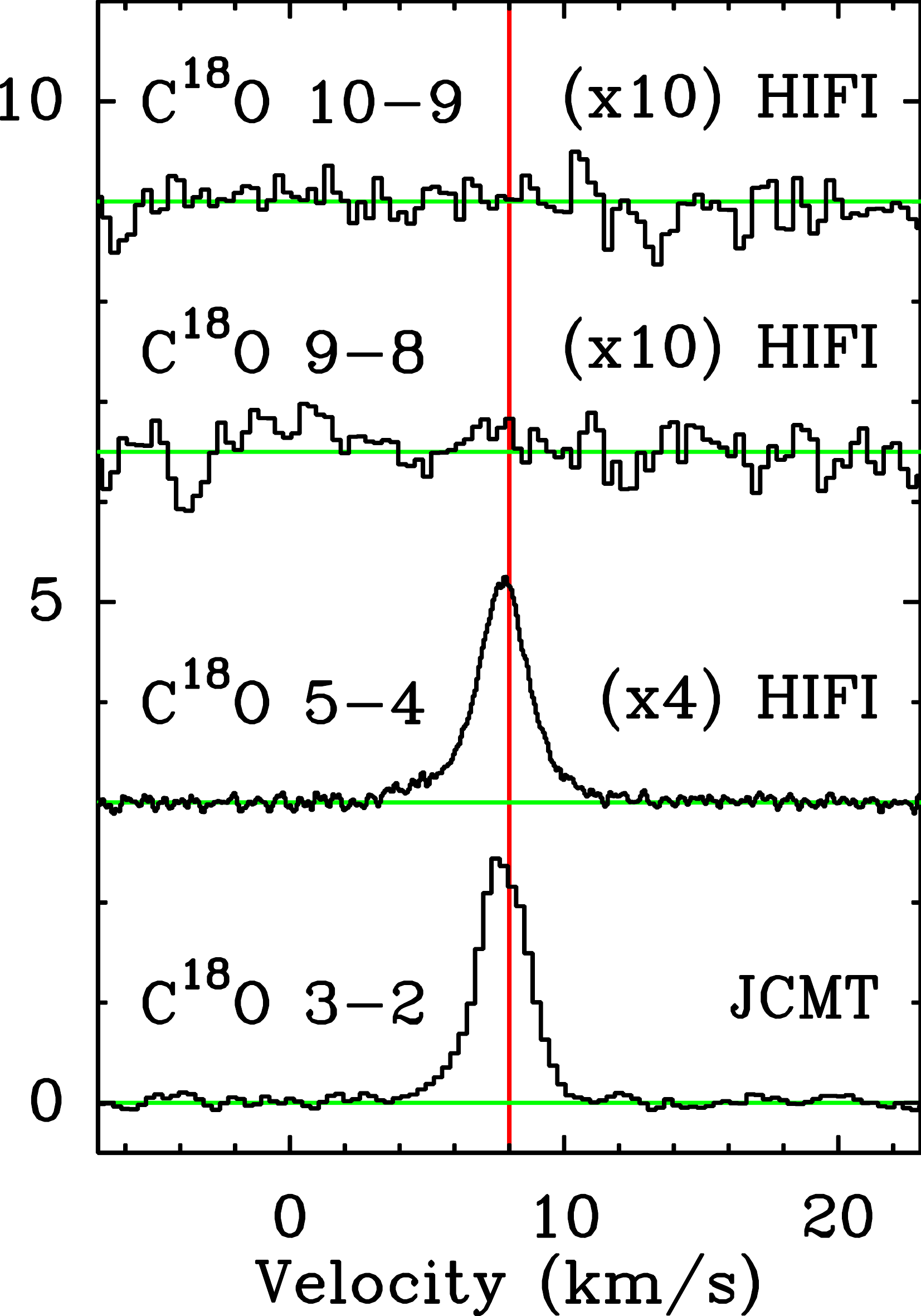}
    \caption{\small Observed $^{12}$CO, $^{13}$CO, and C$^{18}$O transitions for SerSMM4.}
    \label{fig:linesSMM4}
\end{figure*}

\begin{table*}[!ht]
\caption{Observed line intensities for SerSMM4 in all observed transitions.}
\normalsize
\begin{center}
\begin{tabular}{l l l r r r r r r r r r}
\hline \hline
Mol.  & Transition & Telescope & Efficiency & $\int T_{\rm MB} \mathrm{d}V$ & $T_{\mathrm{peak}}$ & $rms$ \\
 &  & &   $\eta$ &[K km s$^{-1}$] & [K] &   [K]\\
\hline
CO        & 2--1 & JCMT-RxA       & 0.69 & 168.38\phantom{0} & 24.80\phantom{0}  & 0.10\phantom{0} \\
          & 3--2 & JCMT-HARPB     & 0.63 & 108.35\phantom{0} & 15.48\phantom{0}  & 0.022 \\
          & 6--5 & APEX-CHAMP$^+$ & 0.52 & 159.61\phantom{0} & 27.56\phantom{0}  & 0.23\phantom{0} \\
          & 7--6 & APEX-CHAMP$^+$ & 0.40 & 110.08\phantom{0} & 19.39\phantom{0}  & 0.63\phantom{0} \\
          &10--9 & {\it Herschel}-HIFI\tablefootmark{a}      & 0.64 & 40.24\phantom{0} & 4.98\phantom{0}  & 0.13\phantom{0} \\
$^{13}$CO & 3--2 & JCMT-HARPB     & 0.63 & 26.08\phantom{0}  & 8.10\phantom{0}   & 0.048 \\
          & 6--5 & APEX-CHAMP$^+$ & 0.52 & 10.65\phantom{0}  & 2.61\phantom{0}   & 0.15\phantom{0} \\
          & 8--7 & {\it Herschel}-HIFI\tablefootmark{b}      & 0.75 & 1.62\phantom{0} & 0.31\phantom{0} & 0.009 \\
          & 10--9& {\it Herschel}-HIFI\tablefootmark{b}      & 0.74 & 0.52\phantom{0} & 0.10\phantom{0}  & 0.020 \\
C$^{18}$O & 3--2 & JCMT-HARPB     & 0.63 & 6.02\phantom{0}   & 2.52\phantom{0} & 0.061 \\
          & 5--4 & {\it Herschel}-HIFI\tablefootmark{b} & 0.76 & 1.41\phantom{0} & 0.53\phantom{0}  & 0.010 \\
          & 9--8 & {\it Herschel}-HIFI\tablefootmark{b} & 0.74 & $<$0.055 & \dots\phantom{0}  & 0.024 \\
          &10--9 & {\it Herschel}-HIFI\tablefootmark{b} & 0.74 & $<$0.048 & \dots\phantom{0}  & 0.021 \\
\hline 
\end{tabular}
\end{center}
\tablefoot{
\tablefoottext{a}{Only H-polarization observation is used.}
\tablefoottext{b}{H- and V-polarization observations averaged.}
}
\label{tbl:linesSMM4}
\end{table*}

\newpage

\onecolumn
\subsection{SMM3}
\begin{figure*}[htb]
    \centering
    \includegraphics[scale=0.3]{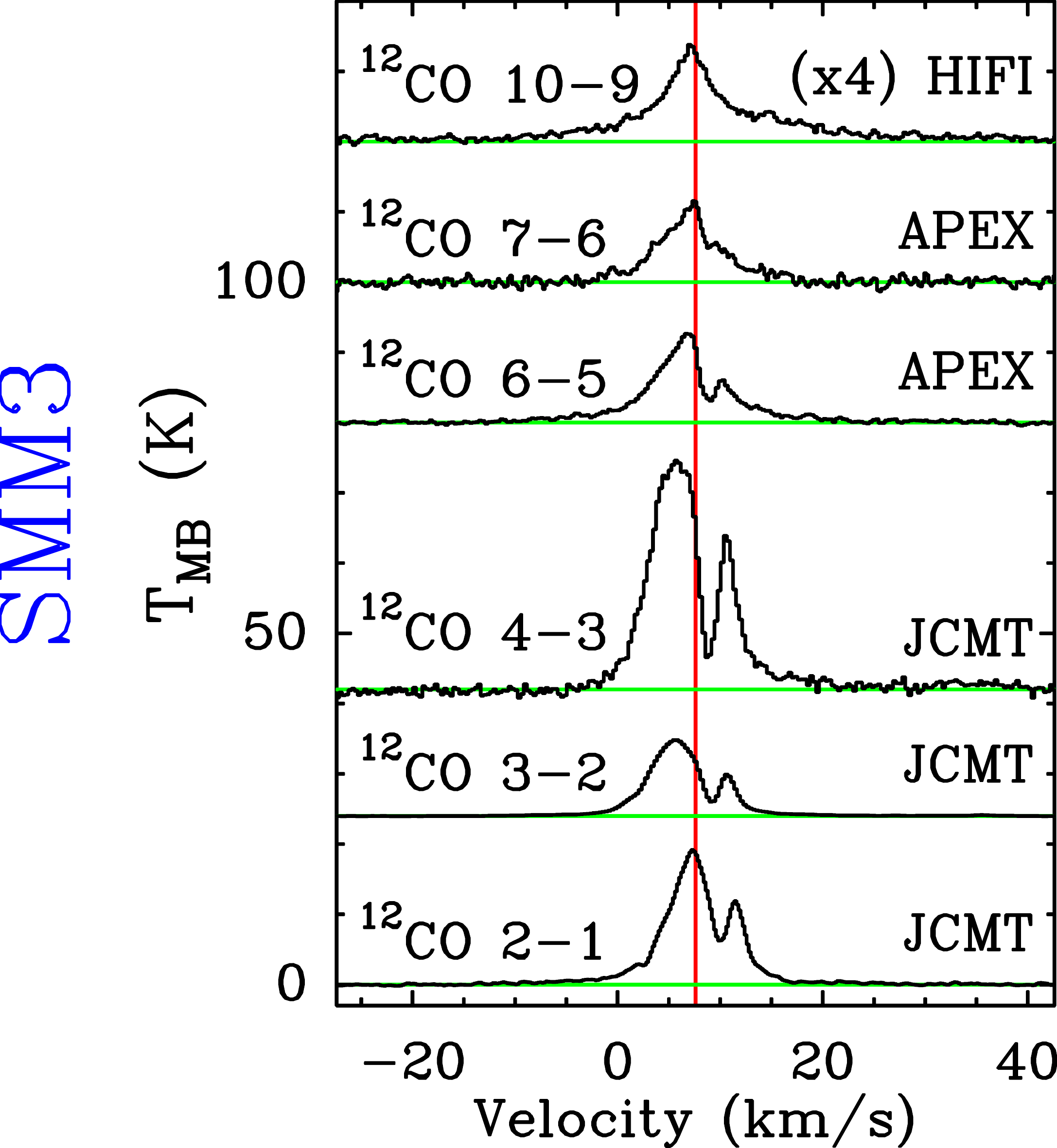}
    \includegraphics[scale=0.3]{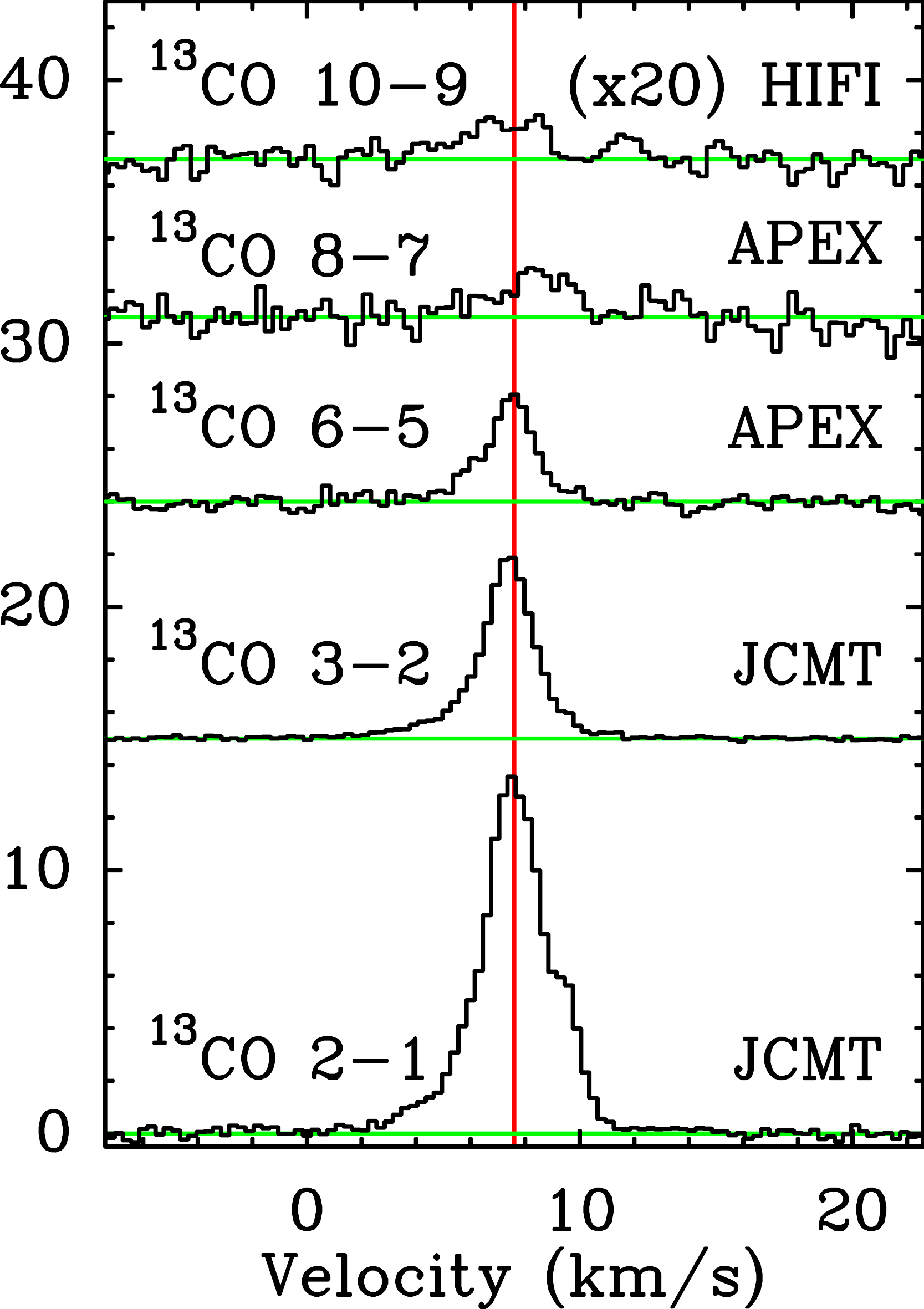}
    \includegraphics[scale=0.3]{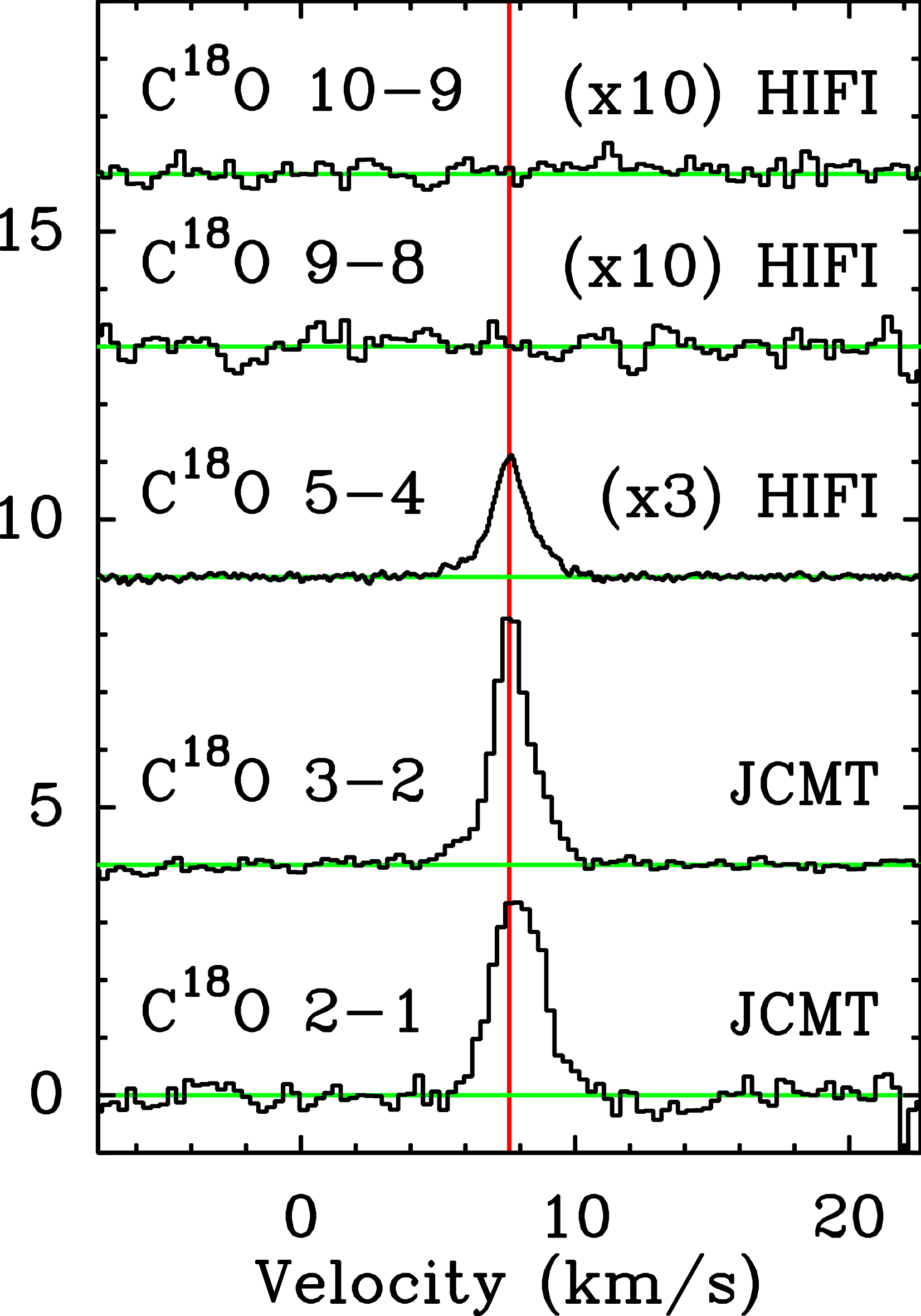}
    \caption{\small Observed $^{12}$CO, $^{13}$CO, and C$^{18}$O transitions for SerSMM3.}
    \label{fig:linesSMM3}
\end{figure*}

\begin{table*}[!ht]
\caption{Observed line intensities for SerSMM3 in all observed transitions.}
\normalsize
\begin{center}
\begin{tabular}{l l l r r r r r r r r r}
\hline \hline
Mol.  & Transition & Telescope & Efficiency & $\int T_{\rm MB} \mathrm{d}V$ & $T_{\mathrm{peak}}$ & $rms$ \\
 &  & &   $\eta$ &[K km s$^{-1}$] & [K] &   [K]\\
\hline
CO        & 2--1 & JCMT-RxA         & 0.69   &  132.99\phantom{0} & 19.15\phantom{0} & 0.11\phantom{0} \\
          & 3--2 & JCMT-HARPB       & 0.63   &  78.14\phantom{0}  & 10.83\phantom{0} & 0.034 \\
          & 4--3  & JCMT            & 0.38   &  242.95\phantom{0} & 32.85\phantom{0} & 0.79\phantom{0} \\
          & 6--5 & APEX-CHAMP$^+$   & 0.52   &  97.56\phantom{0}  & 12.73\phantom{0} & 0.22\phantom{0} \\
          & 7--6  & APEX-CHAMP$^+$  & 0.40   &  74.26\phantom{0}  & 11.54\phantom{0} & 0.55\phantom{0} \\
          &10--9 & {\it Herschel}-HIFI\tablefootmark{a} & 0.64   & 35.71\phantom{0} &  3.49\phantom{0}  & 0.095 \\
$^{13}$CO & 2--1 & JCMT-RxA         & 0.74   &  44.50\phantom{0}  & 13.65\phantom{0} & 0.18\phantom{0} \\
          & 3--2 & JCMT-HARPB       & 0.63   &  18.30\phantom{0}  & 7.06\phantom{0}  & 0.072 \\
          & 6--5 & APEX-CHAMP$^+$   & 0.52   &  9.59\phantom{0}   & 4.12\phantom{0}  & 0.26\phantom{0} \\
          & 8--7  & APEX-CHAMP$^+$  & 0.49   &  5.59\phantom{0}   & 2.28\phantom{0}  & 0.61\phantom{0} \\
          & 10--9& {\it Herschel}-HIFI\tablefootmark{a} & 0.74   & 0.33\phantom{0}  &  0.092 & 0.02\phantom{0} \\
C$^{18}$O & 2--1 & JCMT-RxA         & 0.69   &  8.01\phantom{0}   & 3.44\phantom{0}  & 0.32\phantom{0} \\
          & 3--2 & JCMT-HARPB       & 0.63   &  4.95\phantom{0}   & 2.63\phantom{0}  & 0.060 \\
          & 5--4 & {\it Herschel}-HIFI\tablefootmark{b} & 0.76   & 1.29\phantom{0}  &  0.70\phantom{0}  & 0.012 \\
          & 9--8 & {\it Herschel}-HIFI\tablefootmark{a} & 0.74   & $<$0.043         & \dots\phantom{0}  & 0.019 \\
          &10--9 & {\it Herschel}-HIFI\tablefootmark{a} & 0.74   & $<$0.032        & \dots\phantom{0}  & 0.014 \\
\hline 
\end{tabular}
\end{center}
\tablefoot{
\tablefoottext{a}{H- and V-polarization observations averaged.}
\tablefoottext{b}{Only H-polarization observation is used.}
}
\label{tbl:linesSMM3}
\end{table*}

\newpage
\clearpage

\onecolumn
\subsection{L723mm}
\begin{figure*}[htb]
    \centering
    \includegraphics[scale=0.3]{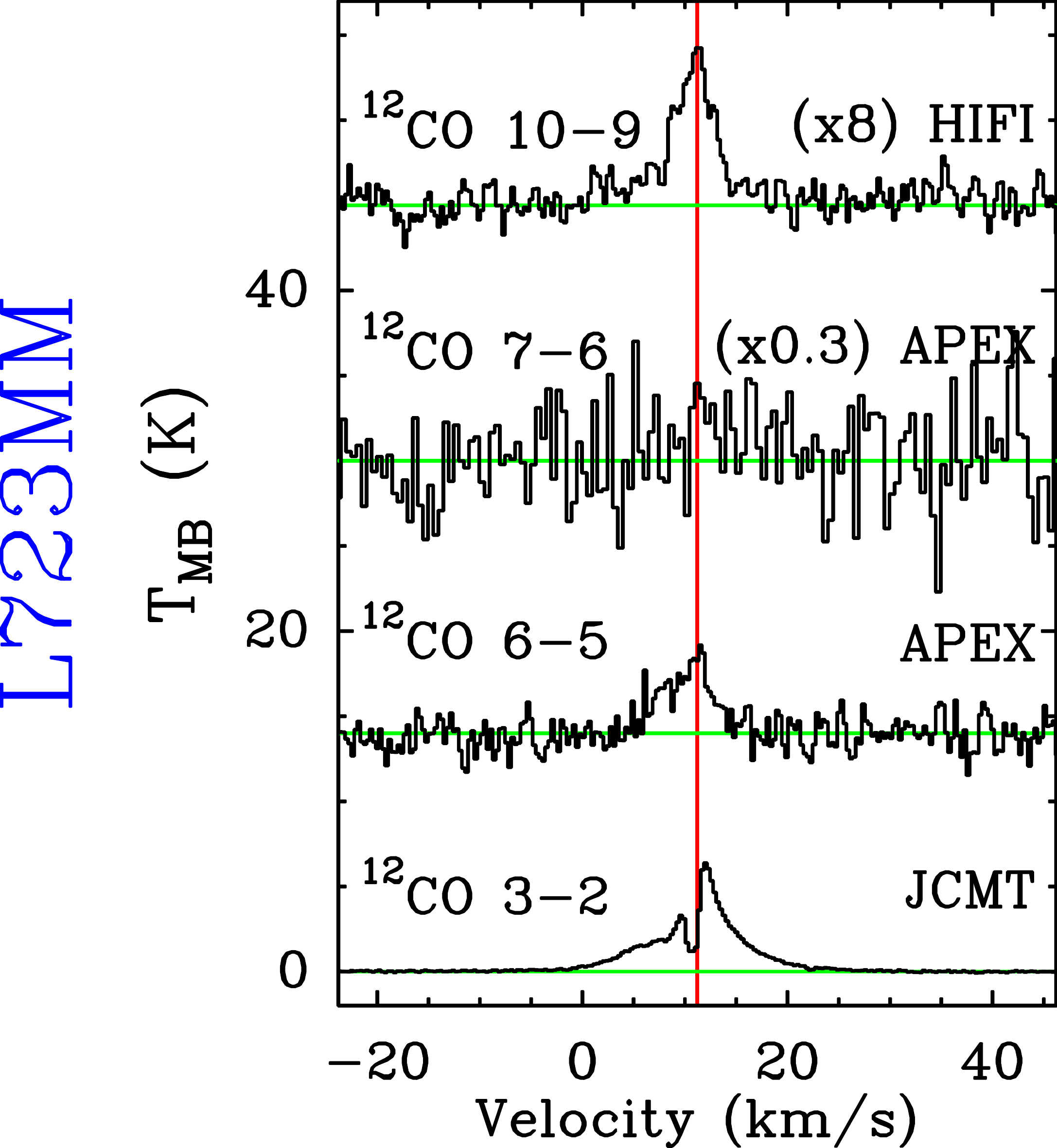}
    \includegraphics[scale=0.3]{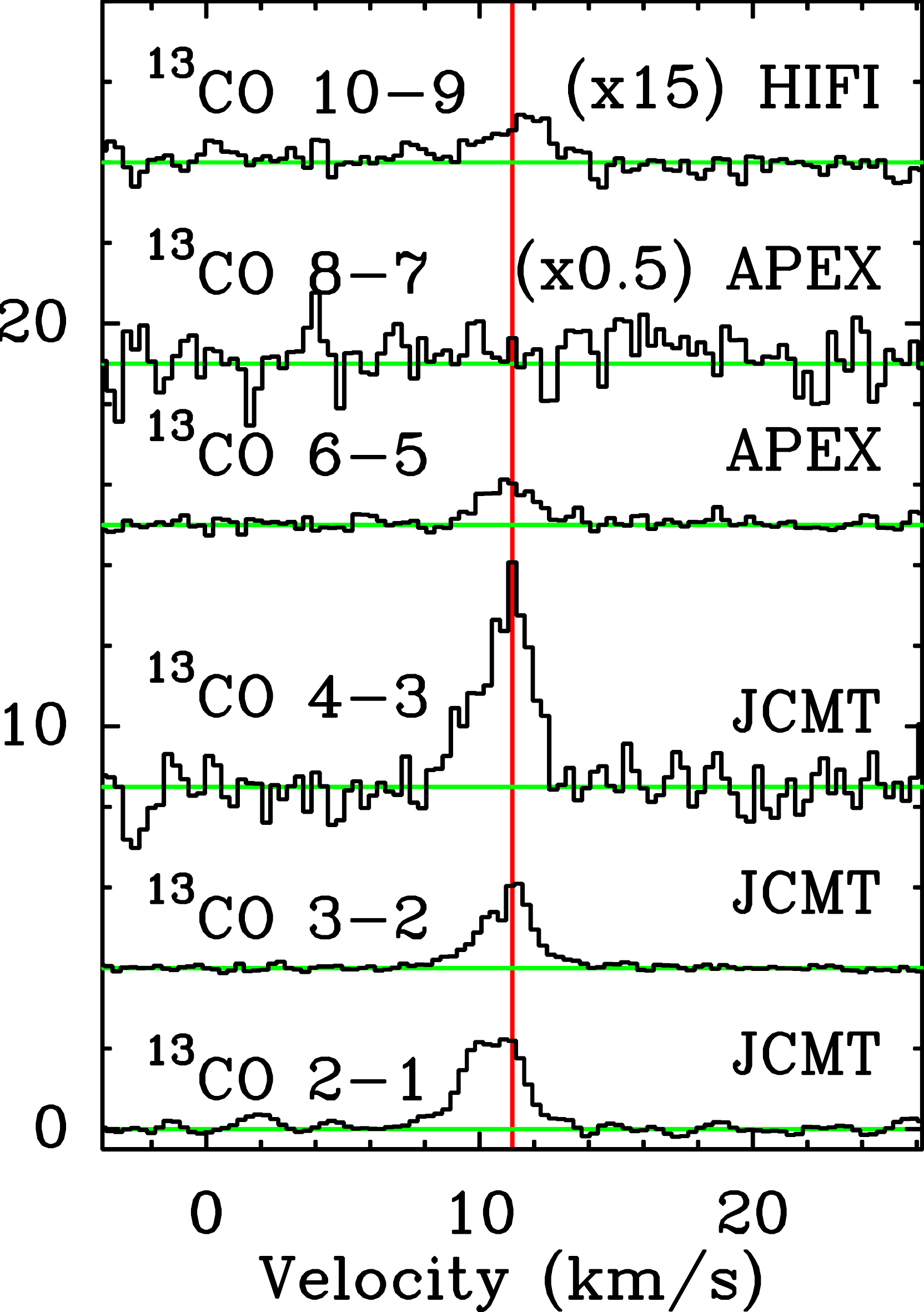}
    \includegraphics[scale=0.3]{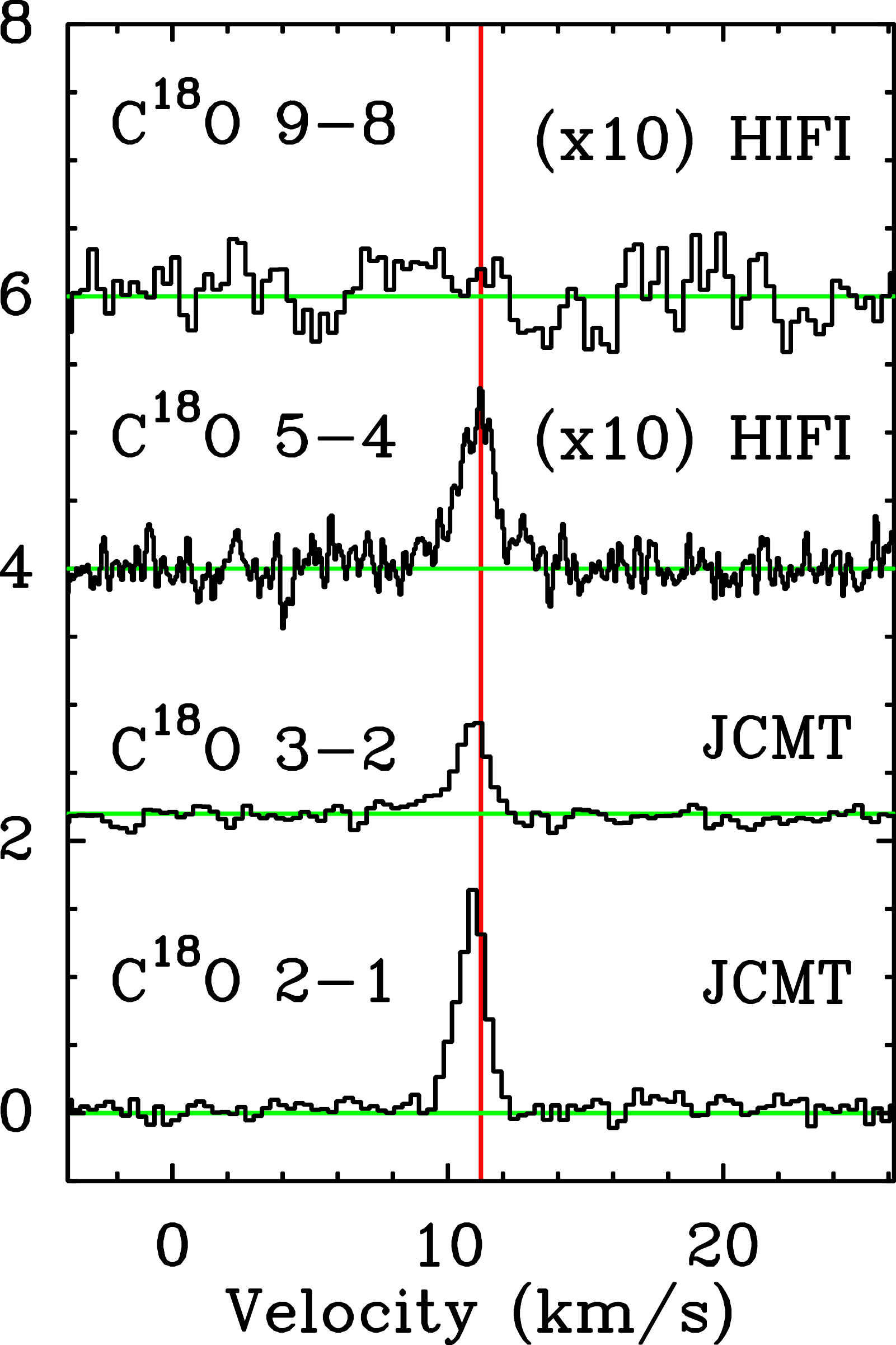}
    \caption{\small Observed $^{12}$CO, $^{13}$CO, and C$^{18}$O transitions for L723mm.}
    \label{fig:linesL723mm}
\end{figure*}

\begin{table*}[!ht]
\caption{Observed line intensities for L723mm in all observed transitions.}
\normalsize
\begin{center}
\begin{tabular}{l l l r r r r r r r r r}
\hline \hline
Mol.  & Transition & Telescope & Efficiency & $\int T_{\rm MB} \mathrm{d}V$ & $T_{\mathrm{peak}}$ & $rms$ \\
 &  & &   $\eta$ &[K km s$^{-1}$] & [K] &   [K]\\
\hline
CO        & 3--2 & JCMT-HARPB     & 0.63   &  39.54\phantom{0}  &  6.39\phantom{0}  & 0.052 \\
          & 6--5 & APEX-CHAMP$^+$ & 0.52   &  22.02\phantom{0}  &  5.44\phantom{0}  & 1.11\phantom{0} \\
          & 7--6 & APEX-CHAMP$^+$ & 0.49   &  11.00\phantom{0}  &  6.00\phantom{0}  & 3.81\phantom{0} \\
          &10--9 & {\it Herschel}-HIFI\tablefootmark{a} & 0.64 & 6.65\phantom{0} & 1.17\phantom{0} & 0.085 \\
$^{13}$CO & 2--1 & JCMT-RxA       & 0.74   &  7.09\phantom{0}   &  2.23\phantom{0}  & 0.10\phantom{0} \\
          & 3--2 & JCMT-HARPB     & 0.63   &  4.79\phantom{0}   &  2.18\phantom{0}  & 0.08\phantom{0} \\
          & 4--3 & JCMT\tablefootmark{b}   & 0.38   &  10.81\phantom{0} &  5.92\phantom{0}  & 0.76\phantom{0} \\
          & 6--5 & APEX-CHAMP$^+$ & 0.45   &  3.33\phantom{0}   &  1.25\phantom{0}  & 0.18\phantom{0} \\
          & 8--7 & APEX-CHAMP$^+$ & 0.49   &  1.94\phantom{0}   &  4.67\phantom{0}  & 1.55\phantom{0} \\
          & 10--9& {\it Herschel}-HIFI\tablefootmark{a} & 0.74 &  0.21\phantom{0}  &  0.13\phantom{0}  & 0.025 \\
C$^{18}$O & 2--1 & JCMT-RxA       & 0.69   &  2.16\phantom{0}   &  1.75\phantom{0}  & 0.077 \\
          & 3--2 & JCMT-HARPB     & 0.63   &  0.96\phantom{0}   &  0.70\phantom{0} & 0.094 \\
          & 5--4 & {\it Herschel}-HIFI\tablefootmark{c} & 0.76 &  0.20\phantom{0}  &  0.14\phantom{0} & 0.019 \\
          & 9--8 & {\it Herschel}-HIFI\tablefootmark{a} & 0.74 &  $<$0.043         & \dots\phantom{0} & 0.019 \\
\hline 
\end{tabular}
\end{center}
\tablefoot{
\tablefoottext{a}{H- and V-polarization observations averaged.}
\tablefoottext{b}{Taken in 11$\arcsec$ beam.}
\tablefoottext{c}{Only H-polarization observation is used.}
}
\label{tbl:linesL723mm}
\end{table*}

\newpage

\onecolumn
\subsection{B335}
\begin{figure*}[!htb]
    \centering
    \includegraphics[scale=0.3]{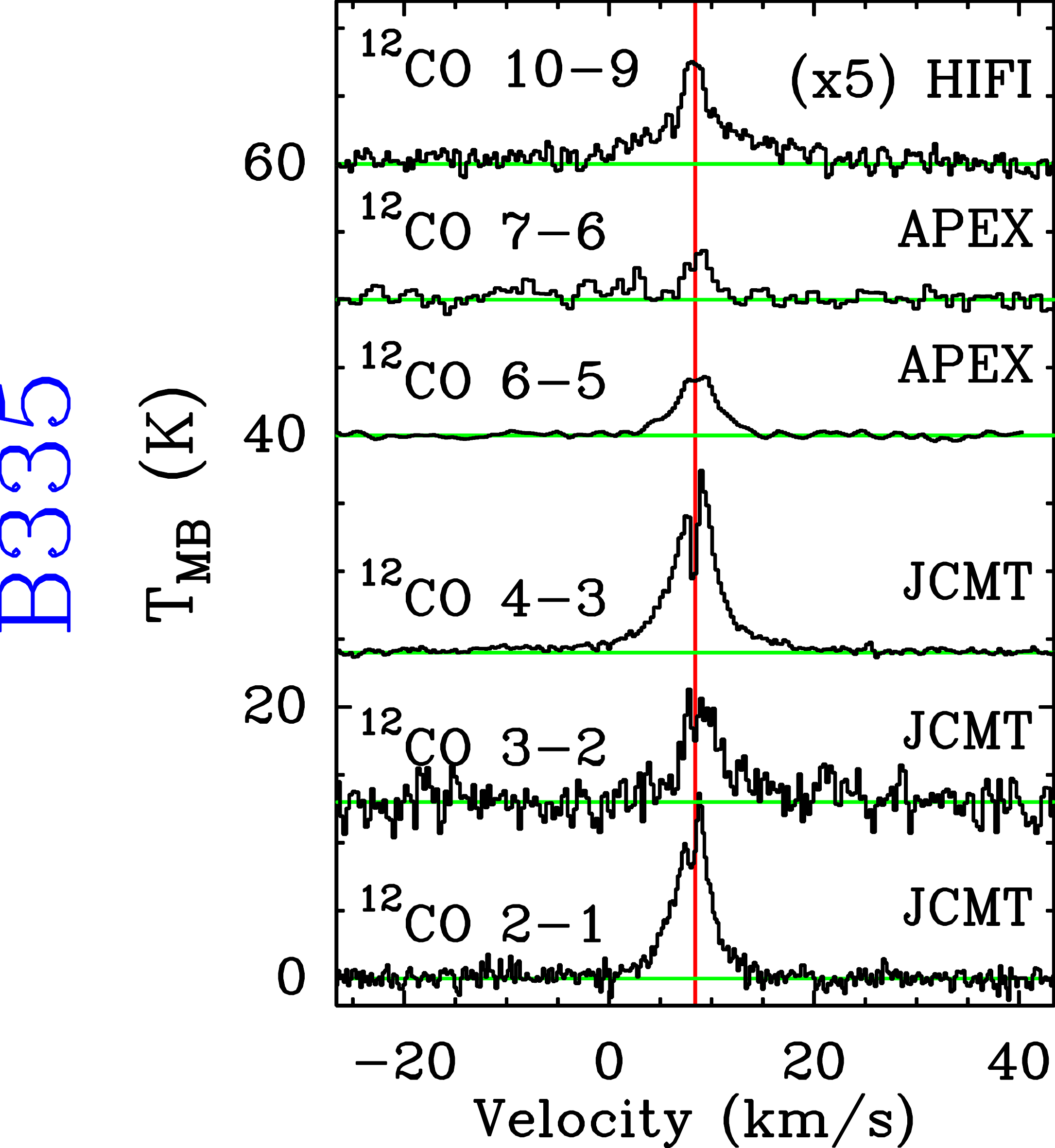}
    \includegraphics[scale=0.3]{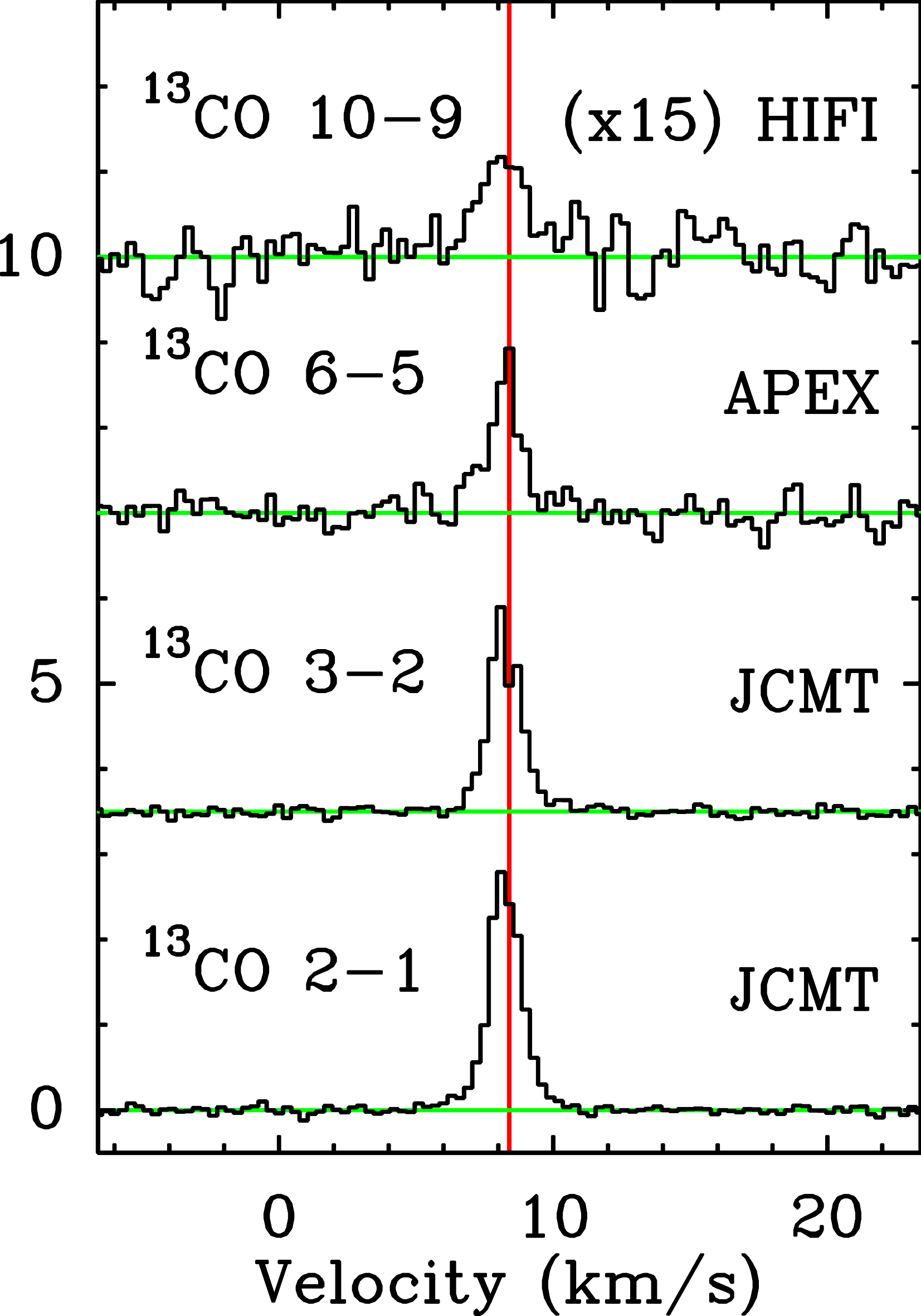}
    \includegraphics[scale=0.3]{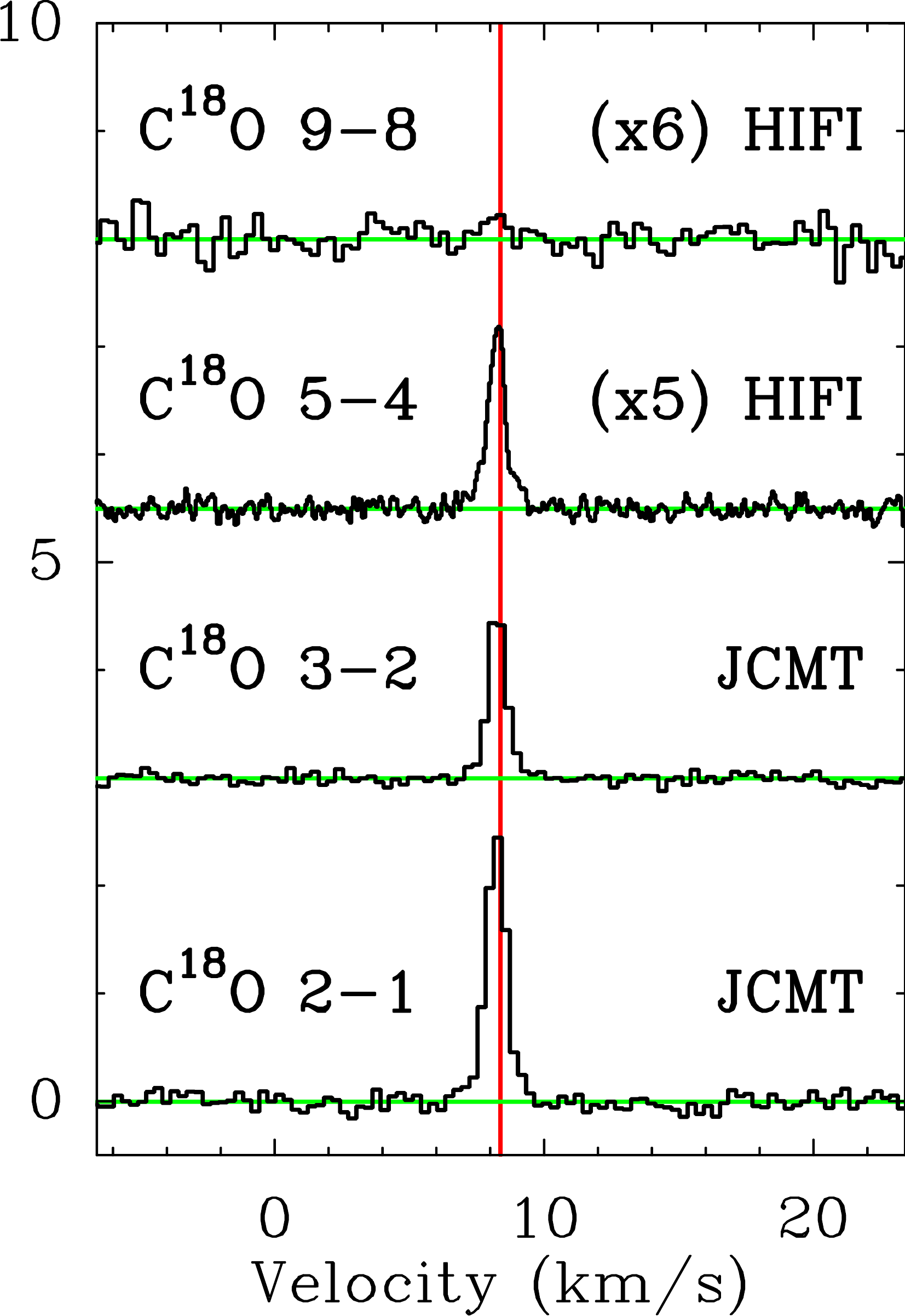}
    \caption{\small Observed $^{12}$CO, $^{13}$CO, and C$^{18}$O transitions for B335.}
    \label{fig:linesB355}
\end{figure*}

\begin{table*}[ht]
\caption{Observed line intensities for B335 in all observed transitions.}
\normalsize
\begin{center}
\begin{tabular}{l l l r r r r r r r r r}
\hline \hline
Mol.  & Transition & Telescope & Efficiency & $\int T_{\rm MB} \mathrm{d}V$ & $T_{\mathrm{peak}}$ & $rms$ \\
 &  & &   $\eta$ &[K km s$^{-1}$] & [K] &   [K]\\
\hline
CO        & 2--1 & JCMT-RxA       & 0.69   &  56.15\phantom{0}  &  13.63\phantom{0} & 0.51\phantom{0} \\
          & 3--2 & JCMT-HARPB     & 0.63   &  38.33\phantom{0}  &  8.69\phantom{0}  & 1.49\phantom{0} \\
          & 4--3 & JCMT           & 0.38   &  68.67\phantom{0}  &  13.41\phantom{0} & 0.15\phantom{0} \\
          & 6--5 & APEX-CHAMP$^+$ & 0.52   &  25.89\phantom{0}  &  4.41\phantom{0}  & 0.23\phantom{0} \\
          & 7--6 & APEX-CHAMP$^+$ & 0.49   &  21.88\phantom{0}  &  3.61\phantom{0}  & 0.58\phantom{0} \\
          &10--9 & {\it Herschel}-HIFI\tablefootmark{a} & 0.64 &  12.49\phantom{0} &  1.50\phantom{0}  & 0.13\phantom{0} \\
$^{13}$CO & 2--1 & JCMT-RxA       & 0.74   &  4.27\phantom{0}   &  3.23\phantom{0}  & 0.051 \\
          & 3--2 & JCMT-HARPB     & 0.63   &  3.40\phantom{0}   &  2.95\phantom{0}  & 0.070 \\
          & 6--5 & APEX-CHAMP$^+$ & 0.48   &  2.56\phantom{0}   &  2.13\phantom{0}  & 0.21\phantom{0} \\
          & 10--9&  {\it Herschel}-HIFI\tablefootmark{b}       & 0.74 & 0.19\phantom{0} & 0.08\phantom{0} & 0.019 \\
C$^{18}$O & 2--1 & JCMT-RxA       & 0.69   &  2.40\phantom{0}   &  2.54\phantom{0}  & 0.10\phantom{0} \\
          & 3--2 & JCMT-HARPB     & 0.63   &  1.39\phantom{0}   &  1.90\phantom{0}  & 0.062 \\
          & 5--4 & {\it Herschel}-HIFI\tablefootmark{a} & 0.76 &  0.29\phantom{0}  &  0.32\phantom{0}  & 0.012 \\
          & 9--8 & {\it Herschel}-HIFI\tablefootmark{b} & 0.74 &  $<$0.057         & \dots\phantom{0}  & 0.025 \\
\hline 
\end{tabular}
\end{center}
\tablefoot{
\tablefoottext{a}{Only H-polarization observation is used.}
\tablefoottext{b}{H- and V-polarization observations averaged.}
}
\label{tbl:linesB335}
\end{table*}

\newpage

\onecolumn
\subsection{L1157}
\begin{figure*}[htb]
    \centering
    \includegraphics[scale=0.3]{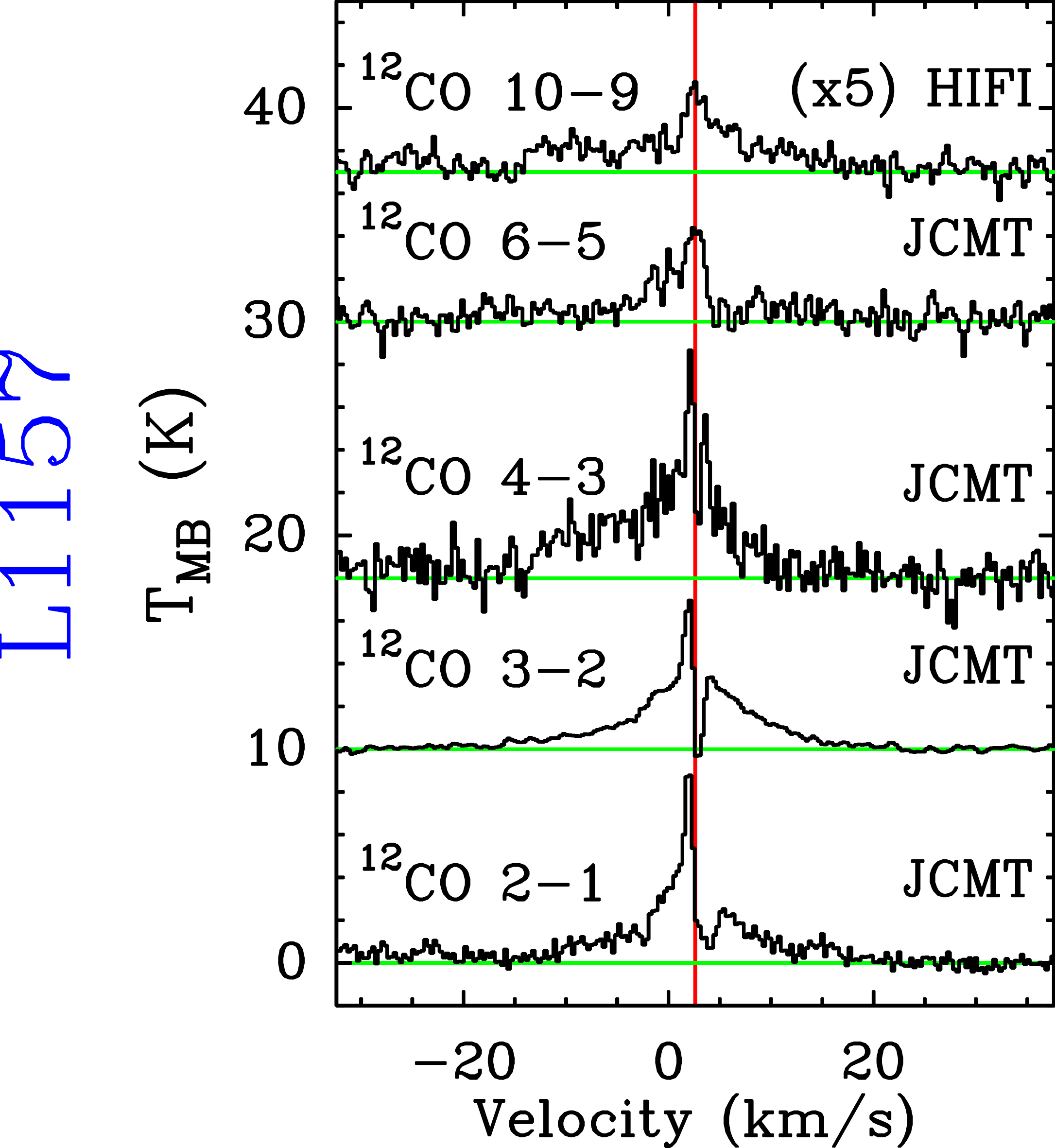}
    \includegraphics[scale=0.3]{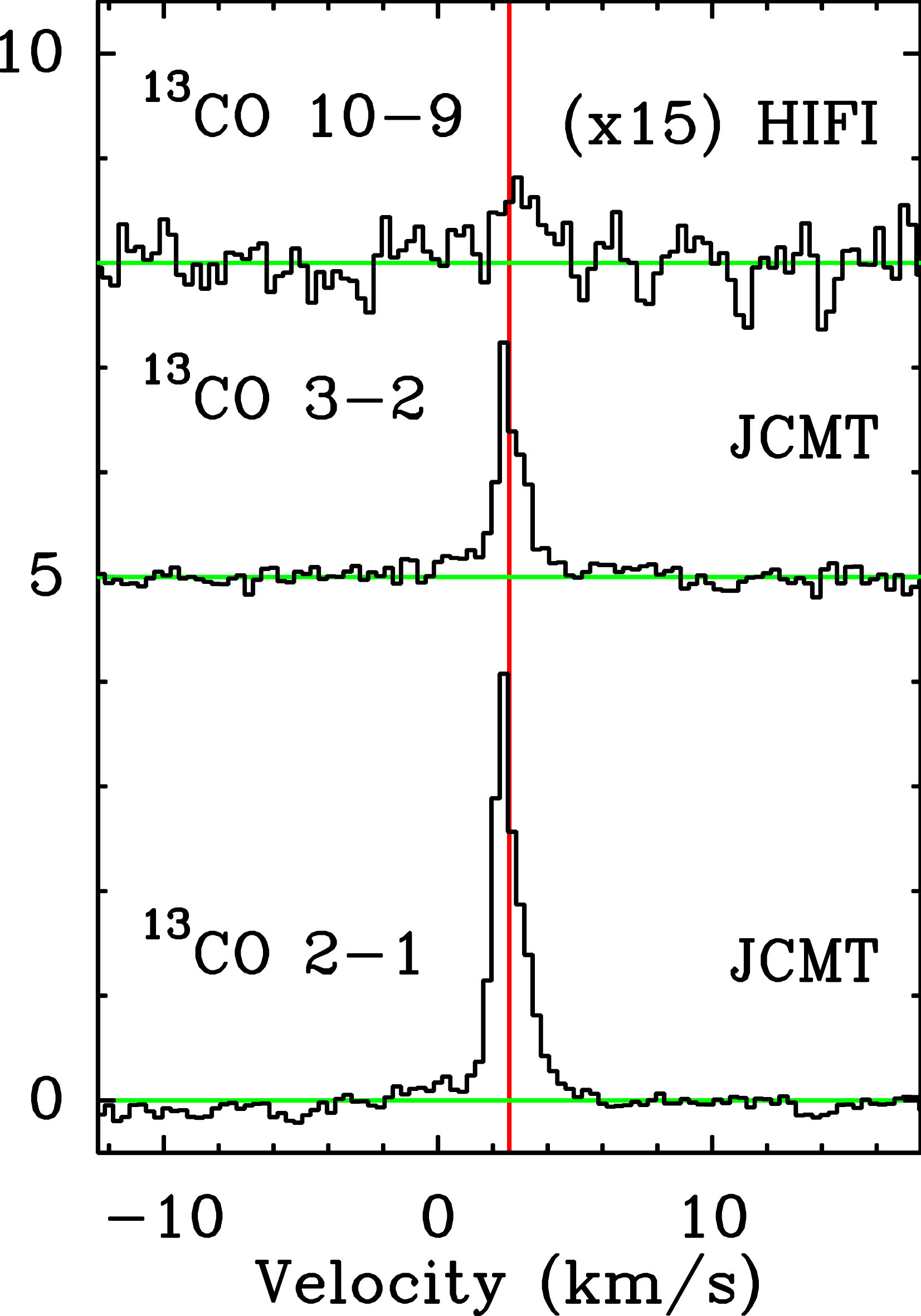}
    \includegraphics[scale=0.3]{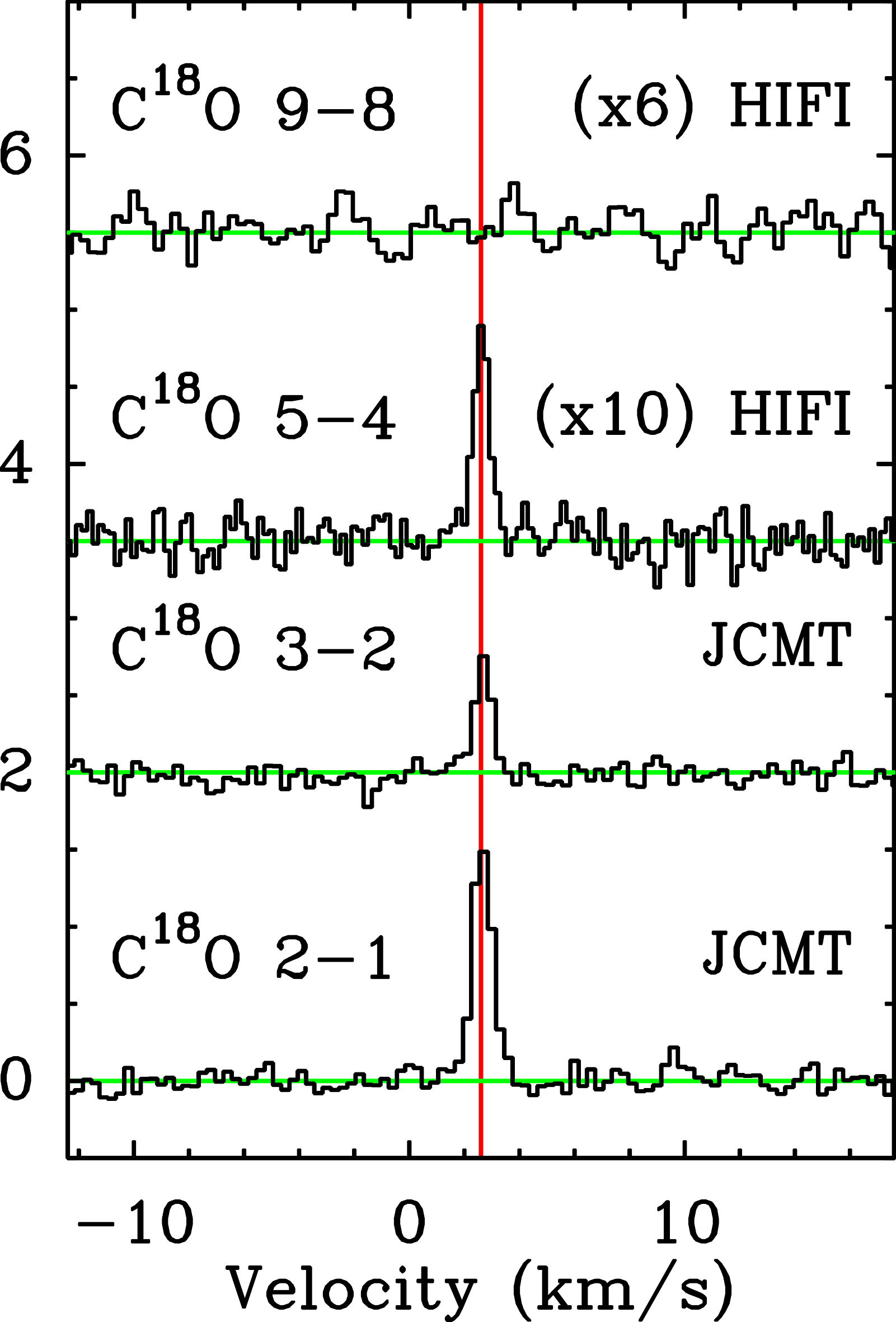}
    \caption{\small Observed $^{12}$CO, $^{13}$CO, and C$^{18}$O transitions for L1157.}
    \label{fig:linesL1157}
\end{figure*}

\begin{table*}[!ht]
\caption{Observed line intensities for L1157 in all observed transitions.}
\normalsize
\begin{center}
\begin{tabular}{l l l r r r r r r r r r}
\hline \hline
Mol.  & Transition & Telescope & Efficiency & $\int T_{\rm MB} \mathrm{d}V$ & $T_{\mathrm{peak}}$ & $rms$ \\
 &  & &   $\eta$ &[K km s$^{-1}$] & [K] &   [K]\\
\hline
CO        & 2--1 & JCMT-RxA      & 0.69  &  47.92 &  8.77  & 0.28\phantom{0} \\
          & 3--2 & JCMT-HARPB    & 0.63  &  48.22 &  7.41  & 0.10\phantom{0} \\
          & 4--3 & JCMT          & 0.38  &  78.15 & 10.67  & 0.88\phantom{0} \\
          &10--9 & {\it Herschel}-HIFI\tablefootmark{a}     & 0.64 &  8.94 &  0.85  & 0.15\phantom{0} \\
$^{13}$CO & 2--1 & JCMT-RxA      & 0.74  &  5.10  &  4.34  & 0.10\phantom{0} \\
          & 3--2 & JCMT-HARPB    & 0.63  &  3.11  &  2.57  & 0.11\phantom{0} \\
          & 10--9& {\it Herschel}-HIFI\tablefootmark{b} & 0.74 &  0.33 &  0.10  & 0.027 \\
C$^{18}$O & 2--1 & JCMT-RxA      & 0.69  &  1.46  &  1.50 & 0.094 \\
          & 3--2 & JCMT-HARPB    & 0.63  &  0.58           &  0.83  & 0.090 \\
          & 5--4 & {\it Herschel}-HIFI\tablefootmark{b}    & 0.76 &  0.11 &  0.14  & 0.014 \\
          & 9--8 & {\it Herschel}-HIFI\tablefootmark{a}    & 0.74 &  $<$0.05 & \dots\phantom{0}  & 0.022 \\
\hline 
\end{tabular}
\end{center}
\tablefoot{
\tablefoottext{a}{H- and V-polarization observations averaged.}
\tablefoottext{b}{Only H-polarization observation is used.}
}
\label{tbl:linesL1157}
\end{table*}

\newpage

\onecolumn
\subsection{L1489}
\begin{figure*}[htb]
    \centering
    \includegraphics[scale=0.3]{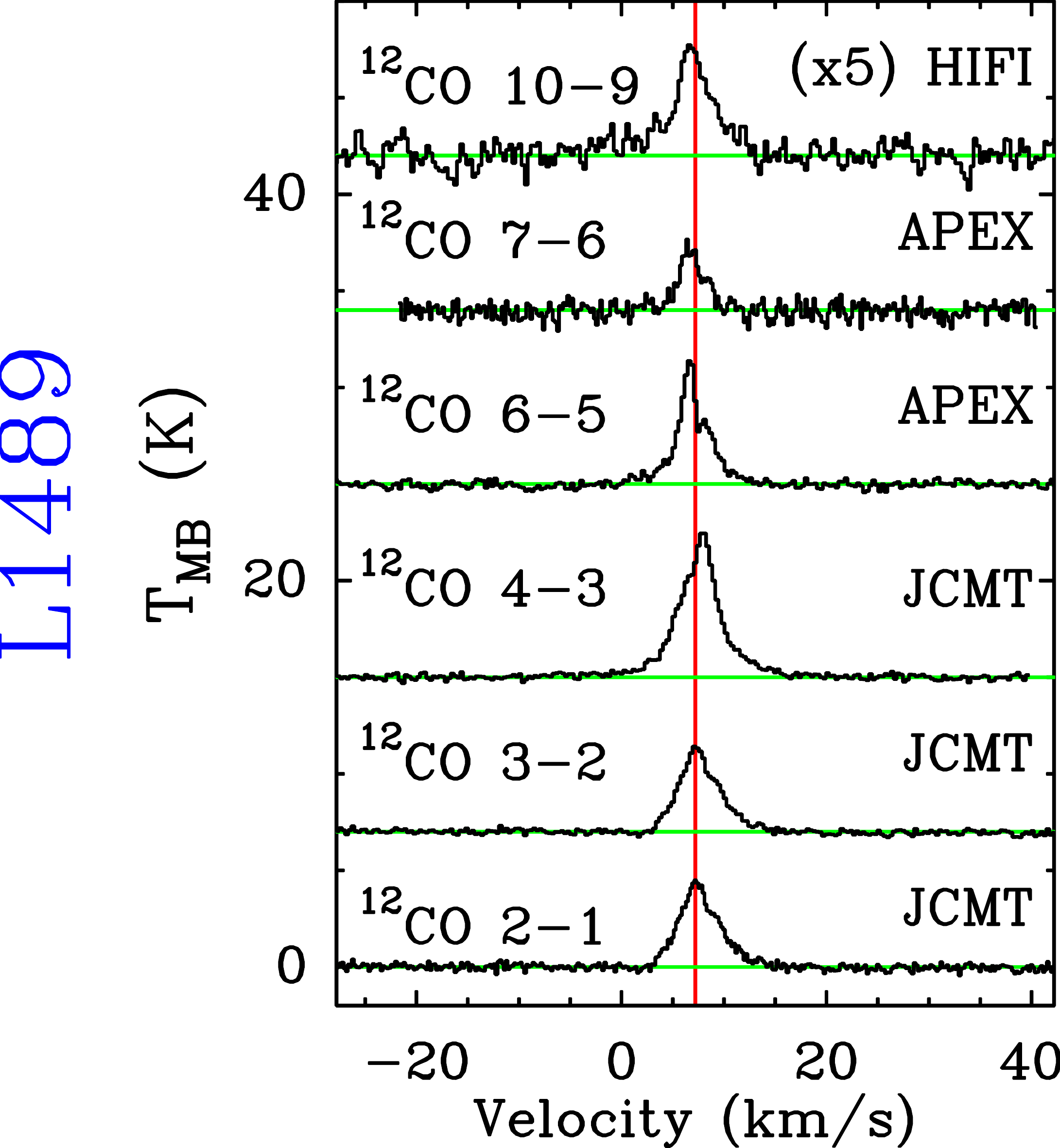}
    \includegraphics[scale=0.3]{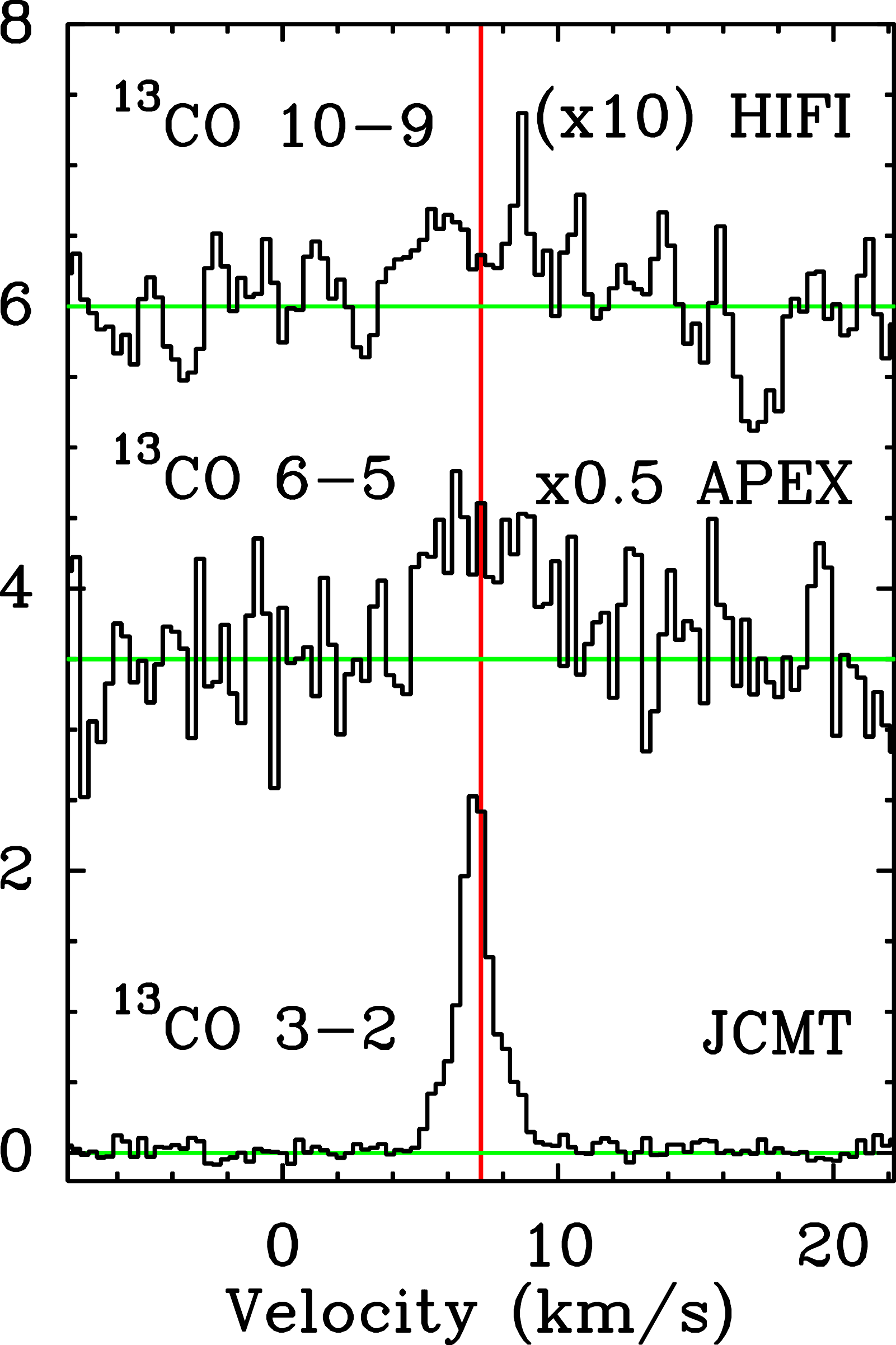}
    \includegraphics[scale=0.3]{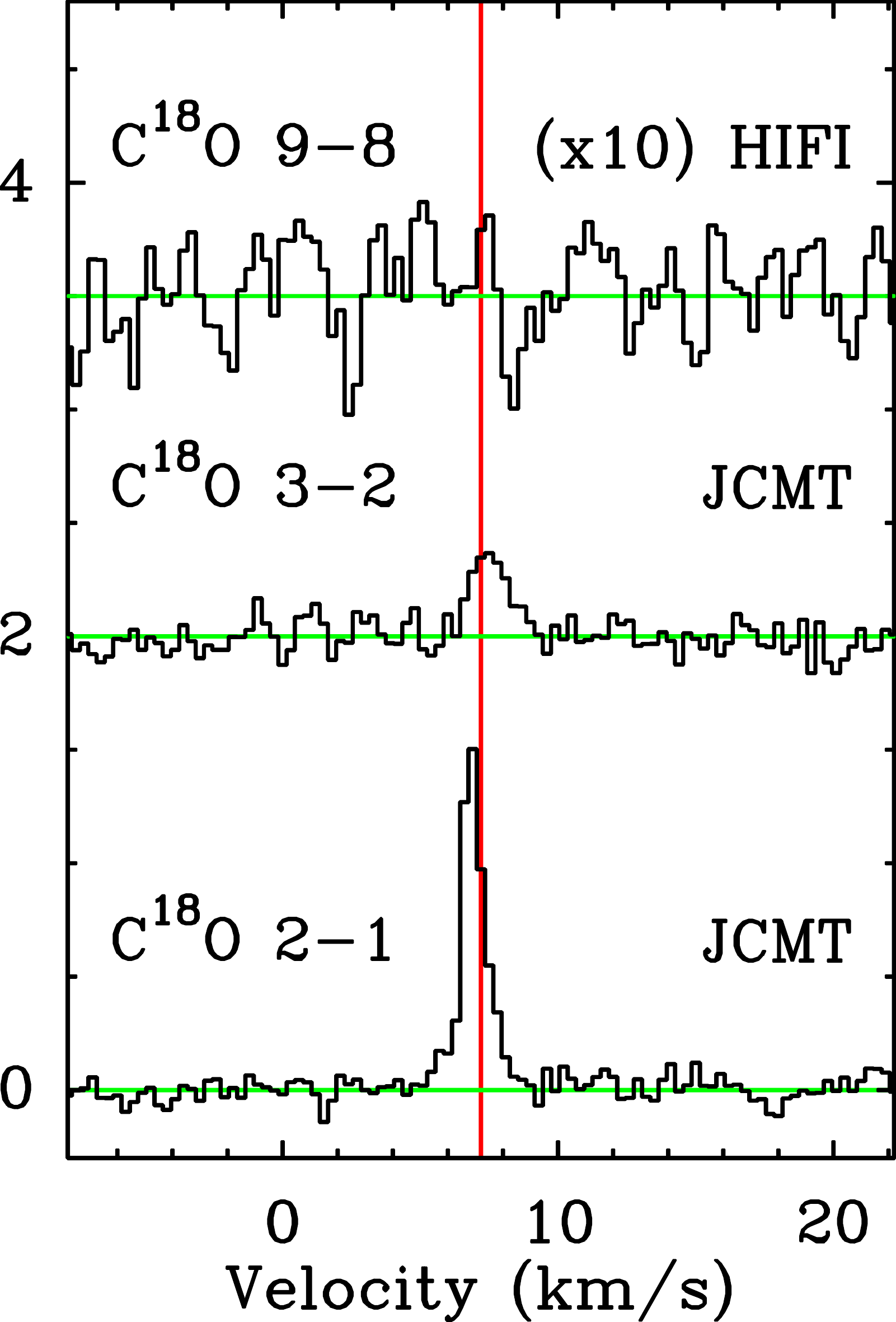}
    \caption{\small Observed $^{12}$CO, $^{13}$CO, and C$^{18}$O transitions for L1489.}
    \label{fig:linesL1489}
\end{figure*}

\begin{table*}[!ht]
\caption{Observed line intensities for L1489 in all observed transitions.}
\normalsize
\begin{center}
\begin{tabular}{l l l r r r r r r r r r}
\hline \hline
Mol.  & Transition & Telescope & Efficiency & $\int T_{\rm MB} \mathrm{d}V$ & $T_{\mathrm{peak}}$ & $rms$ \\
 &  & &   $\eta$ &[K km s$^{-1}$] & [K] &   [K]\\
\hline
CO        & 2--1 & JCMT-RxA        & 0.69 & 19.87\phantom{0} & 4.49\phantom{0}  & 0.15\phantom{0} \\
          & 3--2 & JCMT-HARPB      & 0.63 & 13.04\phantom{0} & 4.00\phantom{0}  & 0.49\phantom{0} \\
          & 4--3 & JCMT            & 0.38 & 33.80\phantom{0} & 7.14\phantom{0}  & 0.15\phantom{0} \\
          & 6--5 & APEX-CHAMP$^+$  & 0.45 & 20.87\phantom{0} & 6.48\phantom{0}  & 0.16\phantom{0} \\
          & 7--6 & APEX-CHAMP$^+$  & 0.49 & 9.06\phantom{0}  & 3.65\phantom{0}  & 0.37\phantom{0} \\
          &10--9 & {\it Herschel}-HIFI\tablefootmark{a} & 0.64 & 6.20\phantom{0} & 1.16\phantom{0}  & 0.11\phantom{0} \\
$^{13}$CO & 3--2 & JCMT-HARPB      & 0.63 & 4.45\phantom{0}  & 2.70\phantom{0}  & 0.082 \\
          & 6--5 & APEX-CHAMP$^+$  & 0.45 & 8.22\phantom{0}  & 3.77\phantom{0}  & 1.29\phantom{0} \\
          & 10--9& {\it Herschel}-HIFI\tablefootmark{a}     & 0.74   &  0.28\phantom{0}  &  0.11\phantom{0} & 0.029 \\
C$^{18}$O & 2--1 & JCMT-RxA        & 0.69 & 1.72\phantom{0}  & 1.62\phantom{0}  & 0.071 \\
          & 3--2 & JCMT-HARPB      & 0.63 & 0.67\phantom{0} & 0.44\phantom{0} & 0.10\phantom{0} \\
          & 9--8 & {\it Herschel}-HIFI\tablefootmark{a}     & 0.74  &  $<$0.055 & \dots\phantom{0}  & 0.024 \\
\hline 
\end{tabular}
\end{center}
\tablefoot{
\tablefoottext{a}{H- and V-polarization observations averaged.}
}
\label{tbl:linesL1489}
\end{table*}

\newpage
\clearpage

\onecolumn
\subsection{L1551IRS5}
\begin{figure*}[htb]
    \centering
    \includegraphics[scale=0.3]{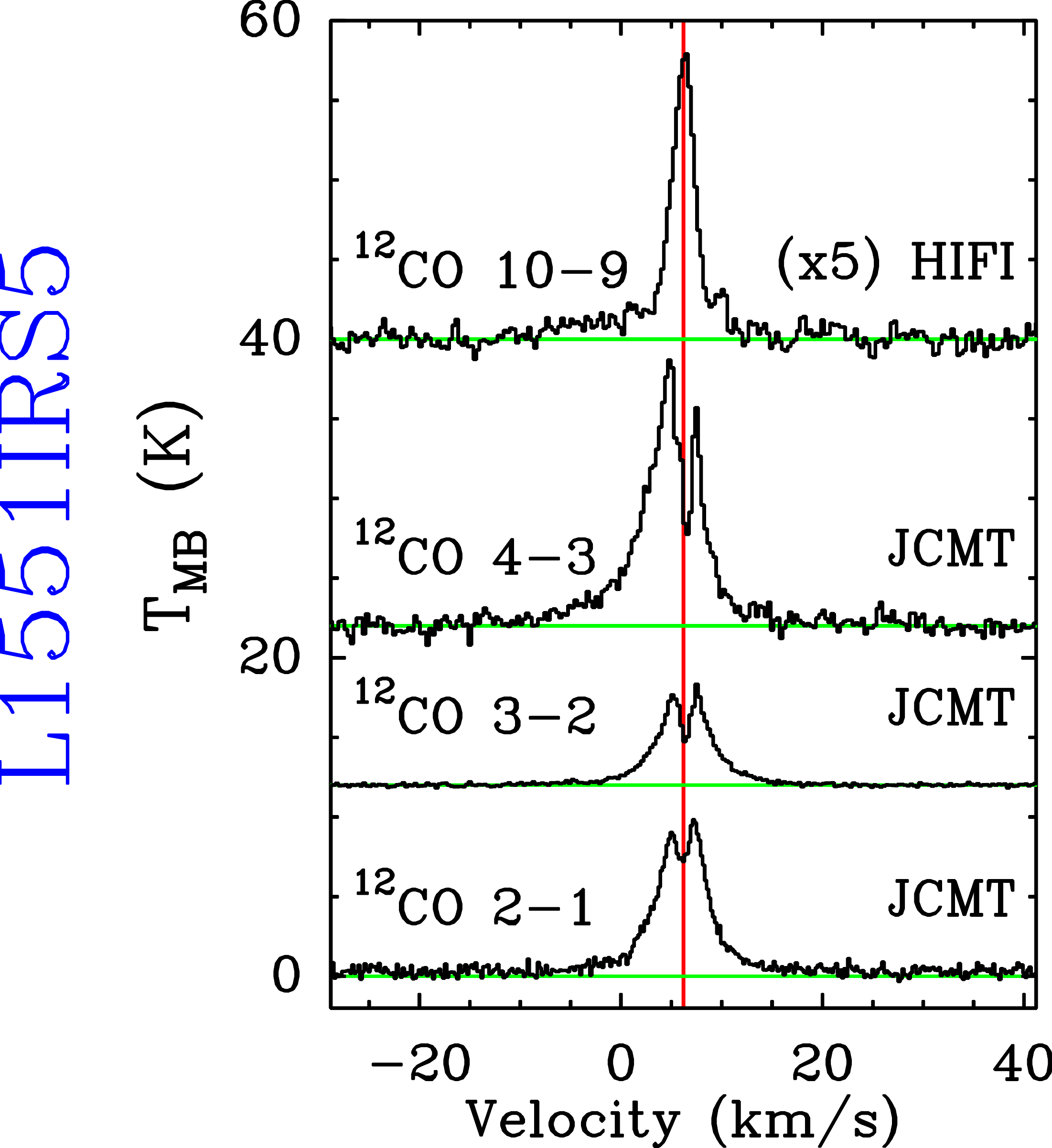}
    \includegraphics[scale=0.3]{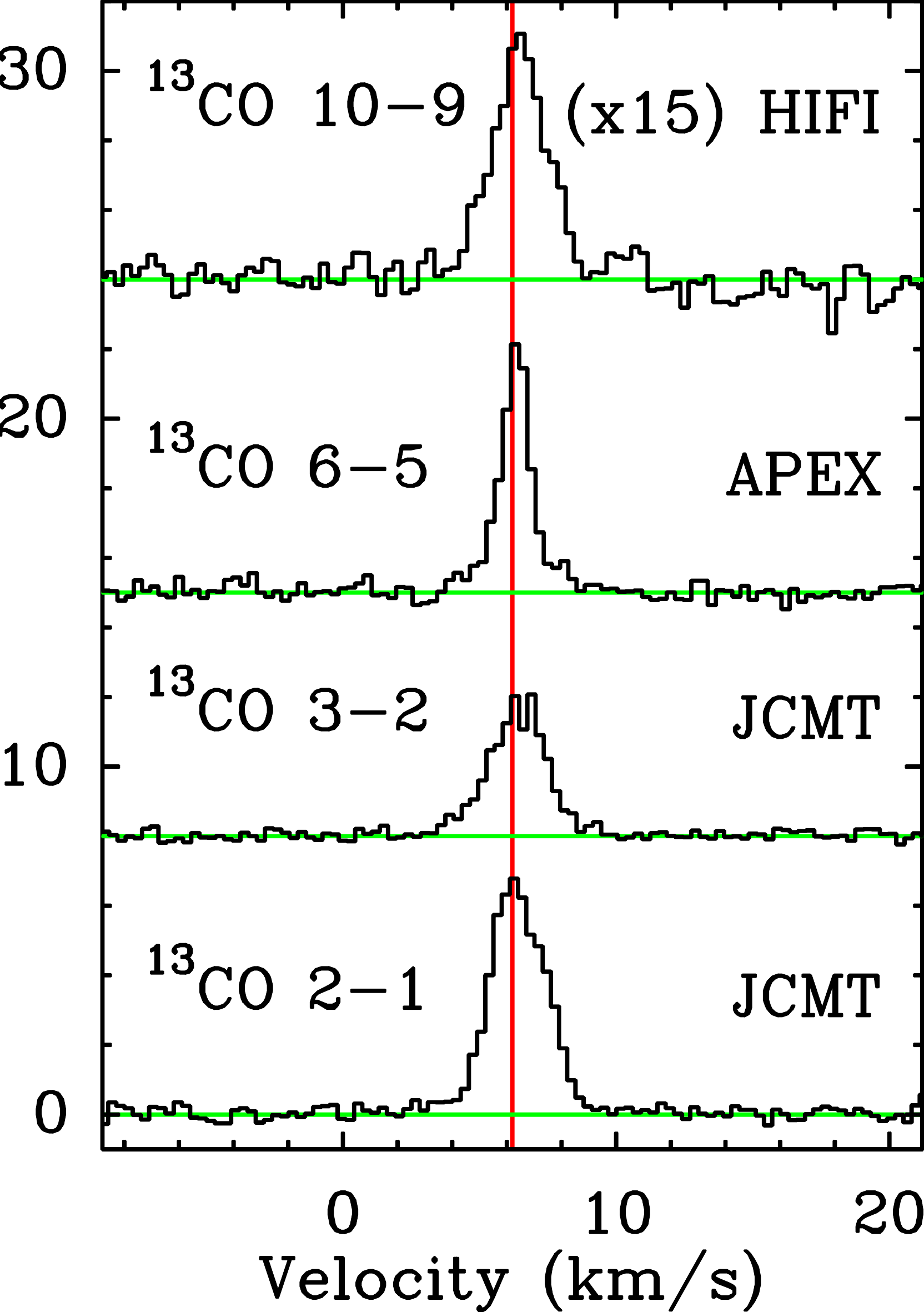}
    \includegraphics[scale=0.3]{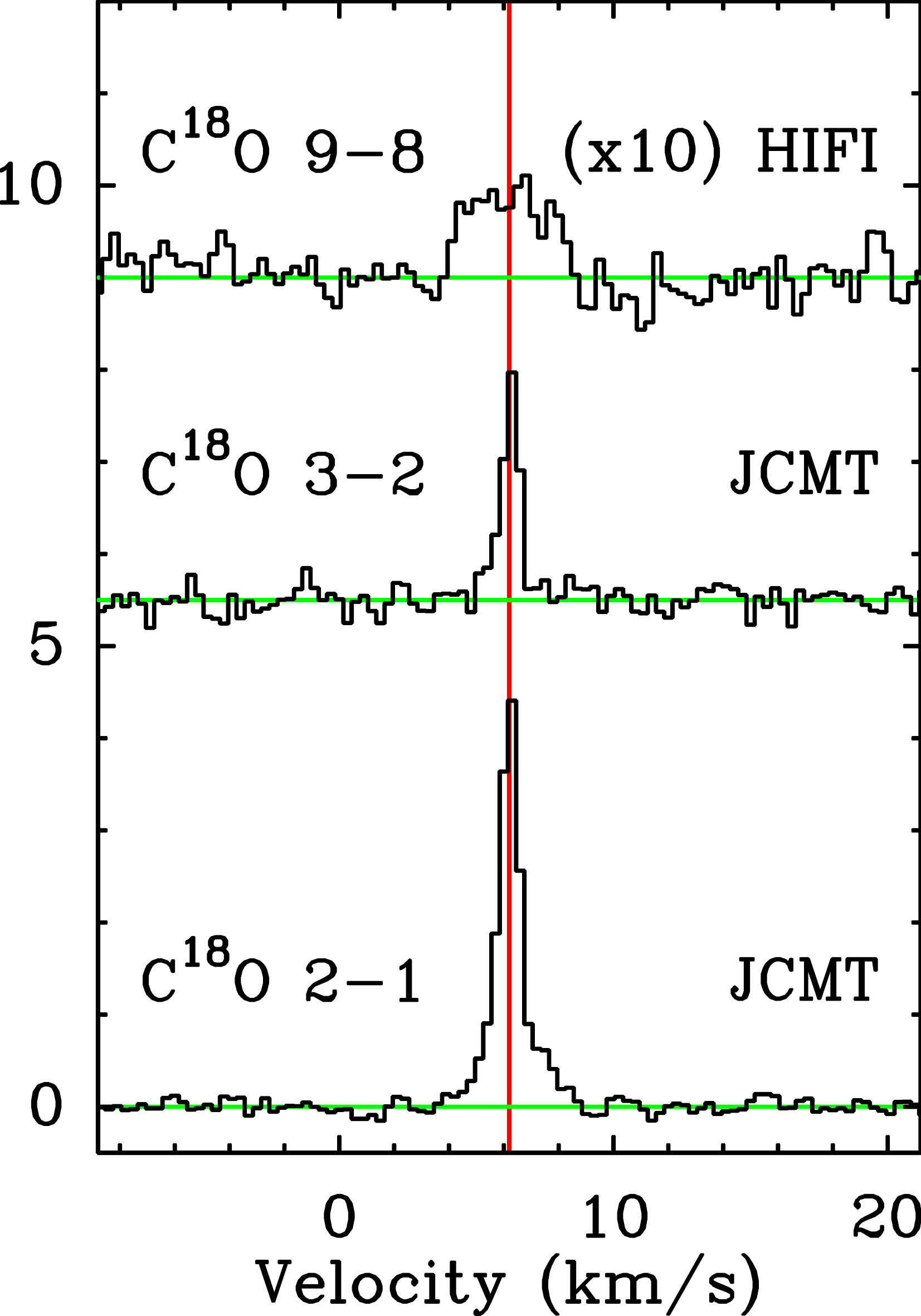}
    \caption{\small Observed $^{12}$CO, $^{13}$CO, and C$^{18}$O transitions for L1551IRS5.}
    \label{fig:linesL1551IRS5}
\end{figure*}

\begin{table*}[!ht]
\caption{Observed line intensities for L1551IRS5 in all observed transitions.}
\normalsize
\begin{center}
\begin{tabular}{l l l r r r r r r r r r}
\hline \hline
Mol.  & Transition & Telescope & Efficiency & $\int T_{\rm MB} \mathrm{d}V$ & $T_{\mathrm{peak}}$ & $rms$ \\
 &  & &   $\eta$ &[K km s$^{-1}$] & [K] &   [K]\\
\hline
CO        & 2--1 & JCMT-RxA             & 0.69 &  71.52\phantom{0} & 9.82\phantom{0}  & 0.27\phantom{0} \\
          & 3--2 & JCMT-HARPB           & 0.63 &  37.52\phantom{0} & 6.68\phantom{0}  & 0.11\phantom{0} \\
          & 4--3 & JCMT\tablefootmark{a}& 0.38 & 108.13\phantom{0} & 16.69\phantom{0} & 0.45\phantom{0} \\
          &10--9 & {\it Herschel}-HIFI\tablefootmark{b} & 0.64 & 14.73\phantom{0} & 3.59\phantom{0} & 0.13\phantom{0} \\
$^{13}$CO & 2--1 & JCMT-RxA             & 0.74 &  17.86\phantom{0} & 7.04\phantom{0}  & 0.25\phantom{0} \\
          & 3--2 & JCMT-HARPB           & 0.63 &  10.23\phantom{0} & 4.82\phantom{0}  & 0.18\phantom{0} \\
          & 6--5 & APEX-CHAMP$^+$       & 0.48 &  10.50\phantom{0} & 7.47\phantom{0}  & 0.24\phantom{0} \\
          & 10--9& {\it Herschel}-HIFI\tablefootmark{b} & 0.74    & 1.38\phantom{0}  &  0.49\phantom{0}  & 0.028 \\
C$^{18}$O & 2--1 & JCMT-RxA             & 0.69 &  5.26\phantom{0}  & 4.62\phantom{0}  & 0.076 \\
          & 3--2 & JCMT-HARPB           & 0.63 &  2.29\phantom{0}  & 2.58\phantom{0}  & 0.18\phantom{0} \\
          & 9--8 & {\it Herschel}-HIFI  & 0.74 &  0.22\phantom{0}  & 0.12\phantom{0}  & 0.025 \\
\hline 
\end{tabular}
\end{center}
\tablefoot{
\tablefoottext{a}{Taken in 11$\arcsec$ beam.}
\tablefoottext{b}{H- and V-polarization observations averaged.}
}
\label{tbl:linesL1551IRS5}
\end{table*}

\newpage

\onecolumn
\subsection{TMR1}
\begin{figure*}[htb]
    \centering
    \includegraphics[scale=0.3]{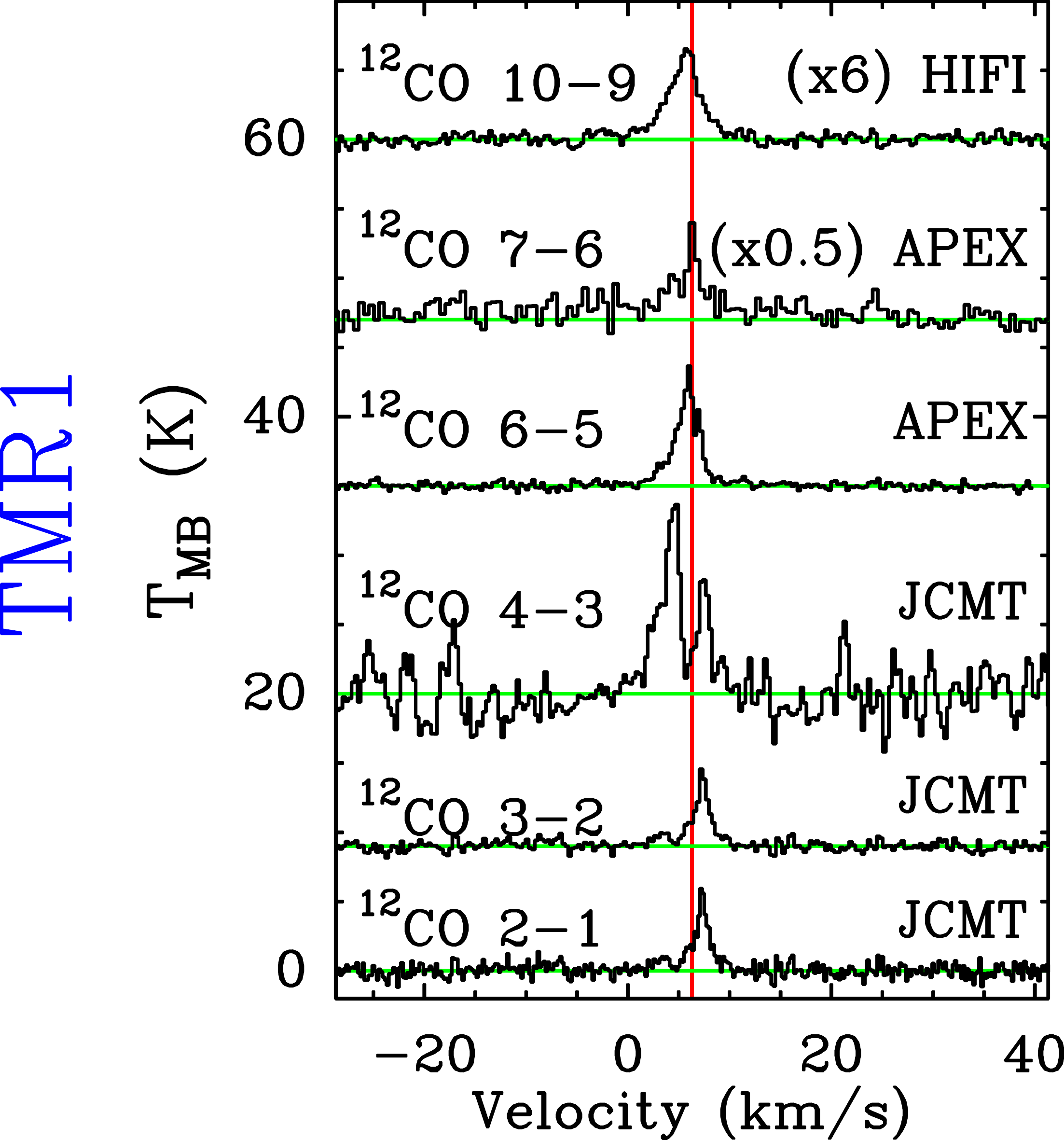}
    \includegraphics[scale=0.3]{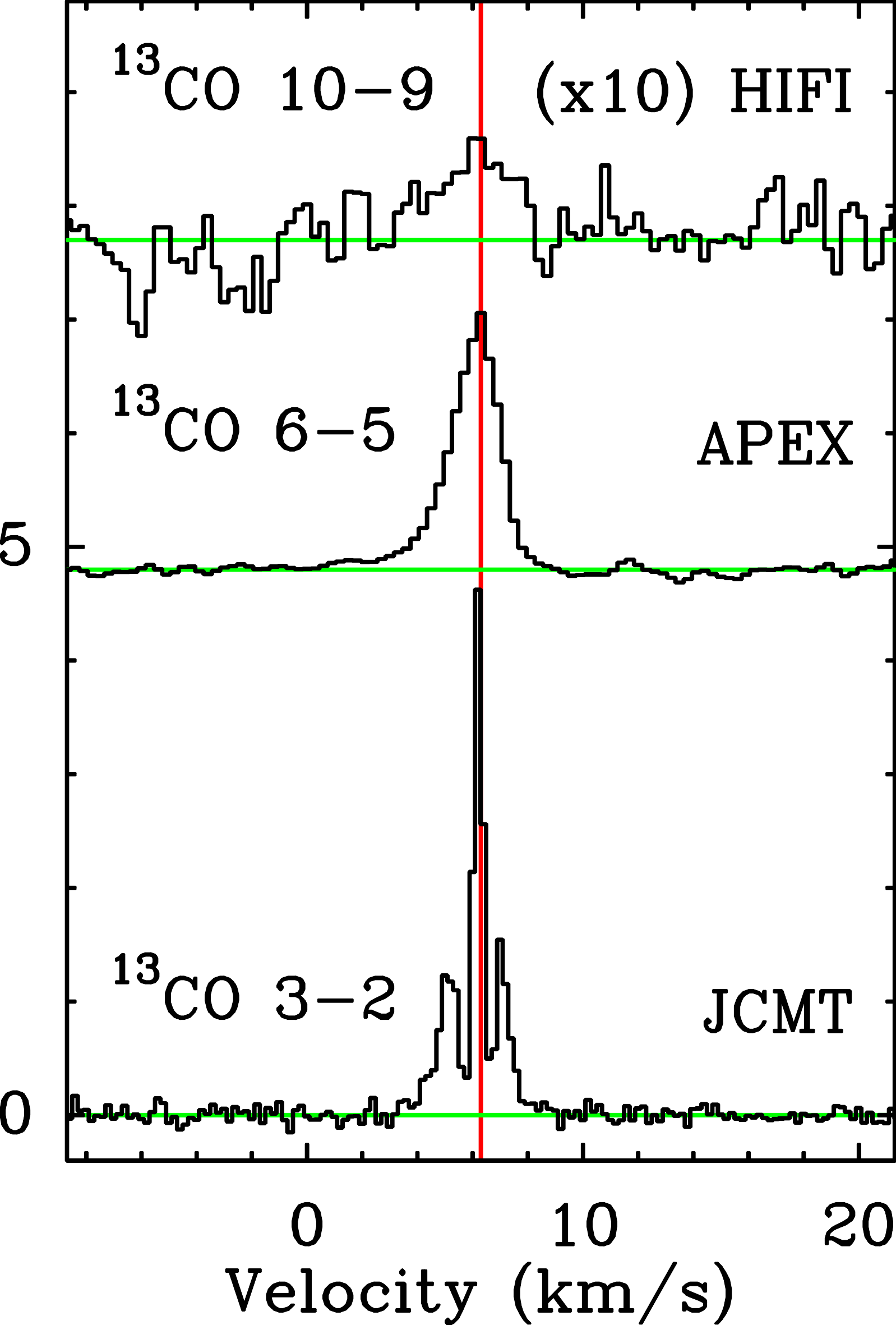}
    \includegraphics[scale=0.3]{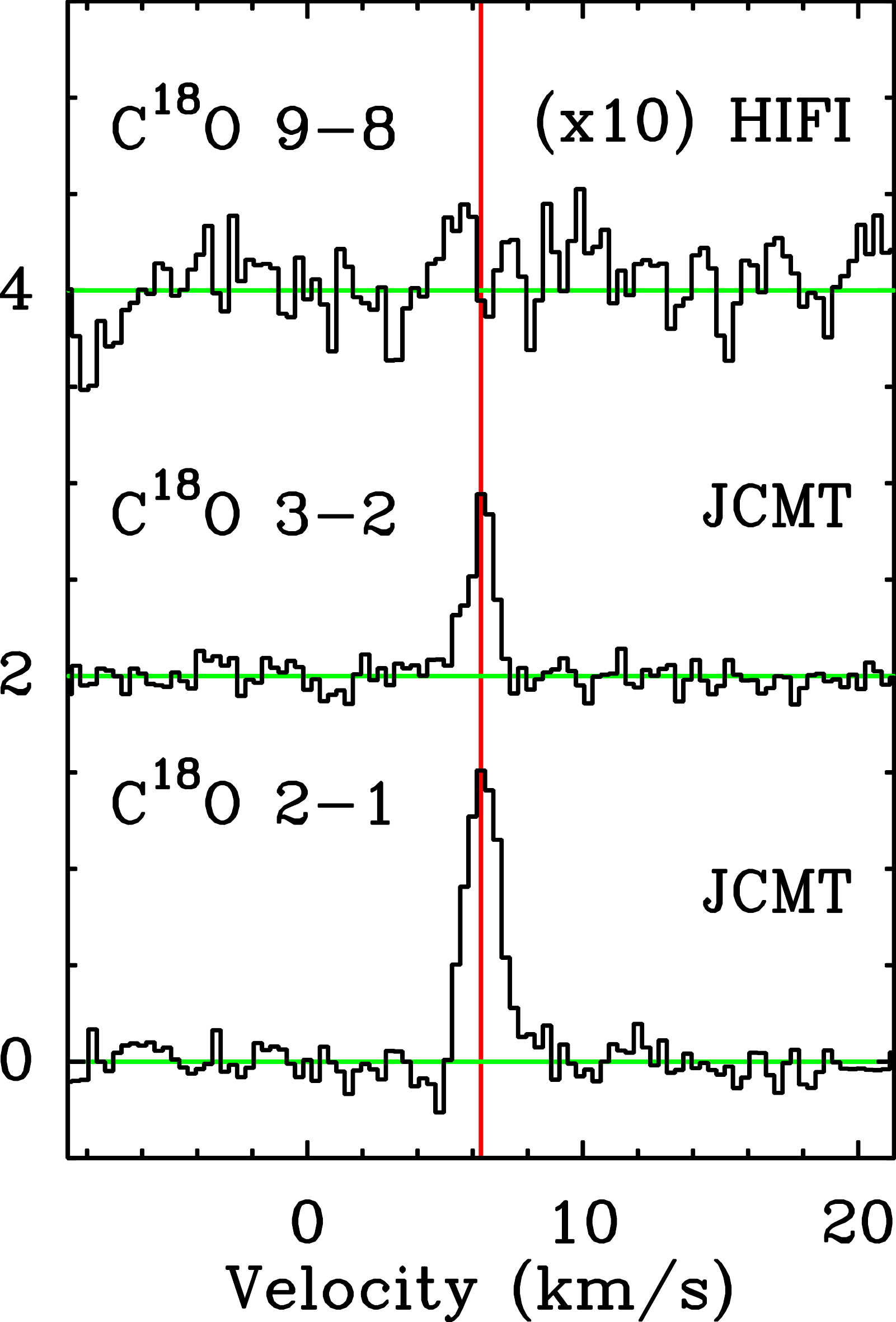}
    \caption{\small Observed $^{12}$CO, $^{13}$CO, and C$^{18}$O transitions for TMR1.}
    \label{fig:linesTMR1}
\end{figure*}

\begin{table*}[!ht]
\caption{Observed line intensities for TMR1 in all observed transitions.}
\normalsize
\begin{center}
\begin{tabular}{l l l r r r r r r r r r}
\hline \hline
Mol.  & Transition & Telescope & Efficiency & $\int T_{\rm MB} \mathrm{d}V$ & $T_{\mathrm{peak}}$ & $rms$ \\
 &  & &   $\eta$ &[K km s$^{-1}$] & [K] &   [K]\\
\hline
CO        & 2--1 & JCMT-RxA             & 0.69 &  11.94\phantom{0} &  5.95\phantom{0}  & 0.41\phantom{0} \\
          & 3--2 & JCMT-HARPB           & 0.63 &  8.55\phantom{0}  &  4.61\phantom{0}  & 0.088 \\
          & 4--3 & JCMT\tablefootmark{a}& 0.38 &  40.78\phantom{0} &  15.23\phantom{0} & 3.26\phantom{0} \\
          & 6--5 & APEX-CHAMP$^+$       & 0.45 &  22.65\phantom{0} &  9.15\phantom{0}  & 0.32\phantom{0} \\
          & 7--6 & APEX-CHAMP$^+$       & 0.42 &  21.89\phantom{0} &  8.92\phantom{0}  & 0.81\phantom{0} \\
          &10--9 & {\it Herschel}-HIFI\tablefootmark{b} & 0.64 & 9.18\phantom{0} & 2.19\phantom{0} & 0.13\phantom{0} \\
$^{13}$CO & 3--2 & JCMT-HARPB           & 0.63 &  4.27\phantom{0}  &  4.62\phantom{0}  & 0.057 \\
          & 6--5 & APEX-CHAMP$^+$       & 0.45 &  4.79\phantom{0}  &  2.00\phantom{0}  & 0.032 \\
          & 10--9& {\it Herschel}-HIFI\tablefootmark{b} & 0.74    &  0.49\phantom{0}  &  0.11\phantom{0}  & 0.027 \\
C$^{18}$O & 2--1 & JCMTRxA              & 0.69 &  2.20\phantom{0}  &  1.61\phantom{0}  & 0.11\phantom{0} \\
          & 3--2 & JCMT-HARPB           & 0.63 &  1.13\phantom{0}  &  1.22\phantom{0}  & 0.091 \\
          & 9--8 & {\it Herschel}-HIFI\tablefootmark{b}  & 0.74   &  $<$0.052 &  \dots\phantom{0}  & 0.023 \\
\hline 
\end{tabular}
\end{center}
\tablefoot{
\tablefoottext{a}{Taken in 11$\arcsec$ beam.}
\tablefoottext{b}{H- and V-polarization observations averaged.}
}
\label{tbl:linesTMR1}
\end{table*}

\newpage
\clearpage

\onecolumn
\subsection{TMC1A}
\begin{figure*}[htb]
    \centering
    \includegraphics[scale=0.3]{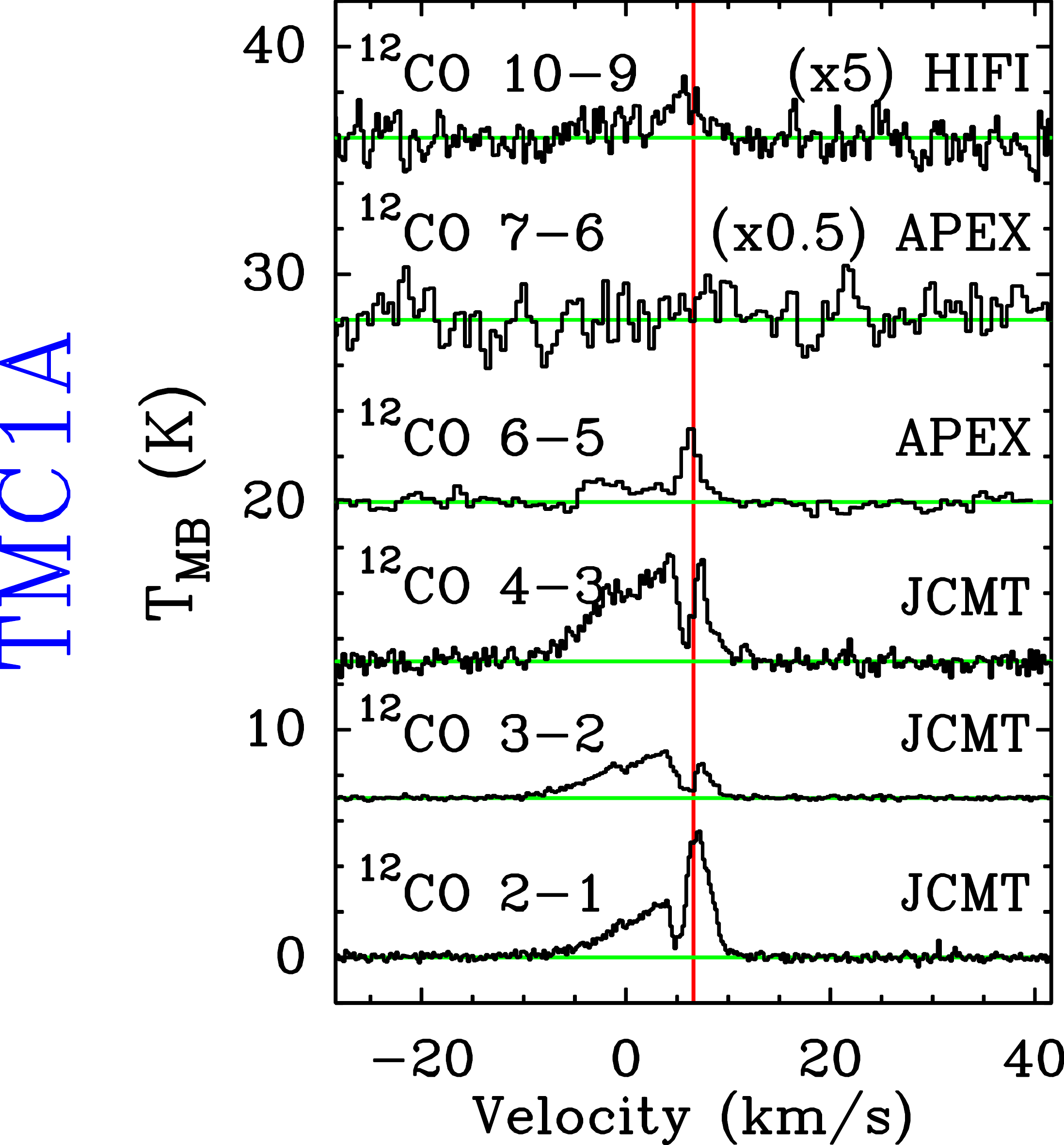}
    \includegraphics[scale=0.3]{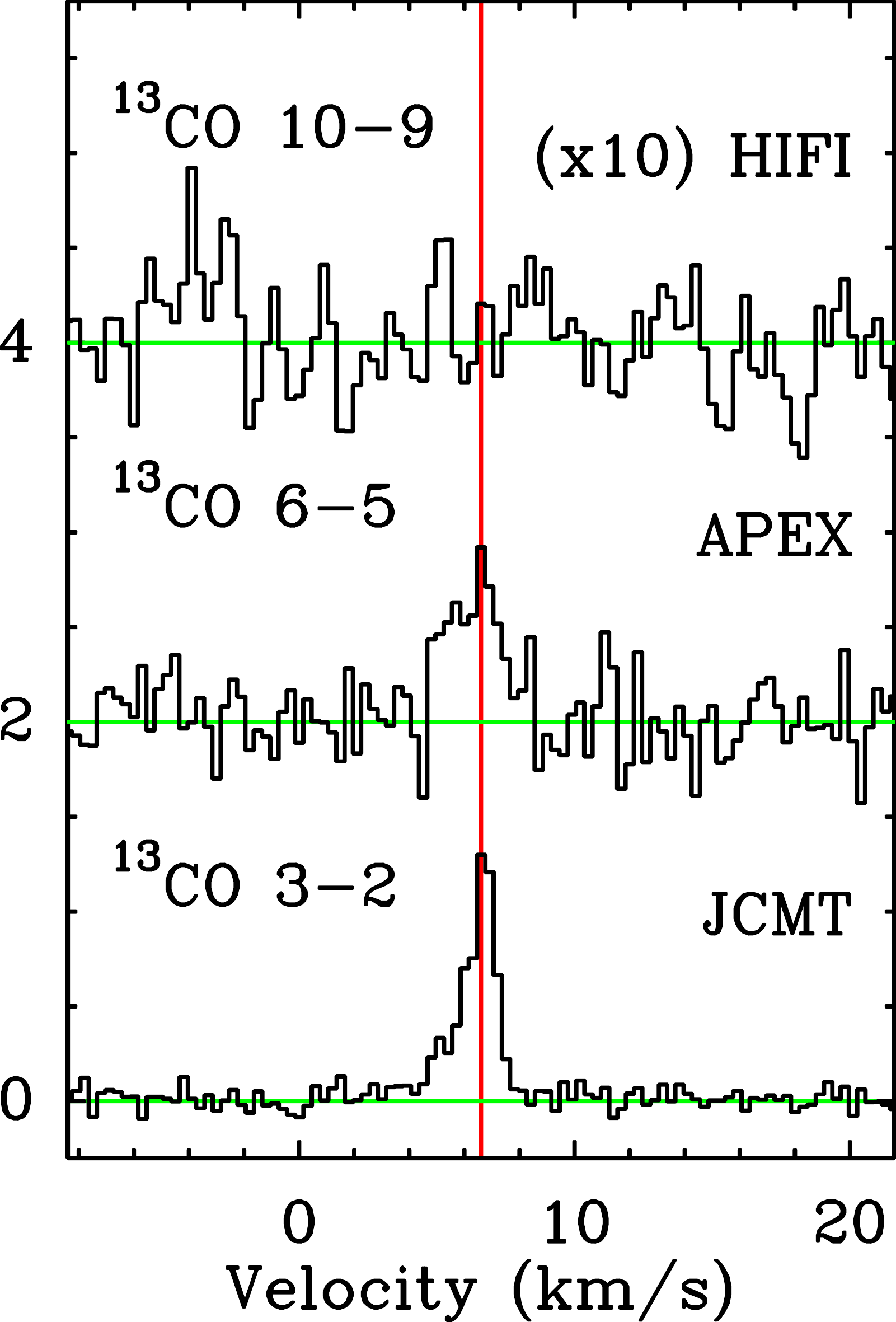}
    \includegraphics[scale=0.3]{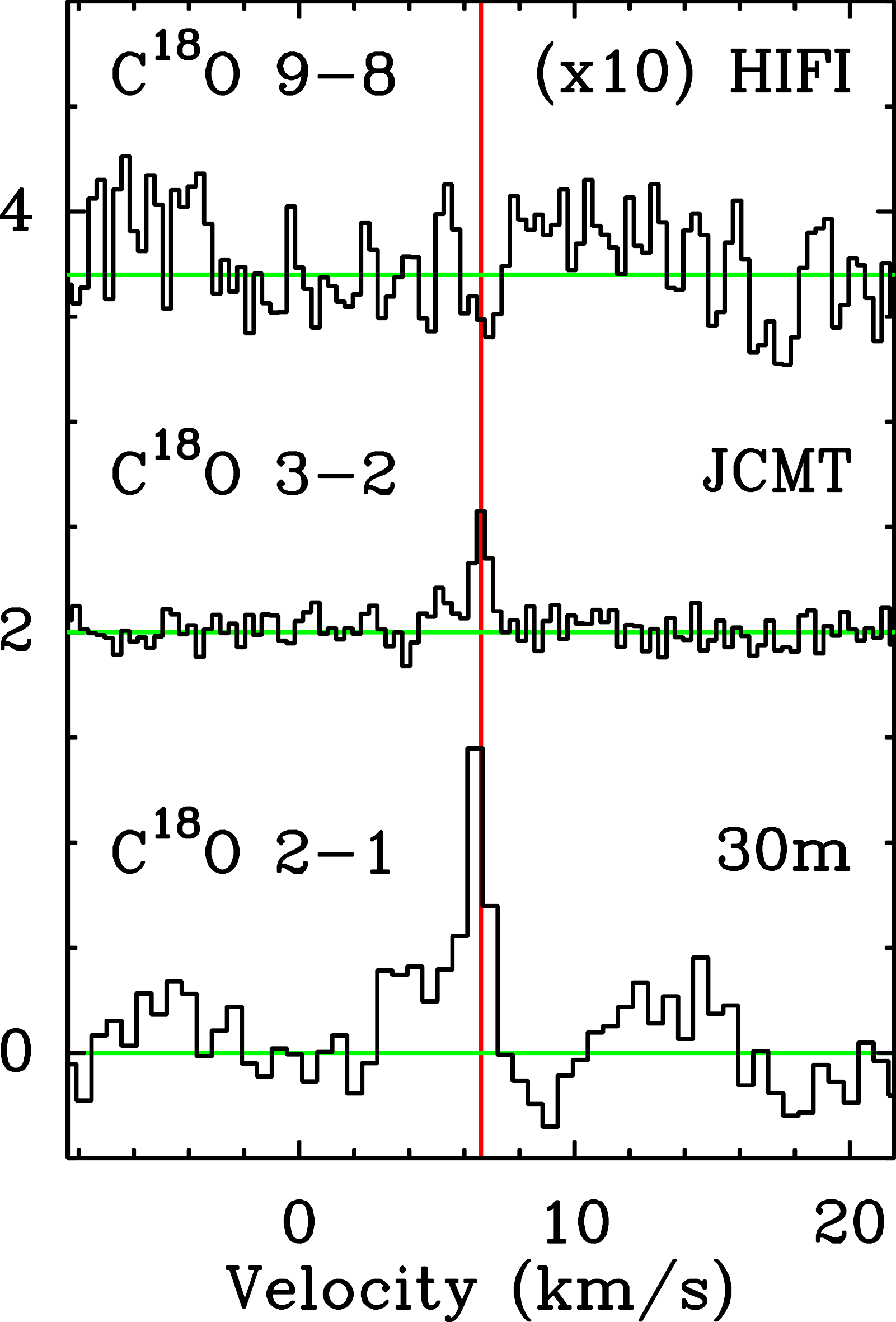}
    \caption{\small Observed $^{12}$CO, $^{13}$CO, and C$^{18}$O transitions for TMC1A.}
    \label{fig:linesTMC1A}
\end{figure*}

\begin{table*}[!ht]
\caption{Observed line intensities for TMC1A in all observed transitions.}
\normalsize
\begin{center}
\begin{tabular}{l l l r r r r r r r r r}
\hline \hline
Mol.  & Transition & Telescope & Efficiency & $\int T_{\rm MB} \mathrm{d}V$ & $T_{\mathrm{peak}}$ & $rms$ \\
 &  & &   $\eta$ &[K km s$^{-1}$] & [K] &   [K]\\
\hline
CO        & 2--1 & JCMT-RxA              & 0.69 & 30.32\phantom{0} & 5.55\phantom{0} & 0.19\phantom{0} \\
          & 3--2 & JCMT-HARPB            & 0.63 & 19.10\phantom{0} & 2.10\phantom{0} & 0.089 \\
          & 4--3 & JCMT\tablefootmark{a} & 0.38 & 42.46\phantom{0} & 4.82\phantom{0} & 0.46\phantom{0} \\
          & 6--5 & APEX-CHAMP$^+$        & 0.52 & 11.68\phantom{0} & 3.21\phantom{0} & 0.24\phantom{0} \\
          & 7--6 & APEX-CHAMP$^+$        & 0.49 & 4.13\phantom{0}  & 2.07\phantom{0} & 0.81\phantom{0} \\
          &10--9 & {\it Herschel}-HIFI\tablefootmark{b} & 0.64 & 1.37\phantom{0} & 0.52\phantom{0} & 0.13\phantom{0} \\
$^{13}$CO & 3--2 & JCMT-HARPB            & 0.63 & 2.18\phantom{0}  & 1.43\phantom{0} & 0.073 \\
          & 6--5 & APEX-CHAMP$^+$        & 0.48 & 1.76\phantom{0}  & 1.09\phantom{0} & 0.25\phantom{0} \\
          & 10--9& {\it Herschel}-HIFI\tablefootmark{b} & 0.74    & $<$0.10\phantom{0} &  \dots\phantom{0} & 0.031 \\
C$^{18}$O & 2--1 & IRAM~30m              & 0.59 & 1.97\phantom{0}  & 2.40\phantom{0} & 0.33\phantom{0} \\
          & 3--2 & JCMT-HARPB            & 0.63 & 0.62\phantom{0}  & 0.61\phantom{0} & 0.11\phantom{0} \\
          & 9--8 & {\it Herschel}-HIFI\tablefootmark{c} & 0.74    & $<$0.052         & \dots\phantom{0} & 0.023 \\
\hline 
\end{tabular}
\end{center}
\tablefoot{
\tablefoottext{a}{Taken in 11$\arcsec$ beam.}
\tablefoottext{b}{Only H-polarization observation is used.}
\tablefoottext{c}{H- and V-polarization observations averaged.}
}
\label{tbl:linesTMC1A}
\end{table*}

\newpage

\onecolumn
\subsection{TMC1}
\begin{figure*}[htb]
    \centering
    \includegraphics[scale=0.3]{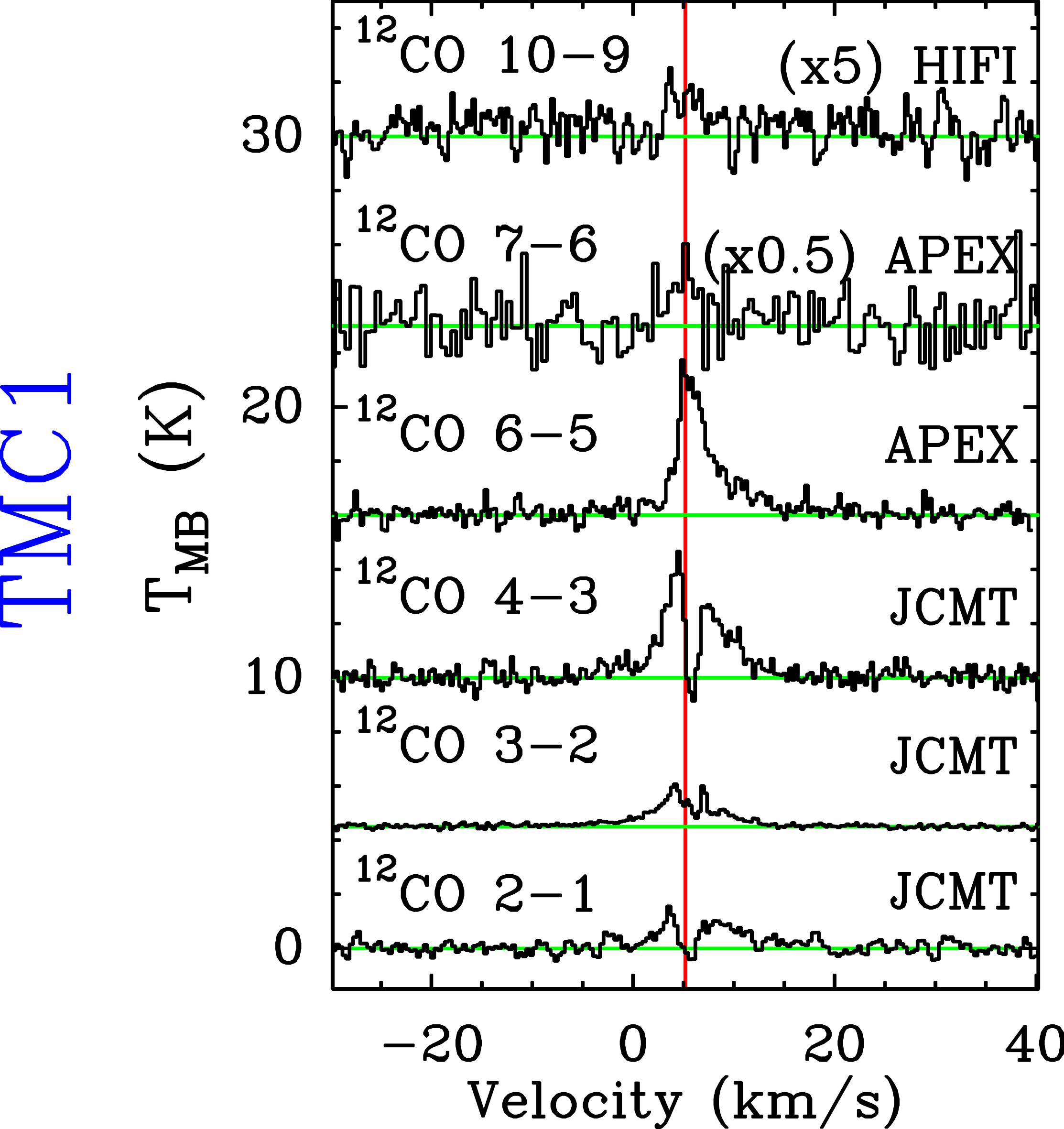}
    \includegraphics[scale=0.3]{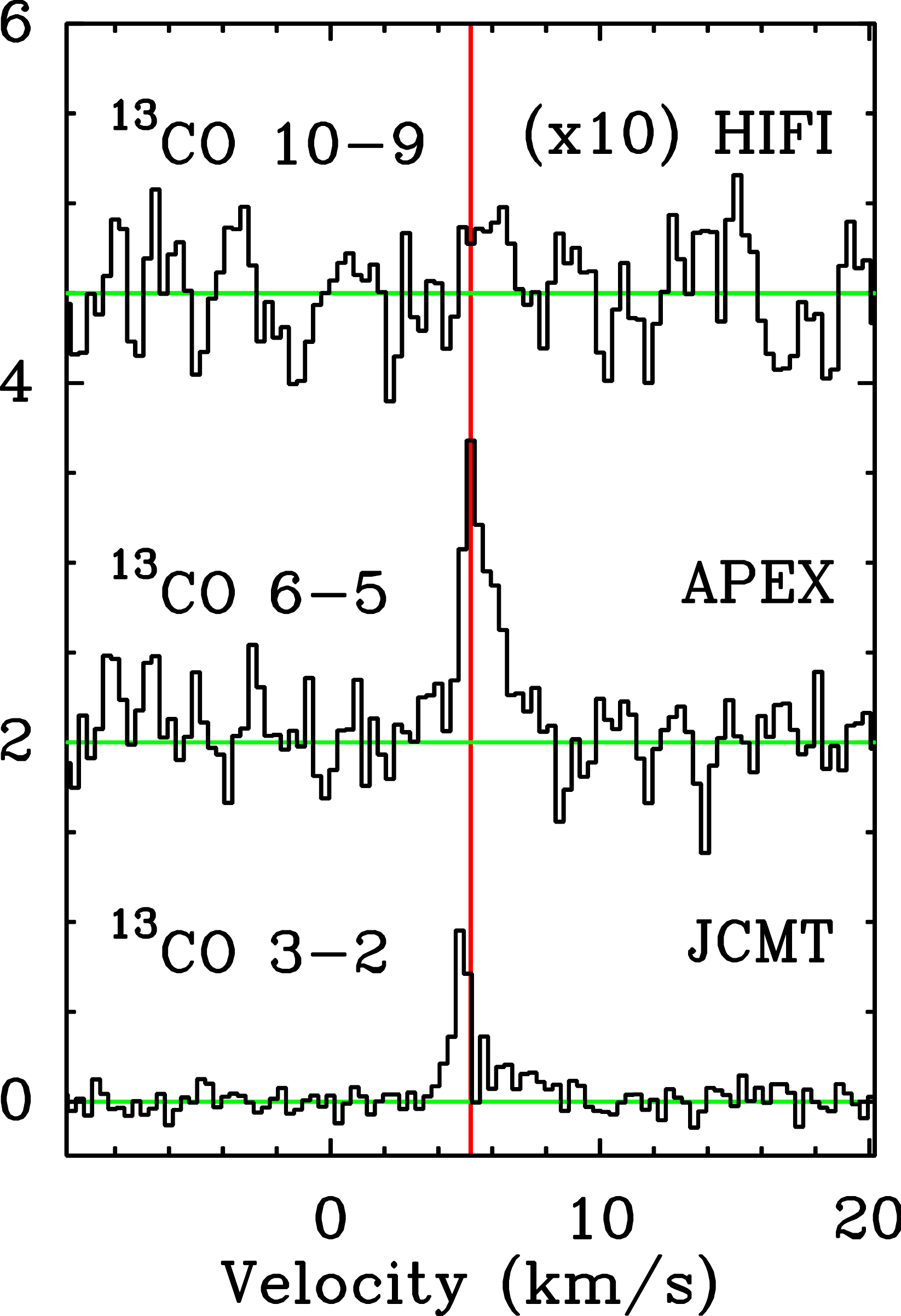}
    \includegraphics[scale=0.3]{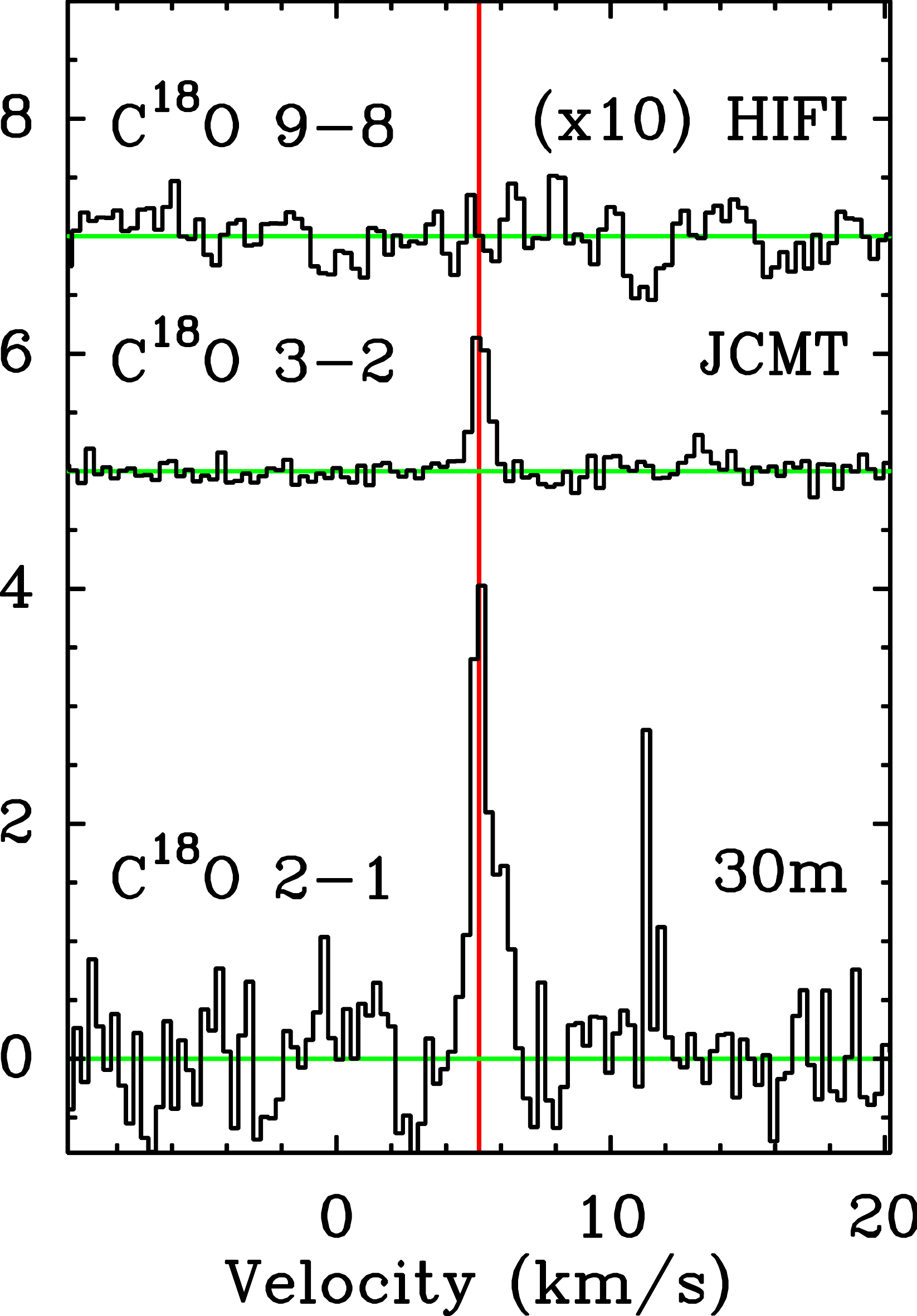}
    \caption{\small Observed $^{12}$CO, $^{13}$CO, and C$^{18}$O transitions for TMC1.}
    \label{fig:linesTMC1}
\end{figure*}

\begin{table*}[!ht]
\caption{Observed line intensities for TMC1 in all observed transitions.}
\normalsize
\begin{center}
\begin{tabular}{l l l r r r r r r r r r}
\hline \hline
Mol.  & Transition & Telescope & Efficiency & $\int T_{\rm MB} \mathrm{d}V$ & $T_{\mathrm{peak}}$ & $rms$ \\
 &  & &   $\eta$ &[K km s$^{-1}$] & [K] &   [K]\\
\hline
CO        & 2--1 & JCMT-RxA             & 0.69 & 9.06\phantom{0}  &  1.56\phantom{0}  & 0.21\phantom{0} \\
          & 3--2 & JCMT-HARPB           & 0.63 & 9.77\phantom{0}  &  2.05\phantom{0}  & 0.10\phantom{0} \\
          & 4--3 & JCMT\tablefootmark{a}& 0.38 & 22.55\phantom{0} &  4.69\phantom{0}  & 0.44\phantom{0} \\
          & 6--5 & APEX-CHAMP$^+$       & 0.52 & 26.58\phantom{0} &  5.98\phantom{0}  & 0.36\phantom{0} \\
          & 7--6 & APEX-CHAMP$^+$       & 0.49 & 11.09\phantom{0} &  4.53\phantom{0}  & 0.92\phantom{0} \\
          &10--9 & {\it Herschel}-HIFI\tablefootmark{b} & 0.64 & 2.87\phantom{0}  & 0.49\phantom{0} & 0.15\phantom{0} \\
$^{13}$CO & 3--2 & JCMT-HARPB           & 0.63 & 1.16\phantom{0}  &  1.09\phantom{0}  & 0.091 \\
          & 6--5 & APEX-CHAMP$^+$       & 0.48 & 2.38\phantom{0}  &  1.81\phantom{0}  & 0.24\phantom{0} \\
          & 10--9& {\it Herschel}-HIFI\tablefootmark{c} & 0.74   &  $<$0.10\phantom{0} &  \dots\phantom{0} & 0.033 \\
C$^{18}$O & 2--1 & IRAM 30m             & 0.59 & 4.17\phantom{0} & 4.10\phantom{0} & 0.41\phantom{0}\\ 
          & 3--2 & JCMT-HARPB           & 0.63 & 0.94\phantom{0}  &  1.35\phantom{0}  & 0.12\phantom{0} \\
          & 9--8 & {\it Herschel}-HIFI\tablefootmark{c} & 0.74   &  $<$0.048 &  \dots\phantom{0}  & 0.021 \\
\hline 
\end{tabular}
\end{center}
\tablefoot{
\tablefoottext{a}{Taken in 11$\arcsec$ beam.}
\tablefoottext{b}{Only H-polarization observation is used.}
\tablefoottext{c}{H- and V-polarization observations averaged.}
}
\label{tbl:linesTMC1}
\end{table*}

\newpage

\onecolumn
\subsection{HH46}
\begin{figure*}[htb]
    \centering
    \includegraphics[scale=0.3]{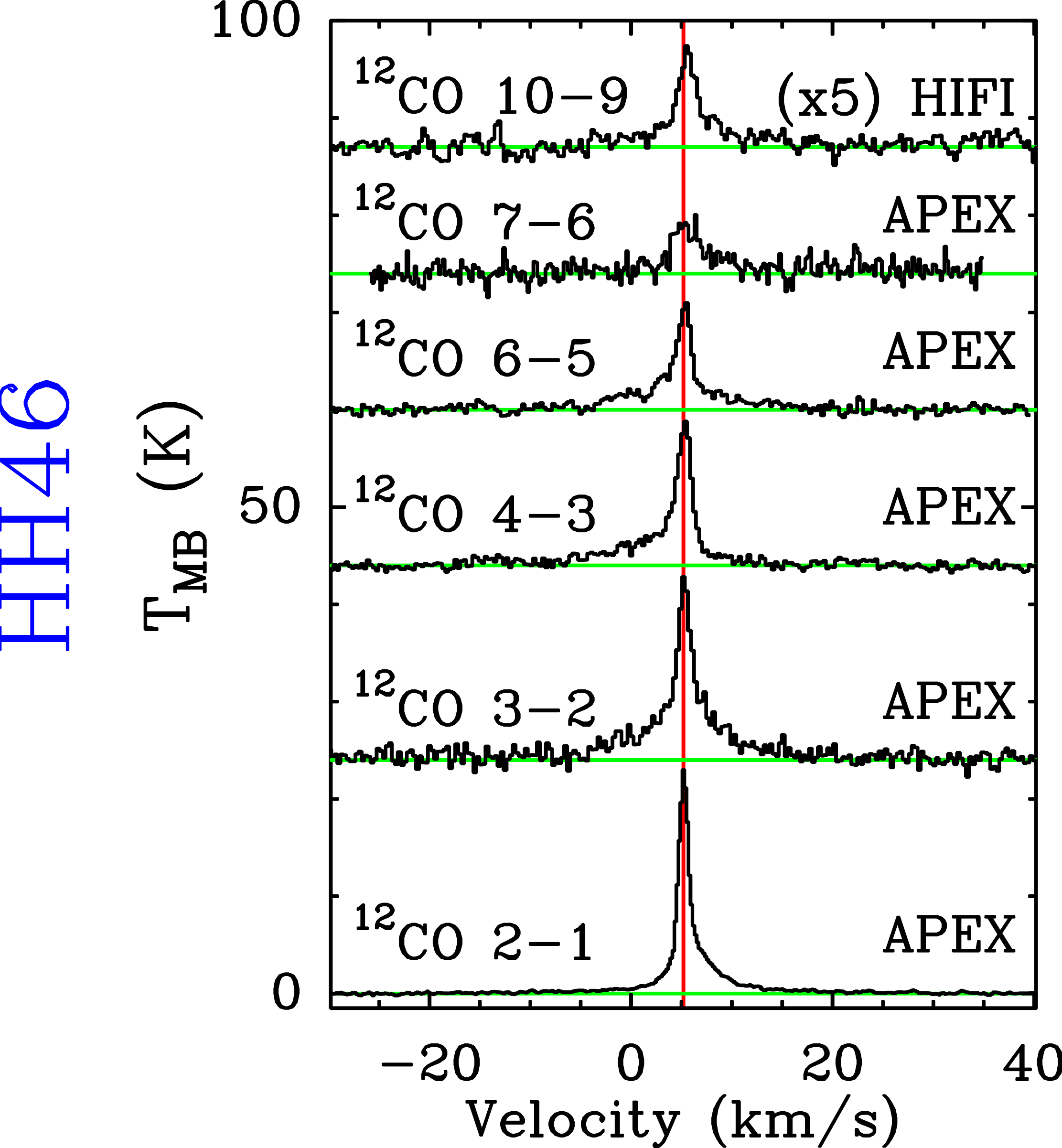}
    \includegraphics[scale=0.3]{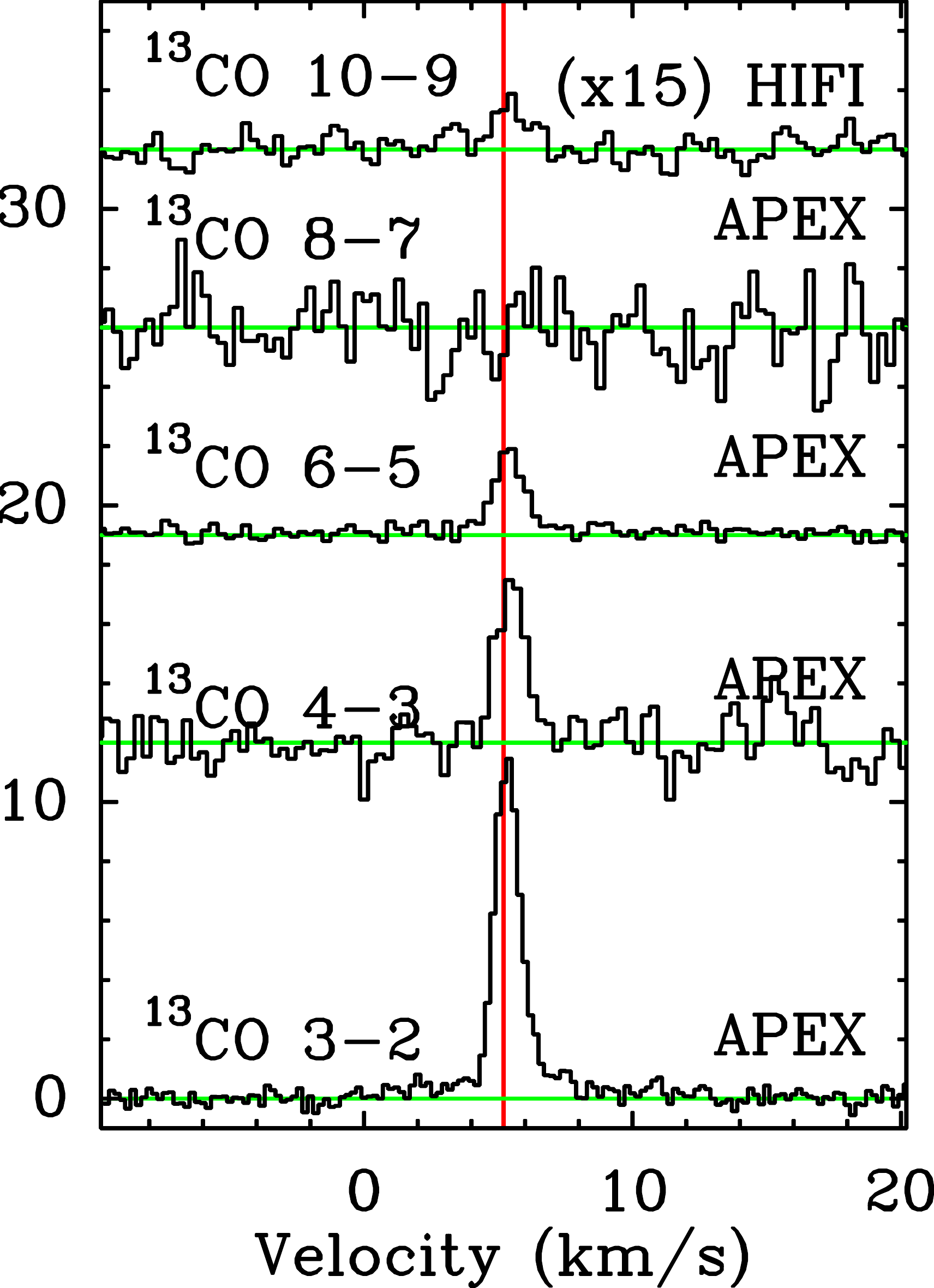}
    \includegraphics[scale=0.3]{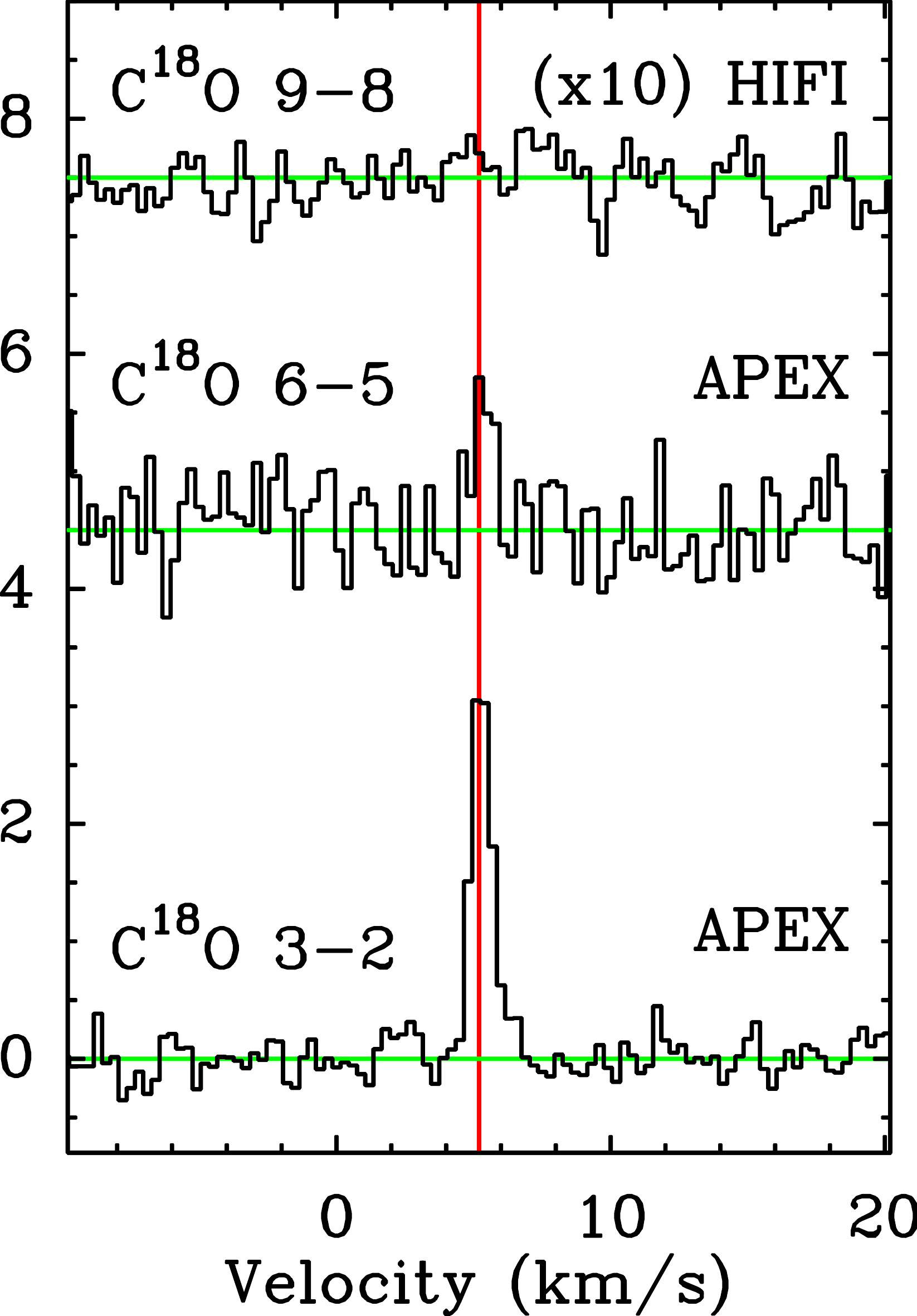}
    \caption{\small Observed $^{12}$CO, $^{13}$CO, and C$^{18}$O transitions for HH46.}
    \label{fig:linesHH46}
\end{figure*}

\begin{table*}[!ht]
\caption{Observed line intensities for HH46 in all observed transitions.}
\normalsize
\begin{center}
\begin{tabular}{l l l r r r r r r r r r}
\hline \hline
Mol.  & Transition & Telescope & Efficiency & $\int T_{\rm MB} \mathrm{d}V$ & $T_{\mathrm{peak}}$ & $rms$ \\
 &  & &   $\eta$ &[K km s$^{-1}$] & [K] &   [K]\\
\hline
CO        & 2--1 & APEX           & 0.73   & 53.56\phantom{0}   &  22.98\phantom{0}  & 0.08\phantom{0} \\
          & 3--2 & APEX           & 0.73   & 81.87\phantom{0}   &  18.78\phantom{0}  & 0.72\phantom{0} \\
          & 4--3 & APEX           & 0.70   & 42.81\phantom{0}   &  14.61\phantom{0}  & 0.49\phantom{0} \\
          & 6--5 & APEX-CHAMP$^+$ & 0.45   & 45.15\phantom{0}   &  11.84\phantom{0}  & 0.47\phantom{0} \\
          & 7--6 & APEX-CHAMP$^+$ & 0.49   & 22.85\phantom{0}   &  6.05\phantom{0}   & 0.89\phantom{0} \\
          &10--9 & {\it Herschel}-HIFI\tablefootmark{a} & 0.64  &  8.21\phantom{0}   & 2.12\phantom{0} & 0.14\phantom{0} \\
$^{13}$CO & 3--2 & APEX           & 0.73   & 17.52\phantom{0}   &  11.89\phantom{0}  & 0.32\phantom{0} \\
          & 4--3 & APEX\tablefootmark{b}   & 0.70 & 8.04\phantom{0} & 6.00\phantom{0} & 1.08\phantom{0} \\
          & 6--5 & APEX-CHAMP$^+$ & 0.45   & 5.94\phantom{0}    &  3.12\phantom{0}  & 0.23\phantom{0} \\
          & 8--7 & APEX-CHAMP$^+$ & 0.42   & $<$1.07\phantom{0} &  \dots\phantom{0} & 0.39\phantom{0} \\
          & 10--9& {\it Herschel}-HIFI\tablefootmark{a} & 0.74  &  0.25\phantom{0}   &  0.18\phantom{0} & 0.041 \\
C$^{18}$O & 3--2 & APEX--2a       & 0.70   &  3.20\phantom{0}   &  \dots \phantom{0}  & 0.10\phantom{0} \\
          & 6--5 & APEX-CHAMP$^+$ & 0.56   &  1.27\phantom{0}   &  1.50\phantom{0}  & 0.50\phantom{0} \\
          & 9--8 & {\it Herschel}-HIFI\tablefootmark{c} & 0.74  &  $<$0.057 & \dots\phantom{0} & 0.025 \\
\hline 
\end{tabular}
\end{center}
\tablefoot{
\tablefoottext{a}{Only H-polarization observation is used.}
\tablefoottext{b}{Taken in 11$\arcsec$ beam.}
\tablefoottext{c}{H- and V-polarization observations averaged.}
}
\label{tbl:linesHH46}
\end{table*}

\newpage

\onecolumn
\subsection{DK Cha}
\begin{figure*}[htb]
    \centering
    \includegraphics[scale=0.3]{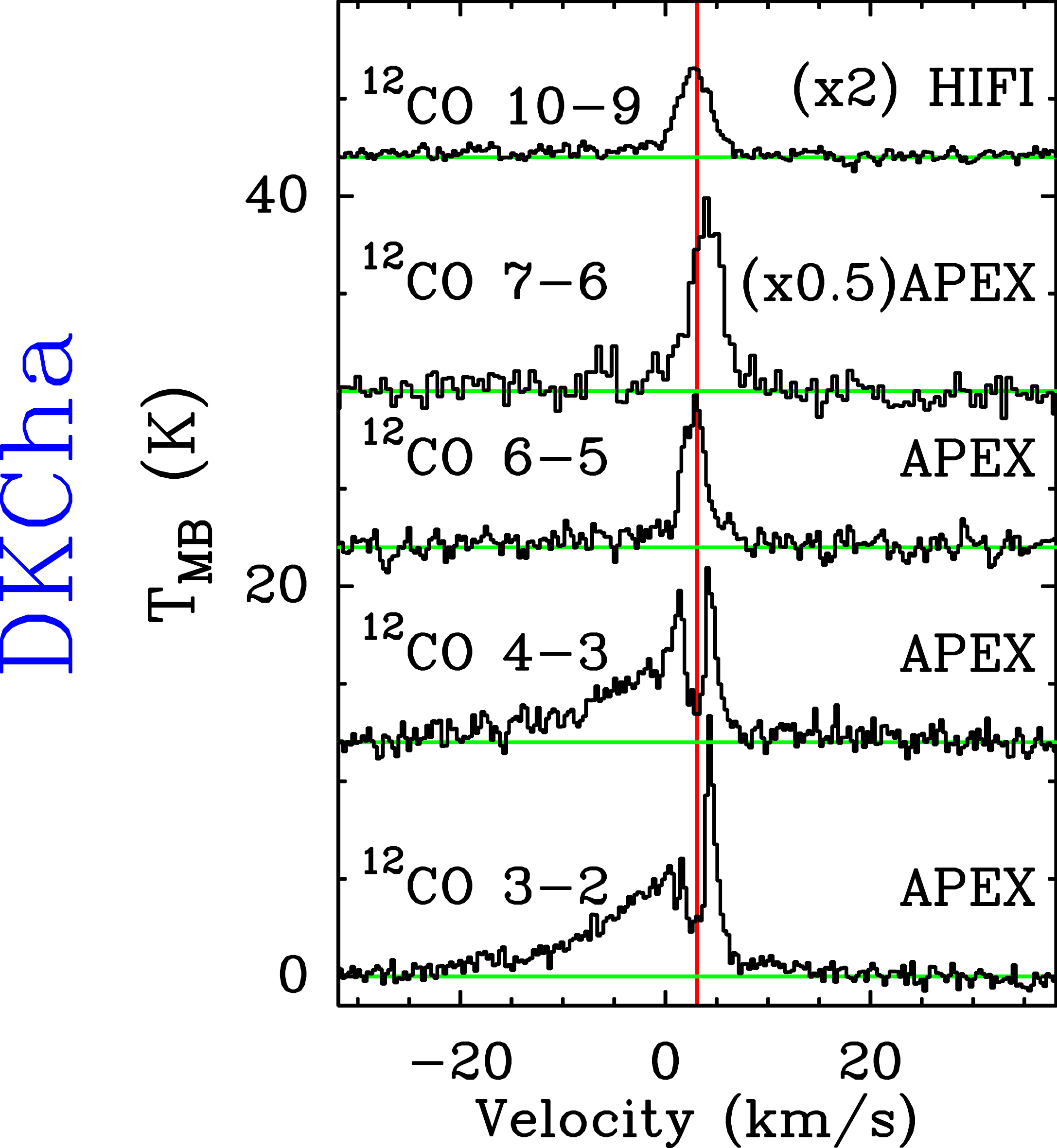}
    \includegraphics[scale=0.3]{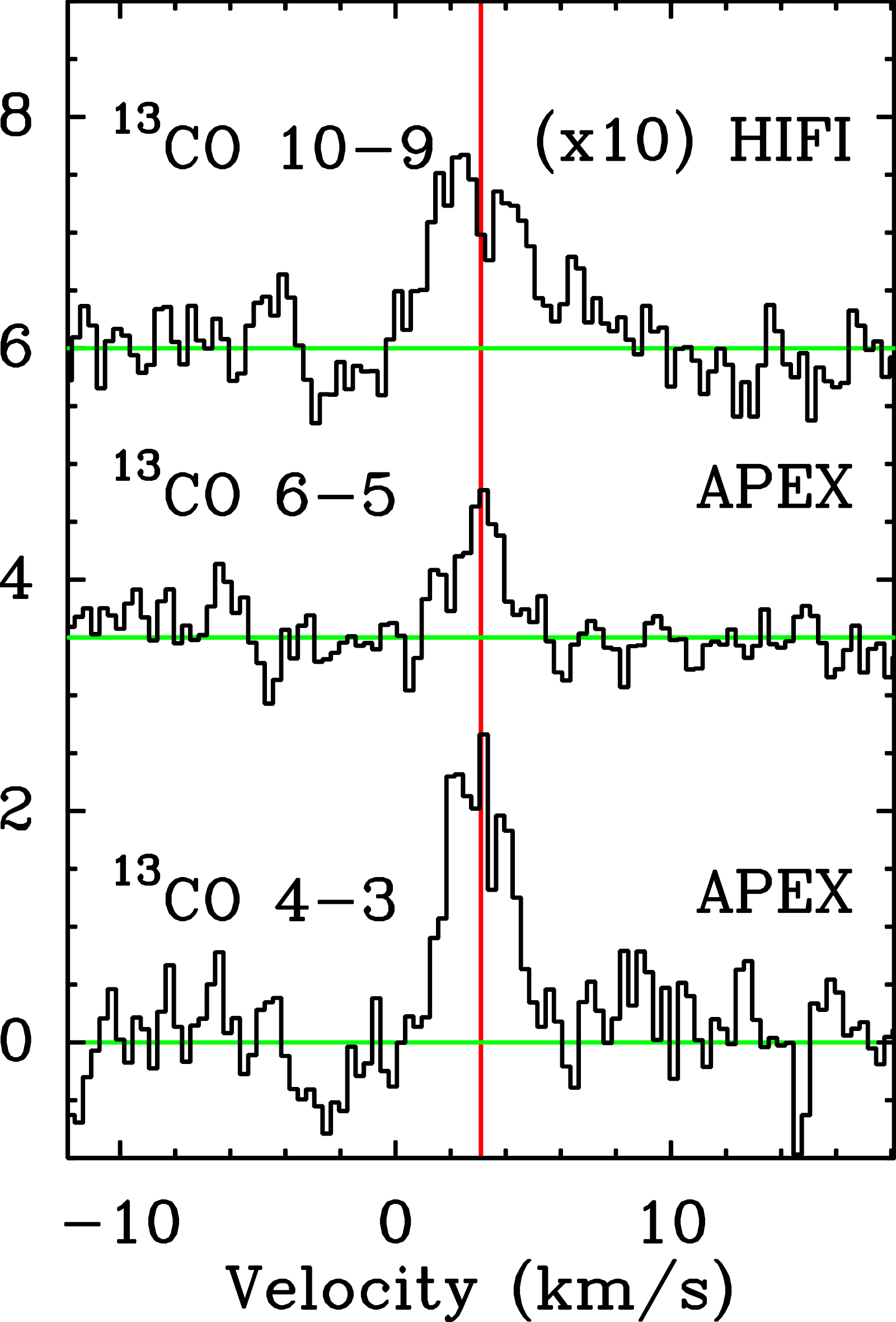}
    \includegraphics[scale=0.3]{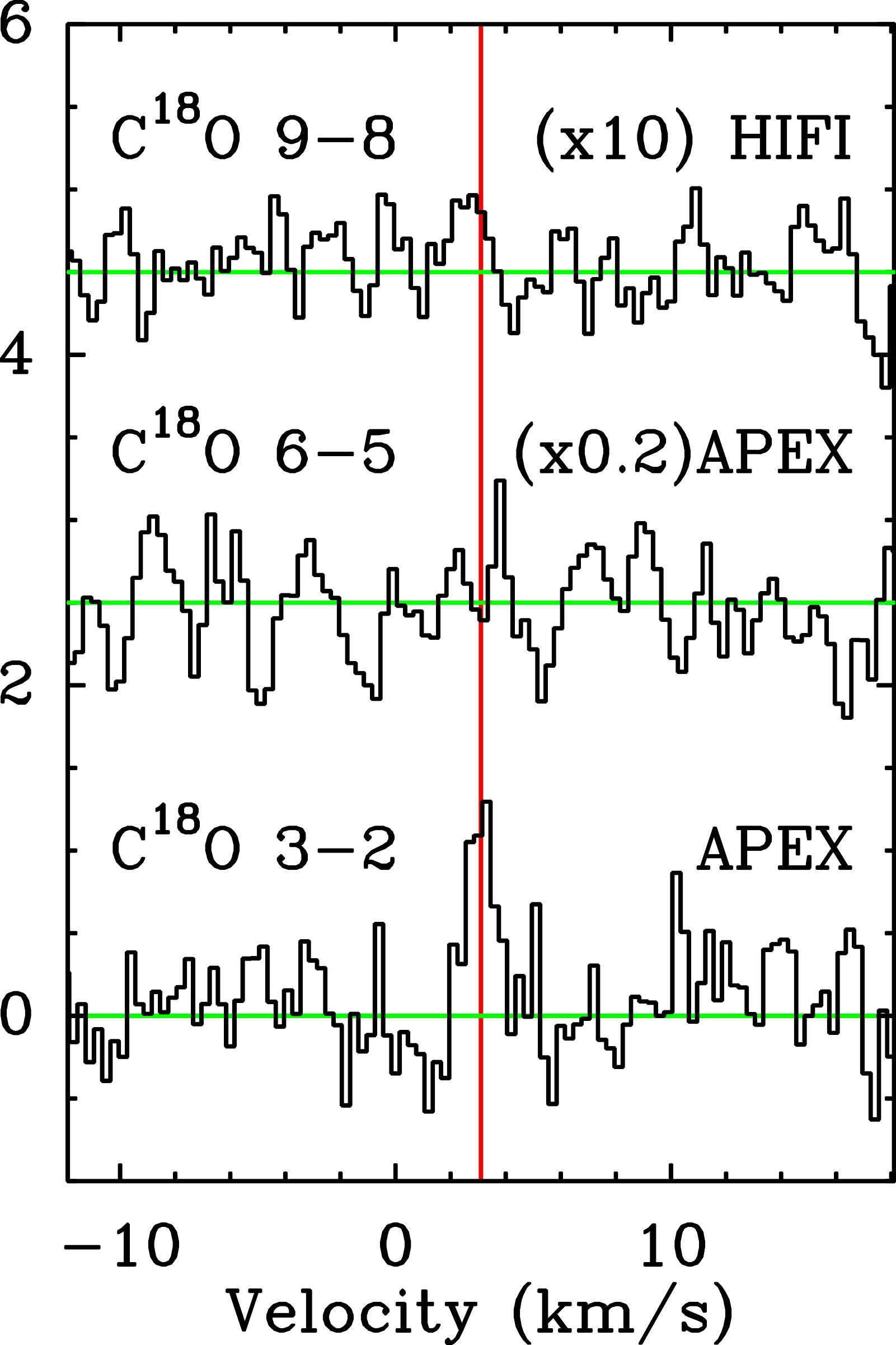}
    \caption{\small Observed $^{12}$CO, $^{13}$CO, and C$^{18}$O transitions for DK Cha.}
    \label{fig:linesDKCha}
\end{figure*}

\begin{table*}[!ht]
\caption{Observed line intensities for DK~Cha in all observed transitions.}
\normalsize
\begin{center}
\begin{tabular}{l l l r r r r r r r r r}
\hline \hline
Mol.  & Transition & Telescope & Efficiency & $\int T_{\rm MB} \mathrm{d}V$ & $T_{\mathrm{peak}}$ & $rms$\\
 &  & &   $\eta$ &[K km s$^{-1}$] & [K] &   [K]\\
\hline
CO        & 3--2 & APEX           & 0.73   &  79.98\phantom{0}   &  13.35\phantom{0}  & 0.38\phantom{0} \\
          & 4--3 & APEX           & 0.65   &  71.77\phantom{0}   &  10.22\phantom{0}  & 0.49\phantom{0} \\
          & 6--5 & APEX-CHAMP$^+$ & 0.45   &  28.02\phantom{0}   &  8.21\phantom{0}   & 0.57\phantom{0} \\
          & 7--6 & APEX-CHAMP$^+$ & 0.42   &  84.86\phantom{0}   &  20.21\phantom{0}  & 1.89\phantom{0} \\
          &10--9 & {\it Herschel}-HIFI\tablefootmark{a} & 0.64 & 11.40\phantom{0} & 2.31\phantom{0} & 0.12\phantom{0} \\
$^{13}$CO & 3--2 & APEX\tablefootmark{b}   & \dots &  6.70\phantom{0} & \dots\phantom{0}  & \dots\phantom{0} \\
          & 4--3 & APEX           & 0.65   &  6.54\phantom{0}    &  3.06\phantom{0}   & 0.44\phantom{0} \\
          & 6--5 & APEX-CHAMP$^+$ & 0.45   &  2.19\phantom{0}    &  1.37\phantom{0}   & 0.25\phantom{0} \\
          & 8--7 & APEX-CHAMP$^+$ & 0.49   &  0.84\phantom{0}    &  \dots \phantom{0}& 0.90\phantom{0} \\
          & 10--9& {\it Herschel}-HIFI\tablefootmark{c}  & 0.74 &  0.82\phantom{0}   &  0.19\phantom{0}  & 0.028 \\
C$^{18}$O & 3--2 & APEX           & 0.70    &  1.41\phantom{0}    &  1.55\phantom{0}  & 0.40\phantom{0} \\
          & 9--8 & {\it Herschel}-HIFI\tablefootmark{a}  & 0.74 &  $<$0.057 & \dots\phantom{0} & 0.025 \\
\hline 
\end{tabular}
\end{center}
\tablefoot{
\tablefoottext{a}{H- and V-polarization observations averaged.}
\tablefoottext{b}{van Kempen et al. (2006)}
\tablefoottext{c}{Only H-polarization observation is used.}
}
\label{tbl:linesDKCha}
\end{table*}

\newpage
\clearpage

\onecolumn
\subsection{GSS30IRS1}
\begin{figure*}[htb]
    \centering
    \includegraphics[scale=0.3]{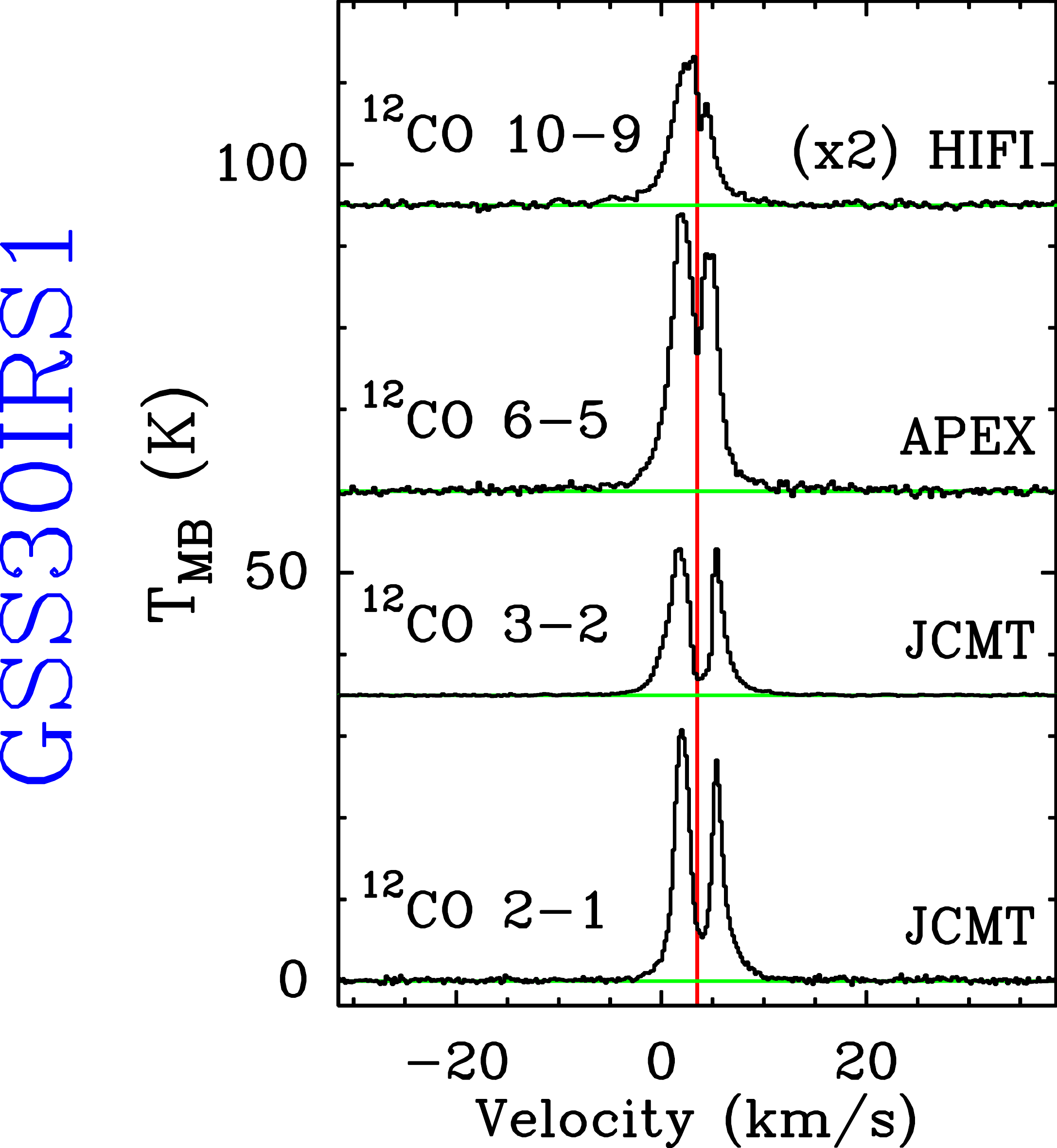}
    \includegraphics[scale=0.3]{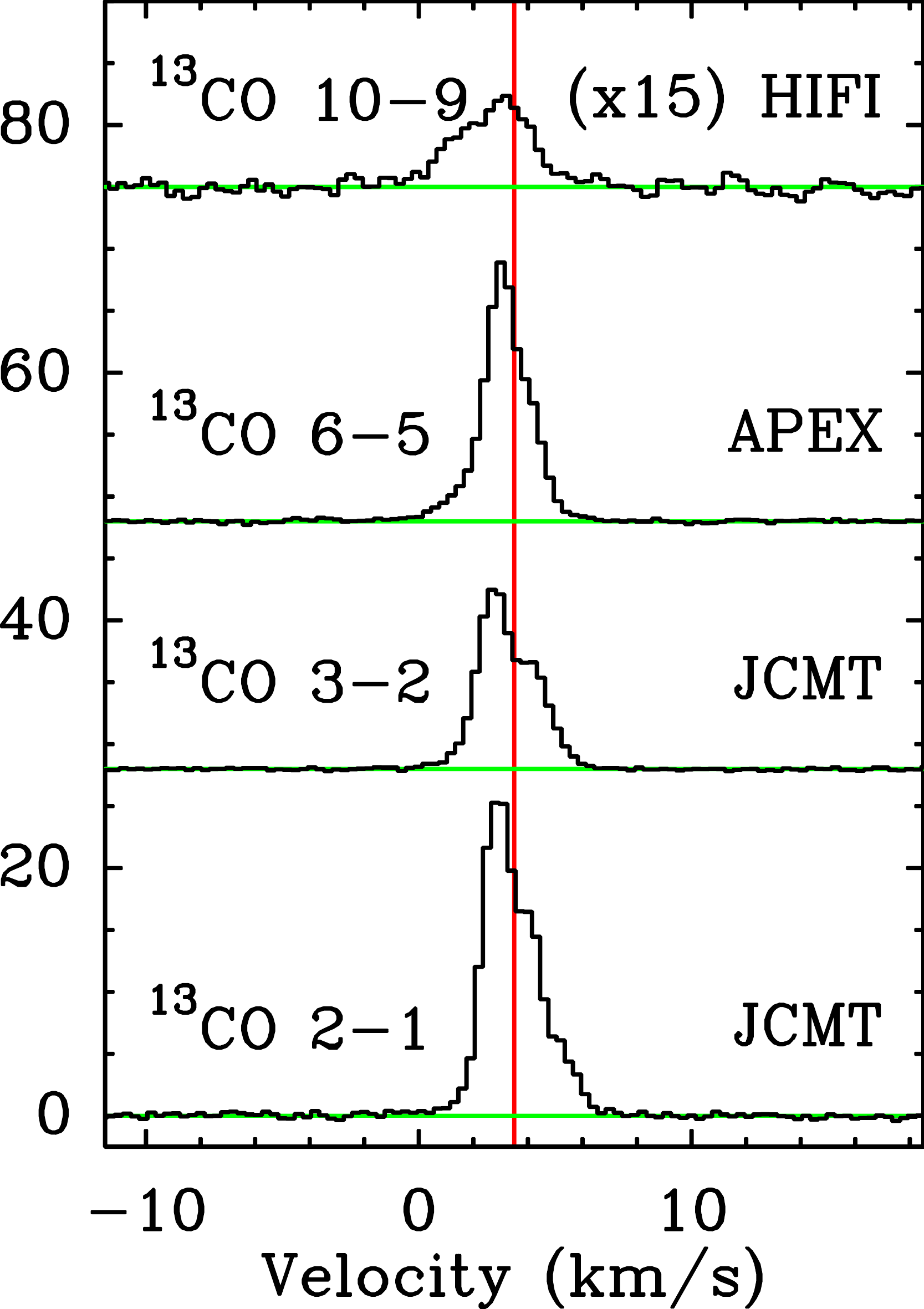}
    \includegraphics[scale=0.3]{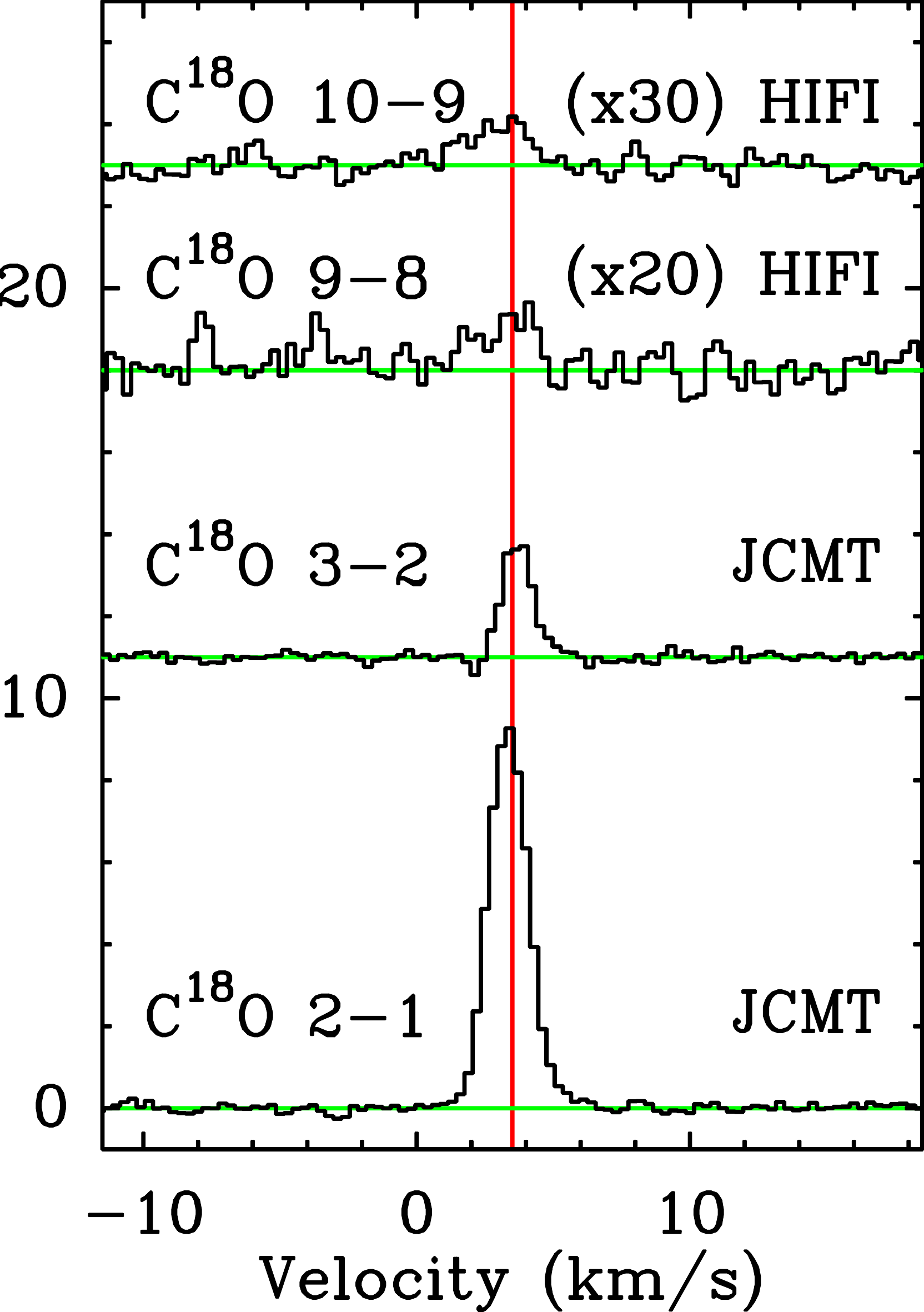}
    \caption{\small Observed $^{12}$CO, $^{13}$CO, and C$^{18}$O transitions for GSS30IRS1.}
    \label{fig:linesGSS30IRS1}
\end{figure*}

\begin{table*}[!ht]
\caption{Observed line intensities for GSS30IRS1 in all observed transitions.}
\normalsize
\begin{center}
\begin{tabular}{l l l r r r r r r r r r}
\hline \hline
Mol.  & Transition & Telescope & Efficiency & $\int T_{\rm MB} \mathrm{d}V$ & $T_{\mathrm{peak}}$ & $rms$ \\
 &  & &   $\eta$ &[K km s$^{-1}$] & [K] &   [K]\\
\hline
CO        & 2--1 & JCMT-RxA         & 0.69   &  114.80\phantom{0} &  30.76\phantom{0} & 0.23\phantom{0} \\
          & 3--2 & JCMT-HARPB       & 0.63   &  77.95\phantom{0}  &  18.90\phantom{0} & 0.079 \\
          & 6--5 & APEX-CHAMP$^+$   & 0.45   &  172.38\phantom{0} &  34.48\phantom{0} & 0.41\phantom{0} \\
          &10--9 & {\it Herschel}-HIFI\tablefootmark{a} & 0.64   &  41.78\phantom{0} & 9.28\phantom{0}  & 0.15\phantom{0} \\
$^{13}$CO & 2--1 & JCMT-RxA         & 0.74   &  58.94\phantom{0}  &  26.30\phantom{0} & 0.25\phantom{0} \\
          & 3--2 & JCMT-HARPB       & 0.63   &  34.33\phantom{0}  &  14.87\phantom{0} & 0.12\phantom{0} \\
          & 6--5 & APEX-CHAMP$^+$   & 0.48   &  41.64\phantom{0}  &  21.37\phantom{0} & 0.13\phantom{0} \\
          & 10--9& {\it Herschel}-HIFI\tablefootmark{a} & 0.74   &  2.06\phantom{0}  & 0.50\phantom{0} & 0.032 \\
C$^{18}$O & 2--1 & JCMT-RxA         & 0.69   &  17.09\phantom{0}  &  9.45\phantom{0}  & 0.12\phantom{0} \\
          & 3--2 & JCMT-HARPB       & 0.63   &  3.73\phantom{0}   &  2.81\phantom{0}  & 0.12\phantom{0} \\
          & 9--8 & {\it Herschel}-HIFI\tablefootmark{b} & 0.74   &  0.21\phantom{0}  & 0.093 & 0.023 \\
          &10--9 & {\it Herschel}-HIFI\tablefootmark{b} & 0.74   &  0.089   &  0.042  & 0.009 \\
\hline 
\end{tabular}
\end{center}
\tablefoot{
\tablefoottext{a}{Only H-polarization observation is used.}
\tablefoottext{b}{H- and V-polarization observations averaged.}
}
\label{tbl:linesGSS30IRS1}
\end{table*}

\newpage
\clearpage

\onecolumn
\subsection{Elias29}
\begin{figure*}[htb]
    \centering
    \includegraphics[scale=0.3]{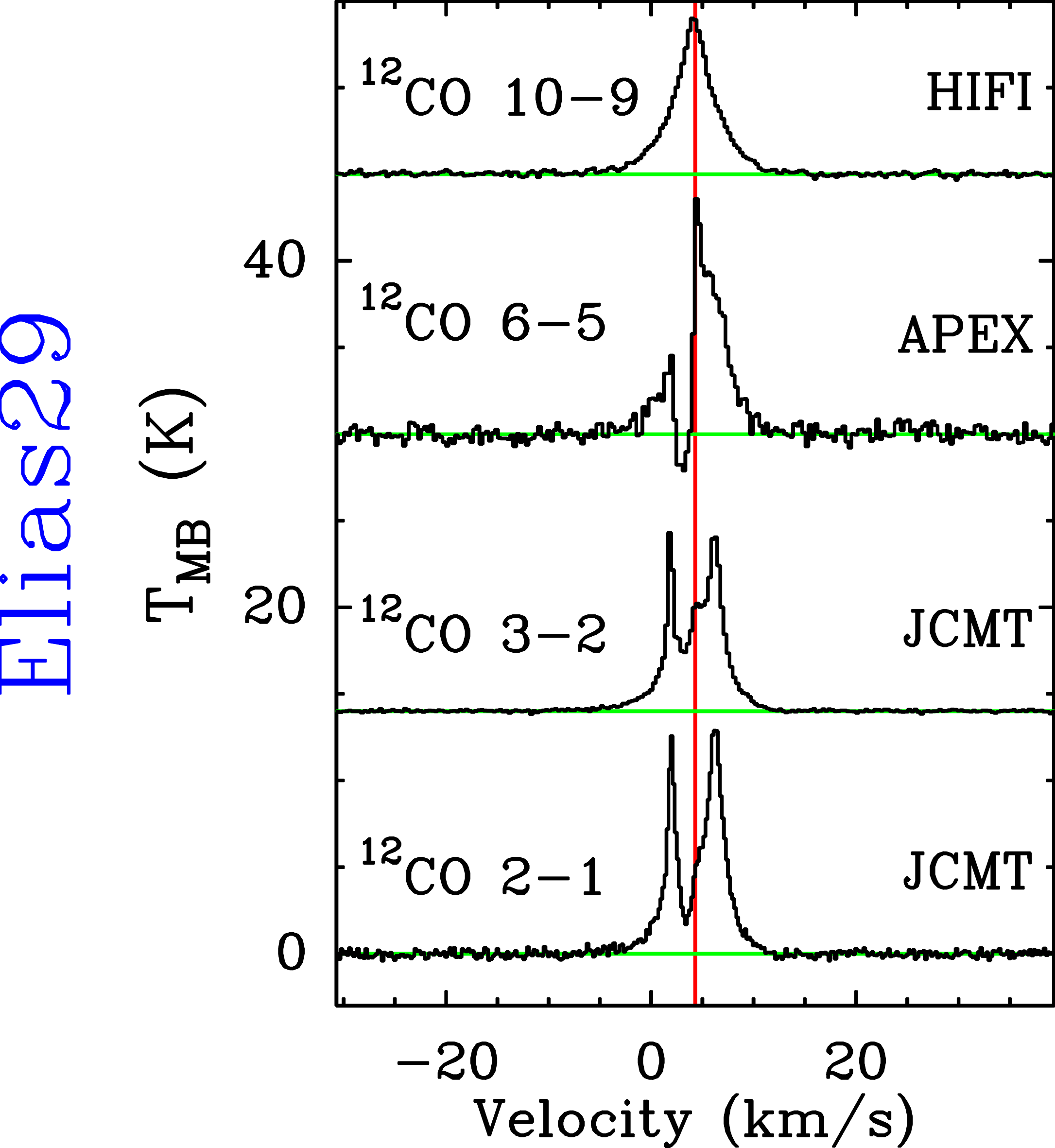}
    \includegraphics[scale=0.3]{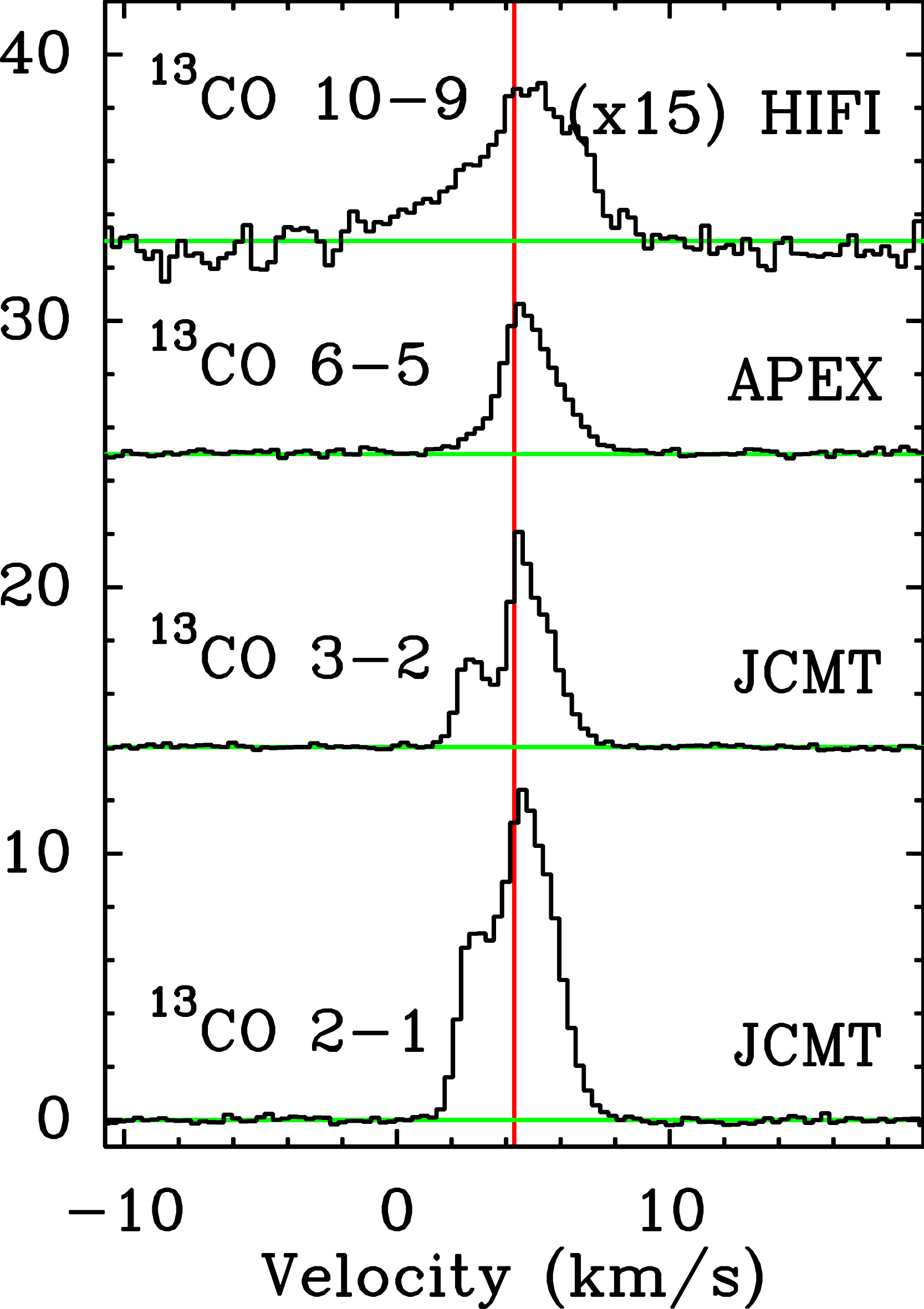}
    \includegraphics[scale=0.3]{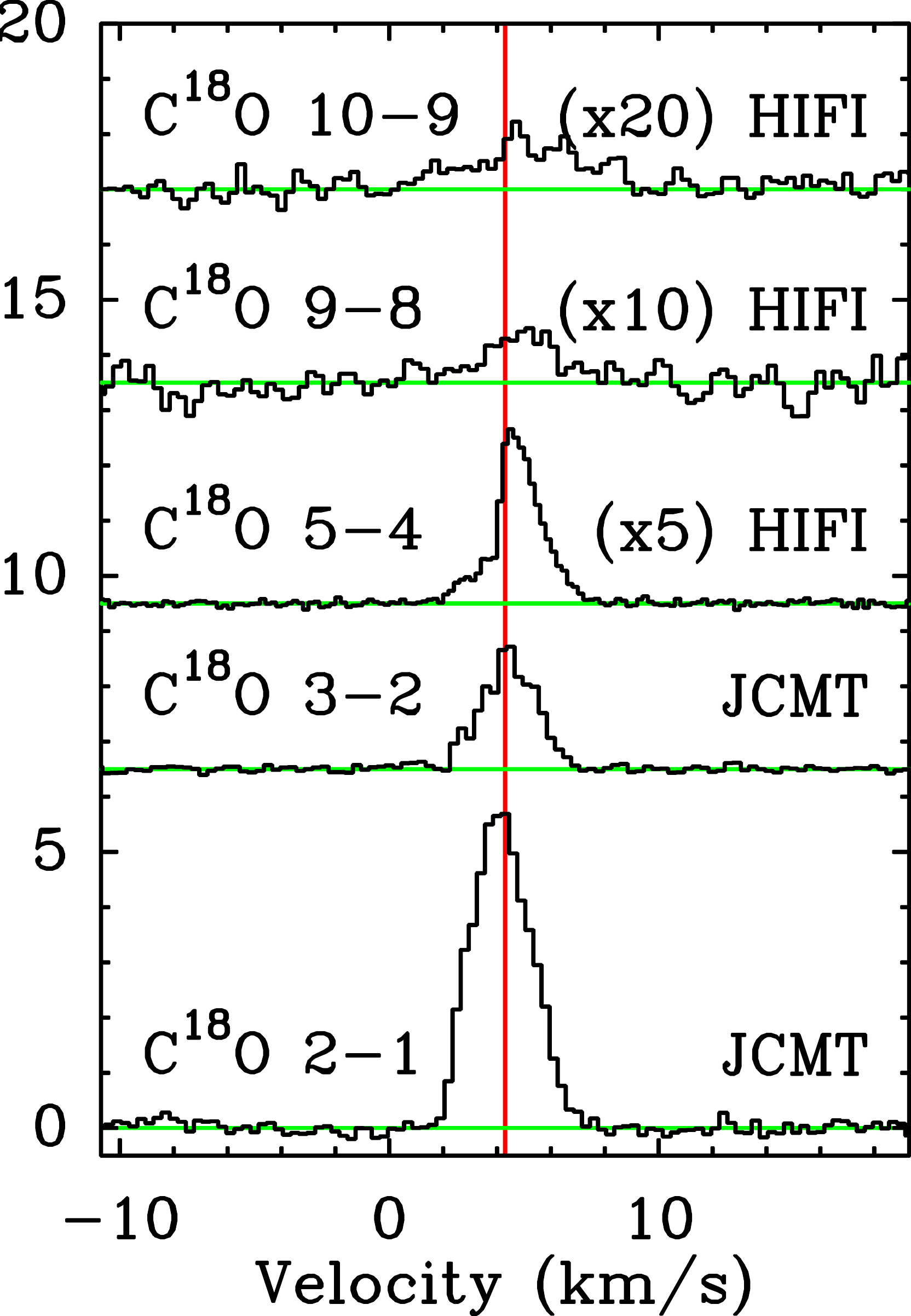}
    \caption{\small Observed $^{12}$CO, $^{13}$CO, and C$^{18}$O transitions for Elias29.}
    \label{fig:linesElias29}
\end{figure*}

\begin{table*}[!ht]
\caption{Observed line intensities for Elias29 in all observed transitions.}
\normalsize
\begin{center}
\begin{tabular}{l l l r r r r r r r r r}
\hline \hline
Mol.  & Transition & Telescope & Efficiency & $\int T_{\rm MB} \mathrm{d}V$ & $T_{\mathrm{peak}}$ & $rms$ \\
 &  & &   $\eta$ &[K km s$^{-1}$] & [K] &   [K]\\
\hline
CO        & 2--1 & JCMT-RxA        & 0.69   &  53.74 & 12.88 & 0.18\phantom{0} \\
          & 3--2 & JCMT-HARPB      & 0.63   &  48.35 & 11.14 & 0.069 \\
          & 6--5 & APEX-CHAMP$^+$  & 0.48   &  45.08 & 14.55 & 0.48\phantom{0} \\
          &10--9 & {\it Herschel}-HIFI\tablefootmark{a} & 0.64 & 45.86 &  9.15  & 0.11\phantom{0} \\
$^{13}$CO & 2--1 & JCMT-RxA        & 0.74   &  37.38 & 12.55 & 0.17\phantom{0} \\
          & 3--2 & JCMT-HARPB      & 0.63   &  18.71 &  8.25 & 0.10\phantom{0} \\
          & 6--5 & APEX-CHAMP$^+$  & 0.48   &  14.62 &  5.68 & 0.13\phantom{0} \\
          & 10--9& {\it Herschel}-HIFI\tablefootmark{a} & 0.74 & 1.72  &  0.39  & 0.031 \\
C$^{18}$O & 2--1 & JCMT-RxA        & 0.69   &  15.69 & 5.89  & 0.15\phantom{0} \\
          & 3--2 & JCMT-HARPB      & 0.63   &  5.27  & 2.53  & 0.066 \\
          & 5--4 & {\it Herschel}-HIFI\tablefootmark{a} & 0.76 & 1.26 &  0.60  & 0.010 \\
          & 9--8 & {\it Herschel}-HIFI\tablefootmark{a} & 0.74 & 0.34 &  0.10  & 0.027 \\
          &10--9 & {\it Herschel}-HIFI\tablefootmark{b} & 0.74 & 0.23 &  0.07  & 0.010 \\
\hline 
\end{tabular}
\end{center}
\tablefoot{
\tablefoottext{a}{H- and V-polarization observations averaged.}
\tablefoottext{b}{Only H-polarization observation is used.}
}
\label{tbl:linesElias29}
\end{table*}

\newpage

\onecolumn
\subsection{Oph~IRS63}
\begin{figure*}[htb]
    \centering
    \includegraphics[scale=0.3]{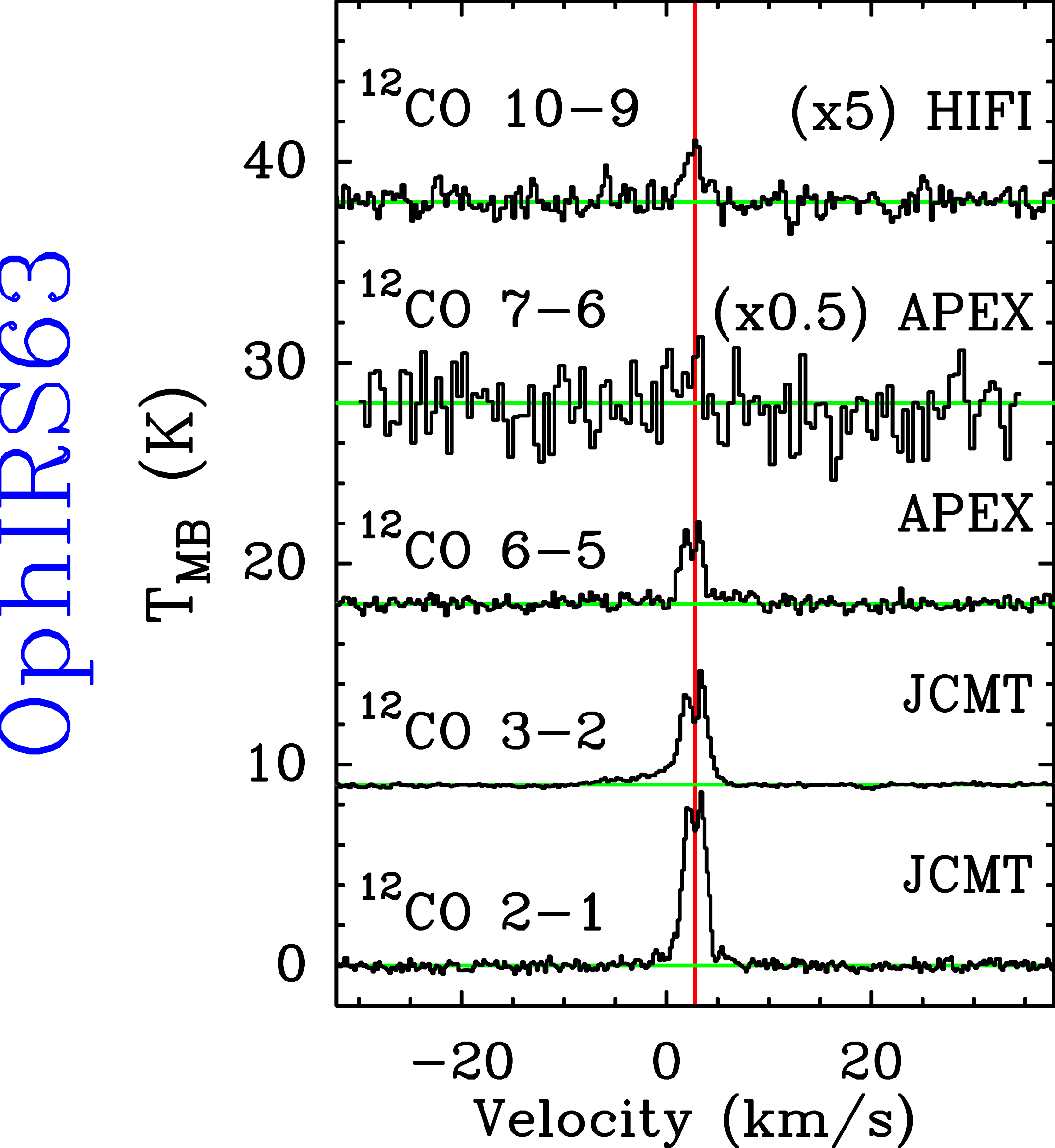}
    \includegraphics[scale=0.3]{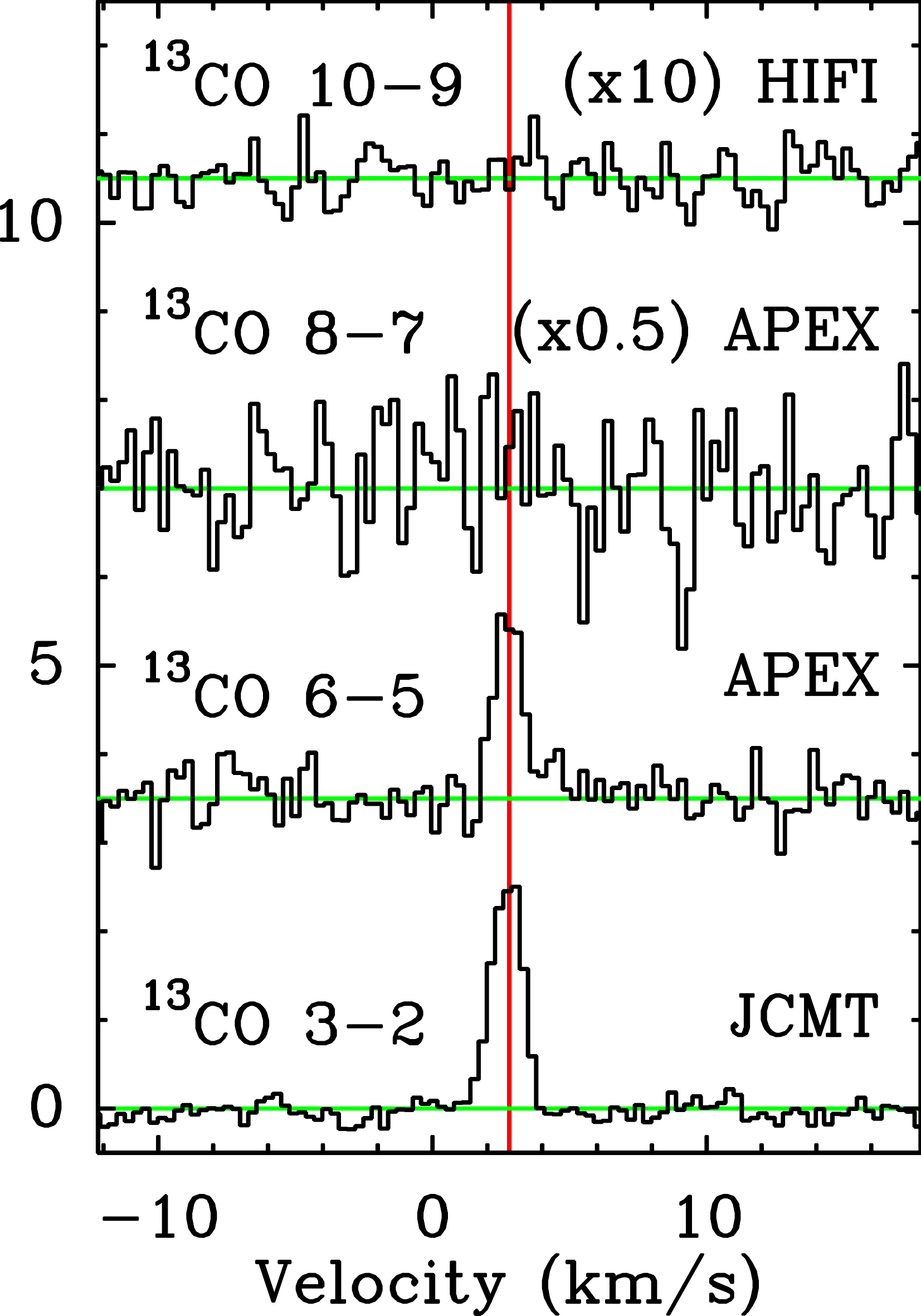}
    \includegraphics[scale=0.3]{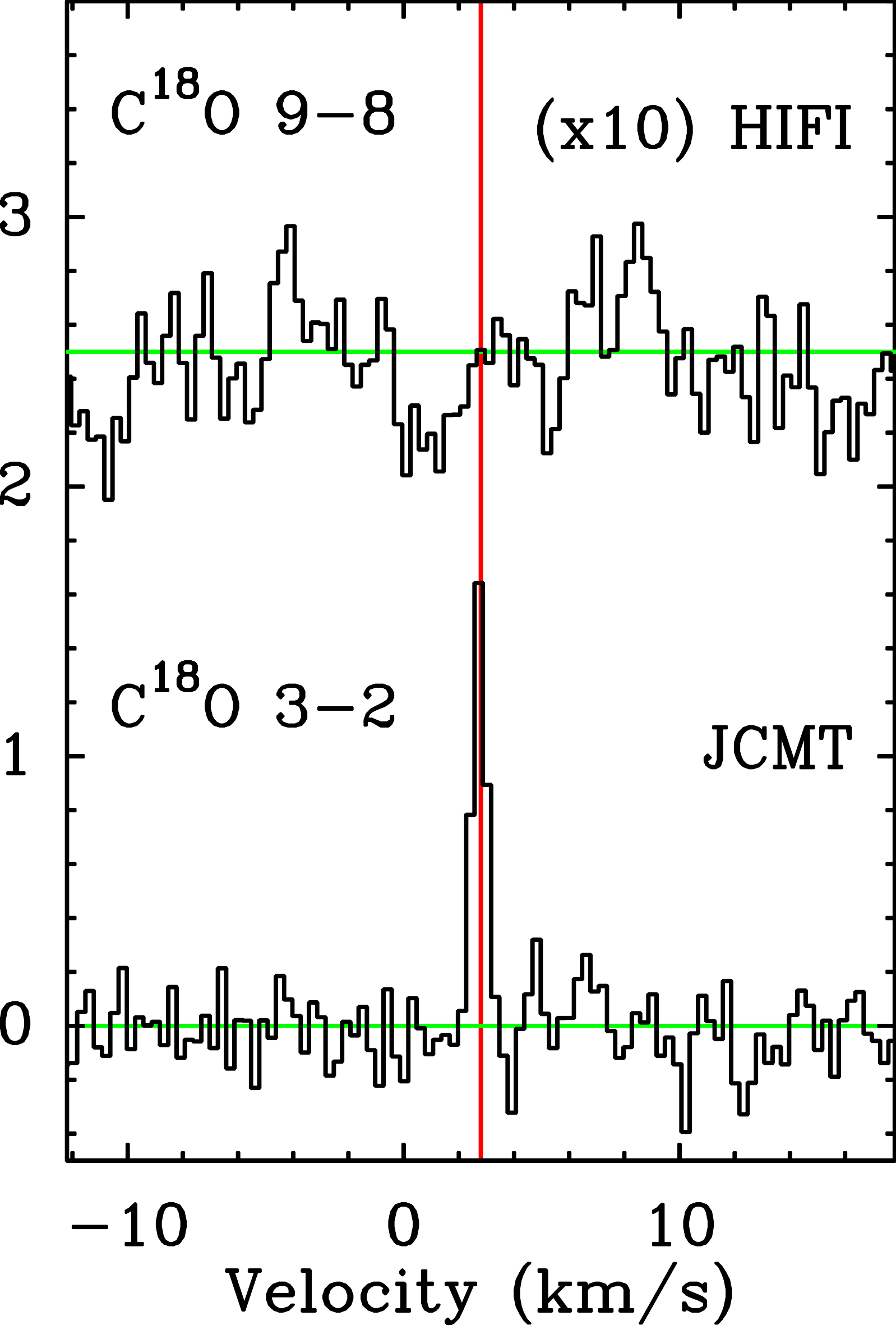}
    \caption{\small Observed $^{12}$CO, $^{13}$CO, and C$^{18}$O transitions for Oph~IRS63.}
    \label{fig:linesOphIRS63}
\end{figure*}

\begin{table*}[!ht]
\caption{Observed line intensities for OphIRS63 in all observed transitions. }
\normalsize
\begin{center}
\begin{tabular}{l l l r r r r r r r r r}
\hline \hline
Mol.  & Transition & Telescope & Efficiency & $\int T_{\rm MB} \mathrm{d}V$ & $T_{\mathrm{peak}}$ & $rms$ \\
 &  & &   $\eta$ &[K km s$^{-1}$] & [K] &   [K]\\
\hline
CO        & 2--1 & JCMT-RxA       & 0.69   &  23.27\phantom{0}  &  8.64  & 0.19\phantom{0} \\
          & 3--2 & JCMT-HARPB     & 0.63   &  16.47\phantom{0}  &  6.06  & 0.094 \\
          & 6--5 & APEX-CHAMP$^+$ & 0.48   &  11.11\phantom{0}  &  4.36  & 0.32\phantom{0} \\
          & 7--6 & APEX-CHAMP$^+$ & 0.48   &  8.63\phantom{0}   &  4.77  & 1.09\phantom{0} \\
          &10--9 & {\it Herschel}-HIFI\tablefootmark{a} & 0.64 &  1.15\phantom{0}  & 0.62  & 0.13\phantom{0} \\
$^{13}$CO & 3--2 & JCMT-HARPB     & 0.63   &  3.58\phantom{0}   &  2.67  & 0.13\phantom{0} \\
          & 6--5 & APEX-CHAMP$^+$ & 0.45   &  3.02\phantom{0}   &  2.51  & 0.35\phantom{0} \\
          & 8--7 & APEX-CHAMP$^+$ & 0.49   &  $<$4.26\phantom{0} & \dots\phantom{0} & 1.68\phantom{0} \\
          & 10--9& {\it Herschel}-HIFI\tablefootmark{a} & 0.74 &  $<$0.066 & \dots\phantom{0}  & 0.026 \\
C$^{18}$O & 3--2 & JCMT-HARPB     & 0.63   &  0.94\phantom{0}   & 1.73  & 0.17\phantom{0} \\
          & 9--8 & {\it Herschel}-HIFI\tablefootmark{a} & 0.74 &  $<$0.048 & \dots\phantom{0}  & 0.021 \\
\hline 
\end{tabular}
\end{center}
\tablefoot{
\tablefoottext{a}{H- and V-polarization observations averaged.}
}
\label{tbl:linesOphIRS63}
\end{table*}

\newpage
\clearpage

\onecolumn
\subsection{RNO91}
\begin{figure*}[htb]
    \centering
    \includegraphics[scale=0.3]{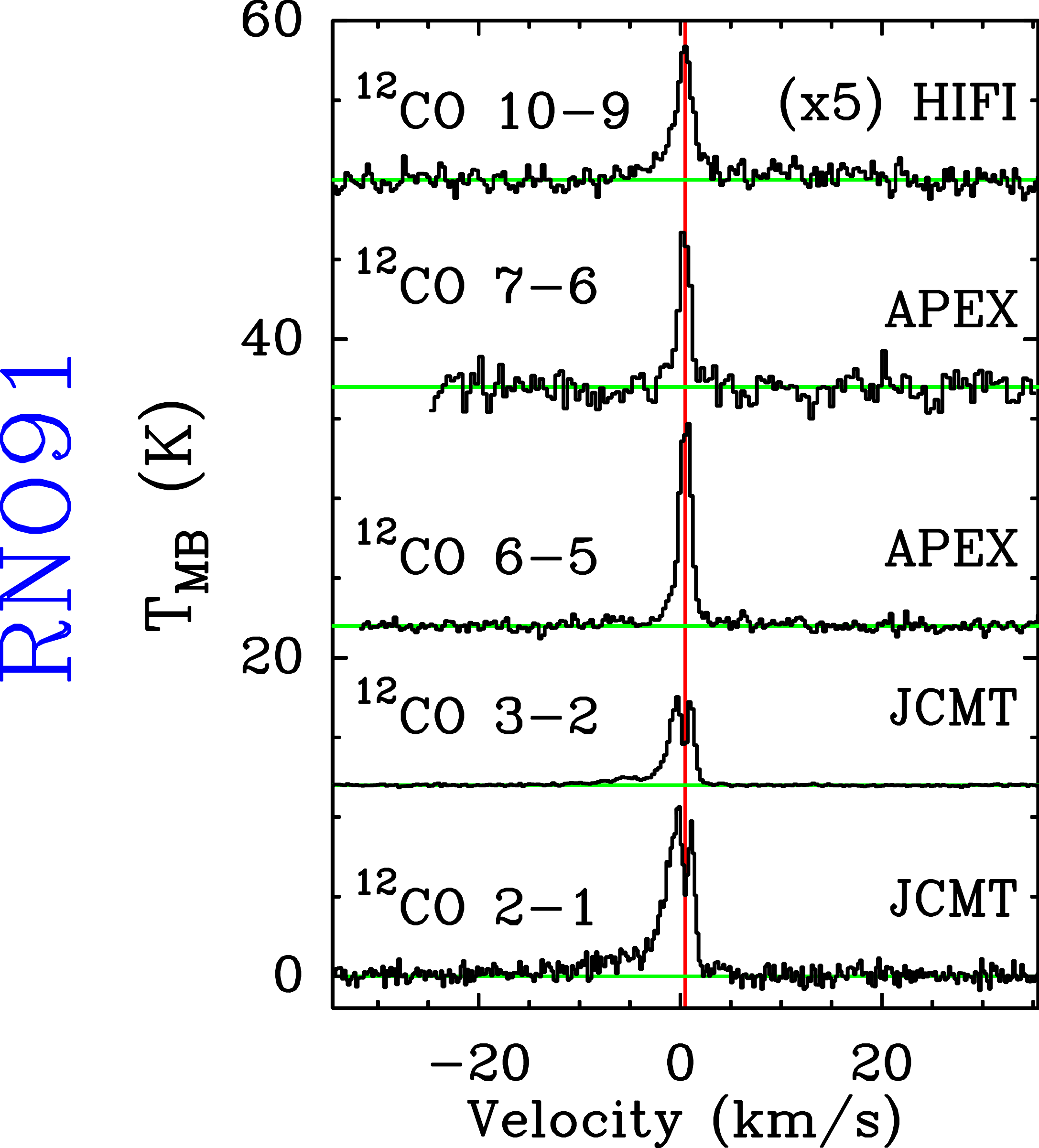}
    \includegraphics[scale=0.3]{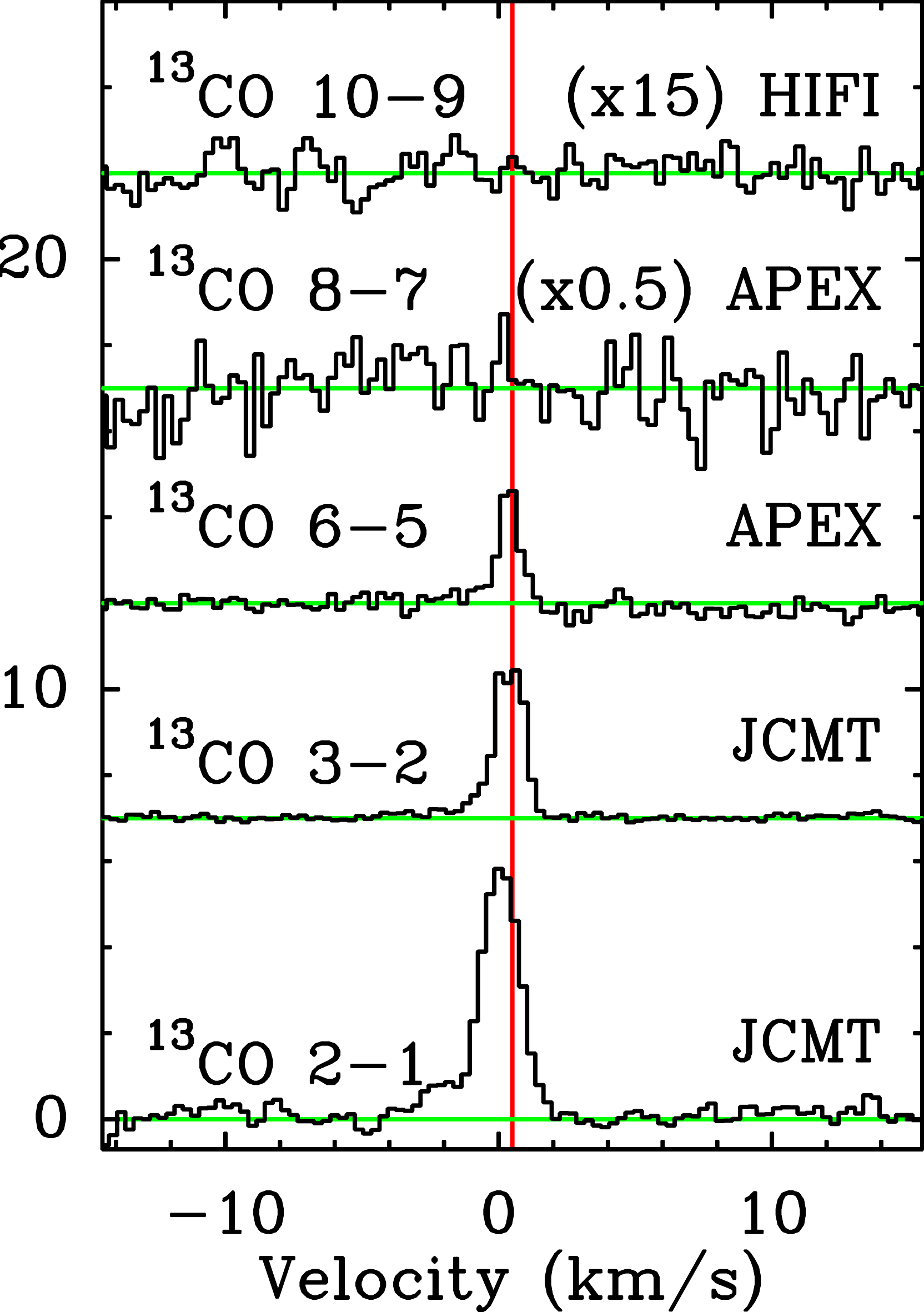}
    \includegraphics[scale=0.3]{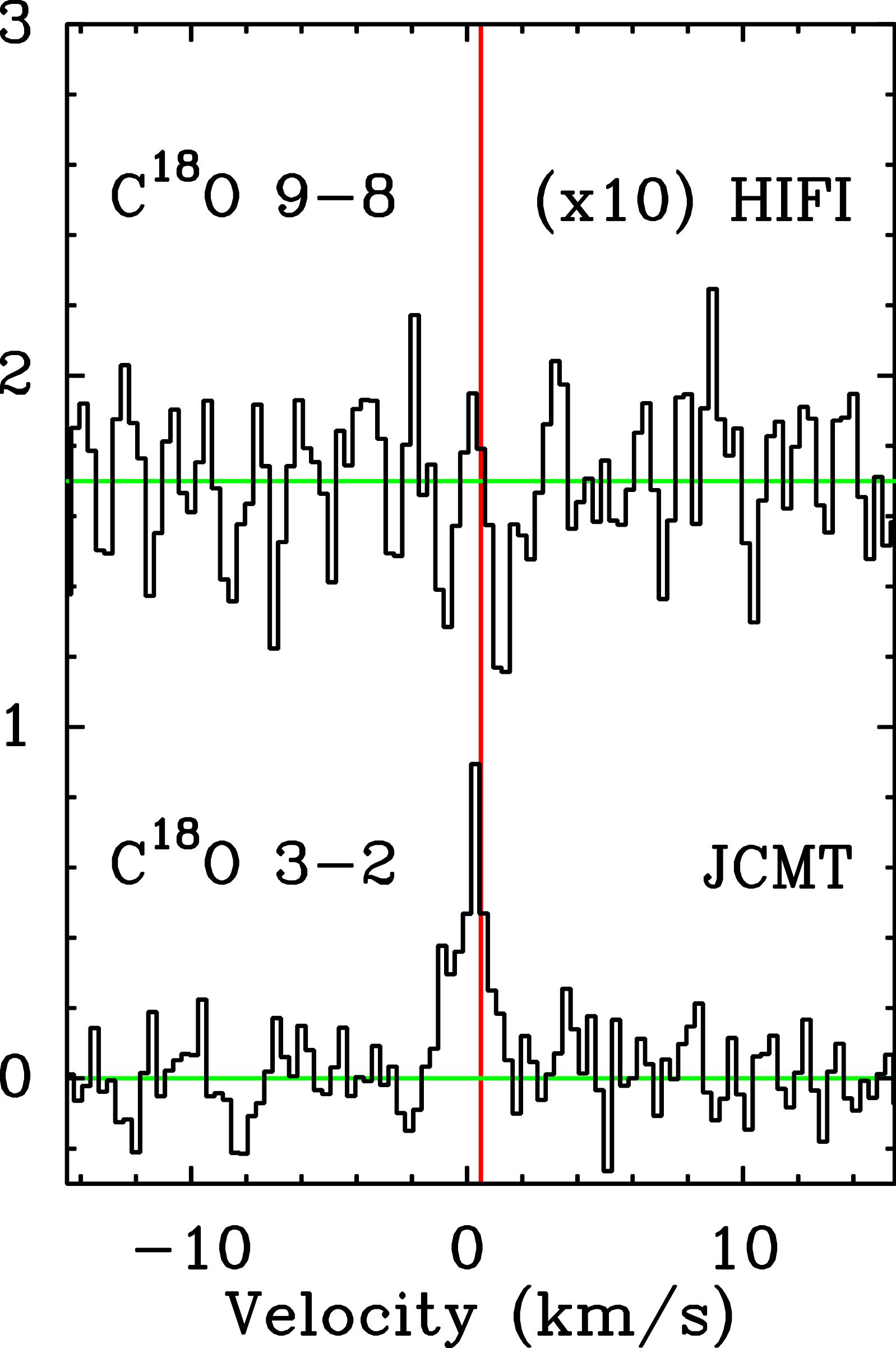}
    \caption{\small Observed $^{12}$CO, $^{13}$CO, and C$^{18}$O transitions for RNO91.}
    \label{fig:linesRNO91}
\end{figure*}

\begin{table*}[!ht]
\caption{Observed line intensities for RNO91 in all observed transitions. }
\normalsize
\begin{center}
\begin{tabular}{l l l r r r r r r r r r}
\hline \hline
Mol.  & Transition & Telescope & Efficiency & $\int T_{\rm MB} \mathrm{d}V$ & $T_{\mathrm{peak}}$ & $rms$ \\
 &  & &   $\eta$ &[K km s$^{-1}$] & [K] &   [K]\\
\hline
CO        & 2--1 & JCMT-RxA       & 0.69   &  40.47\phantom{0}  &  10.64 & 0.42\phantom{0} \\
          & 3--2 & JCMT-HARPB     & 0.63   &  16.75\phantom{0}  &   6.19 & 0.064 \\
          & 6--5 & APEX-CHAMP$^+$ & 0.48   &  25.96\phantom{0}  &  13.45 & 0.33\phantom{0} \\
          & 7--6 & APEX-CHAMP$^+$ & 0.49   &  14.87\phantom{0}  &  11.06 & 1.15\phantom{0} \\
          &10--9 & {\it Herschel}-HIFI\tablefootmark{a} & 0.64 &  5.19\phantom{0}  & 1.69 & 0.10\phantom{0} \\
$^{13}$CO & 2--1 & JCMT-RxA       & 0.74   &  11.77\phantom{0}  &  5.87  & 0.39\phantom{0} \\
          & 3--2 & JCMT-HARPB     & 0.63   &  5.38\phantom{0}   &  3.92  & 0.08\phantom{0} \\
          & 6--5 & APEX-CHAMP$^+$ & 0.45   &  2.43\phantom{0}   &  2.98  & 0.28\phantom{0} \\
          & 8--7 & APEX-CHAMP$^+$ & 0.49   &  1.55\phantom{0}   &  4.58   & 1.77\phantom{0} \\
          & 10--9& {\it Herschel}-HIFI\tablefootmark{a} & 0.74  &  $<$0.072 & \dots\phantom{0} & 0.028 \\
C$^{18}$O & 3--2 & JCMT-HARPB     & 0.63   &  1.13\phantom{0}   &  1.06   &  0.14\phantom{0} \\
          & 9--8 & {\it Herschel}-HIFI\tablefootmark{a} & 0.74  &  $<$0.052 & \dots\phantom{0} & 0.023 \\
\hline 
\end{tabular}
\end{center}
\tablefoot{
\tablefoottext{a}{H- and V-polarization observations averaged.}
}
\label{tbl:linesRNO91}
\end{table*}

\newpage

\section{Herschel-HIFI observation IDs}
\begin{table}[!ht]
\caption{{\it Herschel} Obsids for related observations}
\small
\begin{center}
\begin{tabular}{l c c c c c c c }
\hline \hline
Source 	& $^{12}$CO~10--9 & $^{13}$CO~10--9	& $^{13}$CO~8--7 & C$^{18}$O~5--4 & C$^{18}$O~9--8 & C$^{18}$O~10--9 & \\
\hline
L1448-MM	& 1342203253	& 	1342201803 	&  	 \dots 	   & 	1342203186	& 	1342203182	& 	1342201802	& \\
IRAS2A		& 1342191701	& 	1342191657 	&  1342225937  & 	1342192206	& 	1342191606	& 	1342215968	& \\
IRAS4A		& 1342191721	& 	1342191656 	&  1342225938  & 	1342192207	& 	1342191605	& 	1342249014	& \\
IRAS4B		& 1342191722	& 	1342191655 	&  1342225940  & 	1342192208	& 	1342191604	& 	1342249851	& \\
L1527		& 1342203256	&	1342216335	& 	 \dots 	   & 	1342203188	& 	1342203156	& 	 \dots 		& \\
Ced110IRS	& 1342201734	& 	1342200765 	& 	 \dots 	   & 	 \dots 		& 	1342201756	& 	 \dots 		& \\
BHR71		& 1342201732	&	1342200764	& 	 \dots 	   & 	1342200755	& 	1342215915	& 	 \dots 		& \\
IRAS 15398	& 1342214446	& 	1342214414 	& 	 \dots 	   & 	1342266008 	& 	1342203165	& 	 \dots 		& \\
L483	    & 1342217730	& 	1342207374 	&	\dots	   & 	1342207582	& 	1342218213	& 	1342207375	& \\
Ser SMM1	& 1342207701	& 	1342207379 	&  1342229782  & 	1342194463	& 	1342194994	& 	1342207378	& \\
Ser SMM4	& 1342207700	& 	1342207380 	&  1342229782  & 	1342194464	& 	1342194993	& 	1342207381	& \\
Ser SMM3	& 1342207699	& 	1342207377 	& 	 \dots 	   & 	1342207580	& 	1342207658	& 	1342207376	& \\
L723-MM		& 1342210152	&	1342210168	& 	 \dots 	   & 	1342219172 	& 	1342210041	& 	 \dots 		& \\
B335		& 1342230175	&	1342219248	& 	 \dots 	   &	1342219182	& 	1342219217	& 	 \dots 		& \\
L1157		& 1342198346	&	1342200763	& 	 \dots 	   & 	1342199077	& 	1342197970	& 	 \dots 		& \\
\hline 	 	 	 	
L1489		& 1342203254	& 	1342203938 	& 	 \dots 	 & 	 \dots 		& 	1342203158	& 	 \dots 		& \\
L1551-IRS5	& 1342203258	& 	1342203940 	& 	 \dots 	 & 	 \dots 		& 	1342203153	& 	 \dots 		& \\
TMR1		& 1342225917	& 	1342203937 	& 	 \dots 	 & 	 \dots 		& 	1342203157	& 	 \dots 		& \\
TMC1A		& 1342225916	& 	1342215969 	& 	 \dots 	 & 	 \dots 		& 	1342203154	& 	 \dots 		& \\
TMC1		& 1342203255	& 	1342216336 	& 	 \dots 	 & 	 \dots 		& 	1342203155	& 	 \dots 		& \\
HH46		& 1342222281	& 	1342194785 	& 	 \dots 	 & 	 \dots 		& 	1342195041	& 	 \dots 		& \\
DKCha		& 1342201733	& 	1342201590 	& 	 \dots 	 & 	 \dots 		& 	1342201755	& 	 \dots 		& \\
GSS30 IRS1	& 1342214442	& 	1342214413 	& 	 \dots 	 & 	 \dots 		& 	1342203163	& 	 1342250604	& \\
Elias29		& 1342214443	& 	1342214408 	& 	 \dots 	 & 	 1342266143	& 	1342203162	& 	 1342249849	& \\
Oph IRS63	& 1342214441	& 	1342214407 	& 	 \dots 	 & 	 \dots 		& 	1342203164	& 	 \dots 		& \\
RNO91		& 1342214440	& 	1342214406 	& 	 \dots 	 & 	 \dots 		& 	1342204512	& 	 \dots 		& \\
\hline 
\end{tabular}
\end{center}
\label{tbl:HerschelObsids}
\end{table}

\end{document}